\documentclass{aa}  
\usepackage{graphicx, threeparttable,epstopdf}
\usepackage{natbib}
\usepackage {xcolor}
\usepackage{txfonts}
\newcommand{\lya}{Ly$\alpha$}
\usepackage[above]{placeins}

\usepackage[breaklinks,colorlinks,citecolor=blue,linkcolor=blue,urlcolor=blue]{hyperref}
\begin{document} 

   \title{ Metal enrichment and evolution in four $z>6.5$ quasar sightlines observed with {\em JWST}/NIRSpec
   }

   \author{L. Christensen   
          \inst{1,2} \thanks{e-mail: lichrist@nbi.ku.dk}
          \and
          P. Jakobsen\inst{1,2}
          \and
          C. Willott\inst{3}
          \and
          S. Arribas\inst{4}
          \and
          A. Bunker\inst{5}
          \and
          S. Charlot\inst{6}
          \and
          R. Maiolino\inst{7,8,9}
          \and
          M. Marshall\inst{3}
          \and
          M. Perna\inst{4}
          \and 
          H. {\"U}bler\inst{7,8}
          }

   \institute{
Cosmic Dawn Center (DAWN), University of Copenhagen, Denmark 
\and
Niels Bohr Institute, University of Copenhagen, Jagtvej 128, DK-2200 N, Copenhagen, Denmark 
\and
National Research Council of Canada, Herzberg Astronomy \& Astrophysics Research Centre, 5071 West Saanich Road, Victoria, BC V9E 2E7, Canada 
\and
Centro de Astrobiolog\'{\i}a (CAB),
 CSIC-INTA, Ctra. de Ajalvir km 4, Torrej\'on de Ardoz, E-28850, Madrid, Spain
\and
Department of Physics, University of Oxford, Denys Wilkinson Building, Keble Road, Oxford OX1 3RH, UK
\and
Sorbonne Universit\'e, CNRS, UMR 7095, Institut d'Astrophysique de Paris, 98 bis bd Arago, 75014 Paris, France
\and
Kavli Institute for Cosmology, University of Cambridge, Madingley Road, Cambridge, CB3 0HA, UK
\and
Cavendish Laboratory, University of Cambridge, 19 JJ Thomson Avenue, Cambridge, CB3 0HE, UK
\and
Department of Physics and Astronomy, University College London,
Gower Street, London WC1E 6BT, UK
}

   \date{Received 12 September 2023; accepted 11 October 2023}

  \abstract
   {We present JWST/NIRSpec $R\!\simeq\!2700$ spectra of four high-redshift quasars: VDES J0020--3653 ($z=6.860$), DELS J0411--0907 ($z=6.825$), UHS J0439+1634 ($z=6.519$) and ULAS J1342+0928 ($z=7.535$). The exquisite data quality, signal-to-noise ratio of 50--200, and large $0.86\!~\mu{\rm m}\le \lambda \le 5.5\!~\mu{\rm m}$ spectral coverage allows us to identify between 13 and 17 intervening and proximate metal absorption line systems in each quasar spectrum, with a total number of 61 absorption-line systems detected at $2.42<z<7.48$ including the highest redshift intervening \ion{O}{i}\,$\lambda$1302 and \ion{Mg}{ii} systems at $z=7.37$ and $z=7.44$. We investigate the evolution of the metal enrichment in the epoch of reionization (EoR) at $z>6$ and find:  
{\it i})  A continued increase of the low-ionization \ion{O}{i}, \ion{C}{ii}, and \ion{Si}{ii} incidence,
{\it ii}) Decreasing high-ionization \ion{C}{iv} and \ion{Si}{iv} incidence with a transition from predominantly high- to low-ionization at $z\approx6.0$, and
{\it iii}) a constant \ion{Mg}{ii} incidence across all redshifts.
The observations support a change in the ionization state of the intergalactic medium in the EoR rather than a change in metallicity. The abundance ratio of [Si/O] in five $z>6$ absorption systems show enrichment signatures produced by low-mass Pop III pair instability supernovae, and possibly Pop III hypernovae.
In the Gunn-Peterson troughs we detect transmission spikes where \lya\ photons can escape. From 22 intervening absorption line systems at $z>5.7$, only a single low-ionization system out of 13 lies within 2000\!~km~s$^{-1}$ from a spike, while four high-ionization systems out of nine lie within $\sim$2000\!~km~s$^{-1}$ from a spike. Most spikes do not have associated metal absorbers close by. This confirms that star-forming galaxies responsible for producing the heavy elements that are transported to the circumgalactic medium via galaxy winds do so in predominantly high-density, neutral environments, while lower density environments are ionized without being polluted by metals at $z\approx6-7$.}
\titlerunning{NIRSpec observations of four $z>6.5$ quasars}
\authorrunning{L. Christensen et al.}

   \maketitle
%

\section{Introduction}
Investigations of high redshift quasars contribute significantly to our knowledge of the physical conditions in the early Universe, including the growth of supermassive black holes (SMBH) and their co-evolution with galaxies, and their effect on the reionization history of the Universe. Quasars also serve as bright beacons to illuminate the baryonic material along their lines of sight since their high luminosity allow their detection to very large distances. Finding high-redshift quasars has been the focus of numerous scientific investigations. From color selection of sources in large imaging surveys \citep[e.g.][]{fan01,willott10,wang19,matsuoka22,banados23}, hundreds of quasars beyond redshift 6 have been discovered and compiled in databases \citep{inayoshi20,fan23}, while beyond redshift 7 only eight quasars have been discovered to date \citep{mortlock11,banados18,matsuoka19a,matsuoka19b,yang19,yang21,wang18,wang21}.

Spectra of quasars allow detailed studies of intervening baryonic material from absorption lines arising either in the environment near quasars as well as along random lines of sight through the Universe, and are invaluable probes to analyse the intergalactic medium (IGM) and the epoch of reionization (EoR). Due to the rapid decrease in quasar number densities compared to galaxies at the highest redshifts, quasars contribute only $<7$\% of the reionizing photon budget at $z\sim6-6.6$ \citep{jiang22}. The neutral IGM will appear as a damping wing at the red part of the \lya\ absorption in quasar spectra \citep{miralda-escude98}, and the shape of this damping wing in quasar spectra at different redshift can be used to compute the neutral gas fraction $X_{\ion{H}{i}}$ as the reionization progresses \citep{banados18,davies18,davies20}. A complete absorption by the IGM in the Gunn-Peterson absorption trough (GP hereafter)\citep{gunn65} detected in $z>6$ quasar spectra \citep{fan01,mortlock11} reveals an increase in the \lya\ optical depth with increasing redshift \citep{becker21}. 

In the otherwise dark regions of the GP troughs, transmission spikes in the IGM start appearing at $z~\lesssim~6$ where ionized regions allow the transmission of \lya\ photons at those redshifts. Such spikes therefore need to lie near sources that produce ionizing radiation. \citet{kakiichi18} and \citet{meyer20} investigated the correlation between \lya\ forest transmission spikes and location of \lya-emitters and galaxies within a distance of 60 Mpc from the line of sight towards $z>6$ quasars, and argue that reionization is primarily driven by a population of galaxies that are fainter than the detected galaxies.

Besides probing the EoR through studies of hydrogen absorption, quasar spectra can be used to measure the evolution of heavy element enrichment with cosmic time from intervening absorption line systems to $z\sim6$ \citep[see][for a review of metal absorption lines]{becker15}. Quasars probe random lines of sight through the Universe, and intersect galaxies either in their circumgalactic medium (CGM) or interstellar medium (ISM) out to the quasar redshift, all of which imprint a spectral signature in the background quasar spectrum.

While the metallicity on the average increases with time, metal absorption lines also reflect the ionization state and are therefore sensitive to the ultraviolet background radiation. Early analyses of the \lya\ forest clouds and their correlations with \ion{C}{iv}$\lambda\lambda$1548,1550 absorption lines revealed that the IGM at $z=3$ is enriched to a metallicity level of $\approx10^{-3}~Z_{\odot}$ \citep{cowie95,ellison00}. While metals seem to be prevalent, absorption line densities\footnote{Line densities are also commonly called `incidence rate' or simply `number densities' in the literature. These terms are identical in their definition.}, defined as the number of absorption line systems detected per unit (comoving) redshift path length interval, can be used to map the evolution of the metal enrichment and ionization level. For example, \ion{C}{iv} absorber line densities remain roughly constant with redshift out to $z\sim4$ and then show a steep drop at $z>5$ \citep{becker11,codoreanu18,dodorico22,davies23}. A similar trend is found for the other high-ionization line \ion{Si}{iv} \citep{cooper19,dodorico22}, suggesting a rapid physical change in the ionization state at these redshifts. The change of ionization state rather than a change in metal enrichment is supported from the observation of a change in the \ion{C}{iv}/\ion{C}{ii}\,$\lambda$1334 ratio \citep{simcoe20,davies23}. Similarly, the line density of weak \ion{Mg}{ii}\,$\lambda$2796 absorbers remain constant over a large redshift range of $2<z<7$ \citep{chen17}, whereas the line density of \ion{O}{i}\,$\lambda$1302 increases over the range $3.2<z<6.5$ \citep{becker11,becker19} and is also interpreted as a change in the ionization state at the EoR, where the Universe becomes progressively more neutral.

Bright quasars also allow detailed analyses of individual absorption systems by measuring column densities of various atomic species, similar to studies of $2<z<3$ strong hydrogen  absorption line systems \citep[e.g.][]{prochaska03}. Focusing on the highest redshift absorbers at $z>6$, this epoch approaches the time when the first population of massive stars produce the initial chemical enrichment. However, for intervening absorption lines systems the absence of measurable \ion{H}{i} column densities in the quasars heavily absorbed \lya\ forest or GP troughs makes it impossible to derive absolute metallicities, and only a few studies of relative metallicities at $z>6$ have been performed to date. One example is the $z\sim7.5$ quasar ULAS J1342+0928 whose absorption system at $z=6.84$ with measured iron, carbon and silicon suggests no alpha-element enhancement nor any special signatures of enrichment from very massive stars \citep{simcoe20}. When the absorption system lies close to the quasar redshift, in the so-called associated or proximate absorption systems, detailed metallicity measurements can be performed since the intrinsic \ion{H}{i} column density can be modelled from fits to the \ion{H}{i} damping wing \citep{banados19b,andika22}. However, proximate absorption systems do not trace a random population, but rather probe the high-density environments near the quasars \citep{ellison10}. 

The connection between galaxies in emission and gas in absorption inform us of the transverse metal enrichment in the CGM of galaxies, albeit only along a single line of sight to the background quasar. Hence, combining the observations with simulations will reveal the spatial distribution of metals in the CGM of galaxies and its evolution with redshift \citep[e.g][]{finlator18,peeples19}. At $z\ge5$ overdensities of line emitting galaxies either selected via \lya\ emission \citep{diaz21} or [\ion{O}{iii}] $\lambda$5007 \citep{kashino23} have been found within 200 kpc of the metal absorbers and approximately at the same redshifts. 

Large amounts of ground-based telescope time have in recent years been devoted to spectroscopic studies of $z>6$ quasars \citep[e.g.][]{dodorico23}. However, with the successful launch and commissioning of the James Webb Space Telescope \citep{gardner23,rigby23}, the question arises of what new insights observations of high redshift quasars obtained with JWST's near infrared spectrograph, NIRSpec \citep{jakobsen22}, may bring to bear on the subject matter. 
While the highest spectral resolution possible with NIRSpec of $R\simeq2700$ is only just adequate for intervening absorption line studies, the benefits offered by the instrument's very high sensitivity and wide ($0.81~\mu{\rm m}\!\le\!\lambda\le5.3~\mu{\rm m}$) infrared spectral coverage free of the atmospheric telluric absorption and OH sky emission lines that hamper ground based near-IR observations are self-evident.

In this paper, we present high signal-to-noise ratio $R\simeq2700$ JWST/NIRSpec spectra of four well-studied quasars at $6.5<z<7.5$. We focus our analysis exclusively on the intervening absorption line systems detected. The paper is organised as follows. Section~\ref{sect:nirspec_obs} describes the observations. The spectra are presented and analysed in Section~\ref{sect:nirspec_spectra}. The measurements of metal absorption line systems and the redshift evolution of  line densities of atomic transitions in oxygen, carbon, silicon and magnesium, calcium and sodium are discussed in Section~\ref{sect:abs_metal}. In Section~\ref{sect:abundances} we compute metal abundance patterns and compare with the enrichment of Population III and Pop II massive star Supernova explosions, and in Section~\ref{sect:trans_igm} we compare the metal absorption system redshifts with transmission spikes seen in the quasar GP troughs.  In Section~\ref{sect:discussion} we present a discussion and conclude in Section~\ref{sect:conclusions}. Throughout the paper we assume a flat cosmological model with $H_0$=67.4~km\!~s$^{-1}$\!~Mpc$^{-1}$ and $\Omega_m$=0.315 \citep{planck18}.

\section{NIRSpec observations}
\label{sect:nirspec_obs}

In this paper we present high signal-to-noise ratio ($S/N\!\simeq\!50-200$), $R\!\simeq\!2700$  NIRSpec spectra spanning $0.86\!~\mu{\rm m} \le \lambda \le 5.3\!~\mu{\rm m}$ of the four high redshift quasars:  ULAS~J1342+0928 \citep[][$z=7.54$, $J_{AB}=20.3$]{banados18}, VDES~J0020--3653 \citep[][$z=6.83$, $J_{AB}=20.4$]{reed19}, DELS~J0411--0907 \citep[a.k.a VHS~J0411--0907,][$z=6.81$, $J_{AB}=20.4$]{pons19,wang19} and UHS~J0439+1634 \citep[][$z=6.51$, $J_{AB}=17.4$]{fan19}. These data were obtained as part of the NIRSpec GTO programs 1219 (PI: N. L\"utzgendorf) and 1222 (PI: C. Willott). 

The NIRSpec Band~I ($0.86~\mu{\rm m}\!\le\!\lambda\le1.88~\mu{\rm m}$) and Band~II ($1.70~\mu{\rm m}\le\lambda\le3.15~\mu{\rm m}$) spectra were taken with the NIRSpec  Fixed Slits together with the G140H/F070LP and G235H/F170LP  grating/filter combinations, whereas the long wavelength Band~III ($2.88~\mu{\rm m}\le\lambda\le5.3~\mu{\rm m}$) portions were obtained employing the NIRSpec Integral Field Unit (IFU) and the G395H/F290LP grating/filter combination (see \cite{jakobsen22} and \cite{boker22} for details of these NIRSpec observing modes). The dates on which the observations were carried out and the position angles of the slits employed are summarized in Table~\ref{tab:ephem}.

   \begin{table}
      \caption{Observation Epochs snd Slit Orientations}
       \centering
         \label{tab:ephem}
         \begin{tabular}{l c r r}
            \hline\hline
            \noalign{\smallskip}
        \quad\quad{Quasar}&  Mode& Date\quad\quad\quad& PA [deg]\\
            \noalign{\smallskip}
            \hline
            \noalign{\smallskip}
 ULAS~J1342+0928& FS& 31 Jan 2023& 66.8\phantom{\enskip}\\ 
                & IFU& 23 Jan 2023& 55.1\phantom{\enskip}\\ 
\noalign{\smallskip}
VDES~J0020--3653& FS& 19 Nov 2022& 187.2\phantom{\enskip}\\ 
                & IFU& 16 Oct 2022& 160.2\phantom{\enskip}\\ 
                & IFU& 17 Nov 2022& 189.2\phantom{\enskip}\\ 
 \noalign{\smallskip}
DELS~J0411--0907& FS& 18 Sep 2022& 50.6\phantom{\enskip}\\ 
                & IFU& 1 Oct 2022& 61.9\phantom{\enskip}\\ 
 \noalign{\smallskip}
UHS~J0439+1634  & FS & 24 Jan 2023& 210.8\phantom{\enskip}\\ 
                & IFU& 18 Jan 2023& 210.4\phantom{\enskip}\\ 
          \noalign{\smallskip}
            \hline
         \end{tabular}
   \end{table}

The Band~I and Band~II Fixed Slit observations all employed 65 group NRSIRS2RAPID \citep{rauscher17} full frame detector sub-integrations. For ULAS\!~J1342+0928, VDES\!~J0020--3653, and  DELS\!~J0411--0907, spectra were  obtained in both the S200A1 and S200A2 fixed slits in order to span the detector gap, with two integrations carried out in each of three nodded positions along the slit for G140H/F070LP (11554\!~s total exposure time) and one integration at each nod in G235H/F170LP (5778\!~s total exposure time). The brighter UHS\!~J0439+1634 was only observed in the S200A1 slit at three nodded integrations in both G140H/F070LP and G235H/F170LP (2889\!~s total exposure time). These data were reduced and combined using the NIRSpec GTO Team pipeline (Carniani et al. in preparation). The resulting flux-calibrated spectra refer to Barycentric vacuum wavelengths sampled at $\Delta\lambda=2.36$\!~{\AA} per wavelength bin in Band~I and $\Delta\lambda=3.96$\!~{\AA} per wavelength bin in Band~II, which by design is close to the average native pixel sampling for these dispersers.

One noteworthy aspect of the shortest wavelength Band~I G140H observations is that they were taken with the F070LP order-separation filter, rather than the default F100LP filter. This was done to include wavelengths in the Gunn-Peterson troughs below emitted Ly$\alpha$ down to $0.813\!~\mu$m with the S200A1 slit and down to $0.858\!~\mu$m with the S200A2 slit. This is permissible in this case since there is little or no light from the quasar emerging in the trough and therefore no second order light (beyond that of the sky background, which is subtracted out in the reduction) to contaminate the primary first order spectrum at wavelengths $\lambda>1.26\!~\mu$m. However this choice did require tricking the pipeline into reducing the full Band~I spectrum in two steps. First by running it for the F070LP filter giving the correctly reduced spectrum over $0.85\!~\mu{\rm m}\le\lambda\le 1.26\!~\mu{\rm m}$ and then running it a second time pretending that the F100LP filter was used and then scaling the resulting red end of the spectrum covering the $1.0\!~\mu{\rm m}\le\lambda \le 1.88\!~\mu{\rm m}$  region by the known ratio of the transmission of the F070LP and F100LP filters. The latter correction amounts to less than 4\% in flux.

The Band~III G395H/F290LP IFU observations of ULAS\!~J1342+0928, VDES\!~J0020--3653, and  DELS\!~J0411--0907 employed 25 group NRSIRS2 sub-integrations at 6 dithered IFU positions (11029\!~s total integration time) and UHS\!~J0439+1634 18 group NRSIRS2 integrations at 8 dithered positions (10620\!~s total integration time). The IFU observations of VDES\!~J0020--3653, and  DELS\!~J0411--0907 are presented in \cite{marshall23}, who along with \cite{perna23} also provide the details of how the IFU data were reduced. Here we merely employ the one dimensional G395H/F290LP spectra extracted from the four IFU data cubes over a 1\farcs0 diameter aperture centered on the quasars to supplement the corresponding Band~I and II Fixed Slit spectra of primary interest. The resulting Band~III spectra are sampled at $\Delta\lambda=6.65$\!~{\AA} per wavelength bin.

\begin{figure*}
    \centering  
    \begin{minipage}{\textwidth}
        \centering
        \includegraphics[bb = 80 40 1110 650, clip, width=0.98\hsize]{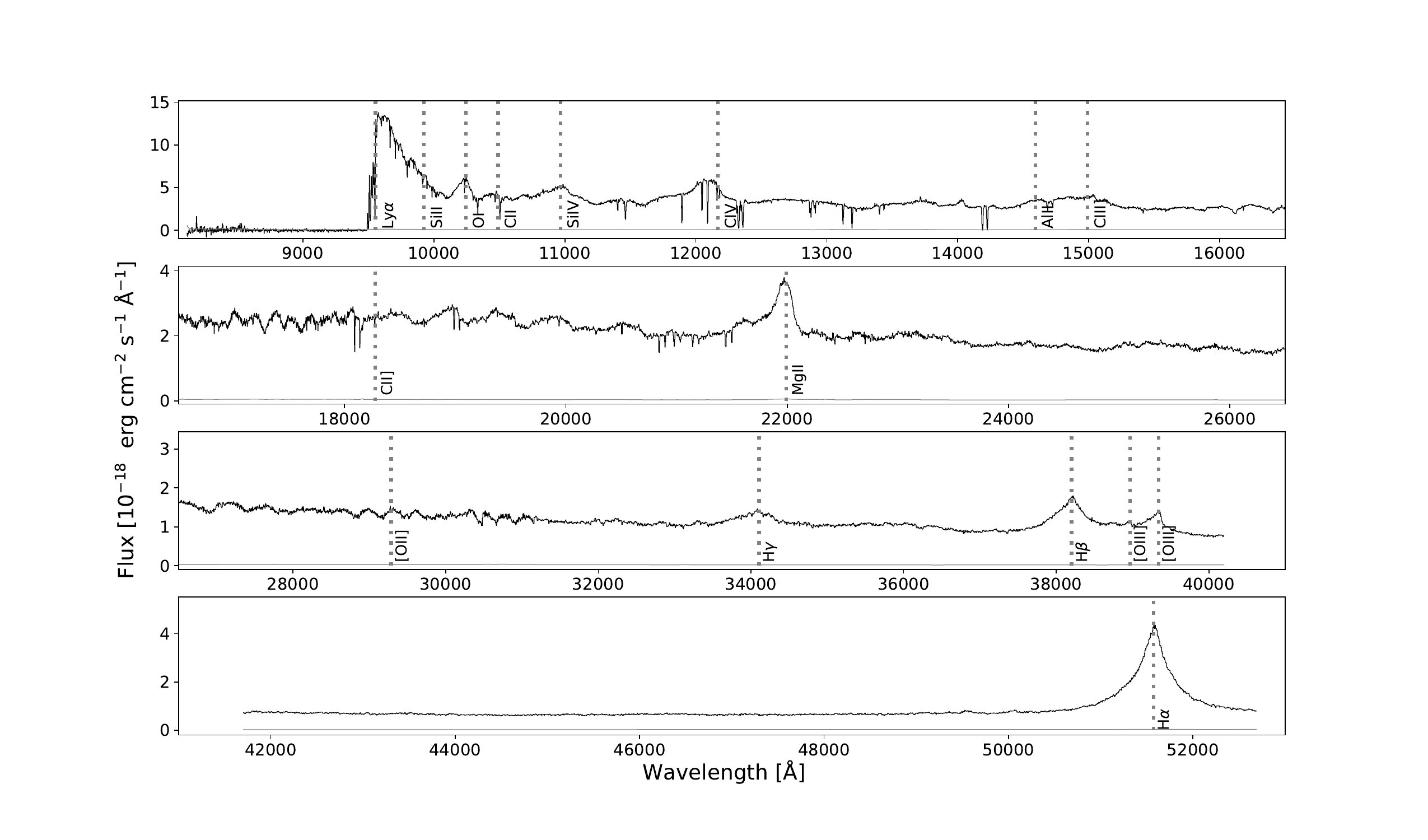}  
       \caption{NIRSpec fixed slit ($\lambda < 31160$ {\AA}) and extracted 1D spectrum from the Band~III data ($\lambda > 31160$ {\AA}) of VDES~J0020--3653. The modulation of the flux, which is particularly pronounced around $\sim$18000 and $\sim$30000 {\AA} here and also the other quasar spectra, is caused by the coarse sampling of the spectrum described in Sect.~\ref{sect:nirspec_obs}.       
       The associated error spectrum is shown in gray. With the very high $S/N$ ratio of the spectrum, the error spectrum appears virtually at the zero-flux level. }
    \label{fig:J0020_all}
    \bigskip
    \end{minipage}
    \begin{minipage}{\textwidth}
        \centering
        \includegraphics[bb = 80 40 1110 650, clip, width=0.98\hsize]{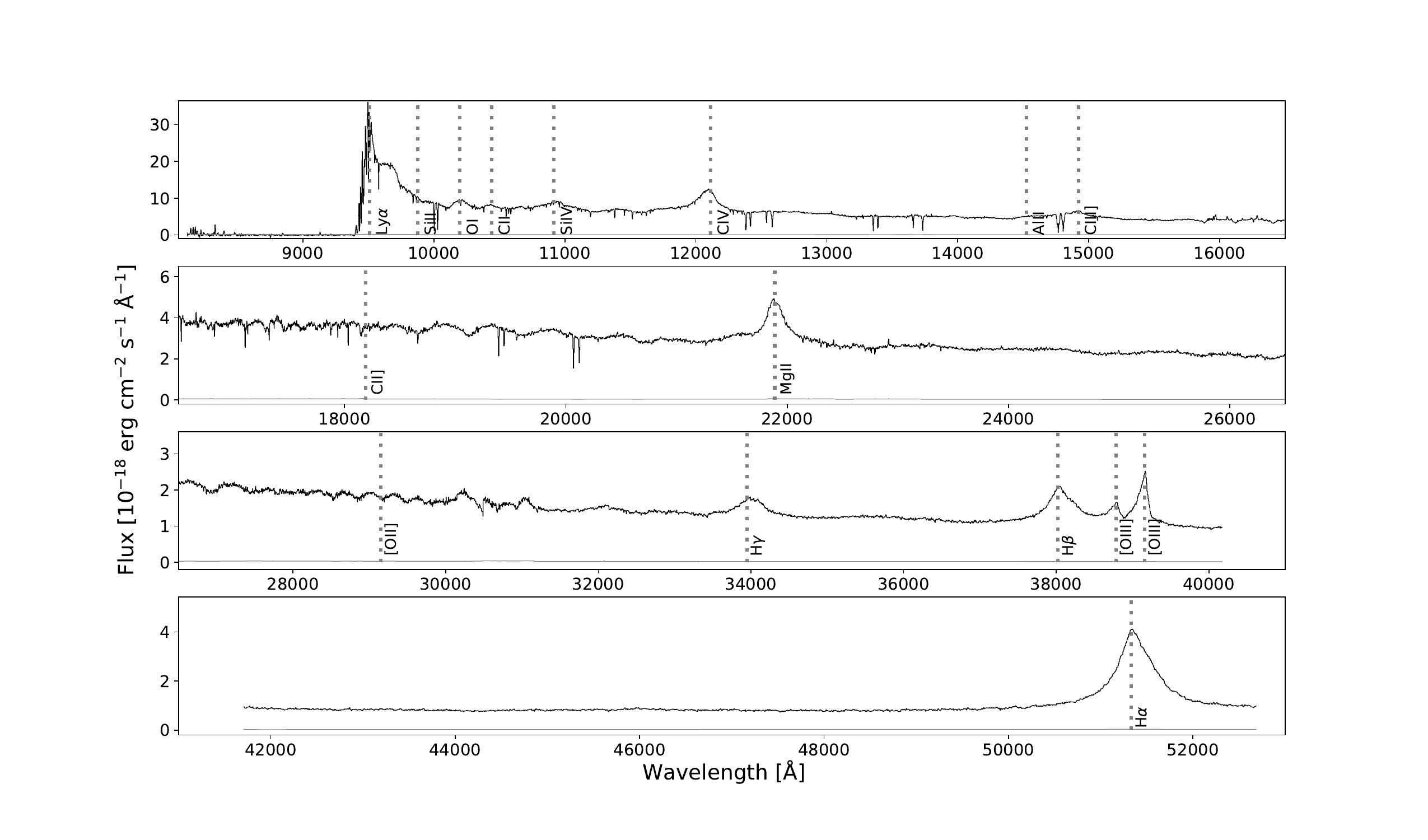}
       \caption{NIRSpec fixed slit  and extracted 1D spectrum from Band~III data ($\lambda > 31160$ {\AA}) of DELS J0411--0907. 
       }
    \label{fig:J0411_all}
\end{minipage}
\end{figure*}

\begin{figure*}[h!]
\begin{minipage}[c]{0.98\textwidth}
\centering
   \includegraphics[bb = 80 35 1110 650, clip, width=0.96\hsize]{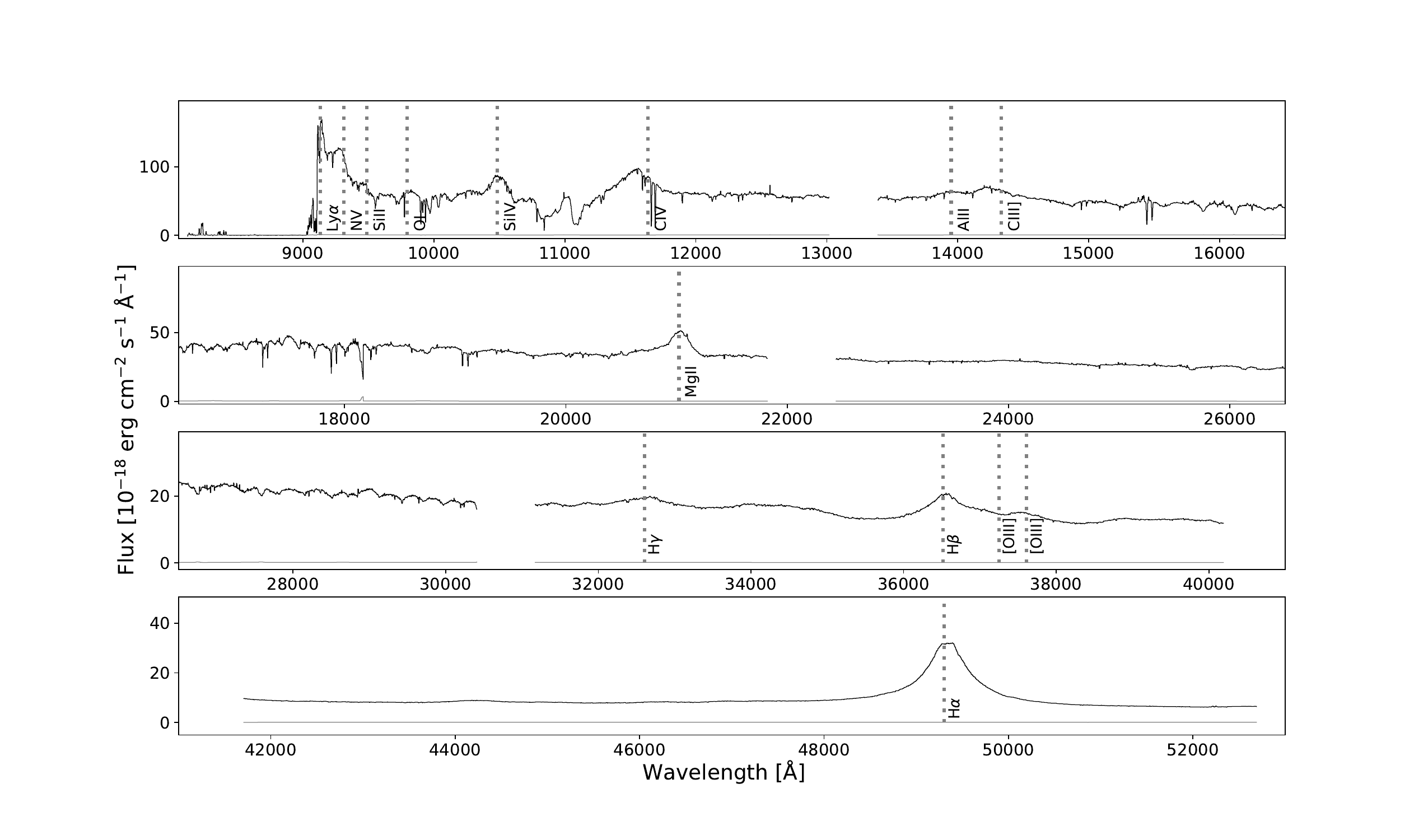}
       \caption{NIRSpec fixed slit and 1D spectrum from Band~III data ($\lambda > 31160$ {\AA}) of UHS J0439+1634. The spectrum shows clear features from intrinsic broad absorption lines (BAL), particularly from \ion{C}{iv}, but also at wavelengths between \lya\ and \ion{Si}{iv}, the underlying quasar continuum suggests an absorbed fraction when extending power-law function continuum fit obtained from wavelengths redwards of the \ion{C}{iii}] line.
    \vspace*{0.3cm}       }
    \label{fig:J0439_all}
\end{minipage}

\begin{minipage}[c]{0.98\textwidth}
\centering
   \includegraphics[bb = 80 40 1110 650, clip, width=0.96\hsize]{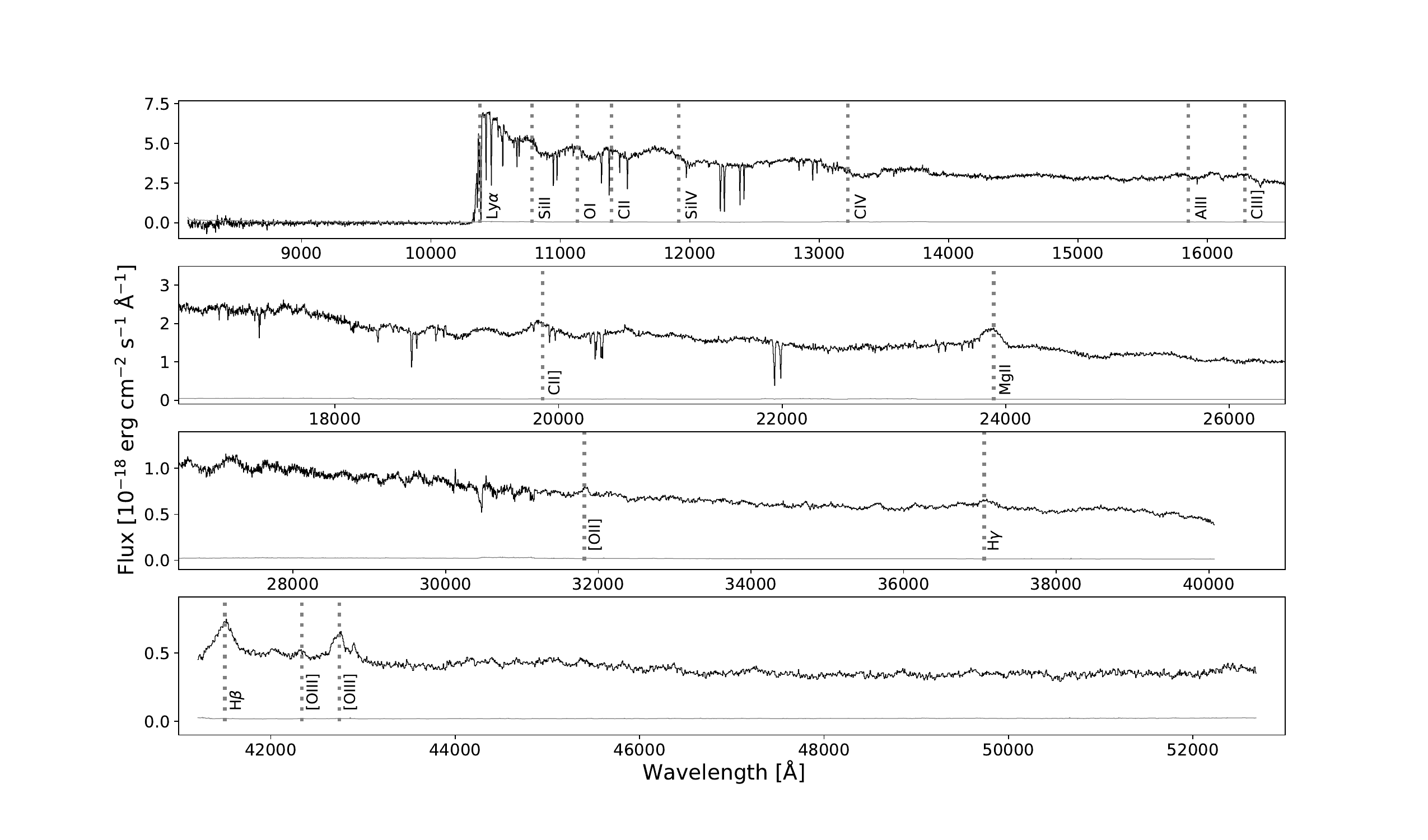}
       \caption{NIRSpec fixed slit and extracted 1D spectrum from Band~III data ($\lambda > 31700$ {\AA}) of ULAS J1342+0928.}
    \label{fig:J1342_all}
\end{minipage}
\end{figure*}

\begin{figure}
  \resizebox{\hsize}{!}{\includegraphics[bb = 5 1 700 540, clip]{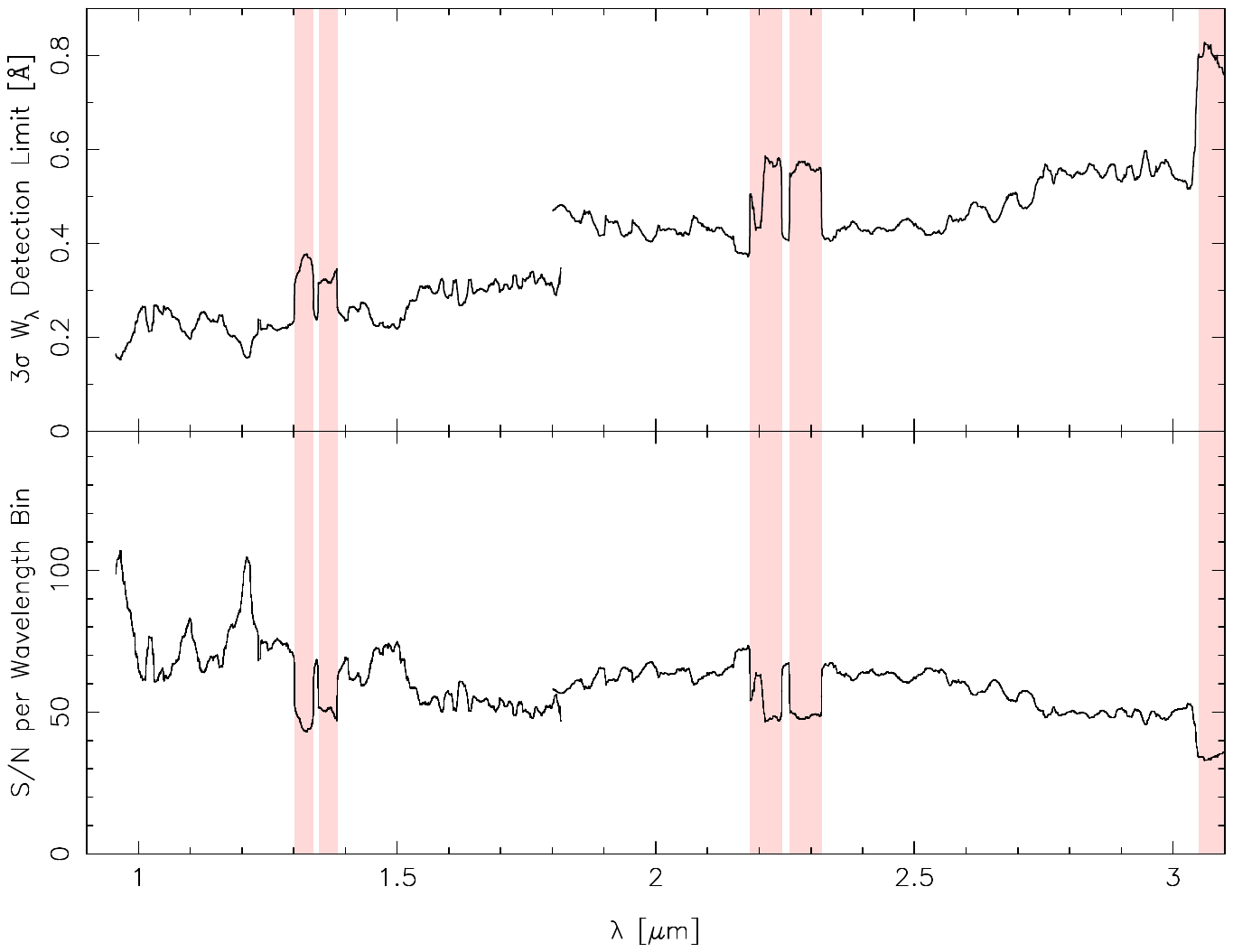}}
       \caption{
{\sl Lower panel:} Signal-to-noise ratio per wavelength bin as a function of wavelength for the combined Band~I and Band~II spectra of VDES\!~J0020--3653. The regions of reduced total exposure caused by the detector gap  are shown as shaded. 
{\sl Upper panel:} Corresponding anticipated three sigma narrow absorption line equivalent width detection limit as a function of wavelength.     
}
    \label{fig:sn_ew}
\end{figure}

The final combined one-dimensional spectra of the quasars along with their $1\sigma$ error spectra are shown in Figures\!~\ref{fig:J0020_all}, \ref{fig:J0411_all}, \ref{fig:J0439_all}, and  \ref{fig:J1342_all}. The signal-to-noise ratios achieved in these four spectra are very high. This is illustrated for the representative case of VDES\!~J0020--3653 in Figure\!~\ref{fig:sn_ew}. The lower panel plots the Band~I and II signal-to-noise ratio per wavelength bin calculated as the ratio of the observed signal divided by the one sigma error spectrum output by the GTO pipeline after median smoothing over a 40 bin window to remove the absorption lines. It is apparent that $S/N\simeq50-100$ per wavelength bin is achieved at nearly all wavelengths. The overall variation with wavelength seen in this plot partly reflects the blaze functions of the two gratings employed and the variation in the continuum flux due to the broad emission lines in the quasar spectrum, but also the fact that not all wavelengths are exposed equally due to the gap between the two NIRSpec detector arrays. When employing the G140H grating with the S200A1 fixed slit the wavelength span $1.3021-1.3391\!~\mu$m falls on the detector gap.  For the S200A2 fixed slits the wavelength range $1.3479-1.3849\!~\mu$m is lost to the gap. Similarly, for the G235H grating respectively $2.1825-2.2445\!~\mu$m and $2.2594-2.3215\!~\mu$m project to the gap for the S200A1 and S200A2 slits. Moreover, with the S200A1 slit, the G235M spectrum falls off the red edge of the detector for $\lambda > 3.0491\!~\mu$m but for $\lambda > 3.1262\!~\mu$m for the S200A2 slit. These unavoidable wavelength regions of reduced $S/N$ only covered in one of the two slit spectra are indicated in Figure\!~\ref{fig:sn_ew}. 

The high $S/N$ ratio observations of the bright ($J_{AB}\simeq20$) quasars presented here represent a radically different and rarer regime of applicability for NIRSpec compared to the perhaps more familiar lower spectral resolution observations of the much fainter ($J_{AB}\simeq26-29$) highest redshift galaxies that the instrument was primarily designed for \citep[cf.][]{curtislake23, bunker23}. It is therefore worth pointing out a few further subtle quirks of NIRSpec revealed by these data that are usually masked at lower $S/N$ ratio. 

\begin{figure}
  \resizebox{\hsize}{!}{\includegraphics[bb = 20 20 670 600, clip]{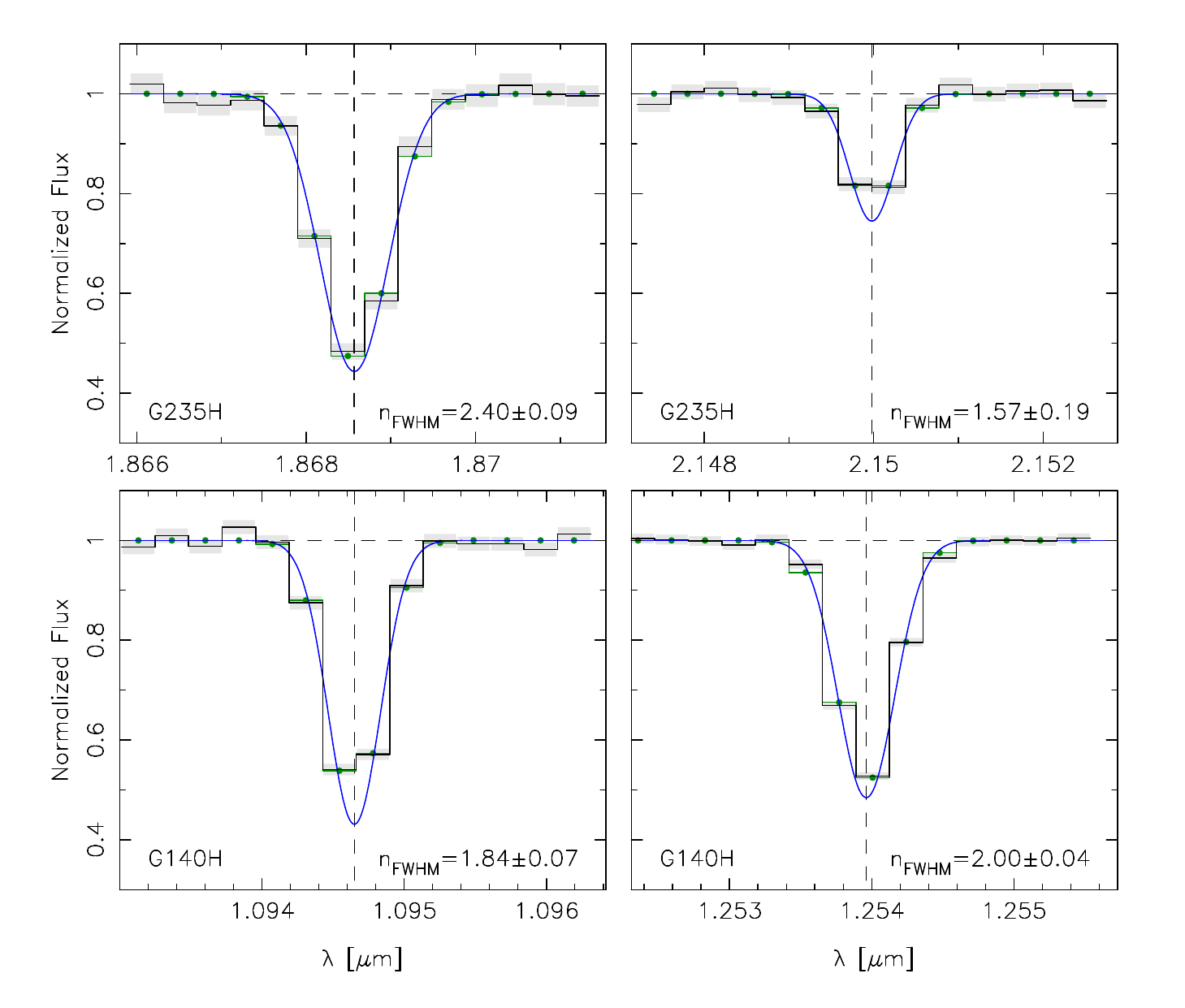}}
       \caption{
Representative examples of normalized line profiles illustrating the coarse wavelength sampling of the NIRSpec spectra. The shaded error bars shown are $\pm1\sigma$. The best-fit Gaussian fits and their FWHM expressed in spectral wavelength bins are also shown.    
       }  
    \label{fig:line_profiles}
\end{figure}

While NIRSpec grating observations of faint high redshift galaxies are invariably severely detector noise limited \citep[cf. Appendix in][]{jakobsen22}, the photon noise in our quasar spectra typically exceeds the total detector read noise and dark current noise by a factor $\simeq4$. Furthermore, the quasar photon signals typically exceeds that of the (zodiacal light dominated) sky background experienced by JWST by a factor $\simeq\!20$. However, one important feature that these quasar spectra still share with their fainter counterparts is that they are rather coarsely sampled both in wavelength and spatially along the slit. As explained in \cite{jakobsen22}, with the aim of optimizing the faint-end sensitivity limit of NIRSpec, the 194\!~mas angular width of the NIRSpec S200A1 and S200A2 Fixed Slits projects on to just under two mean 104\!~mas pixels on the detector. As a result, the apparent line spread function in a given spectrum at a given wavelength will depend on how precisely the slit width is sampled by the between two and three pixels that its monochromatic image happens to project to. This variable sampling issue is further exacerbated by the slit tilt and optical distortion that NIRSpec spectra are subject to, which necessitate that significant resampling of the raw 2D spectra be performed in the NIRSpec reduction pipeline in order to rectify the spectra to a common wavelength and spatial scale that will allow accurate background subtraction and summation of the net quasar signal along the spatial direction. That the final spectra are assembled by co-adding several differently sampled dithered sub-exposures taken at different positions along the slit leads to further smearing of the apparent line spread function. Lastly, the fact that the targets observed here are point sources, whose monochromatic images at the shorter wavelengths are not expected to fill the full width of the slit, adds even further variation in the spectral sampling. The net impact of these effects can be readily seen in Figure\!~\ref{fig:line_profiles} which shows close-ups of four representative narrow absorption lines detected in our final spectra, overlaid by their best fit Gaussian profiles. It is evident that the coarse sampling of NIRSpec spectra causes the apparent widths of the absorption lines to vary randomly at the $\simeq\!1$ wavelength bin level, even among lines having comparable strengths. This stochastic variation in the apparent shape of the pixelated line spread function due to sampling noise clearly precludes performing meaningful detailed profile fits to our unresolved absorption lines. However, it ought not bias measurement of the main parameters of interest --  line equivalent widths and centroid wavelengths -- provided  the error propagation calculation for these measured parameters correctly mirrors the noise level in the resampled spectra.

The top panel of Figure\!~\ref{fig:sn_ew} shows the anticipated $3\sigma$ observed equivalent width sensitivity limit to detecting unresolved absorption lines calculated from the lower $S/N$ curve per the expression 
\begin{equation}
W_\lambda > 3 \frac{ \Delta\lambda }{(S/N)} n_\mathrm{FWHM}^{3/2}
\end{equation}
assuming the FWHM of the NIRSpec line spread function to be a multiple of the wavelength bin size $\delta\lambda=n_\mathrm{FWHM}\Delta\lambda$ with $n_\mathrm{FWHM}$ 
(the number of spectral pixels sampling the line spread function),
conservatively set to $n_\mathrm{FWHM}=2.3$ for both bands. It follows that we anticipate being able to measure absorption lines having observed equivalent widths $W_\lambda>0.2\!~$\!~{\AA} in our spectrum of VDES\!~J0020--3653. The equivalent $S/N$ and line sensitivities achieved for DELS\!~J0411--0907 and ULAS\!~J1342+0928 are comparable to those of VDES\!~J0020--3653 as shown in Figure\!~\ref{fig:sn_ew}. However, our spectrum of the considerably brighter UHS\!~J0439+1634 reaches $S/N\simeq90-200$, albeit with actual gaps in the spectrum at the missing S200A1 wavelengths.

We have not attempted to quantify the $S/N$ achieved in the IFU-based Band~III G395H extensions to our spectra. One reason for this is that the STScI pipeline used to reduce the IFU data outputs an estimate of the error spectrum that significantly underestimates the actual noise level in the data. This does not affect our results, since we only detect a single \ion{Na}{i} absorption feature in the Band~III G395H data  (Section\!~\ref{sect:cana}). 

A final consequence of the coarse sampling and optical distortion that NIRSpec spectra are subject to is worth mentioning. The NIRSpec GTO pipeline employs a strict photon conserving (pixel projection weighting) resampling scheme when rectifying the native 2D spectra to a common wavelength and spatial scale. Nonetheless, in the case of a point source, the coarse sampling of the native 2D spectrum gives rise to Moire-style numerical fringing in the resampled spectrum along any  resampled spatial row \citep[cf.][]{smith07}. The amplitude of this unavoidable numerical fringing (or `wiggle') at any given wavelength is effectively averaged out provided the net signal is integrated over a sufficient number of spatial rows large enough to capture most of the light from the target. In the spectra presented here the GTO pipeline performed this summation over a fixed five spatial rows. This is adequate to integrate down the fringing at wavelengths $\lambda<2.8\!~\mu$m, but gradually fails at longer wavelengths and can be seen in our Band~II G235H spectra due to the FWHM of the PSF increasing linearly with wavelength in the diffraction-limited NIRSpec, causing a varying amount of light to be missed in the five row summation. This numerical fringing effect is particularly pronounced in NIRSpec IFU data \citep{marshall23,perna23}, but is less obvious in our 1D Band~III spectra since these were extracted by summing over a large number of spaxels. In any event, since the effect causes a modulation of the entire spectrum it should not affect measured equivalent widths and line centroids as long as the local continuum is set to follow the modulation (Section~\ref{sect:normalisation}).

\section{Analysis of the spectra}
\label{sect:nirspec_spectra}

The full one-dimensional spectra of the quasars are presented in Figures ~\ref{fig:J0020_all}, \ref{fig:J0411_all}, \ref{fig:J0439_all}, and  \ref{fig:J1342_all}, with emission lines commonly seen in quasar spectra marked.  We find a small shift in the absolute flux levels between the Fixed Slit and the IFU spectra, and have scaled the extracted one-dimensional Band~III spectra by a factor of $\approx0.7$ to match the flux level at the overlapping wavelengths in the Fixed Slit Band~II spectra.

\subsection{Quasar systemic redshifts}

In this paper, we focus on the analysis of intervening absorption line systems, and therefore  re-visit the estimated quasar systemic redshifts in order to distinguish intervening from proximate metal absorbers arising less than 3000 km~s$^{-1}$ from the quasar redshifts. This velocity cut-off is chosen since absorbers nearer than this limit show different physical properties such as metallicities and ionisation levels compared to genuine intervening absorption systems \citep{ellison10}.  UHS J0439+1634 is a broad absorption line (BAL) quasar, and therefore could potentially affect its environment to even larger velocities of $\sim$10,000 km~s$^{-1}$. In this spectrum we find two absorption systems that are offset by ~9000 km~s$^{-1}$, and both systems show narrow lines. We choose to include both absorption systems in the statistical analysis having a uniform cut at 3000 km~s$^{-1}$ for the four quasar spectra.

Determining systemic redshifts for quasars is complex because of the dynamical effects that present themselves in quasar spectra. Broad emission lines can be blueshifted by several thousand km~s$^{-1}$ relative to the systemic redshift, and can be composed of multiple components with peaks at distinct wavelengths, reflecting distinct kinematic components in the broad line regions.  With the extended coverage of the rest-frame optical Balmer lines and narrow lines from [\ion{O}{ii}]\,$\lambda\lambda$3727,3729 or [\ion{O}{iii}]\,$\lambda\lambda$4959,5007 afforded by the NIRSpec spectra, more accurate systemic redshifts can be derived. 

Fits to H$\alpha$ lines from the Band~III spectra of VDES J0020--3653 and DELS J0411--0907 and redshift measurements are presented in \citet{marshall23}. The asymmetric [\ion{O}{iii}] line profile of VDES J0020--3653 is caused by excess emission from extended regions spatially offset by $\sim$1 arcsec from the quasar \citep{marshall23}. Our fit to the H$\alpha$ emission line from VDES J0020--3653 is best reproduced by  a combination of a broad Lorentzian and a broad Gaussian profile plus much weaker narrow components of H$\alpha$ and [\ion{N}{ii}] $\lambda$6584 lines from the host galaxy, which gives a redshift of $z=6.8601\pm0.0001$, i.e. slightly higher than $z=6.855$ based on a fit of two Gaussian components reported by \citet{marshall23}.

Fits of Lorentz profiles to H$\alpha$ and H$\beta$ emission from DELS J0411--0907 suggest a systemic redshift of $z=6.8253\pm0.0002$ consistent with the initial redshift reported by \citet{pons19}, and $z=6.8260\pm0.0007$ based on [\ion{C}{ii}] 158 $\mu$m emission \citep{yang21}, while multi Gaussian component fits to [\ion{O}{iii}]\,$\lambda$5007 and H$\alpha$ emission lines suggest a smaller redshift of $z=6.818$ as the systemic one \citep{marshall23}. The difference in the inferred systemic redshift is caused by a narrow component of H$\alpha$ that is blueshifted by $\sim$380 km~s$^{-1}$ relative to the broad H$\alpha$ component \citep{marshall23}.

The [\ion{O}{iii}] $\lambda$5007 line in the spectrum of ULAS J1342+0928 can be fit by a double Gaussian line profile, where the narrow component suggests a lower redshift of $z= 7.5317\pm0.0007$ compared to the centroid of the broad component at $z= 7.5435\pm0.0002$. 
In comparison, a single Gauss component fit to the H$\beta$ emission line provides an adequate fit with $z=7.5353\pm0.0004$. The redshift derived from H$\beta$ is smaller by $211\pm35$ km~s$^{-1}$ compared to $z=7.5413\pm0.0007$ measured from [\ion{C}{ii}] 158~$\mu$m emission from the quasar host \citep{venemans17}.

The \ion{Mg}{ii} line profile in the spectrum of UHS J0439+1634 is asymmetric with a significant blue wing. Fitting the broad line with a Lorentzian profile, we derive a redshift of $z=6.5102\pm0.0005$ consistent with a fit to the same line in a GNIRS spectrum from the Gemini telescope that reveals $z=6.511\pm0.003$ \citep{fan19}. The [\ion{C}{ii}] 158~$\mu$m emission from this object suggests a systemic redshift of  $z=6.5188\pm0.0004$ \citep{yang19}. The H$\alpha$ line is covered by the NIRSpec Band~III data, and its line profile shows a broad component that is well fit by a single Lorentzian profile with narrow emission line components superimposed. The narrow emission lines are recognised as H$\alpha$ and [\ion{N}{ii}] $\lambda\lambda$6548,6584 from the host galaxy, suggesting a redshift of $z=6.5185\pm0.0003$, consistent with that from the [\ion{C}{ii}] 158 $\mu$m emission. 

The systemic quasar redshifts inferred from our spectra and used in the following analysis are summarized in Table~\ref{tab:qso_table}.

\begin{table*}[]
    \centering
    \begin{tabular}{lccc}
\hline\hline
 \noalign{\smallskip}
\quad\quad\enskip Quasar  & $z_{\mathrm{sys, JWST}}$  &  $z_{\mathrm{sys,[\ion{C}{ii}]}}$ &  $J_{\mathrm{AB}}$ \\
\noalign{\smallskip}
\hline
\noalign{\smallskip}
  VDES J0020--3653    &  6.8601$\pm$0.0001  &                   & 20.40$\pm$0.10 \\
  DELS J0411--0907    &  6.8253$\pm$0.0002  & 6.8260$\pm$0.0007 & 20.02$\pm$0.14 \\
  UHS J0439+1634      &  6.5185$\pm$0.0003  & 6.5188$\pm$0.0004 & 17.46$\pm$0.02 \\
  ULAS J1342+0928     &  7.5353$\pm$0.0004  & 7.5413$\pm$0.0007 & 20.30$\pm$0.02\\
\noalign{\smallskip}
\hline
    \end{tabular}
    \medskip
    \caption{Systemic quasar redshifts derived from fits to the Balmer emission lines in the NIRSpec spectra. 
    The third column lists the quasar systemic redshifts derived from [\ion{C}{ii}] 158 $\mu$m emission \citep{venemans17,yang19,yang21}, and the last column the $J$ band magnitude \citep{yang21,reed19}. }
    \label{tab:qso_table}
\end{table*}

\subsection{Normalization of spectra}
\label{sect:normalisation}

For computations of absorption line equivalent widths, we need to determine the level of the continuum emission around the lines. First, we make a global fit of the continuum level across the entire spectrum using the {\tt Astrocook} code \citep{cupani22}, using nodes with a velocity spacing of 1000 km~s$^{-1}$, and reject nodes that deviate at the 5$\sigma$ level within each velocity-window. Some of these outliers are caused by the presence of absorption lines. A spline function is then fit between the remaining node-points. A small window-size is necessary in order to also fit the broad emission lines of the quasars. The spectra in Figures ~\ref{fig:J0020_all}, \ref{fig:J0411_all}, \ref{fig:J0439_all}, and \ref{fig:J1342_all} show strong modulations in the continuum flux level in the fixed slit spectra around 1.8~${\mu}$m and 3.0~${\mu}$m as described in Section~\ref{sect:nirspec_obs}. We therefore choose to manually correct the continuum node points within {\tt Astrocook} before the normalization of the spectra, $f_{\mathrm{norm}}(\lambda)$, and their associated error spectra, $\sigma(f_{\mathrm{norm, \lambda}}$) are computed. The normalized spectra are shown in Appendix~\ref{app:A} with the lines belonging to the identified absorption systems at different redshifts overlaid.

With normalized spectra the observed equivalent widths ($W_{\mathrm{obs}}$) and their uncertainties are computed by 
\begin{equation}
W_{\mathrm{obs}}  = \sum_{\lambda_1}^{\lambda_2}  (1 - f_{\mathrm{norm}}(\lambda)) \Delta\lambda 
\end{equation}
\begin{equation}
\sigma(W_{\mathrm{obs}}) = \sqrt{ \sum_{\lambda_1}^{\lambda_2} \sigma^2(f_{\mathrm{norm, \lambda}}) \Delta\lambda^2}
\end{equation}
where $\Delta\lambda$ is the wavelength bin per pixel in the spectra.

In order to analyse the quasar spectra bluewards of their \lya\ emission, we have to take a different approach to normalize the quasar spectra since the IGM has effectively absorbed most of the emission in the Gunn-Peterson trough at these redshifts. We use the {\tt python} module {\tt PyQSOFit} \citep{guo18} to construct a model of the quasar continuum emission and broad \lya\ emission line. We choose to model the continuum as a power-law function at wavelengths not affected by broad emission lines and superimpose best fits of the quasar \lya\ emission line on top of the continuum, using a set of Gaussian functions with different widths, including both broad and narrow components. 
The observed spectra are divided by the constructed model spectra and the normalized spectra show the relative transmission in the IGM. We use these normalized spectra to explore the connection between the metal absorption line systems and transmission spikes in the IGM in Section ~\ref{sect:trans_igm}.

\section{Metal-line absorption systems}
\label{sect:abs_metal}

In the four quasar spectra we identify metal absorption line systems at several different intervening redshifts as well as in the vicinity of the quasar redshifts. Compared to the IGM where metal ions are typically highly ionized, intervening ISM or CGM lines are typically recognised via low-ionization absorption lines from \ion{O}{i}, \ion{Si}{ii}, \ion{C}{ii}, \ion{S}{ii}, \ion{Al}{ii}, \ion{Zn}{ii}, \ion{Cr}{ii}, \ion{Mn}{ii}, \ion{Fe}{ii}, \ion{Mg}{ii}, and \ion{Mg}{i}, but also may have higher ionization lines from \ion{Al}{iii}, \ion{C}{iv}, and \ion{Si}{iv}. 
We used the atomic line list in \citet{krogager18}, which is compiled from the Vienna Atomic Line Data Base \citep{piskunov95,ryab15}.

The discovery and classification is done through visual inspection searching for systems of lines commonly detected in quasar damped \lya\ systems \citep[e.g.][]{prochaska03} and Gamma-ray burst afterglow spectra that include strong damped \lya\ systems \citep[e.g.][]{christensen11}. Visual identifications, just as machine generated ones, are aided by the presence of line doublets, such as \ion{Mg}{ii}\,$\lambda\lambda$2796,2803 or \ion{C}{iv}\,$\lambda\lambda$1548,1550 whose wavelength separations and fixed relative strengths given by their oscillator strengths are well known. The relative line strengths are different by a factor of two in the optically thin regime, while in the optically thick regime, the line strengths are identical. In addition, absorption lines may have multiple kinematic components and blending with other lines makes the identification of line doublets more complex. In order to identify an absorption line system, we require that at least two or more lines are found for each system such that the identity of the lines and the redshift of the absorption system is uniquely determined. 

In the spectrum of VDES J0020–3653 we identify 16 absorption systems, in DELS J0411–0907 we find 13 systems, in UHS J0439+1634  17 systems, and in ULAS J1342+0928 we find 15 individual absorption systems. In total, 61 identified systems are detected at $2.4\lesssim z\lesssim 7.5$. Appendix~\ref{app:A} presents the four normalized quasar spectra with identified absorption line systems overlaid. The systems and individual absorption lines are presented in both graphical and tabular form in Appendix~\ref{app:B}.

\subsection{Distinguishing \ion{C}{iv} doublets from \ion{O}{i}/\ion{Si}{ii} absorbers} 
We need to pay particular attention to the identification of \ion{C}{iv}\,$\lambda\lambda$1548,1550 doublet and \ion{O}{i}\,$\lambda$1302-\ion{Si}{ii}\,$\lambda$1304 absorbers since the two pairs of transitions have transition wavelength ratios of 0.99834 and 0.99831, respectively. 
These two separations cannot be distinguished at the $R\!~\simeq\!~2700$ spectral resolution of the NIRSpec spectra, and weak systems may not be detected at the signal-to-noise level needed to unambiguously detect the \ion{C}{iv}\,$\lambda\lambda$1548,1550 doublet. To single out potential \ion{O}{i}\,$\lambda$1302-\ion{Si}{ii}\,$\lambda$1304 absorption systems, absorption lines from \ion{Si}{ii}\,$\lambda$1260 and \ion{Si}{ii}\,$\lambda$1526 should be detected too, since they are stronger by a factor of 8.9 and 1.6, respectively, relative to the \ion{Si}{ii}\,$\lambda$1304 line. The \ion{Si}{ii}\,$\lambda$1260 line may, however, fall in the \lya\ forest region and be completely absorbed in the GP troughs. In addition to other silicon transitions, we also search for other associated lines from \ion{C}{ii}\,$\lambda1334$ or \ion{Mg}{ii}\,$\lambda\lambda$2796,2803. When none of these lines are detected, we classify the absorber as a \ion{C}{iv} system.

\subsection{Comparison with known absorption systems }
\label{sect:compare_abs}
As surveys detect more and more high-redshift quasars, an increasing number of studies of intervening absorption lines have been performed. The recent XQR-30 survey is based on a sample of 30 quasars at $5.8<z<6.6$ observed with VLT/X-shooter at higher-resolution \citep{dodorico23}. The only quasar in common with the NIRSpec sample is UHS J0439+1634, where \citet{davies23b} identify a total of 43 absorption line systems from $z=2.2-6.48$. All of the 17 systems we detect, apart from the \ion{Mg}{ii} doublet at $z=6.208$, are also detected in \citet{davies23b}. In contrast, in our lower spectral resolution data we are not able to clearly identify and detect the other absorption line systems. In particular, \citet{davies23} find several weak \ion{C}{iv} doublets in the range from $z=5.0-5.3$ and also weak rest-frame equivalent width ($W_r<0.05$~{\AA}) \ion{Mg}{ii} absorbers at $z\approx2.3-5$ that we cannot confirm in the NIRSpec data. The non-confirmation is either due to blending with other stronger absorption lines from different redshifts, or the fact that only one of the two lines in the doublet can be detected in the NIRSpec data, while the other component is clearly ruled out from the knowledge of the fixed line strengths between the two doublet lines. We are also not able to confirm several of the reported \ion{Si}{iv} doublets. We note that around $z\sim5$ \ion{C}{iv} falls in a wavelength region affected by (weaker) telluric absorption lines and also strong sky emission lines, that \citet{davies23b} do, however, correct for in their analysis.

Turning to absorption line systems at $z>5$, \citet{cooper19} report two systems at $z=5.936$ and 6.178 towards DELS J0411--0907, and at $z=5.889$, 6.271 and 6.843 towards ULAS J1342+0928. These strong systems are all clearly detected in the NIRSpec spectra. In addition, we also detect a few weaker $z>5$ \ion{Mg}{ii} or \ion{C}{iv} doublets towards DELS J0411--0907, whereas towards ULAS J1342+0928, we also find several additional low-ionization systems at $z>7$ identified through multiple lines. ULAS J1342+0928 was observed with both X-shooter at the VLT and with FIRE at the Magellan Telescope at a higher spectral resolution as presented in a detailed case study of the $z=6.84$ absorption system in \citet{simcoe20}. 
We retrieved the processed 2-dimensional X-shooter spectra Advanced Data Products from the ESO archive in order to verify whether the high-redshift absorption line systems detected in the NIRSpec data are also present in the higher resolution spectra. For the higher redshift system at $z=7.368$, the X-shooter data reveals a clear detection of \ion{Si}{ii}\,$\lambda$1260 whereas \ion{Mg}{ii} falls in a noisy region of the spectrum and is not detected. None of the absorption lines from the $z=7.443$ or $z=7.476$ systems can be detected in the X-shooter data, consistent with the less sensitive detection-limit in the ground-based data.

\subsection{Line densities}
Theoretically we expect a detection limit of absorption lines observed equivalent widths at $W_{\mathrm{obs}} =0.2-0.5$ {\AA} as described in Sect.~\ref{sect:nirspec_obs} and Fig.~\ref{fig:sn_ew}. With the wide redshift range of the absorption systems, this implies that the detection limit on the rest-frame equivalent width \(W_r = W_{\mathrm{obs}}/(1+z_{\mathrm{abs}})\) depends on the absorber redshift. The typical detection limit of $W_r$ is 0.025--0.03 {\AA}. When comparing line densities with those of from the literature, we adopt the same $W_r$ cut-off.

To compute line densities, we combine all tables in Appendix~\ref{app:B}, extract the individual absorption transitions of interest, e.g.~\ion{C}{iv}\,$\lambda$1548, \ion{O}{i}\,$\lambda$1302, or \ion{Mg}{ii}\,$\lambda$2796 and compute their $W_r$. Line densities from metal-absorption lines should represent random lines of sight through the Universe, and we therefore exclude absorbers at redshifts closer than 3000 km~s$^{-1}$ from the quasar systemic redshift, since these absorption lines represent proximate systems that may be influenced by the quasars.

\begin{figure}
\centering

   \includegraphics[bb= 40 20 800 520, clip, width=\hsize]{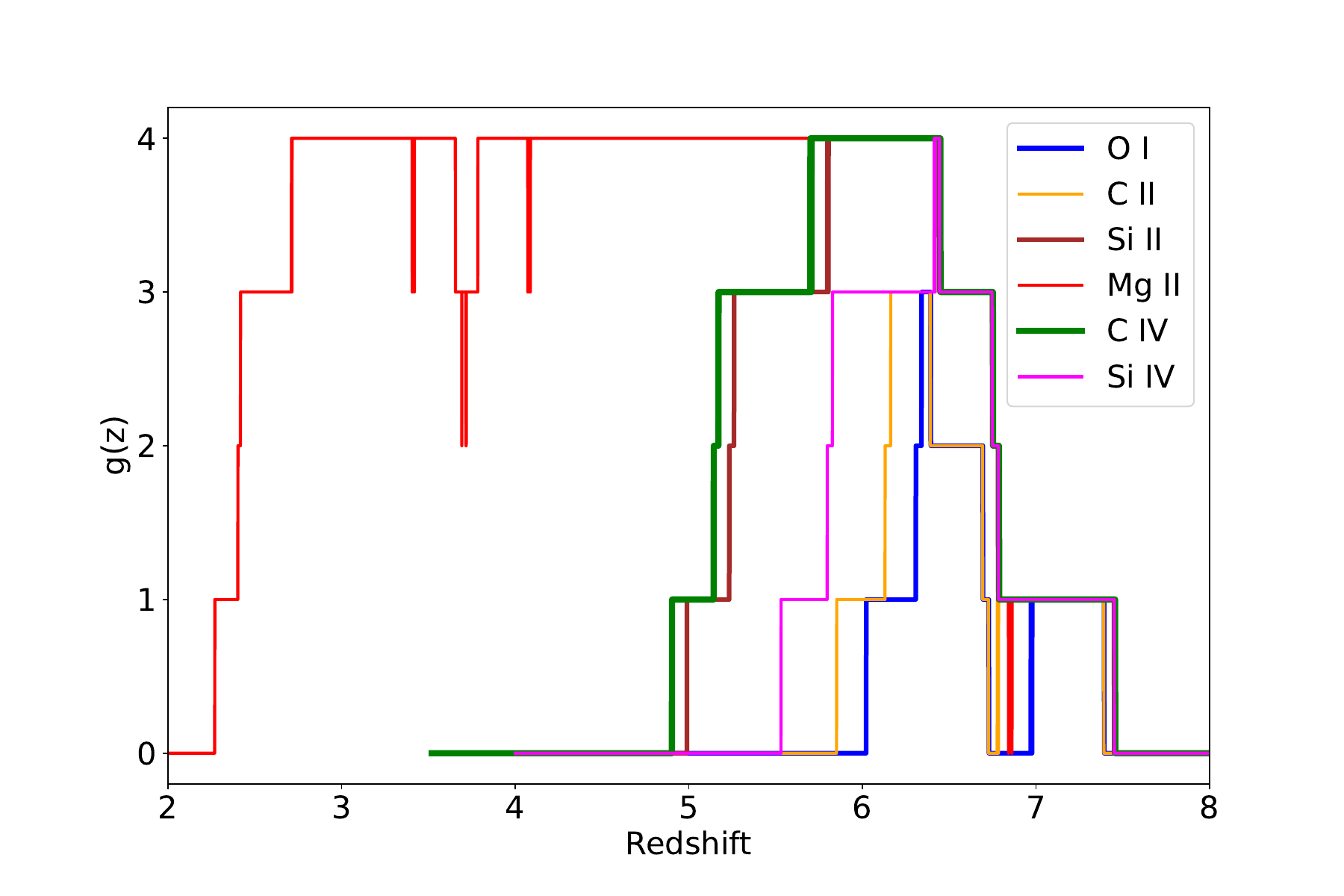}
       \caption{Sensitivity functions, $g(z)$, for the coverage of \ion{O}{i}\,$\lambda$1302, \ion{C}{ii}\,$\lambda$1334, \ion{Si}{ii}\,$\lambda$1526, \ion{C}{iv}\,$\lambda$1548, \ion{Si}{iv}\,$\lambda$1393, and \ion{Mg}{ii}\,$\lambda$2796. The colored lines illustrate how many of the quasar spectra contribute to the search of metal absorption lines as a function of redshift. }  
    \label{fig:sens}
\end{figure}

We compute the redshift path length, which is defined as the full spectral range in each spectrum where it would be possible to detect the chosen absorption line. For each quasar we determine the minimum and maximum redshift where a chosen absorption line can be discovered, excluding regions in the spectra where there are gaps or where the signal-to-noise ratio is low. The sum of this gives the sensitivity function, $g(z)$, which presents for each redshift the number of lines of sight probing the absorption line of interest with an observed $W_{\mathrm{obs}}>0.3$ {\AA}. The result is presented in Fig.~\ref{fig:sens} which illustrates that the NIRSpec observations in this study are sensitive to intervening \ion{O}{i}\,$\lambda$1302 and \ion{C}{ii}\,$\lambda$1334 absorbers at $6 \lesssim z \lesssim 7.5$, \ion{Si}{ii}\,$\lambda$1526 and \ion{C}{iv} absorbers at $5 \lesssim z \lesssim 7.5$, \ion{Mg}{ii} at $2.2\lesssim z\lesssim 7.5$, and \ion{Si}{iv}\,$\lambda$1393 at $5.5\lesssim z\lesssim 7.5$.

By integrating the sensitivity function one can compute the line density $dn/dz$, defined as the number of absorbers $n$ within a chosen redshift interval $ dz$. However, when analysing the line density over a very large redshift range it is conventional to instead analyse the comoving density \citep{bahcall69} and use the comoving absorption path length $X$ defined as: 
\begin{equation}
    dX = \frac{(1+z)^2}{\sqrt{\Omega_M (1+z)^3+\Omega_{\Lambda}}} dz.
\end{equation}

The line density is then computed as
\begin{equation}
    dn/dX =  \frac{\sum_{z_1}^{z_2}n_i} {\int_{z_1}^{z_2} g(z) dX },
\end{equation}
where $n_i$ is the number of absorbers of the chosen species in quasar number $i$, summed over the redshift interval between $z_1$ and $z_2$ in all the quasar spectra. The denominator is the integral of the absorption path length in that redshift interval.

To derive uncertainties of the line densities we assume that the distribution of absorbers is Poissonian in nature. For the small number statistics relevant in this work we use the upper- and lower 68\% confidence intervals tabulated in \citet{gehrels86}.

\subsection{Absorption line density redshift evolution}
Among the many different atomic species detected in the four quasar spectra, in this work we focus on the previously well-studied ones among the low-ionization and high-ionization lines.

For each atomic transition we compute the line density as a function of redshift and compare the JWST data with line densities at lower redshifts from the literature. Additionally, we compare the line densities with the predicted evolution of both low- and high-ionizations column density distribution from simulations in the redshift range $5<z<8$ (Huscher et al.in prep.). The simulations are 
based on updates to the  {\sc Technicolor Dawn}  cosmological hydrodynamic simulation \citep{finlator18,finlator20}. By probing random lines of sight through the simulation volumes from redshifts $z=5$ to $z=10$, one can identify various absorption lines \citep[e.g.][]{doughty19,doughty23} and compute their equivalent widths and column densities. Compared to previous simulations, the model of Huscher et al. includes predictions of equivalent widths and column densities of other atomic transitions for the modelled absorbers, allowing us to compare with the observations of a range of elements in different redshift bins from $z=5$ to $z=10$. For comparison with the NIRSpec observations, we restrict our attention to the relevant redshift interval $5<z<7.5$.

\subsubsection{\ion{O}{i}\,$\lambda$1302 absorbers}

We detect seven \ion{O}{i}\,$\lambda$1302 absorbers listed in Table~\ref{table:oisys} within the total absorber path length interval  $\Delta X = 7.68$. The highest redshift absorption system at $z=7.476$ lies close to the ULAS J1342+0928 quasar redshift at $z=7.535$, and is considered proximate according to the cut-off of at 3000 km~s$^{-1}$, and is not included in the analysis of the statistics of line densities. The two other \ion{O}{i} systems at $z>7$ represent the highest redshift detections of neutral oxygen absorption along the line of sight to high-redshift quasars reported to date. In all cases, when \ion{O}{i} is detected, we also detect more than one other line at the same redshift, so we can be confident to clearly identify the \ion{O}{i} absorption systems (See Appendix~\ref{app:B}). 

Figure~\ref{fig:dndx} shows the line density of \ion{O}{i} absorbers with $W_r>0.05$ {\AA} as a function of redshift for the new NIRSpec measurement compared with values at lower redshifts from \citet{becker19}. The line density for the highest redshift point has a significant uncertainty, implying that even though the line density shows an increase from $z=6$ to $z=7$ the values are also consistent with being constant.

For \ion{O}{i} as well as for other low-ionization line species, the model predicts an increase in line density with increasing redshift up to $z\sim7$ and at higher redshifts ($z>7$) the line density decreases. The reason for this behaviour is a combination of a decreasing metallicity of the CGM with increasing redshift, while the turnover from $z=7\rightarrow6$ is caused by the increasing level of ionization of the CGM with time \citep{doughty19}. With increasing cosmic time, \ion{O}{i} absorbers have a smaller covering fraction as the clouds' outskirts become increasingly ionized while the total metallicity of oxygen and its distribution in the CGM of galaxies does increase with time \citep{doughty19}.

\begin{table}
    \centering
    \caption{\ion{O}{i}\,$\lambda$1302 systems }
    \label{table:oisys}  
    \begin{tabular}{llc}   
    \hline \hline
\noalign{\smallskip}
 \quad\quad\enskip   Quasar &   \enskip$z_{\mathrm{abs}}$ & $W_{\mathrm{rest}}$ (\AA)\\
\noalign{\smallskip}
    \hline
\noalign{\smallskip}
 VDES J0020--3653 &  6.4535  &  0.091$\pm$0.007\\
 VDES J0020--3653 &  6.5625  &  0.062$\pm$0.008\\
 VDES J0020--3653 &  6.6690  &  0.115$\pm$0.010\\
 UHS J0439+1634   &  6.2880  &  0.119$\pm$0.006\\
ULAS J1342+0928   &  7.3680  &  0.061$\pm$0.011\\
ULAS J1342+0928   &  7.4430  &  0.031$\pm$0.010\\
ULAS J1342+0928   &  7.4760$^\dagger$  &  0.045$\pm$0.010\\
\noalign{\smallskip}
\hline                           
\end{tabular}
\begin{tablenotes}
\item $^\dagger$The highest redshift system at $z=7.4760$ lies $\approx$2000 km~s$^{-1}$ from the quasar redshift and is considered as a proximate system, and therefore not included in the computation of line densities.
\end{tablenotes}
\end{table}

\begin{figure*}
\centering
   \includegraphics[bb = 50 350 530 700, clip, width=0.48\hsize]{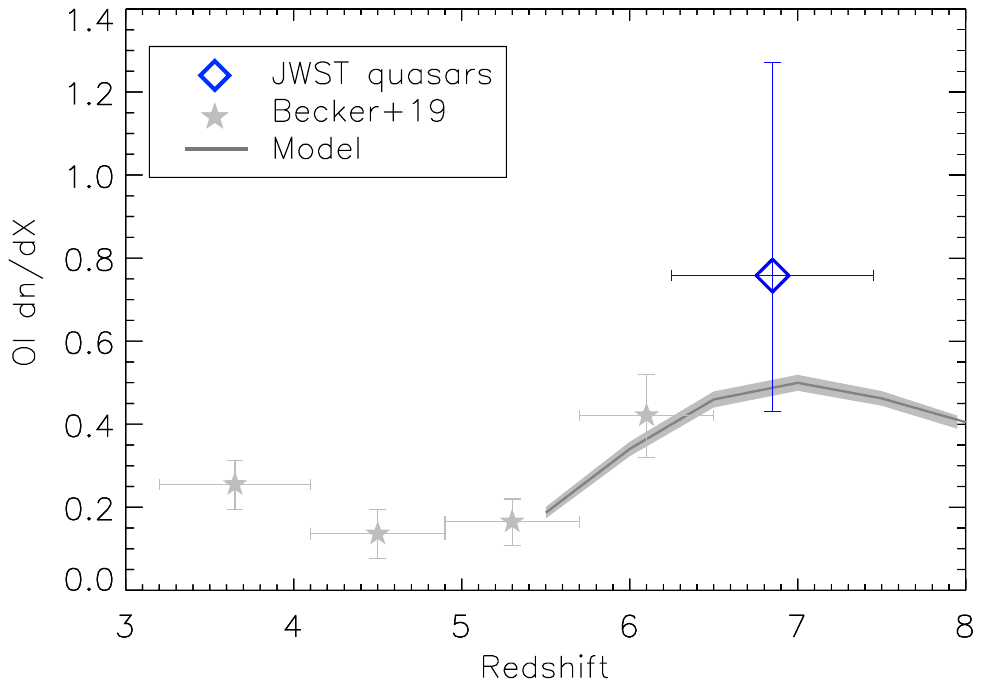}
   \includegraphics[bb = 50 350 530 700, clip, width=0.48\hsize]{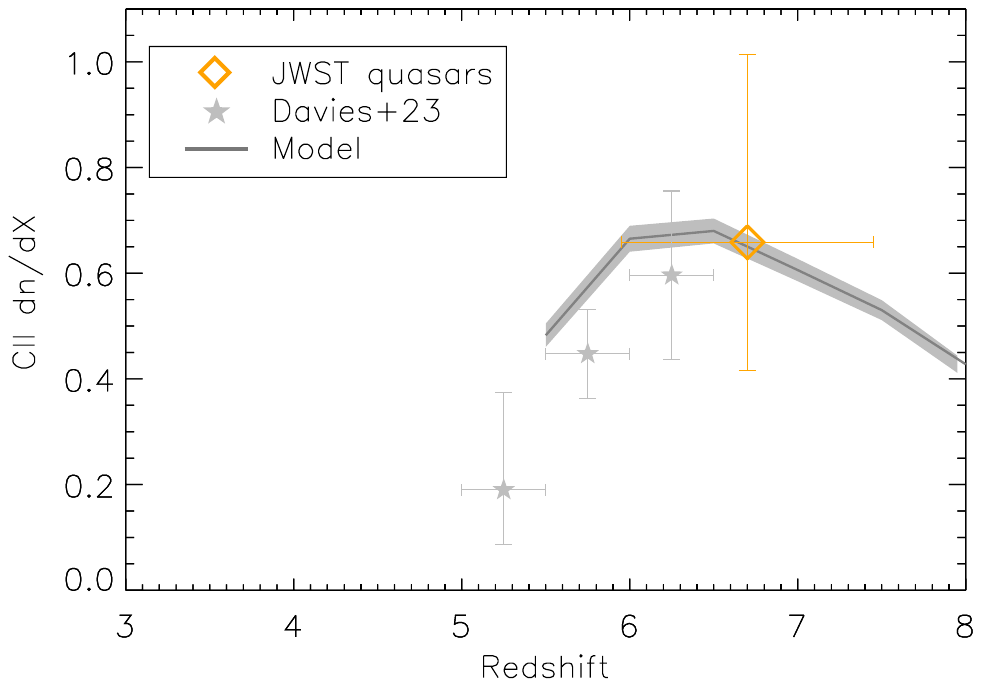}
   \includegraphics[bb = 50 350 530 700, clip, width=0.48\hsize]{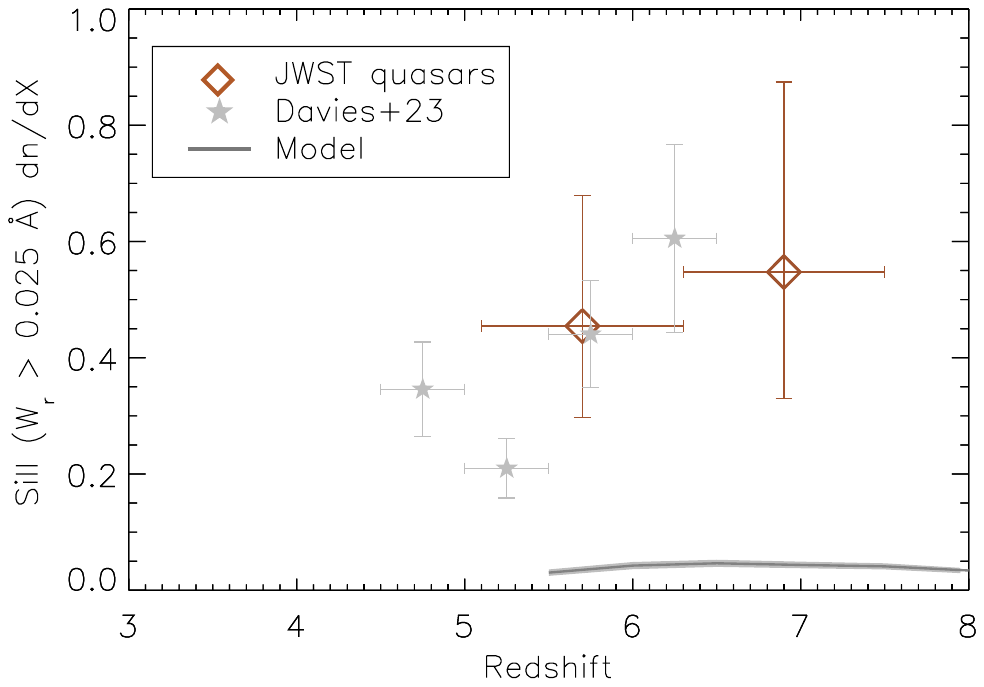}
   \includegraphics[bb = 50 350 530 700, clip, width=0.48\hsize]{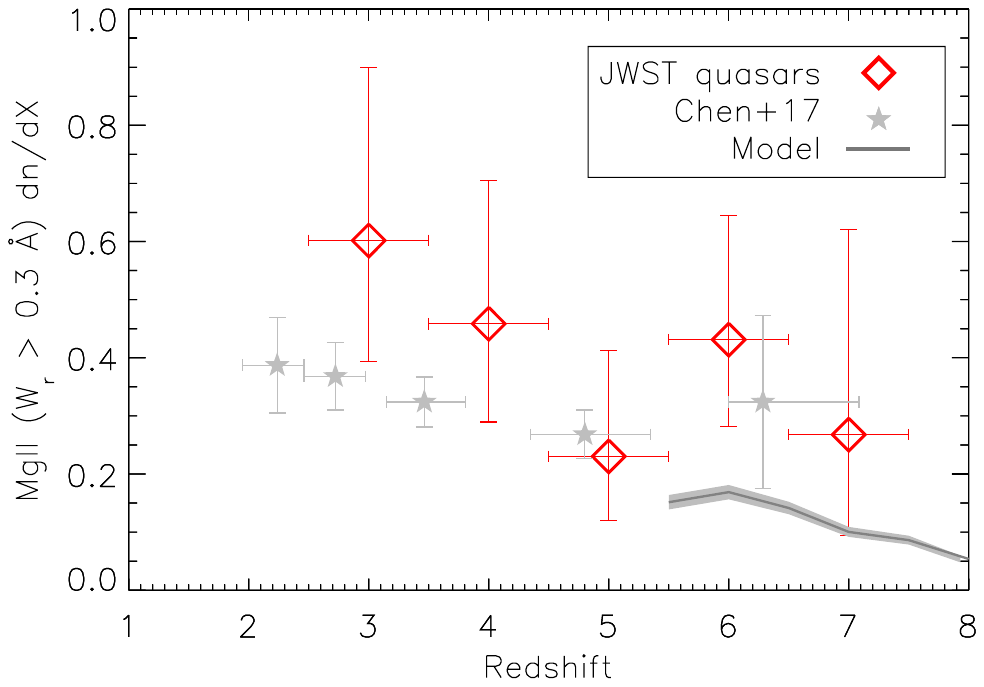}
   \caption{Line densities as a function of redshift for low-ionization lines.
    In all panels, colored symbols represent the JWST data, where error bars for the line densities 
    represent 68\% confidence intervals for small-number Poisson uncertainties \citep{gehrels86}.
    The line densities are compared to predictions from numerical models (Huscher et al. in prep.) computed from the same selection of $W_r$ as from the NIRSpec observations.
      {\it Upper left panel:}  \ion{O}{i}\,$\lambda$1302 absorption line densities for systems with $W_r > 0.05$~{\AA} showing an increase in line density at $z>6.5$ from the NIRSpec data compared to lower redshifts \citep{becker19}.
     {\it Upper right panel:} \ion{C}{ii}\,$\lambda$1334 absorption line densities for systems with $W_r > 0.03$~{\AA} compared to lower redshift values derived from the catalog in \citet{davies23b}.
     {\it Lower left panel:} \ion{Si}{ii}\,$\lambda$1526 absorption line densities for systems with $W_r> 0.025$~{\AA} compared to that derived from the absorption database in \citet{davies23b}. The model prediction at $dn/dX\approx0.04$ under-produces the \ion{Si}{ii} line densities by a factor of $\approx$12. 
       {\it Lower right panel:} \ion{Mg}{ii}\,$\lambda$2796 absorption line densities for systems with $W_r > 0.3$~{\AA}. Compared to the \ion{Mg}{ii} study in \citep{chen17}, the NIRSpec data suggest that the line density could be constant or have a shallow decrease with increasing redshifts in agreement with the model prediction.
       }  
    \label{fig:dndx}
\end{figure*}

\subsubsection{ \ion{C}{ii}\,$\lambda$1334 absorbers}
In the JWST spectra we identify eight intervening \ion{C}{ii}\,$\lambda$1334 absorption systems listed in Table~\ref{tab:ciisys} over a full path length interval of $\Delta X = 11.10$ at $5.9<z<7.5$. Due to the available search path, some of the \ion{C}{ii} lines do not have detected \ion{O}{i} absorption, since these lie blueward of the quasar \lya\ wavelength. The upper right hand panel in Fig.~\ref{fig:dndx} illustrates the \ion{C}{ii} line densities with $W_r> 0.03$~{\AA}.

In comparison, \citet{becker19} find that most of their \ion{O}{i} absorbers also have \ion{C}{ii} lines, but this is primarily caused by selection since to identify \ion{O}{i} absorption lines, the presence of other lines (e.g from \ion{C}{ii}) at the same redshift is required. The line density of \ion{C}{ii} at $z<5$ therefore shows a redshift evolution similar to that of \ion{O}{i}. Using the absorption line catalog of carbon absorbers in \citet{davies23b}, we compute the \ion{C}{ii} line densities in three redshift bins at $5.0<z<6.5$ as illustrated in Fig.~\ref{fig:dndx}. Including the line density at $z>6$ from the NIRSpec data we find a general trend of an increasing line density with redshift.

\begin{table}
    \caption{\ion{C}{ii}\,$\lambda$1334 systems }
    \label{tab:ciisys}  
    \centering
    \begin{tabular}{llc}   
    \hline \hline
    \noalign{\smallskip}
\quad\quad\enskip    Quasar &   \enskip$z_{\mathrm{abs}}$ & $W_{\mathrm{rest}}$ (\AA)\\
    \noalign{\smallskip}
    \hline
    \noalign{\smallskip}
VDES J0020--3653 & 6.4535 &   0.103$\pm$0.010\\  
VDES J0020--3653 & 6.6690 &   0.168$\pm$0.010\\   
DELS J0411--0907  & 6.1774 &   0.215$\pm$0.006\\
UHS J0439+1634   & 5.9280 &   0.008$\pm$0.007\\
UHS J0439+1634   & 6.2880 &   $<$0.086$\pm$0.007$^\dagger$\\
ULAS J1342+0928  & 6.8427 &   0.626$\pm$0.011\\
ULAS J1342+0928  & 7.3680 &   $<$0.063$\pm$0.010$^\dagger$\\
ULAS J1342+0928  & 7.4430 &   0.072$\pm$0.009\\
    \noalign{\smallskip}
\hline                           
\end{tabular}
\begin{tablenotes}
    \item $^\dagger$The upper limit to the equivalent width is quoted because the line is blended with another absorption line at a different redshift (See Appendix~\ref{app:B}).
\end{tablenotes}
\end{table}

\subsubsection{\ion{Si}{ii}\,$\lambda$1526 absorbers}

Several \ion{Si}{ii} transitions can be detected in the spectra with UV transitions at rest-frame wavelengths at 1260, 1304, 1526, and 1808~{\AA}. Due to the wide range of oscillator strengths and possible contamination with other absorption lines, not all absorption lines are detected for any given system. When deriving the line density of \ion{Si}{ii} it is convenient to use a line which is both a strong transition and also has a large redshift path length. The 1808~{\AA} line has the longest redshift path length, but it is the weakest transition of the four, whereas the strongest 1260~{\AA} line can only be detected in a very short path length interval because its wavelength is close to \lya. Therefore we choose to investigate the line density based on the 1526~{\AA} line, which is also a strong transition, and has a larger redshift path length interval.

In the JWST spectra we identify 12 intervening \ion{Si}{ii} absorption systems listed in Table~\ref{tab:siiisys} over a full path length interval of $\Delta X = 29.0$ at $5<z<7.45$. We compare the inferred line density with that derived from the database in \citet{davies23b} in the lower left hand panel Figure~\ref{fig:dndx}.
It is apparent that the evolution of the \ion{Si}{ii} line density could either be constant or show a mild increase with redshift to $z\approx 7$.

In the models of Huscher et al. in prep., rest-frame equivalent widths are computed for the \ion{Si}{ii}\,$\lambda$1260 transition. For non-saturated lines, the relation between two lines in the linear part of the curve of growth scale as \( W_{r_1}/ W_{r_2} = \lambda_1^2 f_1 /(\lambda_2^2 f_2) \), where $\lambda_1$ and $\lambda_2$  are the rest-frame wavelength of the two transitions and $f_1$ and $f_2$ are the oscillator strengths. We can therefore convert the estimated \ion{Si}{ii} $\lambda$1260 line densities to that for \ion{Si}{ii}\,$\lambda$1526. The model predicts a constant line density of $dn/dX\approx0.04$ from $z=5$ to $z=8$ for absorbers with $W_r<0.025$~{\AA}, which is smaller by a factor of $\sim$\!~12 compared to the NIRSpec observations. Alternatively, the model predicts \ion{Si}{ii} equivalent widths that are a factor of 2.5 smaller than the observed ones.

 \begin{table}
 \centering
    \caption{\ion{Si}{ii}\,$\lambda$1526 systems }
    \label{tab:siiisys}
    \begin{tabular}{llc}   
    \hline \hline
\noalign{\smallskip}
\quad\quad\enskip  Quasar &   \enskip$z_{\mathrm{abs}}$ & $W_{\mathrm{rest}}$ (\AA)\\
\noalign{\smallskip}
    \hline
\noalign{\smallskip}
VDES J0020--3653&   5.4695 &   0.028$\pm$0.011\\
VDES J0020--3653&   5.7910 &   0.074$\pm$0.014\\
VDES J0020--3653&   6.4535 &   0.028$\pm$0.012\\
VDES J0020--3653&   6.5625 &   0.059$\pm$0.012\\
VDES J0020--3653&   6.6690 &   0.092$\pm$0.013\\
DELS J0411--0907 &   5.9350 &   0.143$\pm$0.010\\
DELS J0411--0907 &   6.1774 &   0.127$\pm$0.008\\
UHS J0439+1634  &   5.8190 &   0.047$\pm$0.006\\
UHS J0439+1634  &   6.2840 &   0.286$\pm$0.009\\
ULAS J1342+0928 &   6.8427 &   0.186$\pm$0.012\\
ULAS J1342+0928 &   7.3680 &   $< 0.03^\dagger$  \\
ULAS J1342+0928 &   7.4430 &   0.018$\pm$0.010\\
\noalign{\smallskip}
\hline                           
\end{tabular}
\begin{tablenotes}
 \item   $^\dagger$The  \ion{Si}{ii}\,$\lambda$1526 line is not detected in the spectrum, but the system is nevertheless included in the analysis since both \ion{Si}{ii}\,$\lambda$1260 and \ion{Si}{ii}\,$\lambda$1304 are detected at this redshift.    
\end{tablenotes}
\end{table}

\subsubsection{\ion{Mg}{ii}\,$\lambda$2796 absorbers}
The classical \ion{Mg}{ii}\,$\lambda\lambda$2796,2803 doublet is one of the most commonly studied atomic transitions in quasar spectra \citep[e.g.][]{bergeron91,churchill99}. It is easily distinguishable in absorption because of its known line separation and the fact that the 2796 line has twice the oscillator strength as the 2803 line. The line doublet enters the optical range at $z=0.3$ and can be traced up to $z\sim7$ in ground-based near-IR spectra \citep{matejek12,chen17}. In comparison to ground-based observations of \ion{Mg}{ii} absorbers at $z>2.5$ that are heavily affected by the presence of sky emission- and telluric absorption lines, the NIRSpec spectra are much cleaner, allowing easy visual identification of \ion{Mg}{ii} absorbers even at the highest redshifts. It is therefore not surprising that we find some previously undetected absorption systems in these well-studied high-redshift quasars.

The 3$\sigma$ detection limit of the $W_{\mathrm{obs}}$ for \ion{Mg}{ii} systems in the NIRSpec spectra is 0.3~{\AA}. Above this limit, we find 49 \ion{Mg}{ii} absorbers from $z=2.4$ to $z=7.45$ including the highest redshift detection of an intervening absorption system to date (see Appendix~\ref{app:B}). As for the other lines, absorption line systems that lie closer than 3000 km~s$^{-1}$ to the quasar redshift are omitted in the computation of line densities. The highest redshift proximate absorption system at $z=7.476$ is also detected in \ion{Mg}{ii}, and is excluded in the line density computation.

The total redshift path length for \ion{Mg}{ii} absorption systems covered by the four spectra is $\Delta X = 73.3$. To compare with other studies, we focus on the absorbers with $W_r>0.3$ {\AA}, where we detect 30 systems. The lower right hand panel in Fig.~\ref{fig:dndx} illustrates that the line density for \ion{Mg}{ii} absorbers with $W_r>0.3$ {\AA} remains constant or at most exhibits a shallow decrease to the highest redshifts in agreement with previous findings at $z\lesssim6$  \citep{matejek12,chen17,bosman17}. The observed distribution agrees with the model prediction of a shallow decrease in line density with redshift at $z>6$.

\begin{figure}
\centering
   \includegraphics[bb = 70 350 550 700, clip, width=\hsize]{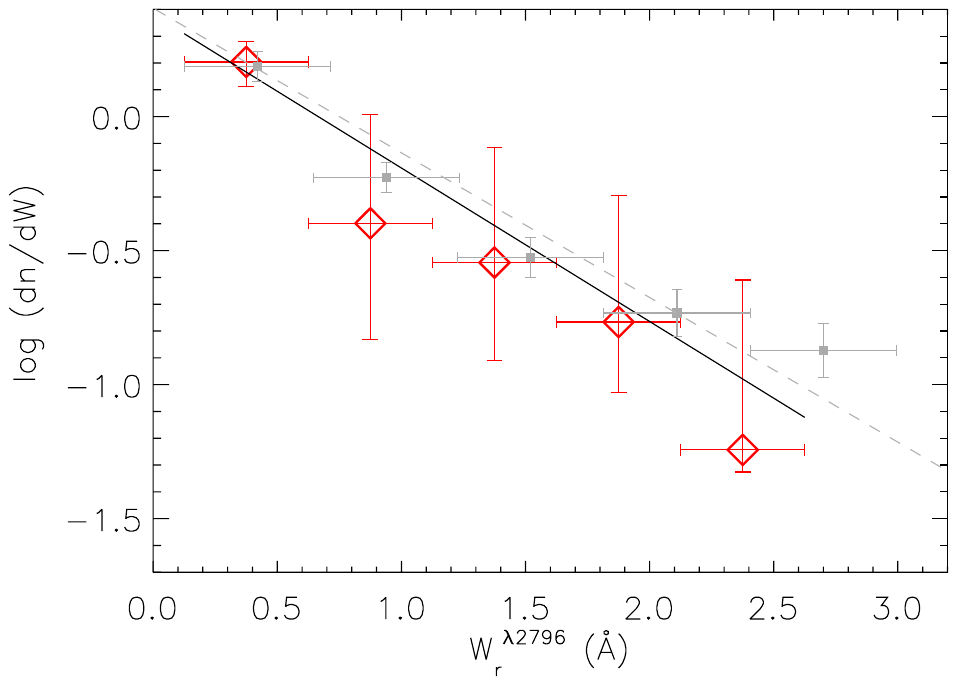}
       \caption{
       Distribution of \ion{Mg}{ii} equivalent widths from all systems across the redshift interval from $z=2.5-7.5$. The best fit is illustrated by the solid line. Compared to much larger quasar surveys at lower redshifts \citep{chen17}, where the dashed line is the best fit, there is no evidence for a different distribution in the  NIRSpec data.} 
    \label{fig:dndw}
\end{figure}

Focusing on all the detected \ion{Mg}{ii} systems including both strong- and weak lines, we derive the rest-frame equivalent width distribution described by a powerlaw function
\begin{equation}
    \frac{dn}{dW}= \frac{n_*}{W*}\exp (-W /W_* )
\end{equation}
where $dW$ is the rest-frame equivalent width increment. We do not include a redshift dependence, since the line density is constant with redshift. The best fit parameters are $W_*$=0.758 and $n_*$=1.82.  Figure~\ref{fig:dndw} presents the distribution compared to a large quasar sample examined in \citet{chen17}, and compared to ground-based observations, we see no evidence for a change in the $W_r$ distribution in the NIRSpec spectra.

\begin{figure}
\centering
   \includegraphics[bb = 50 350 530 700, clip, width=\hsize]{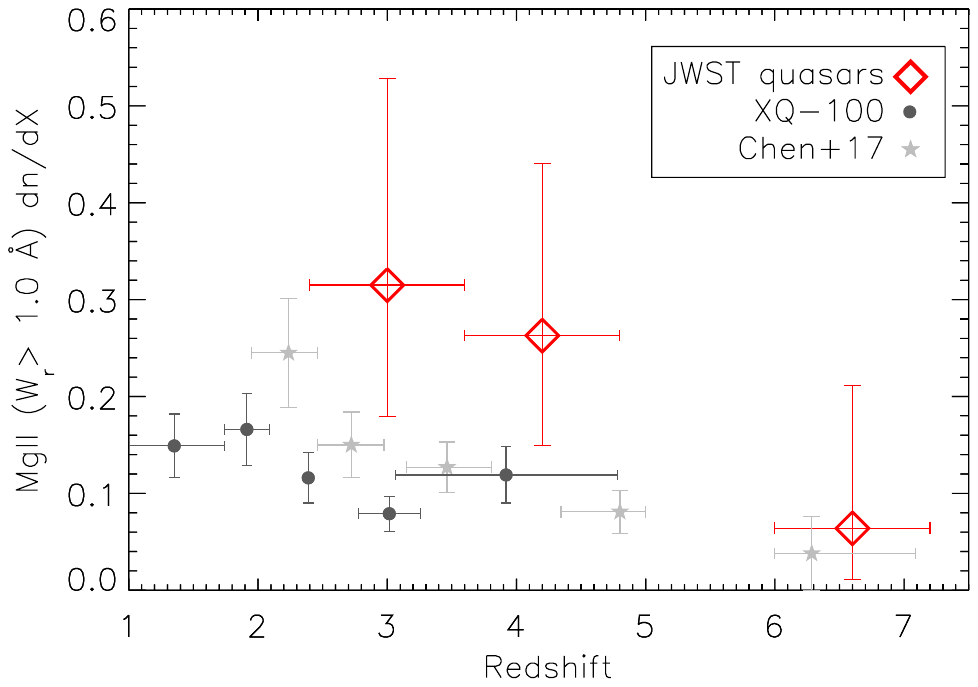}
       \caption{Strong \ion{Mg}{ii}\,$\lambda$2796 absorption line densities for systems with $W_r > 1$ {\AA}.  The JWST data are compared with the strong \ion{Mg}{ii} line densities in a large quasar sample \citep{chen17} and with the XQ-100 survey \citep{christensen17}, and are consistent with a decrease in the line densities to the highest redshifts.  }  
    \label{fig:dndx_mg_strong}
\end{figure}

Strong \ion{Mg}{ii}\,$\lambda$2796 absorption lines with rest-frame $W_r > 1$~{\AA} have received particular attention since these absorbers are found to have the same redshift evolution as that derived from the integrated [\ion{O}{ii}] luminosity of galaxies  and are proposed to trace the cosmic star-formation rate density at $z\approx1$ \citep{prochter06,menard11}. It is therefore relevant to investigate the strongest \ion{Mg}{ii} lines at the highest redshifts with the new NIRSpec data. We identify 11 strong \ion{Mg}{ii}\,$\lambda$2796 absorbers with $W_r >$ 1{\AA} listed in Table~\ref{tab:mgsys}. The line density $dn/dX$ is illustrated in Fig.~\ref{fig:dndx_mg_strong}. Since we only have four JWST spectra, small-number Poisson statistics dominate the uncertainties \citep{gehrels86}, but overall we see that the line density is consistent with other high-redshift quasar and gamma-ray burst afterglow samples that find a decrease in strong $W_r>1$~{\AA} systems at $z>3$ \citep{chen17,christensen17,zou21}. None of the absorbers detected at $z>7$ fall in the strong-line category.

 \begin{table}
 \centering 
    \caption{Strong \ion{Mg}{ii} systems}
    \label{tab:mgsys}  
    \begin{tabular}{llc}   
    \hline\hline
\noalign{\smallskip}
    \quad\quad\enskip Quasar & \enskip$z_{\mathrm{abs}}$ & $W_{\mathrm{rest}}$ (\AA)\\
\noalign{\smallskip}
\hline        
\noalign{\smallskip}
  VDES J0020--3653  &  3.4087   &  2.72$\pm$0.02\\
   VDES J0020--3653  &  3.6040   &  1.39$\pm$0.02\\
   VDES J0020--3653  &  4.0738   &  2.08$\pm$0.02\\
   DELS J0411--0907   &  2.5771   &  1.77$\pm$0.02\\
   DELS J0411--0907   &  3.4290   &  1.79$\pm$0.02\\
   DELS J0411--0907   &  3.7760   &  1.00$\pm$0.02\\
   DELS J0411--0907   &  4.2815   &  2.54$\pm$0.01\\
   UHS J0439+1634   &  3.1700   &  1.31$\pm$0.01\\
   UHS J0439+1634   &  4.5230   &  1.16$\pm$0.01\\
   ULAS J1342+0928  &  3.3758   &  1.48$\pm$0.02\\
   ULAS J1342+0928  &  6.8427   &  1.38$\pm$0.04\\
\noalign{\smallskip}
\hline                           
    \end{tabular}
\end{table}

\subsubsection{\ion{C}{iv} absorbers}
\begin{figure*}
\centering
   \includegraphics[clip, trim=1.5cm 12.5cm 3cm 2.8cm, width=0.48\hsize]{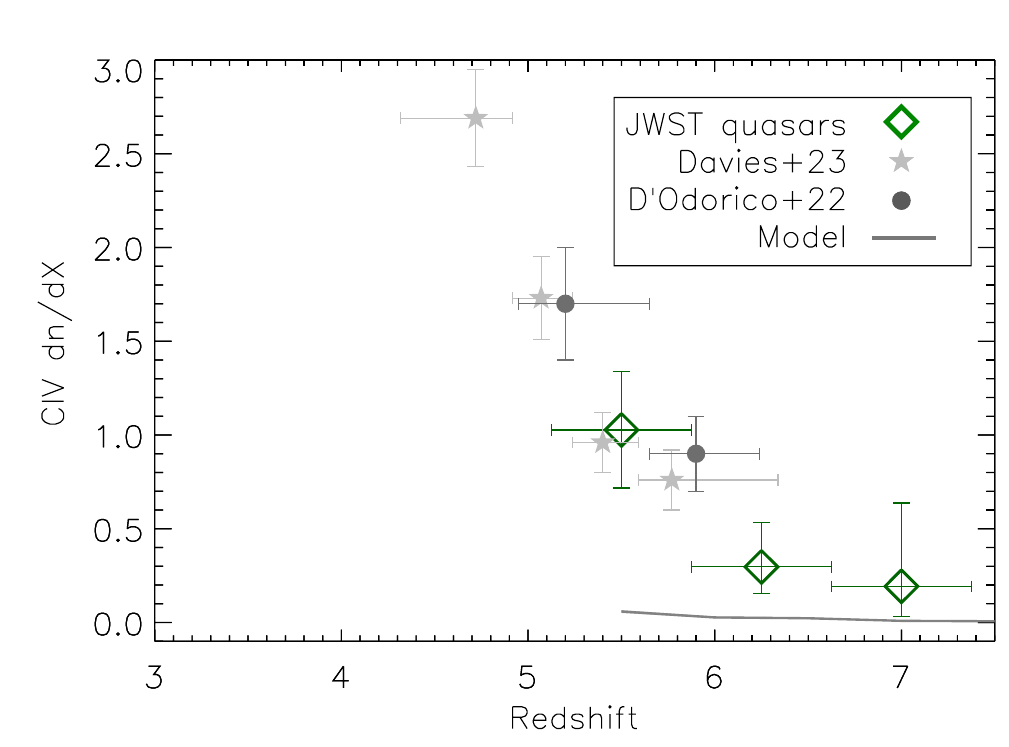}
   \includegraphics[clip, trim=1.5cm 12.5cm 3cm 2.8cm, width=0.48\hsize]{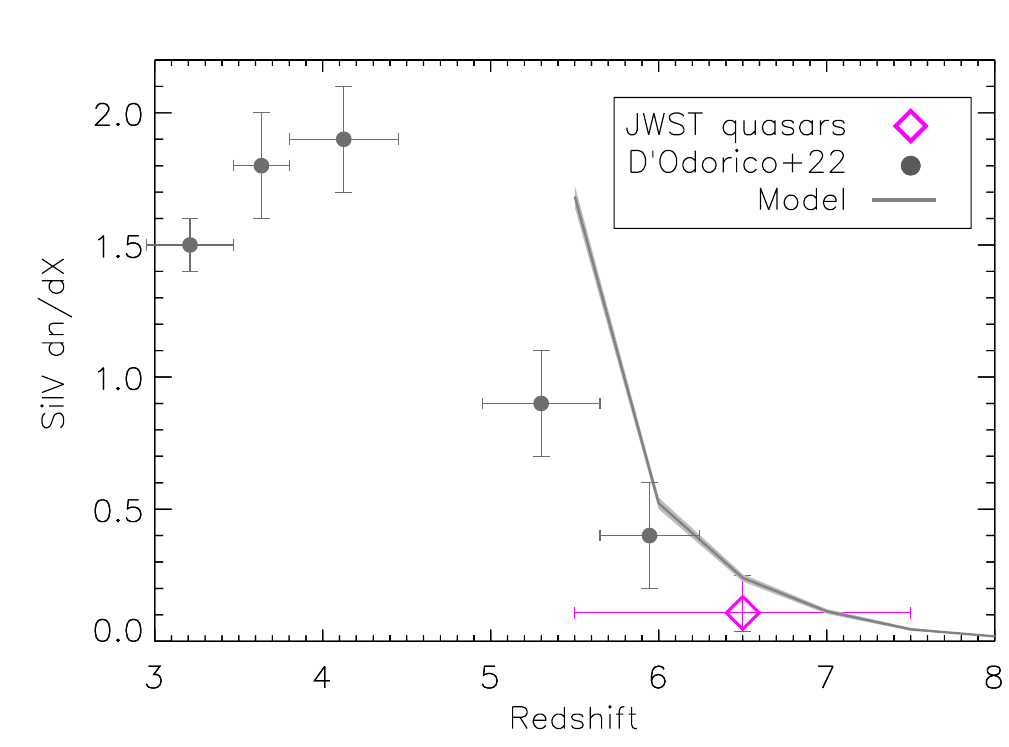}
       \caption{Line densities of high-ionization lines. The left hand panels shows the \ion{C}{iv} absorption line densities with $W_r > 0.05$ {\AA}.
       The \ion{C}{iv} line densities at the highest redshifts from the E-XQR-30 survey  \citep{dodorico22} are illustrated for comparison. The same E-XQR-30 data set was analysed in \citet{davies23} and has been corrected for incompleteness. The {\it right hand panel} shows \ion{Si}{iv} absorption line densities. The \ion{Si}{iv} line densities for systems with $W_r > 0.03$ {\AA} from \citet{dodorico22} are illustrated for comparison.}  
    \label{fig:dndx_high}
\end{figure*}

Turning to the high-ionization transitions, we identify 17 \ion{C}{iv}\,$\lambda\lambda$1548,1550 absorption systems at $5.20< z< 6.86$ in our spectra as listed in Table~\ref{table:civsys}. The total path length interval for \ion{C}{iv} is $\Delta X$ = 30.8. 
Two of the absorbers lie closer than 3000 km~s$^{-1}$ from the quasar systemic redshifts, and are excluded in the computation of absorption line density shown in Figure~\ref{fig:dndx_high}. The NIRSpec data reveal that the \ion{C}{iv} line density continues to decrease with redshift beyond $z>6$. Compared to the observed line densities, the model predicts a much lower line density. This discrepancy was also noted in \citet{dodorico22}, who discussed the mismatching \ion{C}{iv} as a problem in the models in \citet{finlator20}. If models are forced to match the line density of \ion{C}{iv}, then the line densities for \ion{Si}{iv} would be over-predicted.

\begin{table}
    \centering
    \caption{\ion{C}{iv} systems} 
    \label{table:civsys}  
    \begin{tabular}{llc}   
    \hline\hline
\noalign{\smallskip}
    \quad\quad\enskip Quasar & \enskip$z_{\mathrm{abs}}$ & $W_{\mathrm{rest}}$ (\AA)\\
\noalign{\smallskip}
    \hline
\noalign{\smallskip}
    VDES J0020--3653  & 5.2005 &   0.044$\pm$0.008\\
    VDES J0020--3653  & 5.3275 &   0.170$\pm$0.008\\
    VDES J0020--3653  & 5.4699 &   0.228$\pm$0.013\\
    VDES J0020--3653  & 5.7913 &   0.357$\pm$0.014\\
    VDES J0020--3653  & 6.5030 &   0.070$\pm$0.013\\
    VDES J0020--3653  & 6.8550$^\dagger$ &   0.201$\pm$0.007\\
    DELS J0411--0907   & 5.4258 &   0.052$\pm$0.009\\   
    DELS J0411--0907   & 5.9349 &   0.067$\pm$0.009\\
    UHS J0439+1634  &  5.2740 &   0.179$\pm$0.007\\
    UHS J0439+1634  &  5.3945 &   0.808$\pm$0.006\\
    UHS J0439+1634  &  5.3945 &   0.808$\pm$0.006\\
    UHS J0439+1634  &  5.5228 &   0.108$\pm$0.007\\
    UHS J0439+1634  &  5.7360 &   0.123$\pm$0.006\\
    UHS J0439+1634  &  6.2840 &   0.390$\pm$0.007\\
    UHS J0439+1634  &  6.4877$^\dagger$ &  0.238$\pm$0.004\\
    ULAS J1342+0928 &  5.8888 &   0.263$\pm$0.010\\
    ULAS J1342+0928 &  6.7490 &   0.103$\pm$0.010\\
\noalign{\smallskip}
\hline                           
\end{tabular}
\begin{tablenotes}
\item $^\dagger$These absorption systems lie within 3000 km~s$^{-1}$ of the quasar systemic redshift and are excluded from the line density computation.
\end{tablenotes}
\end{table}

\subsubsection{\ion{Si}{iv} absorbers}
In the spectrum of ULAS J1342+0928 we detect a candidate \ion{Si}{iv}\,$\lambda$1393 line at $z=6.476$, which is also detected in \ion{C}{iv}. The spectrum does not clearly reveal the fainter \ion{Si}{iv}\,$\lambda$1402 line. In the spectrum of J0439+1634 we detect a single \ion{Si}{iv} doublet, also reported by \citet{davies23b}, but we note that both transitions in this doublet lie at the wings of two broad absorption line features in the quasar spectrum. These absorbers are listed in Table~\ref{table:siivsys}.

The total path length for \ion{Si}{iv} systems in the JWST spectra is $\Delta X = 18.60$. Fig.~\ref{fig:dndx_high} compares the \ion{Si}{iv} line density from the highest redshift NIRSpec data  to the lower redshift measurements of \citet{dodorico22}. As for the \ion{C}{iv} systems, \ion{Si}{iv} line densities are known to exhibit a rapid decrease in line density at $z>5$ \citep{dodorico22}, which we confirm with our very few detections of  \ion{Si}{iv} absorption lines.

\begin{table}
    \centering
    \caption{\ion{Si}{iv}\,$\lambda$1393 systems }
    \label{table:siivsys}  
    \begin{tabular}{llc}   
    \hline \hline
\noalign{\smallskip}
\quad\quad\enskip    Quasar &   \enskip$z_{\mathrm{abs}}$ & $W_{\mathrm{rest}}$ (\AA)\\
\noalign{\smallskip}
    \hline
\noalign{\smallskip}
UHS J0439+1634  &   6.173 &   0.096$\pm$0.007\\
ULAS J1342+0928 &   6.746 &   0.041$\pm$0.012\\
\noalign{\smallskip}
\hline                           
\end{tabular}
\end{table}

\subsubsection{ \ion{Ca}{ii} and \ion{Na}{i} lines }
\label{sect:cana}

Most of the strong absorption lines arising in the ISM, CGM and IGM are from rest-frame UV lines, but the extended NIR wavelength coverage of the NIRSpec Band~III G395H data allows us to search for rest-frame optical absorption lines associated with any of the absorption systems. 
Since the ionization potential of \ion{Ca}{ii}\,$\lambda\lambda$3934,3969 and \ion{Na}{i}\,$\lambda\lambda$5891,5897 is less than that of hydrogen, these absorption lines arise in neutral regions and are frequently associated with dusty absorbers at lower redshifts detected in quasar spectra ($z\lesssim1$) \citep{heckman00,wild06}. Weak and strong \ion{Ca}{ii} absorbers, with a division at $W_r =0.7$ {\AA}, probe either galaxy disc gas or halo-components, respectively \citep{fang23}.  We detect \ion{Ca}{ii} lines in three absorption systems at $z=$3.17, 3.631, and 3.37, however, the signal is not sufficient to recover both lines in the doublet in any of the systems. Two of the absorbers fall into the weak line category with $W_{r,3934} < 0.7$~{\AA}, while the $z=3.631$ system towards ULAS J1342+0928 is a strong one.

The other rest-frame optical line of interest, the doublet \ion{Na}{i}\,$\lambda\lambda$5891,5897 is covered by the Band~III data for the highest redshift absorbers. However, as discussed in Section~\ref{sect:nirspec_obs}, the statistical errors on the Band~III portions of the spectra are poorly defined. We therefore choose to only focus on absorption line systems that were identified in the Bands~I and II Fixed Slit data, and test whether an associated \ion{Na}{i} doublet is present in any of the $z>3.9$ systems. Only a single high-redshift absorber at $z=5.6814$ towards ULAS J1342+0928 has a potential \ion{Na}{i} absorption line doublet detection.

\subsection{Combined line density redshift evolution}

\begin{figure}
\centering
   \includegraphics[bb= 20 30 800 770, clip, width=\hsize]{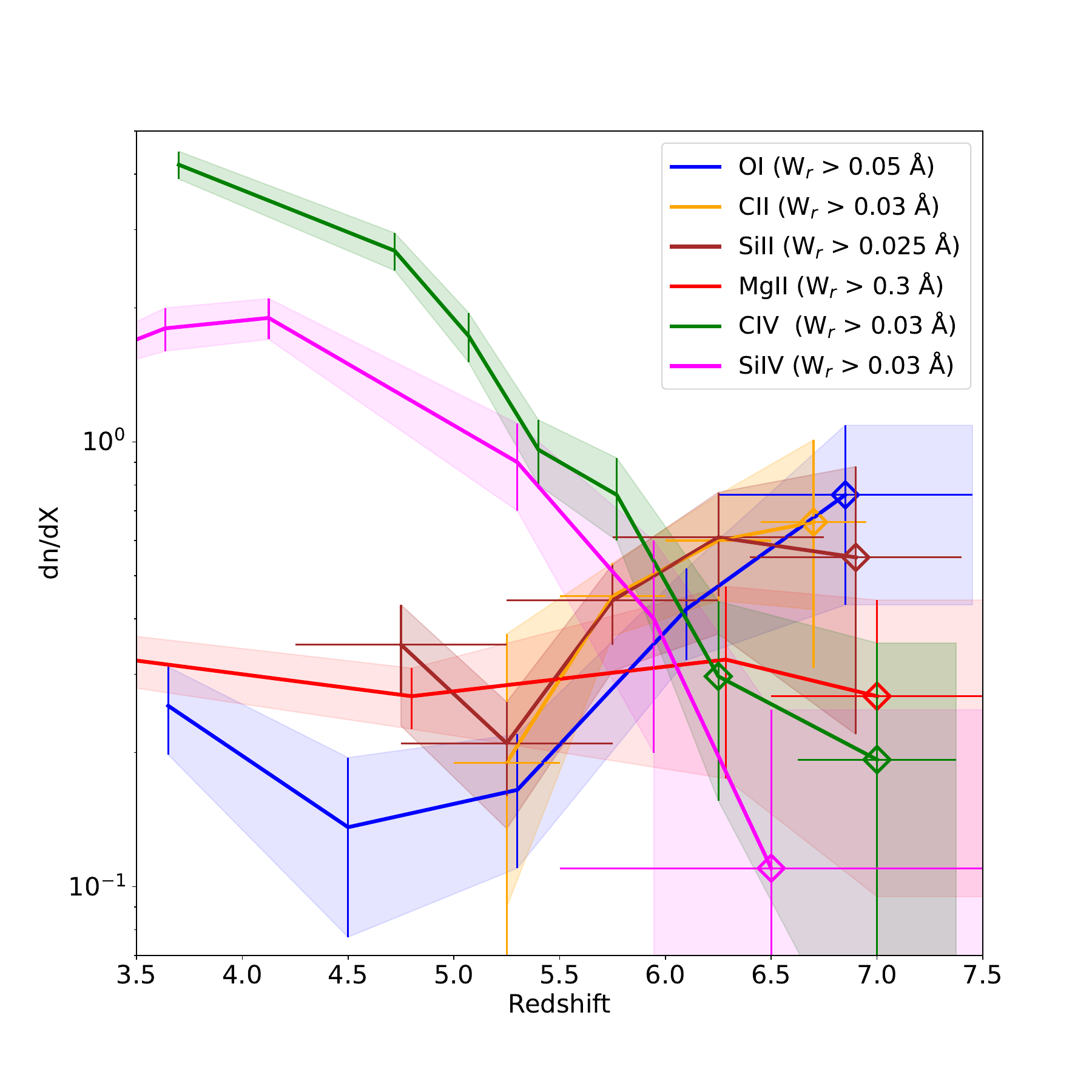}
       \caption{Comparison of the redshift evolution of the line densities of 
       \ion{O}{i}\,$\lambda$1302, \ion{C}{ii}\,$\lambda$1334, \ion{Si}{ii}\,$\lambda$1526, \ion{Mg}{ii}\,$\lambda$2796, \ion{C}{iv}\,$\lambda$1548, 
       and \ion{Si}{iv}\,$\lambda$1393. 
       The highest redshift data points for each of the colored curves represent the NIRSpec analyses in this work, extending from the lower-redshift line densities from \ion{O}{i} \citep{becker19}, \ion{Si}{ii} and \ion{C}{ii} from the catalog in \citet{davies23b}, \ion{C}{iv} \citep{dodorico22,davies23}, \ion{Si}{iv} \citep{dodorico22}, and \ion{Mg}{ii} \citep{chen17}.
       }  
    \label{fig:all_abs}
\end{figure}

To compare the evolution of the atomic species, all the line densities in previous subsections are combined in Fig.~\ref{fig:all_abs}, which shows the evolution with redshifts for different elements derived in the literature with an extension to the highest redshifts ($z>6.0$) detected in the NIRSpec data. All the computed line densities from the JWST data are listed in Table~\ref{tab:absdens}.

Both \ion{C}{iv} and \ion{Si}{iv} experience a drastic drop in line densities continuing at the highest redshifts ($z>6.5$) detected in the NIRSpec data, while the line densities of the low-ionization species \ion{O}{i} and \ion{C}{ii} continue to increase at these redshifts. The evolution of \ion{Si}{ii} from low redshift ($z<5$) is not constrained, and with the large uncertainties from the NIRSpec data, the \ion{Si}{ii} line density could be constant at the redshifts covered in this work, but could also trace the same increase in line density as detected for \ion{O}{i} and \ion{C}{ii}. The transition in line densities of both carbon and silicon from predominantly low-ionization at high redshifts to high-ionization species at low-redshifts  occurs at $z\approx6.0$. In contrast, the line density of \ion{Mg}{ii} remains constant throughout the range from $z=3$ to $z=7$, or may even experience  a shallow decline towards higher redshifts as predicted by the model.

\begin{table*}
    \centering
    \caption{Compilation of line densities for all elements from the NIRSpec data. Column (2) presents the redshift range, (3) the absorption path-length interval, (4) the number of detected absorbers, and (5) the line density.}
    \label{tab:absdens}  
    \begin{tabular}{lcccc}   
    \hline\hline
    \noalign{\smallskip}
        Element  & $z$ & $\Delta X$ &  $n$   & $dn/dX$ \\
        (1) & (2) &  (3) & (4) &  (5) \\
    \noalign{\smallskip}
    \hline
        \noalign{\smallskip}
    \ion{O}{i} $\lambda$1302 ($W_r > 0.03$ {\AA})  &  6.25 -- 7.45  & 6.60  & 6 & $0.91^{+0.54}_{-0.36}$\\
    \ion{O}{i} $\lambda$1302 ($W_r > 0.05$ {\AA})  &  6.25 -- 7.45  & 6.60  & 5 & $0.76^{+0.51}_{-0.33}$\\
    \ion{C}{ii} $\lambda$1334 ($W_r > 0.03$ {\AA}) &  5.95 -- 7.45  & 10.64 & 6 & $0.66^{+0.35}_{-0.24}$\\
    \ion{Si}{ii} $\lambda$1526 ($W_r > 0.025$ {\AA})&  5.1 -- 6.3   & 17.60 & 6 & $0.45^{+0.22}_{-0.16}$\\
    \ion{Si}{ii} $\lambda$1526                     &  6.3 -- 7.5    & 10.96 & 6 & $0.55^{+0.33}_{-0.22}$\\
    \ion{Si}{iv} $\lambda$1394 ($W_r > 0.03$ {\AA})&  5.5 -- 7.5    & 18.60 & 2 & $0.11^{+0.14}_{-0.07}$\\ 
    \ion{C}{iv} $\lambda$1548 ($W_r > 0.05$ {\AA}) &  5.13 -- 5.88  & 10.70 & 11& $1.03^{+0.31}_{-0.31}$\\
    \ion{C}{iv} $\lambda$1548                      &  5.88 -- 6.63  & 13.49 & 4 & $0.30^{+0.23}_{-0.14}$\\
    \ion{C}{iv} $\lambda$1548                      &  6.63 --7.38   &  5.19 & 1 & $0.19^{+0.44}_{-0.16}$\\
    \ion{Mg}{ii} $\lambda$2796 ($W_r > 0.3$ {\AA}) &  2.5 -- 3.5    & 13.30 & 8 & $0.60^{+0.30}_{-0.21}$\\
    \ion{Mg}{ii} $\lambda$2796                     &  3.5 -- 4.5    & 15.27 & 7 & $0.46^{+0.25}_{-0.17}$\\
    \ion{Mg}{ii} $\lambda$2796                     &  4.5 -- 5.5    & 17.36 & 4 & $0.23^{+0.18}_{-0.11}$\\
    \ion{Mg}{ii} $\lambda$2796                     &  5.5 -- 6.5    & 18.56 & 8 & $0.43^{+0.21}_{-0.15}$\\
    \ion{Mg}{ii} $\lambda$2796                     &  6.5 -- 7.5   &   7.49 & 2 & $0.27^{+0.35}_{-0.17}$\\
    \noalign{\smallskip}
\hline                           
    \end{tabular}
\end{table*}

\section{Metal abundances}
\label{sect:abundances}

\subsection{Metal abundance ratios}

The spectral resolution and sampling (Section~\ref{sect:nirspec_obs}) of our NIRSpec spectra is insufficient to accurately model the absorption line profiles via Voigt profile fitting techniques with the aim of deriving column densities for the different ions. However in simple cases, we can extract the same information from the measurements of the rest-frame equivalent widths. In the optically thin regime when the optical depth is $\tau_{\lambda} \ll 1$, the column density $N$ scales linearly with the rest-frame equivalent width \citep{spitzer78}:
\begin{equation}
 N = 1.13 \times 10^{20} \frac{W_r ({\rm\AA})}{f \lambda^2 ({\rm \AA})}  \; \; \; \mathrm{[cm^{-2}]}, 
\end{equation}
where $f$ is the oscillator strength, and $\lambda$ the rest-frame transition wavelength.

We derive the relative abundances assuming that the singly ionized atoms and neutral oxygen are the dominant state of ionization in the absorption systems. This is a reasonable assumption, since none of the systems with \ion{O}{i} absorption show any associated high-ionization transitions. Moreover at the highest redshifts, the metallicities are expected to be low, so the optical depths of the transitions are low too. Finally, the column density ratios are compared to the solar photosphere abundances in \citet{asplund21}. The silicon-to-oxygen and carbon-to-oxygen ratios are computed as:
\begin{equation}
    \mathrm{[Si/O]} = \log \left( 
    \frac{N_{\mathrm{SiII}}}{N_{\mathrm{OI}}} \right) - 
    \log \left( \frac{N_{\mathrm{Si}}} { N_{\mathrm{O}}} \right)_{\odot}
\end{equation}
\begin{equation}
    \mathrm{[C/O]} = \log \left(\frac{N_{\mathrm{CII}}}{N_{\mathrm{OI}}}\right) - 
    \log \left( \dfrac{N_{\mathrm{C}}} { N_{\mathrm{O}}} 
    \right)_{\odot}, 
\end{equation}
where the solar values are
$\log \left( \frac{N_{\mathrm{Si}}} { N_{\mathrm{O}}}\right)_{\odot} = -1.18\pm0.05$
and
$\log \left( \frac{N_{\mathrm{C}}} { N_{\mathrm{O}}}\right)_{\odot} = -0.23\pm0.06$ \citep{asplund21}.

\subsection{Pop III enrichment signatures in [Si/O]}

\begin{figure}[!ht]
\centering
   \includegraphics[bb= 1 1 530 725, clip, width=\hsize]{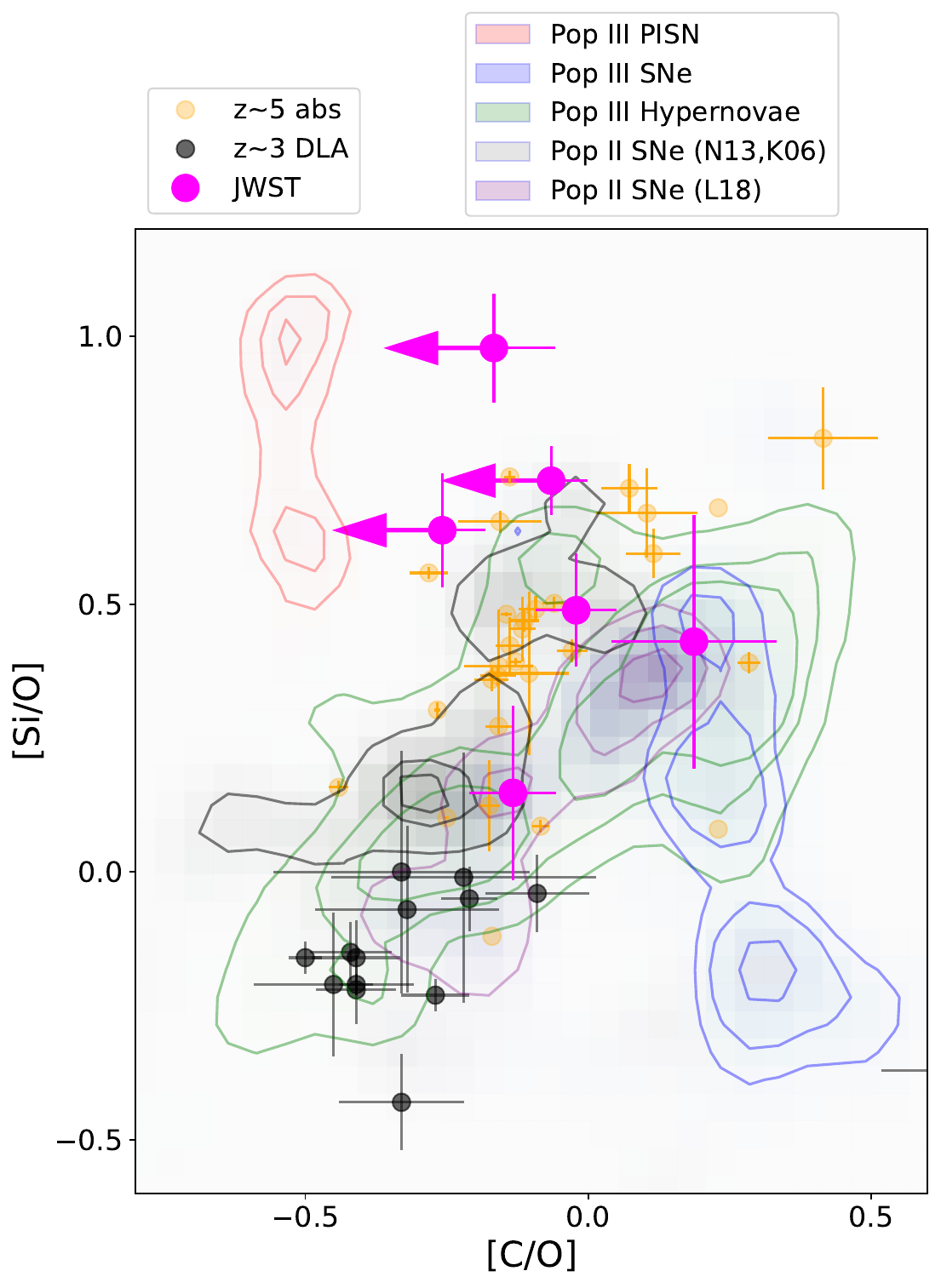}
       \caption{Abundance ratios of [Si/O] versus [C/O]. The background colored map with contours illustrates the metal yields from different stellar explosions colored as follows: {\it Red region: } Pop~III Pair-instability Supernovae (PISN), {\it Green}: Pop~III Hypernovae, {\it Blue}: Pop~III Core-collapse Supernovae, {\it Gray and Purple}: Pop~II Supernovae. Yields from the different stellar populations are scaled assuming a Salpeter IMF. The {\it Magenta circles} illustrate that the NIRSpec absorption systems listed in Table~\ref{tab:metals} have values of [Si/O] similar to Pop III models, although some overlap with the yields of a single 15 $M_{\odot}$ Pop II star. Other intervening absorption-line systems at $4.5<z_{\mathrm{abs}}<6$ \citep{becker19,cooper19} ({\it orange circles}) fall in the same region of the abundance pattern space, whereas the {\it gray circles} representing very metal-poor damped \lya\ absorbers at $z_{\mathrm{abs}}\sim3$ \citep{cooke11,cooke17,welsh19} are also consistent with enrichment of Pop II core collapse SNe \citep[][annotated as K06, N13, and L18]{kobayashi06,nomoto13,limongi18}.
       }  
    \label{fig:abundance}
\end{figure}

To date, no secure discovery of metal-free Pop-III stars has been made and the nature of the first generation of stars remains unknown. Pop-III stars leave behind a signature in the form of distinct metal-abundance patterns created by the stellar explosions of the massive stars. These patterns make their imprint on the chemical abundances of extremely metal-poor halo stars in the Milky Way \citep[e.g.][]{frebel15}, but also absorption line systems at high redshifts can be used to trace the nature of the first stars, since the time span for metal-enrichment of gas in the Universe is small.

In this section we focus on the six systems at $z_{\mathrm{abs}}>6.5$ with identified \ion{O}{i}\,$\lambda$1302 absorption lines in Table~\ref{table:oisys}. Since no high-ionization lines are detected in these absorption systems, they are predominantly neutral and we do not require ionization corrections to derive metal abundance ratios. We examine whether the absorption lines are likely to be intrinsically saturated. Even though the spectra in Figures B1--B61 in Appendix~\ref{app:B} suggest that none of the lines are saturated, at the low spectral resolution some lines could have hidden saturation. Indeed some of the lower-redshift strong absorption lines like the strong \ion{Mg}{ii} lines are clearly saturated. For each of the absorption line systems we construct a conventional curve-of-growth with a best fit Doppler parameter between 20 and 40 km~s$^{-1}$ from unblended \ion{Si}{ii} lines, where two or three separate transitions are detected. For the uncontaminated \ion{O}{i}, \ion{C}{ii}, and \ion{Fe}{ii} lines, we verify that the transitions in $W_r/\lambda$ vs. $Nf\lambda$ fall on a straight line in the linear, non-saturated part of the curve of growth. This confirms that most lines are not saturated and that we can reliably derive the metal abundance ratios listed in Table~\ref{tab:metals}.

Figure~\ref{fig:abundance} shows the abundance ratios in a diagram of [Si/O] vs. [C/O] for the six $z>6$ \ion{O}{i} absorbers in our sample. Since \ion{C}{ii} in two of the six absorbers is partly contaminated by other absorption lines at other redshifts, and one is not detected, we provide only upper limits for three [C/O] ratios. 

To compare the observed abundance pattern with the chemical enrichment from various types of stellar explosions, Figure~\ref{fig:abundance} illustrates the predicted ratios where the combined yields are weighted by the number of stars in a Salpeter initial mass function (IMF), reflecting the combined enrichment pattern generated by a stellar population. Pop III pair-instability SNe (PISN) with initial masses of $140-260$\!~$M_{\odot}$ are illustrated by red coloured contours, and occupy a distinct region with a high [Si/O] and low [C/O] ratio \citep{heger02}. Pop III hypernovae (shown in green contours) and Pop III supernovae (blue contours) with a range of energies and initial masses of $10-100$\!~$M_{\odot}$ also display distinct abundance patterns \citep{heger10}. Hypernovae are defined as having explosion energies of $>10^{52}$ erg, roughly ten times more than that of core-collapse SNe. 

Several theoretical works have computed yields from  metal-enriched Pop II. Using the yields table in \citet{nomoto13}, stars with initial stellar masses between $10-40$\!~$M_{\odot}$  and  5\% --100\% solar metallicity (illustrated in gray color in Fig. \ref{fig:abundance}) produce an abundance pattern that overlaps with that of hypernovae. The yields in \citet{kobayashi06} gives more or less identical results. A more recent yields table that includes Pop II stars with higher initial stellar masses between $13-120$\!~$M_{\odot}$ \citep{limongi18} (purple contours) produce an IMF weighted yield with a higher concentration around [C/O]$\approx$0.1 and [Si/O]$\approx$0.3.

Variations of abundance ratios from Pop III and Pop II stellar yields have been used to infer the enrichment patterns predicted in Gamma-ray burst hosts \citep{ma17}, observed in low-metallicity absorption-line systems \citep{cooke17,welsh19,welsh23}, and used to predict the yields in absorbers at the highest redshifts \citep{jeon19}. Stars in the local Universe also retain chemical signatures originating from the first populations of stars reflecting distinct abundance ratio patterns \citep{vanni23,vanni23b,salvadori23}. See also \citet{ma17}, Vanni et al. in prep and D'Odorico et al. in prep. for abundance ratio diagrams other than [Si/O] and [C/O].

Figure~\ref{fig:abundance} illustrates that also other intervening absorption-line systems at $4.5<z<6$ (orange circles) \citep{becker19,cooper19} fall in the same region as the JWST absorption systems. However, several of these other absorption line systems at $z<5$ have detected \ion{Si}{iv} and \ion{C}{iv} absorption lines that display different kinematical profiles and therefore arise in a different medium than the low-ionization lines \citep{becker19}. In such cases a direct conversion from $W_r$ of \ion{Si}{ii} and \ion{C}{ii} to a total Si and C column density is not valid. In fact, some of the absorption systems at $z<4.2$ in \citet{becker19} have hydrogen column densities $N$(\ion{H}{i})$< 10^{20}$ cm$^{-2}$ for which ionization corrections are necessary, and when these are included, all the absorbers at $3.0\!<\!z\!<\!4.2$ have [Si/O]$<\!0.4$ \citep{saccardi23}. For seven absorption systems at $5<z<6.2$ their [Si/O] ratios without ionization corrections remain close to the solar value \citep{becker12}, suggesting that Pop III enrichment does not dominate these systems.

In higher \ion{H}{i} column density Damped-\lya\ systems (DLAs), where the hydrogen column density is $N$(\ion{H}{i}) $> 2\times 10^{20}$ cm$^{-2}$, the clouds are self shielded, such that the dominant contributions to the metal column densities are from the low-ionization species. Above metallicities of 1\% solar, C,N,O absorption lines are saturated in DLA systems and only lower limits on the column densities can be derived \citep[e.g.][]{berg16}. The gray circles in Fig.~\ref{fig:abundance} illustrate that metal-poor DLAs at $z_{\mathrm{abs}}\sim3$ with metallicities from $10^{-3}$ to $10^{-2}$  $Z_{\odot}$ in oxygen \citep{cooke11,cooke17,welsh19} generally have abundance patterns similar to solar values and consistent with Pop II SNe enrichment patterns, albeit with an overlap from the high-mass Pop III hypernovae enrichment pattern. However, when analysing other elements (Fe and Al), one of the DLA systems has an enrichment pattern which is better matched by Pop~III SNe  \citep{cooke17,welsh23}. 

For the three absorption systems detected by NIRSpec with limits on the carbon abundance the enrichment could possibly be created by Pop III PISN.  Two of the other absorbers show abundance ratios with high values of [Si/O] that are consistent with either enrichment by primarily lower mass ($<15\,M_{\odot}$) Pop III hypernovae or even Pop III core-collapse SNe for one of them. Models predict that the abundance ratio $0.5<$[Si/O]$<1.0$ is produced by energetic Pop~III SNe (with explosion energies $>10^{51}$ erg) from stars with initial stellar masses in the range from $10-15$ $M_{\odot}$ \citep{heger10}. However, also single 15\!~$M_{\odot}$ star Pop II explosions produce a high [Si/O]\!~$\approx\!~0.5$ ratio \citep{woosley95,thielemann96,nomoto06,nomoto13}, so the enrichment by a single Pop II star remains a possibility. We do have to consider that gas clouds seen in absorption may not only be polluted by a single star alone, but by several SNe explosions over a range of stellar masses. When integrating the yields over the IMF, stars with masses in the range from $13-40$\!~$M_{\odot}$ with initial population metallicities less than solar all produce an integrated ratio of [Si/O]$\lesssim0.1$ \citep{woosley95,kobayashi06,nomoto13}.

\begin{table}
    \centering    \caption{Metal abundance ratios of absorption systems}
    \label{tab:metals}  
    \begin{tabular}{llcc}   
    \hline\hline
\noalign{\smallskip}    
\quad\quad\enskip  Quasar  & \enskip$z_{\mathrm{abs}}$ & [Si/O] &  [C/O]  \\
    \noalign{\smallskip}
    \hline
    \noalign{\smallskip}
 ULAS J1342+0928 &  7.3680$^\dagger$ &  0.98$\pm$0.10 & $<$--0.17$\pm$0.11   \\
 ULAS J1342+0928 &  7.4430           &  0.43$\pm$0.24 & ~~~~0.19$\pm$0.15  \\
 VDES J0020–3653 &  6.4535           &  0.15$\pm$0.16 & ~~--0.13$\pm$0.08  \\
 VDES J0020–3653 &  6.5625$^\dagger$ &  0.64$\pm$0.11 & $<$--0.26~~~~~~~~~~  \\
 VDES J0020–3653 &  6.6690           &  0.49$\pm$0.11 & ~~--0.02$\pm$0.07  \\ 
 UHS J0439+1634  &  6.2880$^\dagger$ &  0.73$\pm$0.06 & $<$--0.07$\pm$0.06 \\
\noalign{\smallskip}
\hline                           
    \end{tabular}
    \begin{tablenotes}
        \item $^\dagger$Upper limits for [C/O] are computed because the column densities of carbon lines are overestimated due to line contamination from other systems, or an upper detection limit on \ion{C}{ii} in the case of the $z=6.5625$ system.
    \end{tablenotes}
\end{table}

\section{Transmission spikes in the Gunn-Peterson trough}
\label{sect:trans_igm}
One of the most debated questions arising from studies of the high-redshift Universe in the past decades, is what sources reionized the IGM? To further investigate this, we examine whether the sources associated with metal absorbers have ionized their surroundings.

\begin{figure*}
\centering
   \includegraphics[bb= 100 40 1329 650, clip, width=0.95\hsize]{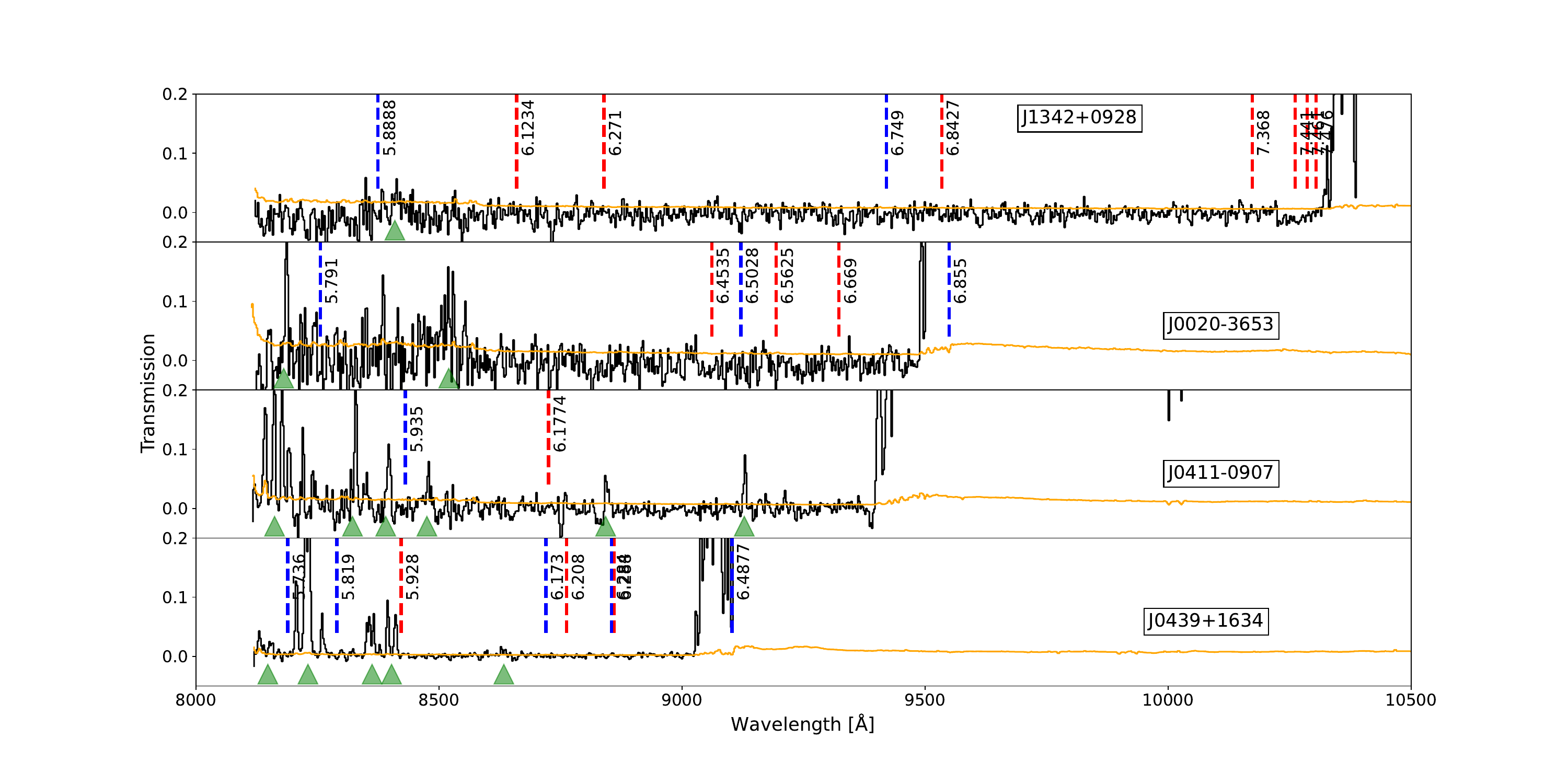}
       \caption{The panels show the relative transmission in the GP trough after normalizing the spectra to a model of the quasar continuum in the \lya\ forest. The individual absorber redshifts are marked where their expected Ly$\alpha$ wavelength falls. Red dashed lines are absorbers that are only low-ionization systems, and blue dashed lines mark absorbers with high-ionization lines from \ion{C}{iv}. Two of the absorbers lie close to the quasar redshift, and their \lya\ wavelengths fall in the proximity region of the quasar spectrum. A single low-ionization absorber at $z=5.928$ is associated with a transmission spike in the intervening IGM illustrated by the green triangles, while four of the high-ionization absorbers lie within $\pm$2000 km~s$^{-1}$ from the IGM transmission spikes at $z\sim5.8$.
       }  
    \label{fig:igm_trans}
\end{figure*}

We investigate the narrow transmission spikes at $z>5.7$ in the \lya\ forest region and in the GP troughs covered by the NIRSpec data using the normalized quasar spectra bluewards of the quasars \lya\ lines described in Sect.~\ref{sect:normalisation}.  To locate narrow transmission spikes we search for peaks with a 3-$\sigma$ prominence above zero, using the normalized spectrum divided by the normalized noise spectrum, i.e. a spectrum of the $S/N$ ratio. We require that the peaks extend over three or more pixels to avoid detections from a single noisy pixel. The identified spikes have relative transmissions of 2--5\% or higher. Figure~\ref{fig:igm_trans} illustrates the location of the \lya\ wavelengths for the identified metal-absorption line system redshifts plotted on the normalized spectrum in the GP trough region of the four spectra. The blue and red colored dashed lines mark high- and low-ionisation absorption systems, respectively. The green triangles mark the spikes in the IGM where there is a significant transmission of \lya\ photons, and Table~\ref{tab:spikes} lists the redshifts of the encountered \lya\ transmission spikes.

We briefly compare the detected IGM transmission spikes with previously published results. \citet{yang20} examine transmission spikes at $z>5.5$ based on ground-based spectra of a large sample of high-redshift quasars including two of the quasars in this paper (DELS J0411--0907 and VDES J0020--3653). They list four spikes in DELS J0411--0907, with three being multiple peaks at an almost identical redshift of $z\sim5.7$. We also detect these four spikes in the NIRSpec data but in addition we also detect spikes at even higher redshift. For VDES J0020--3653, \citet{yang20} do not report spike redshifts. Due to the higher $S/N$ ratio of the JWST spectra and the absence of noise due to telluric features, we can detect weaker spikes in the GP troughs.

Only a single low-ionization absorption system (at $z=5.928$ towards UHS J0439+1634) out of the total of 13 intervening low-ionization absorption systems (red dashed lines in Fig.~\ref{fig:igm_trans}) match the transmission spikes in the IGM to within a velocity offset of 2000 km~s$^{-1}$, whereas four out of nine of the high-ionization system lie within $\approx$2000 km~s$^{-1}$ of transmission spikes, all of which at $z\sim5.8$. Not surprisingly, the two proximate absorption systems in the spectra of UHS J0439+1634 and VDES J0020--3653 are indeed high-ionization systems. Conversely, from the total of 21 absorption systems at $z>5.7$, 16 spikes do not match any metal-absorption system within $\pm$2000 km~s$^{-1}$. The fact that most of the low-ionization systems are not associated with a transmission spike supports the hypothesis that the metal-enrichment arises from star-formation followed by heavy element productions by supernova explosions preferentially in higher density regions that may be the last regions to be reionized \citep{oh02}. On the other hand, half of the high-ionization systems at $5.7 < z < 6.8$ appear in regions of the IGM that are ionized to a degree that allows the escape of \lya\ photons.

 \begin{table}
       \centering
      \caption{Redshifts of transmission spikes in the GP troughs}
         \label{tab:spikes}
         \begin{tabular}{l l }
            \hline\hline
            \noalign{\smallskip}
        \quad\quad{Quasar}& $z_{\mathrm{spikes}}$  \\
            \noalign{\smallskip}
            \hline
            \noalign{\smallskip}
                 ULAS~J1342+0928  & 5.9214 \\ 
             \noalign{\smallskip}
                 VDES J0020--3653 & 5.7333,  6.0125\\ 
             \noalign{\smallskip}
                 DELS J0411--0907 & 5.7178, 5.8497, 5.9059\\
                                &   5.9756, 6.2781, 6.5127\\
             \noalign{\smallskip}
                 UHS J0439+1634   & 5.70232, 5.7702, 5.8787\\
                                  & 5.9117,  6.1017\\
          \noalign{\smallskip}
            \hline
         \end{tabular}
   \end{table}

\section{Discussion}

\label{sect:discussion}

\subsection{A forest of lines or discrete absorption line systems?}
 \cite{oh02} has suggested that as the Universe becomes progressively more neutral, a forest of weak metal lines may appear at redshifts close to the EoR. Analogous to the \ion{H}{i} \lya\ forest arising from neutral clouds of hydrogen in the IGM that becomes more numerous towards higher redshifts, metals in the IGM would also progressively become more neutral. For example an \ion{O}{i}\,$\lambda$1302 forest could give rise to a depression of the quasar continuum at the 10-20\% level \citep{oh02}. Other atoms such as \ion{C}{ii}, or \ion{Si}{ii} have ionization potentials close to that of \ion{H}{i} and have oscillator strengths that also make the lines useful as tracers of the enrichment and the increasing neutral IGM towards the reionization epoch. Likewise, the \ion{Mg}{ii}\,$\lambda\lambda$2796,2803 doublet is easily detected, and could give rise to a continuous \ion{Mg}{ii} forest provided the neutral IGM is globally enriched with metals \citep{hennawi21}. However, ground-based spectra of high redshift quasars have not yet revealed any signature from a \ion{Mg}{ii} forest \citep{tie23}. In the NIRSpec data we detect a line density of \ion{Mg}{ii} that remains constant out to $z=7.5$. \ion{Mg}{ii} lines are identified as discrete absorption systems rather than a global decrease in the continuum in the form of a \ion{Mg}{ii} forest. Also in the case of \ion{O}{i}, we are only able to detect discrete absorbers that likely arise in the CGM or ISM of intervening galaxies. Further analysis of JWST/NIRSpec data using more advanced methods such as auto-correlation techniques may be able to find a metal forest, but that is beyond the scope of the analysis of individual absorption systems in this work. Detecting a \ion{Mg}{ii} metal-line forest with a global decrease in the continuum flux at the few percent level is difficult due to the wiggles in the NIRSpec data described in Section~\ref{sect:nirspec_obs}, but also the broad emission-line composite from \ion{Mg}{ii} and \ion{Fe}{ii} in the quasar spectrum will make it difficult to accurately determine the level of the unabsorbed continuum emission.

\subsection{High-redshift metallicities}
In the absence of measurable \ion{H}{i} column densities, we can only measure relative metallicities. In the redshift interval $2<z<6$ absorption systems display a relatively constant relative abundance pattern in [Si/O] and [C/Fe] with redshift and do not reveal clear signatures of Pop III SNe enrichment \citep{becker12,kulkarni14}, with a couple of possible exceptions from DLA systems at $z\approx3$ \citep{welsh23}. A relevant question is therefore if we have overestimated the [Si/O] or [C/O] ratios for the $z>6$ absorption systems.  In comparison with ground-based spectroscopic studies, the NIRSpec data have  low spectral resolution and could potentially have hidden saturation, or be composed of multiple components rather than a single one assumed for the \ion{O}{i} systems in this work. We argue against the possibility of hidden saturation based on the analysis of the curve of growth for the absorption lines in any of the \ion{O}{i} absorbers where we find that they are consistent with being on the linear, optically thin region of the curve of growth. In agreement with our finding, numerical simulations at $z\sim7$ also find that \ion{O}{i} absorption lines along random sight-lines are optically thin \citep{doughty19}.

Alternatively, multiple absorption components in the observed spectra could have variations in abundance patterns between them, but in order to display a high [Si/O] ratio, some of the components will inevitably have to have an extreme value. Ultimately, we need higher spectral resolution measurements to reveal and decompose the absorption lines in distinct components but this may have to wait until the arrival of high-resolution spectrographs on future extremely large telescopes \citep[e.g.][]{maiolino13}.

Another possibility leading to inaccurate abundance ratios arises if the atoms exist predominantly in ionization states other than \ion{Si}{ii} or \ion{C}{ii} in which case ionization corrections are needed to derive [Si/O] and [C/O] ratios accurately. However, we do not regard this as likely for following reasons. Firstly, the intervening absorption systems are primarily neutral because there are no high-ionization \ion{C}{iv} or \ion{Si}{iv} lines associated with the \ion{O}{i} systems. Secondly, the \ion{O}{i} absorbers are not associated with a transmission spike in the GP troughs at these redshifts, and consequently the environment is predominantly neutral. \ion{O}{i} is charge-exchanged locked with \ion{H}{i}, so \ion{O}{i} is the dominant state of oxygen. Possibly, \ion{Si}{iii} and \ion{C}{iii} could be present, but these atomic states do not have absorption lines in the spectral range covered by the NIRSpec data. However, if that were the case, then the total [Si/O] and [C/O] would be even higher than detected in the high-redshift absorbers. For similar $z\sim6$ absorption systems, \citet{cooper19} argue that the low-ionization systems are equivalents to $z\sim3$ DLA systems in which metallicities derived without ionization corrections are reliable.

\subsection{Ionization state of the CGM}
Our finding that the number density of the high-ionization ions \ion{C}{iv} and \ion{Si}{iv} continues to drop at $z>6.5$ extends the trends observed at slightly lower redshifts \citep{cooper19,dodorico22,davies23}. The disappearing of these absorption lines cannot be attributed to a change in metallicity, since \ion{C}{ii} and \ion{Si}{ii} are frequently detected with column densities that are ten times higher than in the high-ionization states \citep{becker11,bosman17,cooper19}.  Moreover, we find that the \ion{Mg}{ii} line density appears to be constant throughout the redshift range probed out to $z\sim7.5$, or at least only display a modest decrease to the highest redshifts covered,  $z\sim7.5$. This supports the contention that the change in the high-ionization line densities is caused by a softening of the UV radiation field dominated by star-forming galaxies rather than by the harder radiation of active galactic nuclei at these redshifts.

The under-production of \ion{C}{iv} at $z>6$ is known to be a problem in cosmological simulations \citep{keating16,finlator18,doughty18} although interestingly we find that the number density of \ion{C}{ii} absorbers agrees with observations. We detect only two \ion{Si}{iv} absorbers in the JWST spectra, and confirm that the line density of this transition continues to decrease with redshift consistent with model predictions. The fact that the observed line density of \ion{Si}{iv} at $z>5.5$ matches the simulations \citep{dodorico22}, whereas the observed \ion{Si}{ii} line density is higher than in simulations, is not likely due to the hardness of the UV background in the simulations, because the ratio of these lines does not depend on the spectral hardness \citep{doughty18}. In the epoch where reionization has not yet been completed, the low-ionization ions will be the dominant ones, and may be present out to large distances from the galaxies \citep{doughty23}.

\subsection{Environments of the metal absorbers}
To date, not many studies have attempted to match metal-absorption lines to galaxies at $z>6$ primarily due to the lack of instrument sensitivity since the galaxies are very faint. From ground-based observations, targeting \lya\ emitters has recently been the best option due to the increased contrast of the emission lines above the background. By studying correlations of high-ionization \ion{C}{iv} systems at $z\sim5$ with \lya\ emitters \citet{diaz21} found emitters within 200 kpc in projection from the absorbers. The environments of the even higher redshift low-ionization systems are less clear from observations. Using JWST/NIRcam slitless spectra of one $z=6.3$ quasar field \citet{kashino23} identify galaxies with [\ion{O}{iii}] emission lines at $5.3<z<6.2$ and find that most of the detected galaxy redshifts match either high- or low-ionizations absorption-line systems in the quasar spectrum. Conversely, out of the eight intervening absorption line systems, only four of the systems have detected galaxies within 200 kpc of the quasar sight line \citep{kashino23}, where most of the non-detections are from low-ionization absorber systems.

In a survey of eight quasars at $z>6$, \citet{meyer20} detect a weak correlation between transmission spikes and $z\approx6$ \lya-emitters and Lyman-break galaxies (LBGs) within 1--10 proper Mpc in projection along the line of sight, but argue that fainter galaxies below the detection threshold are necessary to drive the reionization. In a similar analysis of a single quasar at $z=6.4$ \citet{kakiichi18} find that three out of six LBGs lie close in redshift space to the IGM transmission spikes, and also conclude that fainter galaxies clustered around the brighter LBGs are needed to drive reionization, but also argue that luminous quasars do contribute to the ionizing flux at the latest stages of the reionization. In a recent analysis of JWST NIRcam slitless data, \citet{kashino23} measure a clear correlation between galaxy redshifts at $5.7<z<6.1$ and transmission spikes in the \lya\ forest, but find that metal-absorption systems are not connected to the transmission spikes. This is consistent with our finding of no clear association of low-ionization absorbers with IGM transmission peaks.

Whether any of the absorption systems have detectable galaxy counterparts nearby can be investigated with the Band~III JWST integral field spectra. The 3 arcsec field of view of the IFU limits the search to within 7--10 kpc in projection from the quasar, which is much smaller than the typical size of 100 kpc of the metal-enriched CGM around high-redshift galaxies \citep{tumlinson17}. The sizes and covering fractions of the metal-enriched CGM depend on the galaxy masses and redshifts, where simulations predict that the median separation of an \ion{O}{i} absorber at $z=7$ to its host galaxy is about 250 $h^{-1}$ comoving kpc or a physical size of about 45 kpc, and that the absorber may probe either the CGM or the ISM of the host \citep{doughty19}. Even though the IFU of NIRSpec has a small field of view, there is a non-negligible chance of detecting the associated galaxy through emission lines at the absorber redshift in the IFS data cube, especially if the strongest absorbers probe the ISM. This association of metal absorption lines with galaxies will be investigated in forthcoming papers.

\section{Summary and conclusions}
\label{sect:conclusions}

We have presented JWST/NIRSpec spectra of four $z>6.5$ quasars covering the wide spectral range  0.8--5.3 $\mu$m. Targeting luminous quasars with magnitudes of $J_{\mathrm{AB}}$=17--20, the data shows an exquisite signal with a $S/N$ ratio of 50--200 per spectral bin allowing for an in-depth analysis of each of the quasar lines of sight. We detect metal-absorption line systems from intervening absorbers and systems associated with the quasars.  In each of the four spectra we identify between 13 and 17 absorption line systems with a total of 61 systems at redshifts ranging from $z=2.4$ to $z=7.5$. Since the line of sight towards the quasars probe random locations, the identifications include detections of intervening metal-enriched gas in the CGM or the ISM of intervening galaxies at the highest redshift to date. Most absorption line systems are identified in multiple atomic transitions due to the high quality of the NIRSpec data, while some weaker systems are only detected through \ion{Mg}{ii} or \ion{C}{iv} doublets.

At the quasar redshifts, the \lya\ transitions of each absorption system fall in the Gunn-Peterson trough, so deriving hydrogen column densities or metallicities based on the JWST data is impossible, and likewise is it impossible to derive the neutral hydrogen gas fractions. Instead, we use low-ionization lines such as \ion{O}{i} as a proxy to examine the changing ionization state of the IGM \citep{oh02}. Rather than detecting a forest of metal lines in the IGM at the end of the EoR, we detect instead a number of discrete \ion{O}{i} (and \ion{Mg}{ii}) absorbers, as logically expected since winds from galaxies are responsible for distributing the heavy elements out into the CGM of the host galaxies. 

From the number statistics of the low-ionization absorbers \ion{O}{i}\,$\lambda$1302, \ion{C}{ii}\,$\lambda$1334, \ion{Si}{ii}\,$\lambda$1526, and \ion{Mg}{ii}\,$\lambda$2796, we derive the redshift evolution of the line densities beyond $z>6$. Combined with other studies of lower redshift quasar spectra \citep{becker19,chen17,davies23b}, we find that the line densities of \ion{O}{i}, \ion{C}{ii}, and \ion{Si}{ii} derived from the NIRSpec data reveal a continued increase beyond $z>6$, whereas \ion{Mg}{ii} shows a redshift evolution that is consistent with being constant or mildly decreasing with increasing redshift. In all atomic species, the number of detected absorbers remains small and the redshift path-length interval is also limited in our NIRSpec data set of just four quasars. Consequently, the Poisson uncertainties are large and  the line density evolution could also be constant to within $\sim1\sigma$ uncertainties for all atomic species. Nevertheless, compared to numerical simulations, the observations follow the predicted redshift dependence for the low-ionization lines at $z=5\rightarrow8$, where the model predicts an increase with increasing redshift to $z\sim7$ due to the increased neutral content and a decreasing UV ionizing background flux. At $z>7$ the model predicts that line densities decrease with redshift as a consequence of the decreasing metal enrichment of the CGM.

For the high-ionization transitions \ion{C}{iv} and \ion{Si}{iv} we derive a decrease in line density beyond $z>6$ following the previously reported rapid decrease seen for these lines from $z=4\rightarrow6$ \citep{dodorico22,davies23}. Again, the drop is consistent with the model prediction. The transitions in line densities from a predominantly low-ionization level to a high-ionization level occurs at $z\approx6.0$ for both carbon and silicon. The observations therefore suggest that there is a continuous change in the ionization state rather than a drastic change in the distribution of metals at $z>6$.

We matched the redshifts of the absorption line systems at $z>5.7$ with spikes of  transmission of \lya\ photons in the GP troughs, and find that half of the intervening absorbers that have high-ionization absorption lines fall within $\pm$2000 km~s$^{-1}$ of the  \lya\ redshift of a matching transmission spike. For low-ionization absorbers on the other hand, only a single system out of 13 intervening absorbers falls within $\pm$2000 km~s$^{-1}$ of a spike. This demonstrates that low-ionization systems are found in environments that are not ionized sufficiently to allow \lya\ photons to escape. Low-ionization metal-absorption line systems therefore likely trace higher density environments in the Universe that are among the last ones to become reionized.

While we are not able to derive absolute metallicities, our relative metallicity ratios are accurate provided that the absorption lines are not saturated, and that ionization corrections are insignificant. We argue that the systems traced through \ion{O}{i} absorption lines are reliable tracers, since in none of the cases do we detect high-ionization lines and they are not associated with IGM transmission spikes. Focusing on silicon, carbon and oxygen atoms, we derive high values of the metal line ratio [Si/O] in the range from 0.5--1.0 in five of the $z>6$ absorption systems. Such high values of [Si/O] are consistent with the enrichment pattern expected from energetic Pop III hypernovae with initial stellar masses from $10-15$\!~$M_{\odot}$, or even from Pop III core-collapse SNe \citep{heger10}. However, explosions of single Pop II stars with a stellar mass of 15\!~$M_{\odot}$ can also produce a high [Si/O] ratio \citep{woosley95,kobayashi06,nomoto13}, so this scenario cannot be ruled out if the absorption system happen to probe a gas cloud enriched by a single star alone. Absorbers with the highest [Si/O] ratios and upper limits on [C/O] could be enriched by Pop III pair instability supernovae. The high [Si/O] ratio and limit on the carbon abundance cannot be explained by the yields produced by Pop II SNe. We therefore suggest that the CGM of the neutral absorption systems at $z>6$ are dominated by the enrichment from Pop III explosions.

In the near future, surveys from LSST and Euclid will detect an increasing number of (fainter) quasars at $z>7$. Also with the increasing depth and completed wide-field surveys with JWST, new detections of quasars at $z>10$ are being made \citep{goulding23,maiolino23}. Follow-up near-IR spectroscopic observations with NIRSpec and future ELT spectrographs will allow to extend this work to even higher redshifts into the first half of the reionization epoch beyond $z>7.5$. Higher resolution spectra compared to the $R\!\simeq\!2700$ NIRSpec observations are needed to fit the absorption line profiles with conventional Voigt fitting techniques to verify metal column densities, and with an increasing number of high redshift quasar spectra the larger path length interval will provide line densities with smaller Poisson uncertainties. Future observations of the fields surrounding the four quasars presented in this paper may be able to identify the galaxies responsible for the intervening absorption line systems and single out the galaxies that are possibly responsible for the Pop III metal-enrichment at $z>6$.

\begin{acknowledgements}

We thank Kristian Finlator for useful discussions, and Ezra Huscher for sharing results from their simulations. We also thank Stefania Salvadori for valuable insights into Pop III enrichment.

L.C. is supported by DFF/Independent Research Fund Denmark, grant-ID 2032--00071.

The Cosmic Dawn Center is funded by the Danish National Research Foundation under grant no. 140.

AJB has received funding from the European Research Council (ERC) under the European Union’s Horizon 2020 Advanced Grant 789056 “First Galaxies”

MP acknowledges support from the research project PID2021-127718NB-I00 of the Spanish Ministry of Science and Innovation/State Agency of Research (MICIN/AEI), and the Programa Atracci\'on de Talento de la Comunidad de Madrid via grant 2018-T2/TIC-11715.

H{\"U} gratefully acknowledges support by the Isaac Newton Trust and by the Kavli Foundation through a Newton-Kavli Junior Fellowship.

The work is based on data obtained by the European Southern Observatory, Paranal, Chile, Program IDs: 098.B-0537, 0100.A-0446 and 0100.A-0898.

This work is based on observations made with the NASA/ESA/CSA James Webb Space Telescope. The data were obtained from the Mikulski Archive for Space Telescopes at the Space Telescope Science Institute, which is operated by the Association of Universities for Research in Astronomy, Inc., under NASA contract NAS 5-03127 for JWST. These observations are associated with GTO programs 1219 and 1222.

\end{acknowledgements}

\bibliographystyle{aa} 
\bibliography{paper}

%
%

\begin{appendix}

\section{Normalised spectra and absorption line systems}
\label{app:A}
For visualisation purposes the normalized Fixed Slit spectra of the quasars are shown here, with all the different intervening and proximate absorption line systems overlaid. Absorption lines are color coded to mark lines belonging to the same redshift absorbers. 
Figures with zoom in on individually detected absorption lines for each system are presented in Appendix~\ref{app:B}.
Note that the \lya~forest is ignored, and we do not include the Band~III data from 3-5.3 $\mu$m, because only a very few absorption lines are detected at these wavelengths.

\begin{figure*}
\centering
   \includegraphics[bb= 90 120 1150 1290, clip, width=\hsize]{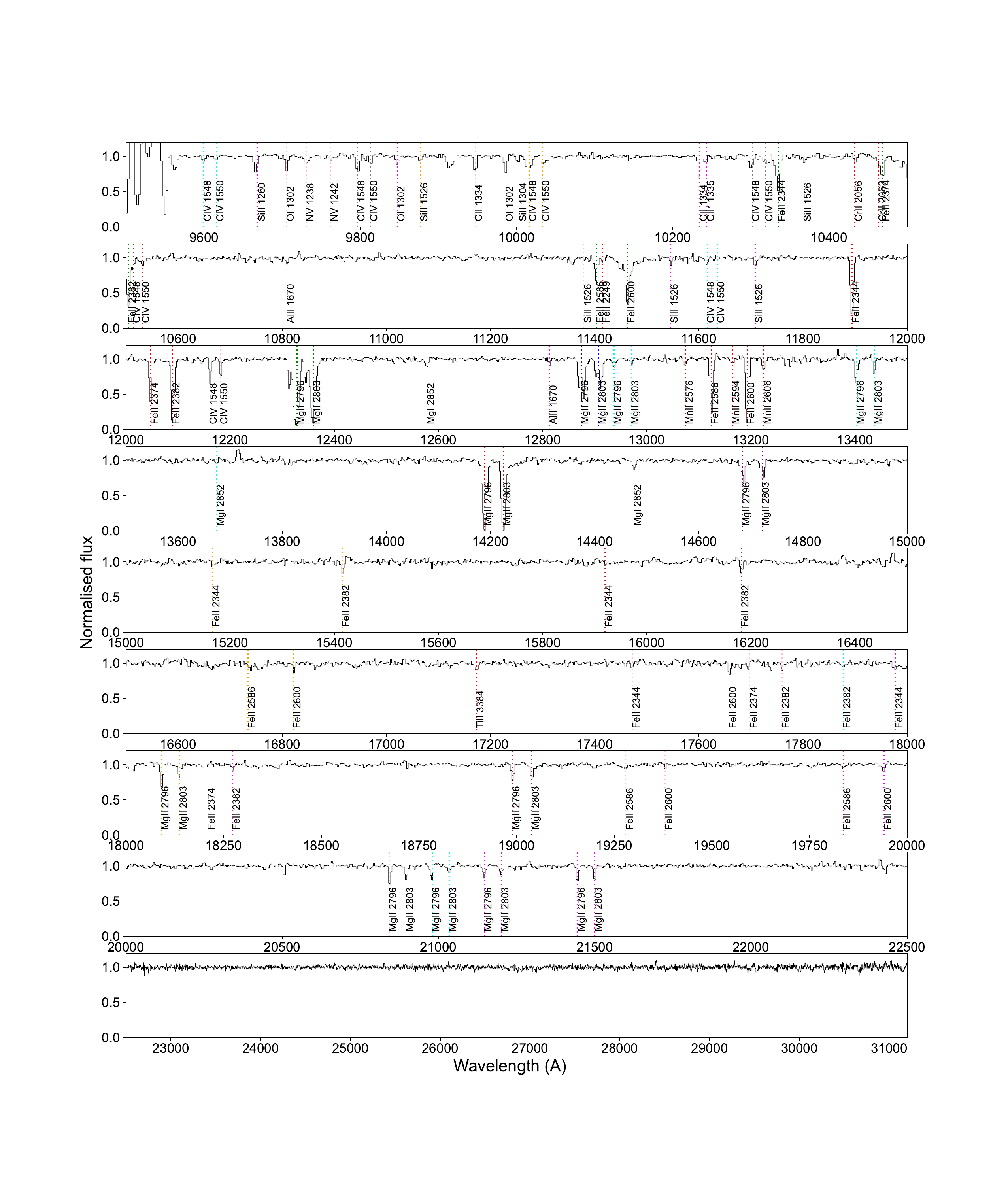}
       \caption{Normalized spectrum of VDES J0020--3653. Absorption line systems are identified via multiple absorption lines at the same redshifts. Here, 16 different absorption systems are detected at $z$ = 3.4087, 3.6040, 3.6265, 3.7930, 4.0738, 4.2510, 5.2005, 5.3277, 5.4695, 5.6540, 5.7910, 6.4535, 6.5028, 6.5625, 6.669, and 6.855. Each absorption system is specified with a distinct colour. The absorber at the highest redshift lies at the quasar redshift. 
       }  
    \label{fig:allspec_0020}
\end{figure*}

\begin{figure*}
\centering
   \includegraphics[bb= 90 120 1150 1290, clip, width=\hsize]{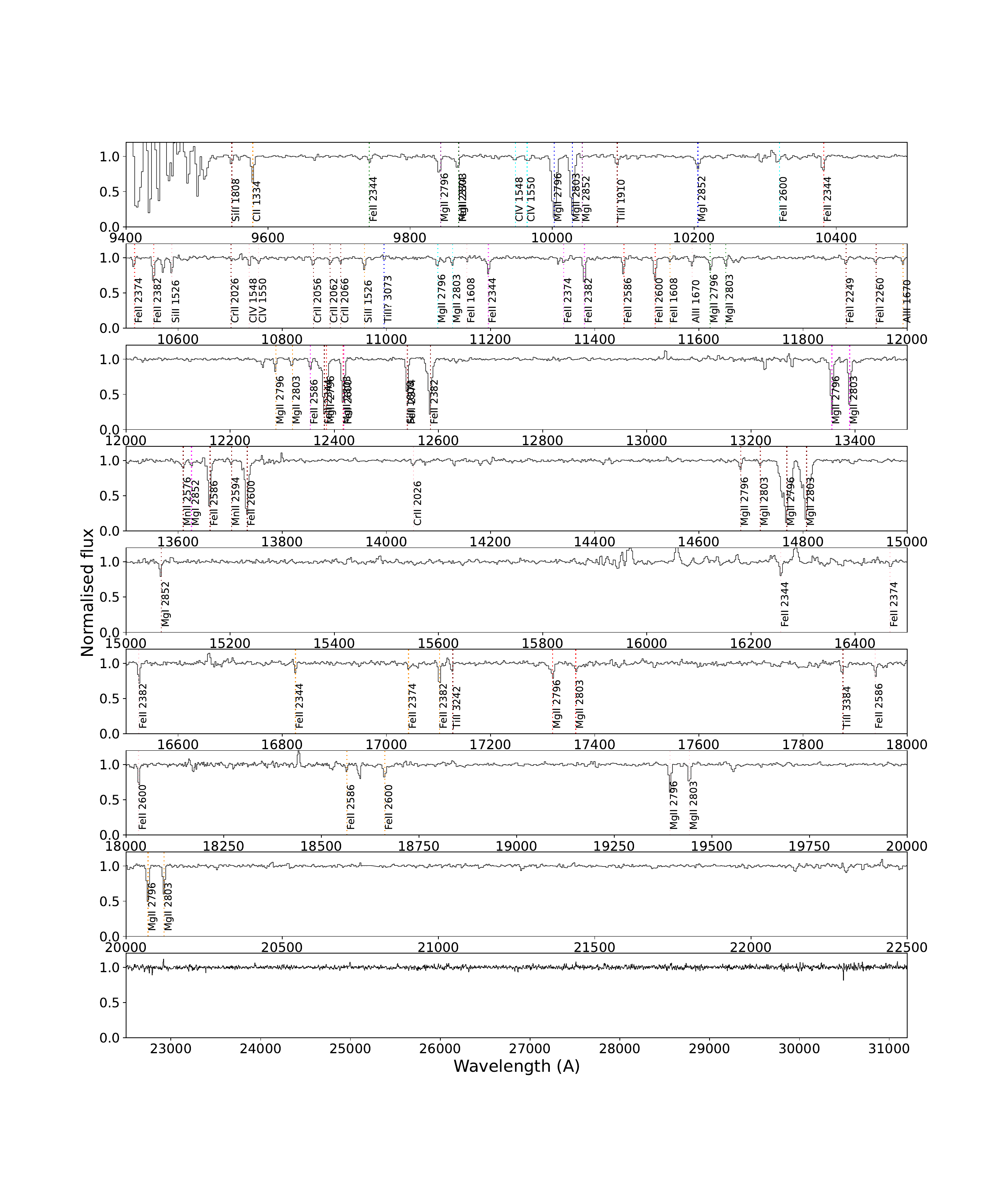}
       \caption{Normalized spectrum of DELS J0411--0907. Absorption line systems are identified via multiple absorption lines at the same redshifts. Here, 13 different absorption systems are detected at $z$ = 2.52, 2.5771, 2.969, 3.156, 3.3943, 3.4290, 3.7760, 4.2498, 4.2815, 5.1935, 5.4258, 5.935, and 6.1774. 
       }  
    \label{fig:allspec_0411}
\end{figure*}

\begin{figure*}
\centering
   \includegraphics[bb= 90 120 1150 1290, clip, width=\hsize]{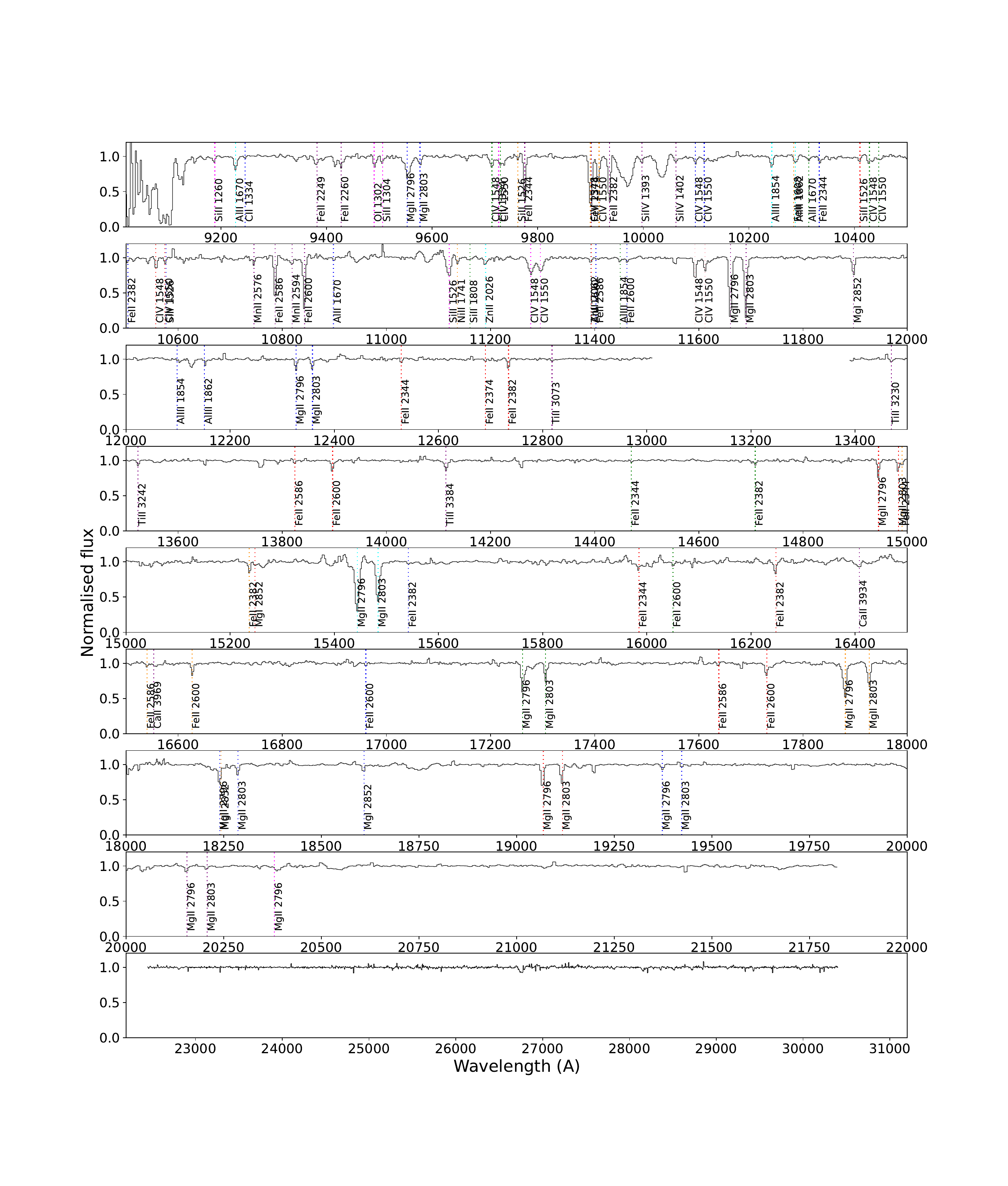}
       \caption{Full spectrum of UHS J0439+1634. The normalization of the quasar continuum emission is not accurate around 10,000--11,000 {\AA}, where the quasar spectrum has strong BAL features.
       Absorption line systems are identified via multiple absorption lines at the same redshifts. Here, 17 different absorption systems are detected at $z$ = 2.4161, 3.170, 3.4081, 4.3445, 4.523, 5.1728, 5.274, 5.3945, 5.5228, 5.736, 5.819, 5.928, 6.028, 6.173, 6.284, 6.288, and 6.4877. The highest redshift absorber is at the quasar redshift.}
    \label{fig:allspec_0439}
\end{figure*}

\begin{figure*}
\centering
   \includegraphics[bb= 90 120 1150 1290, clip, width=\hsize]{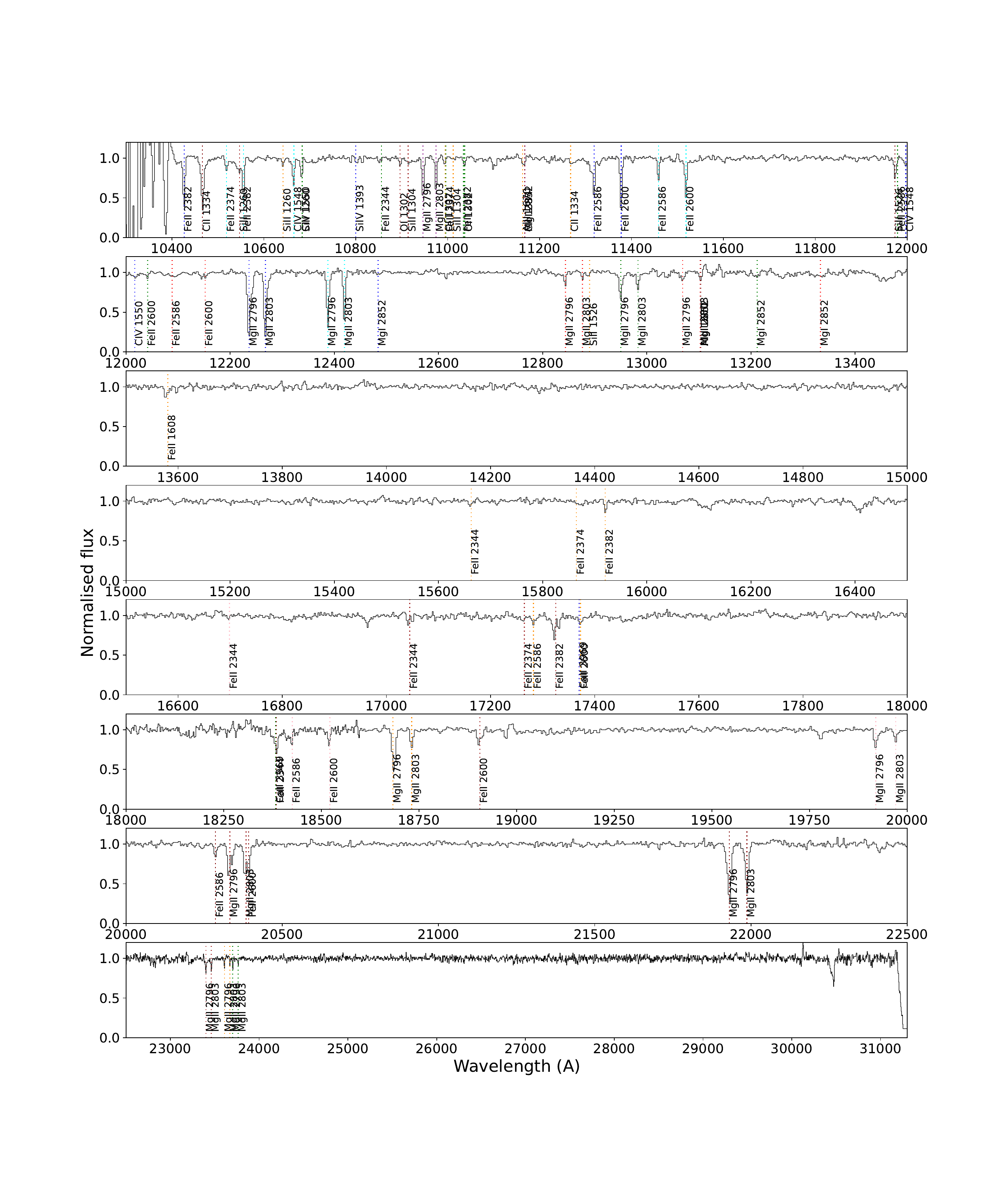}
       \caption{Full spectrum of ULAS J1342+0928. Absorption line systems are identified via multiple absorption lines at the same redshifts. Here, 15 different absorption systems are detected at $z$ = 2.9145, 3.3758, 3.430, 3.593, 3.631, 3.6735, 5.6814, 5.8888, 6.1234, 6.271, 6.749, 6.8427, 7.368, 7.443, and 7.476. The highest redshift absorber is associated with the quasar, being offset by $\approx$2000 km~s$^{-1}$ from the quasar systemic redshift.  }  
    \label{fig:allspec_1342}
\end{figure*}

\newpage
\clearpage 

\section{Individual absorption systems}
\label{app:B}

The appendix presents each of the detected absorption line systems. In each of the quasar spectra we have identified between 13--17 intervening absorption line systems at different redshifts, ranging from lower redshifts ($z=2.4$) to absorbers located at the quasar redshifts. 

In the figures in this appendix all panels illustrate the normalised quasar spectrum around the detected absorption line species within $\pm$600 km~s$^{-1}$ for each of the absorption systems. Flux errors per spectral pixel are overlaid, and the y-axis of each absorption line is scaled to match the depth of the line. Equivalent widths in the observed frame are computed by integrating the lines from within $\pm$200 km~s$^{-1}$, while for the stronger systems that display broader lines, the lines are integrated within $\pm$300 km~s$^{-1}$. The reported wavelengths of the absorption lines in all tables are computed from weighted mean flux values rather than based on absorption line fits. Line profiles for narrow line systems for a single absorption system may appear visually different as explained in Fig.~\ref{fig:line_profiles} in Section~\ref{sect:nirspec_obs}. Some lines are blended or partially contaminated with other identified lines from other absorption systems. We describe in the comments to the tables which lines are blended with a line from another system. Some weak lines or lines towards longer wavelengths are dominated by errors from the complex normalization of the quasar spectrum.

\begin{figure}[h!]
    \centering
   \includegraphics[bb = 1 50 550 750, clip, width=\hsize]{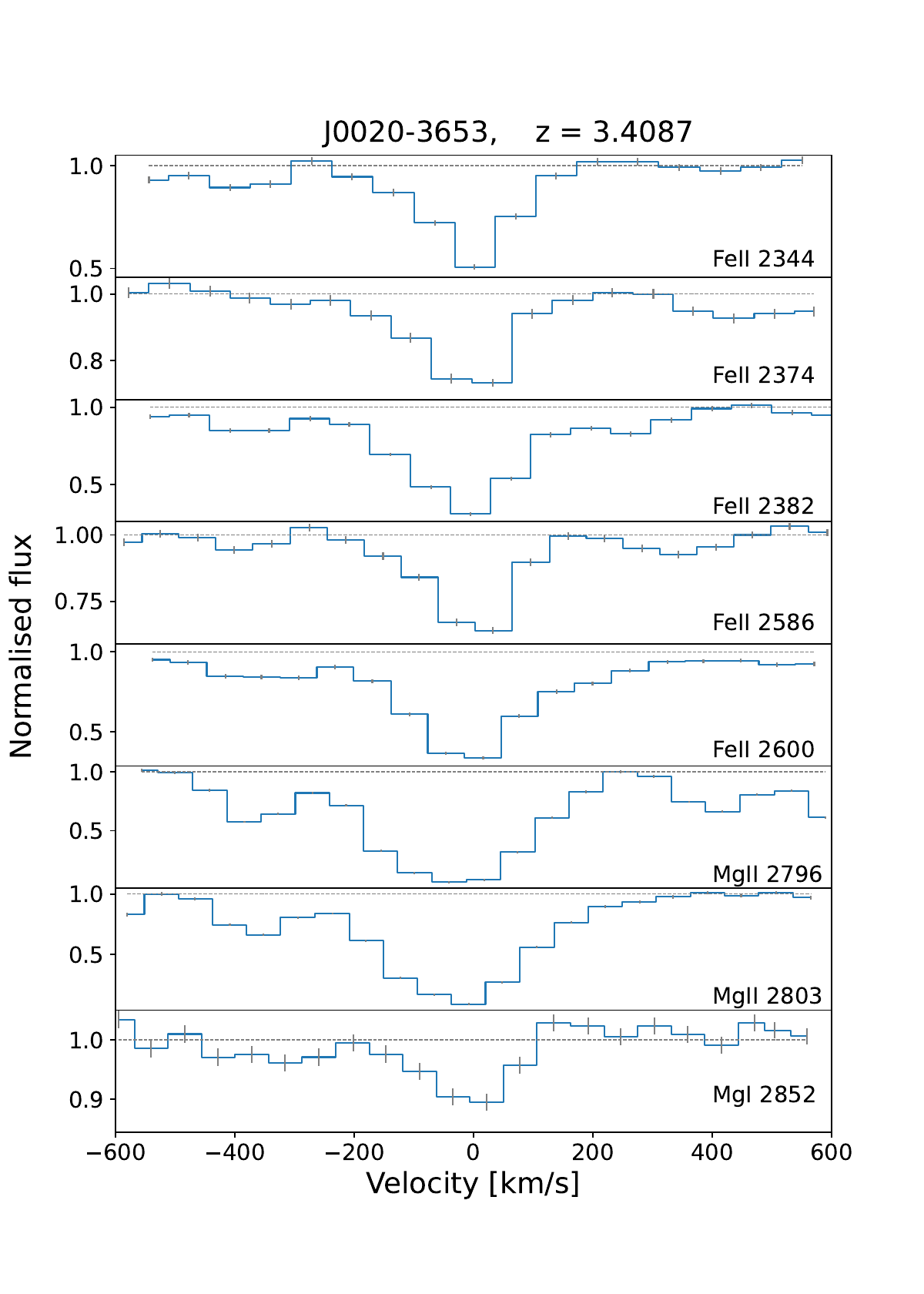}
    \caption{Absorption system at $z=3.4087$ toward J0020--3653. Note that the y-axis range changes according to the strengths of the absorption lines in this and all other panels.}
    \label{fig:J0020_3.4087}
\end{figure}
\begin{table}
\centering  
    \caption{J0020--3653 $z=3.4087$} 
    \label{table:J0020_3.4087}  
    \begin{tabular}{cccc}   
    \hline\hline
\noalign{\smallskip} 
$\lambda_{\mathrm{obs}}$ & ID & $W_{\mathrm{obs}}$  \\   
\noalign{\smallskip}
\hline
\noalign{\smallskip}
10334.94 & \ion{Fe}{ii}\ 2344.21 &  2.821$\pm$0.119\\
10468.29 & \ion{Fe}{ii}\ 2374.46 &  1.916$\pm$0.098$^\dagger$   \\
10504.90 & \ion{Fe}{ii}\ 2382.76 &  6.216$\pm$0.104$^\ddagger$  \\
11403.76 & \ion{Fe}{ii}\ 2586.65 &  2.574$\pm$0.104\\
11463.38 & \ion{Fe}{ii}\ 2600.17 &  7.266$\pm$0.095\\
12328.28 & \ion{Mg}{ii}\ 2796.35 &  12.011$\pm$0.081\\
12359.93 & \ion{Mg}{ii}\ 2803.53 &  11.204$\pm$0.081\\
12577.86 & \ion{Mg}{i}\ 2852.96 &  0.700$\pm$0.107\\
\noalign{\smallskip}
\hline                           
\end{tabular}
\begin{tablenotes}
\item $^\dagger$ Line blended with \ion{Cr}{ii} $\lambda$2062 at $z=4.0738$ 
\item$^\ddagger$ Line partly blended with \ion{C}{iv} $\lambda$1548 at $z=5.791$
\end{tablenotes}
\end{table}

\begin{figure}
   \centering
   \includegraphics[bb= 1 1 537 221, clip, width=\hsize]{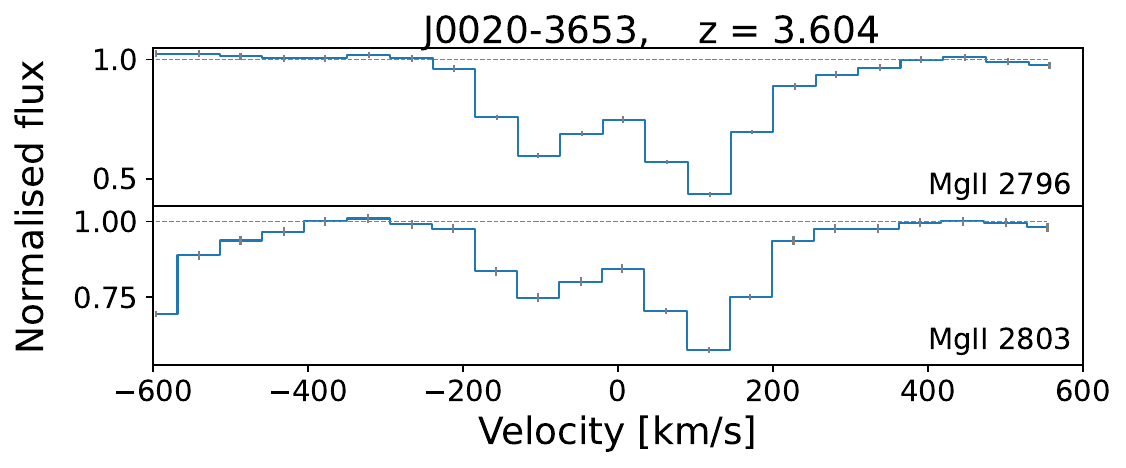}
    \caption{Absorption system at $z=3.604$ toward J0020--3653. }
    \label{fig:J0020_3.604}
\end{figure}
\begin{table}
\centering
\caption{J0020--3653 $z=3.604$} 
\label{table:J0020_3.604}  
\begin{tabular}{cccc}   
\noalign{\smallskip}
\hline\hline
\noalign{\smallskip}
$\lambda_{\mathrm{obs}}$ & ID & $W_{\mathrm{obs}}$ \\   
\noalign{\smallskip} \hline \noalign{\smallskip}     
12874.40 & \ion{Mg}{ii}\ 2796.35 &  6.396$\pm$0.095\\
12907.46 & \ion{Mg}{ii}\ 2803.53 &  4.372$\pm$0.105\\
\noalign{\smallskip}
\hline                           
\end{tabular}
\end{table}

\begin{figure}
   \centering
   \includegraphics[width=\hsize]{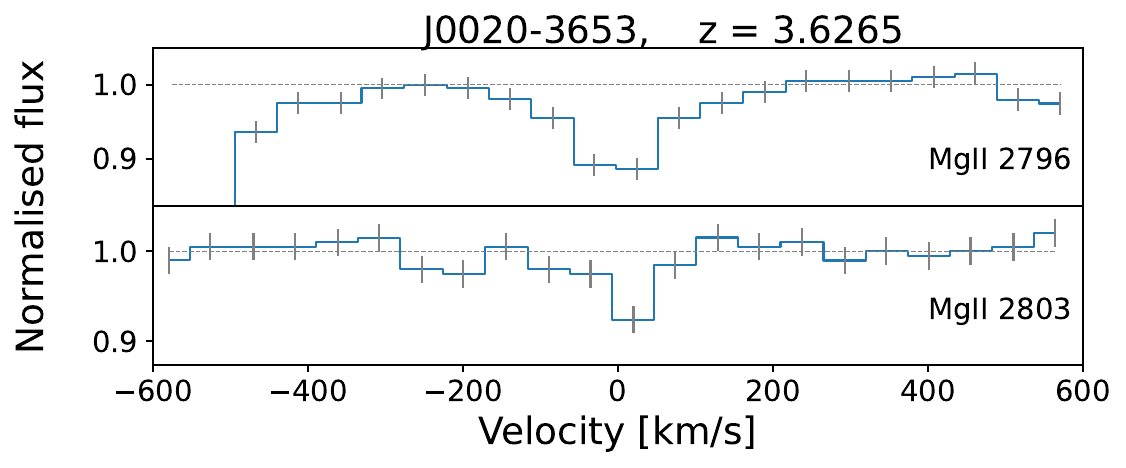}
    \caption{Absorption system at $z=3.6265$ toward J0020--3653. }
    \label{fig:J0020_3.6265}
\end{figure}
\begin{table}[h!]
\centering
\caption{J0020--3653 $z=3.6265$} 
\label{table:J0020_3.6265}  
\begin{tabular}{cccc}   
\hline\hline \noalign{\smallskip}
$\lambda_{\mathrm{obs}}$ & ID & $W_{\mathrm{obs}}$ \\   
\noalign{\smallskip} \hline \noalign{\smallskip}     
12937.32 & \ion{Mg}{ii}\ 2796.35 &  0.872$\pm$0.098\\
12970.54 & \ion{Mg}{ii}\ 2803.53 &  0.320$\pm$0.102\\
\noalign{\smallskip}
\hline                           
\end{tabular}
\end{table}

\FloatBarrier


\begin{figure}
   \centering
   \includegraphics[width=\hsize]{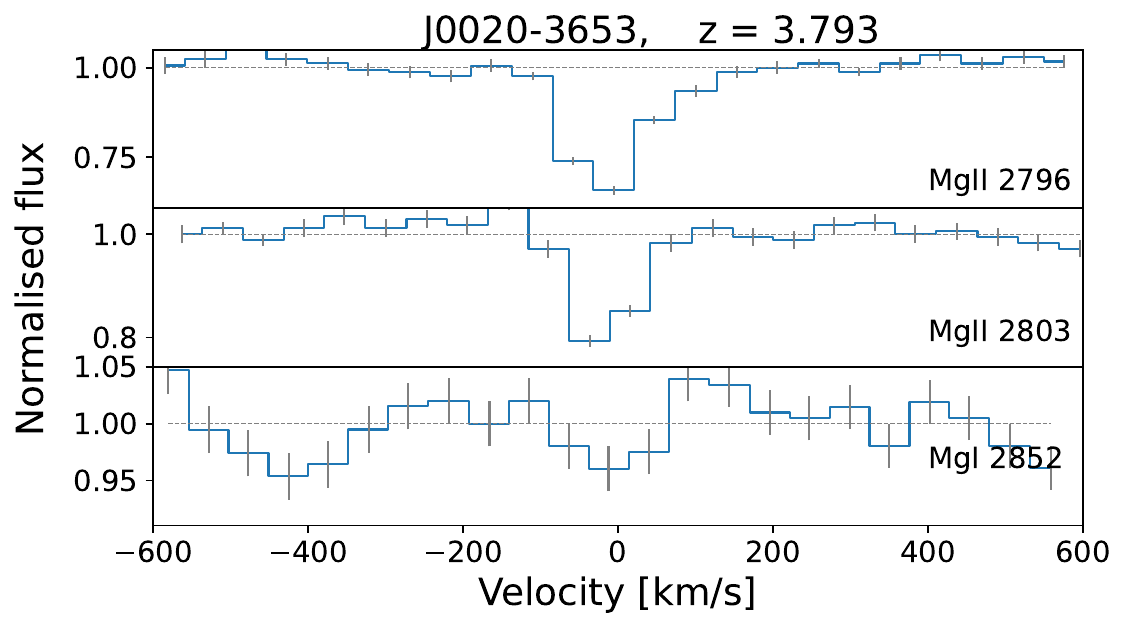}
    \caption{Absorption system at $z=3.793$ toward J0020--3653. }
    \label{fig:J0020_3.793}
\end{figure}
\begin{table}
\centering
\caption{J0020--3653 $z=3.793$} 
\label{table:J0020_3.793}  
\begin{tabular}{cccc}   
\hline\hline \noalign{\smallskip}
$\lambda_{\mathrm{obs}}$ & ID & $W_{\mathrm{obs}}$ \\   
\noalign{\smallskip} \hline \noalign{\smallskip}     
13402.92 & \ion{Mg}{ii}\ 2796.35 &  1.977$\pm$0.070\\
13437.32 & \ion{Mg}{ii}\ 2803.53 &  0.770$\pm$0.089\\
13674.26 & \ion{Mg}{i}\ 2852.96 &  0.190$\pm$0.125\\
\noalign{\smallskip}
\hline                           
\end{tabular}
\end{table}

\begin{figure}
   \centering
   \includegraphics[clip, trim=0.cm 5cm 0cm 5cm, width=\hsize]{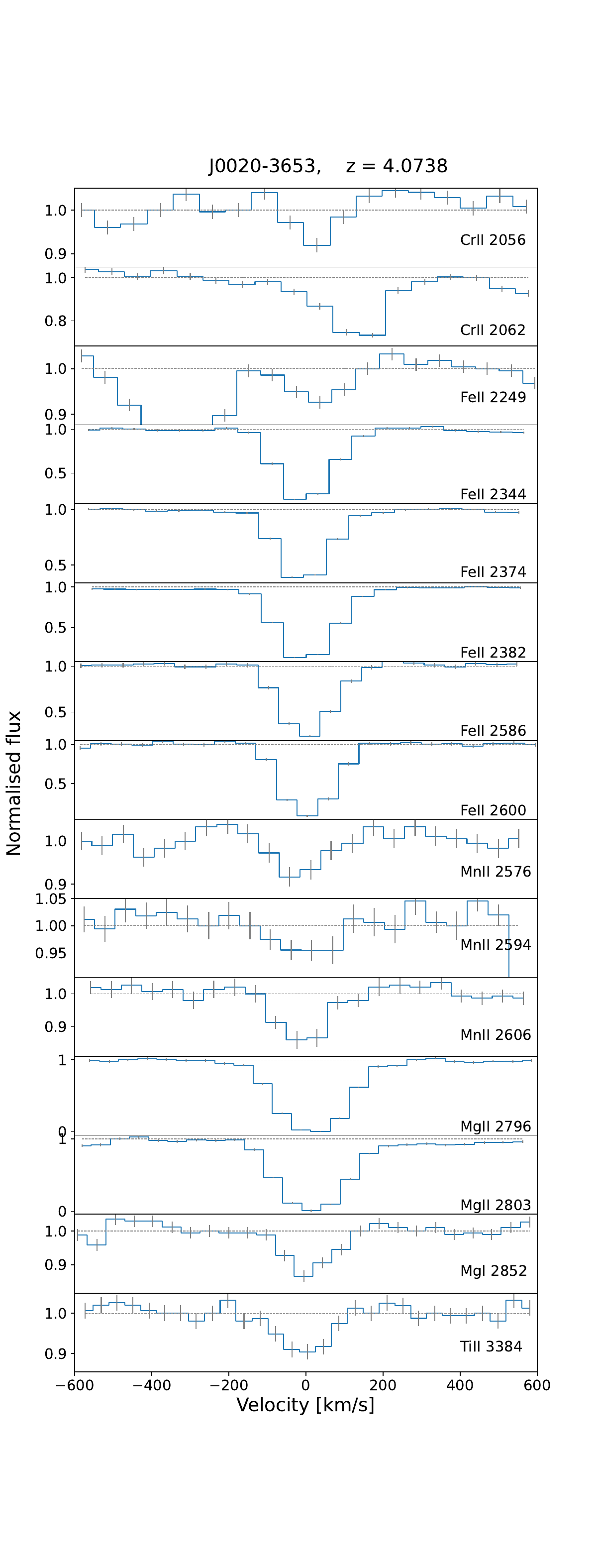}
    \caption{Absorption system at $z=4.0738$ toward J0020--3653. }
    \label{fig:J0020_4.0738}
\end{figure}
\begin{table}
\centering
\caption{J0020--3653 $z=4.0738$} 
\label{table:J0020_4.0738}  
\begin{tabular}{cccc}   
\hline\hline \noalign{\smallskip}
$\lambda_{\mathrm{obs}}$ & ID & $W_{\mathrm{obs}}$ \\   
\noalign{\smallskip} \hline \noalign{\smallskip}     
10433.02 & \ion{Cr}{ii}\ 2056.25 &  0.128$\pm$0.093\\
10463.36 & \ion{Cr}{ii}\ 2062.23 &  1.813$\pm$0.085$^\dagger$ \\
11415.42 & \ion{Fe}{ii}\ 2249.88 &  0.614$\pm$0.093\\
11894.07 & \ion{Fe}{ii}\ 2344.21 &  5.535$\pm$0.073\\
12047.54 & \ion{Fe}{ii}\ 2374.46 &  4.390$\pm$0.053\\ 
12089.67 & \ion{Fe}{ii}\ 2382.76 &  6.750$\pm$0.052\\
13124.14 & \ion{Fe}{ii}\ 2586.65 &  5.113$\pm$0.140\\
13192.76 & \ion{Fe}{ii}\ 2600.17 &  6.237$\pm$0.135\\
13074.56 & \ion{Mn}{ii}\ 2576.88 &  0.265$\pm$0.157\\
13163.97 & \ion{Mn}{ii}\ 2594.50 &  0.289$\pm$0.156\\
13224.65 & \ion{Mn}{ii}\ 2606.46 &  0.866$\pm$0.163\\
14188.13 & \ion{Mg}{ii}\ 2796.35 &  10.553$\pm$0.095\\
14224.56 & \ion{Mg}{ii}\ 2803.53 &  10.212$\pm$0.097\\
14475.37 & \ion{Mg}{i}\ 2852.96 &  0.842$\pm$0.118\\
17173.49 & \ion{Ti}{ii}\ 3384.74 &  0.718$\pm$0.150\\
\noalign{\smallskip}   
\hline                           
\end{tabular}
\begin{tablenotes}
\item  $^\dagger$ Line blended with \ion{Fe}{ii} $\lambda$2374 at $z=3.4087$
\end{tablenotes}
\end{table}

\FloatBarrier

\begin{figure}
   \centering
   \includegraphics[width=\hsize]{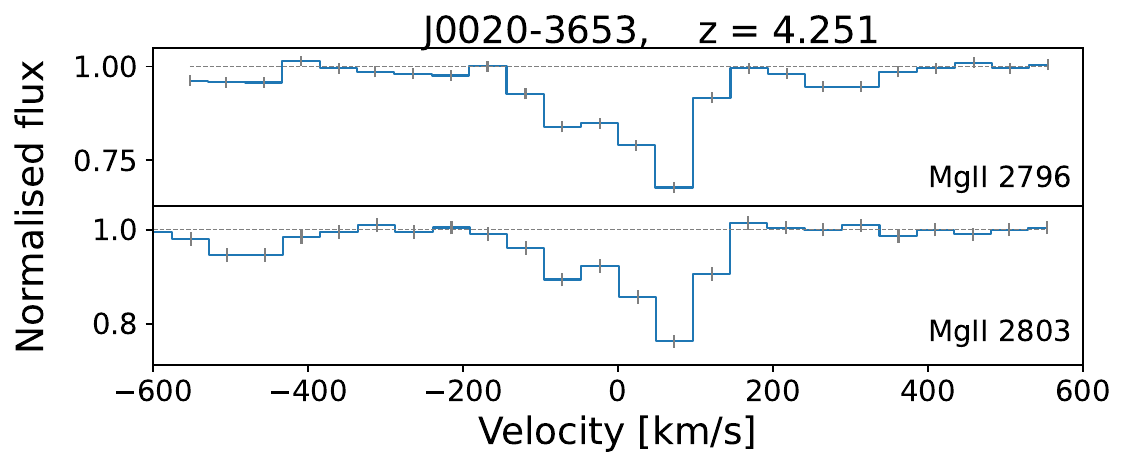}
    \caption{Absorption system at $z=4.251$ toward J0020--3653.}
    \label{fig:J0020_4.251}
\end{figure}
\begin{table}
\centering
\caption{J0020--3653 $z=4.251$} 
\label{table:J0020_4.251}  
\begin{tabular}{cccc}   
\hline\hline \noalign{\smallskip}
$\lambda_{\mathrm{obs}}$ & ID & $W_{\mathrm{obs}}$ \\   
\noalign{\smallskip} \hline \noalign{\smallskip}
14683.64 & \ion{Mg}{ii}\ 2796.35 &  2.392$\pm$0.098\\
14721.34 & \ion{Mg}{ii}\ 2803.53 &  1.630$\pm$0.096\\
\noalign{\smallskip}
\hline                           
\end{tabular}
\end{table}

\FloatBarrier

\begin{figure}
   \centering
   \includegraphics[width=\hsize]{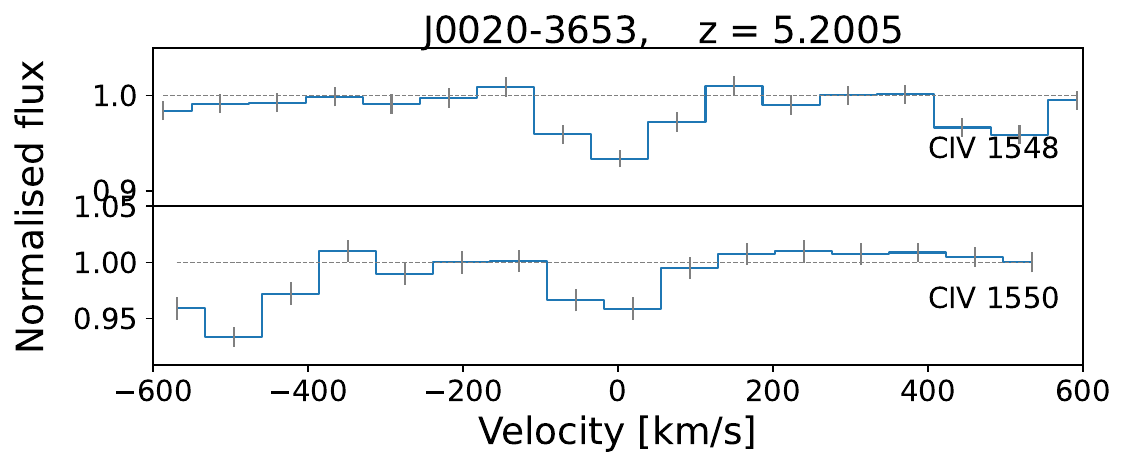}
    \caption{Absorption system at $z=5.2005$ toward J0020--3653.}
    \label{fig:J0020_6.372}
\end{figure}
\begin{table}
\centering
\caption{J0020--3653  $z=5.2005$} 
\label{table:J0020_5.2005}  
\begin{tabular}{cccc}   
\hline\hline \noalign{\smallskip}
$\lambda_{\mathrm{obs}}$ & ID & $W_{\mathrm{obs}}$ \\   
\noalign{\smallskip} \hline \noalign{\smallskip}
9599.58& \ion{C}{iv} 1548.19 & 0.274$\pm$0.052 \\
9615.55& \ion{C}{iv} 1550.77 & 0.168$\pm$0.053 \\
\noalign{\smallskip}
\hline                           
\end{tabular}
\end{table}

\begin{figure}
   \centering
   \includegraphics[width=\hsize]{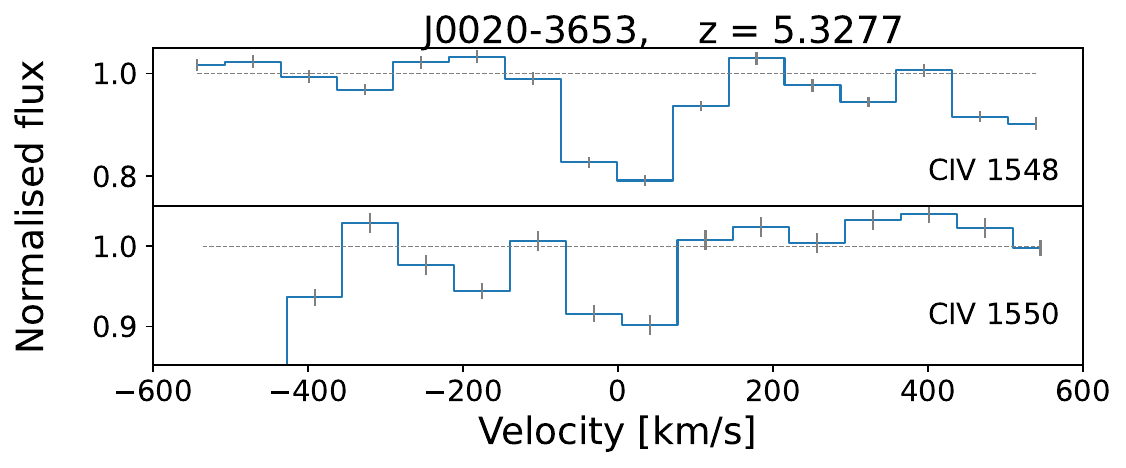}
    \caption{Absorption system at $z=5.3277$ toward J0020--3653. }
    \label{fig:J0020_5.3277}
\end{figure}
\begin{table}
\centering
\caption{J0020--3653 $z=5.3277$} 
\label{table:J0020_5.3277}  
\begin{tabular}{cccc}   
\hline\hline \noalign{\smallskip}
$\lambda_{\mathrm{obs}}$ & ID & $W_{\mathrm{obs}}$ \\   
\noalign{\smallskip} \hline \noalign{\smallskip}
9796.51 & \ion{C}{iv}\ 1548.19 &  1.075$\pm$0.053\\
9812.81 & \ion{C}{iv}\ 1550.77 &  0.396$\pm$0.056\\
\noalign{\smallskip}
\hline                           
\end{tabular}
\end{table}

\begin{figure}
   \centering
   \includegraphics[clip, trim=0.cm 2cm 0cm 3.cm, width=\hsize]{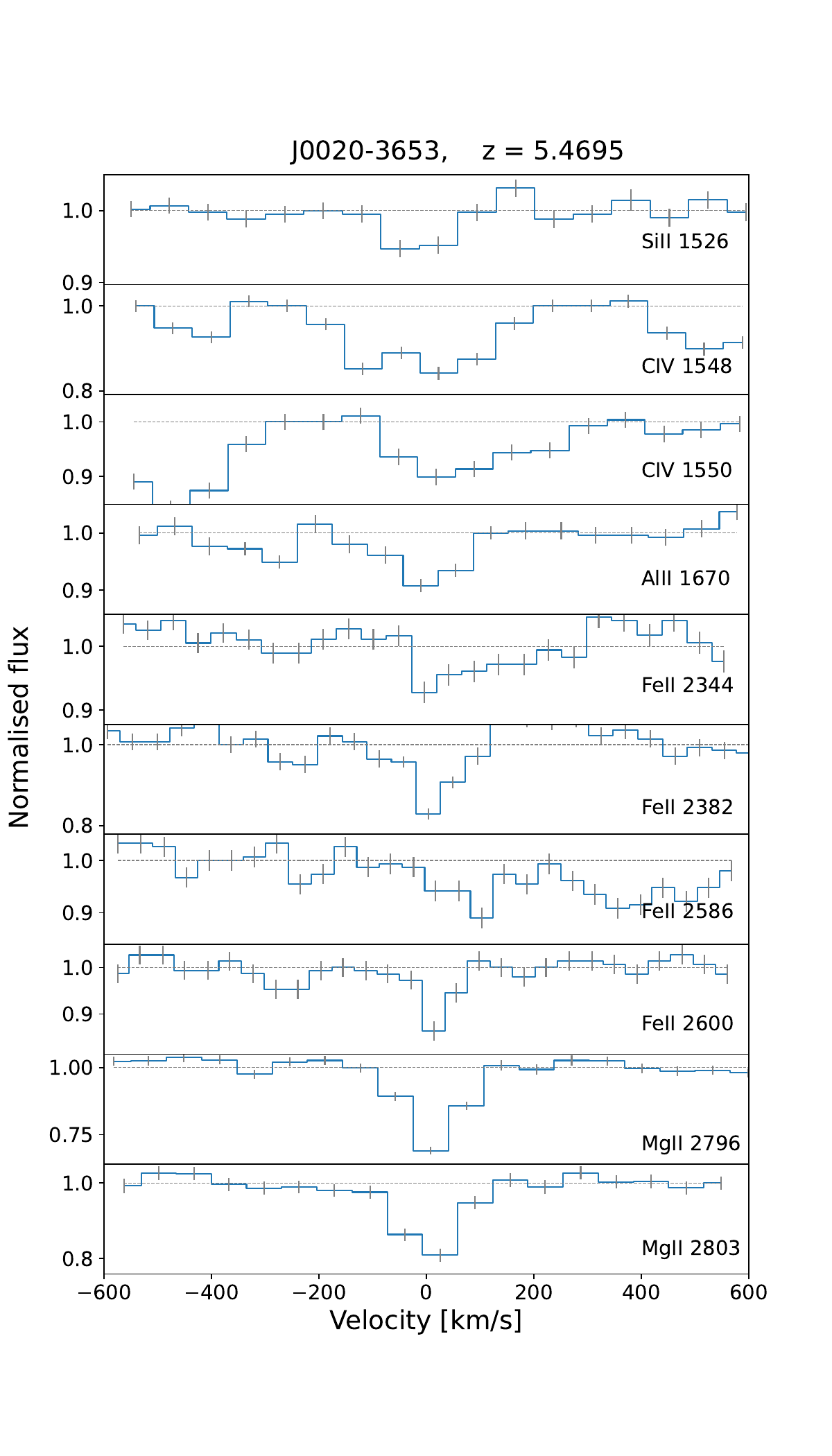}
    \caption{Absorption system at $z=5.4695$ toward J0020--3653. }
    \label{fig:J0020_5.4695}
\end{figure}
\begin{table}
\centering
\caption{J0020--3653 $z=5.4695$} 
\label{table:J0020_5.4695}  
\begin{tabular}{cccc}   
\hline\hline \noalign{\smallskip}
$\lambda_{\mathrm{obs}}$ & ID & $W_{\mathrm{obs}}$ \\   
\noalign{\smallskip} \hline \noalign{\smallskip}
9877.03 & \ion{Si}{ii}\ 1526.71 &  0.181$\pm$0.069\\
10016.05 & \ion{C}{iv}\ 1548.19 &  1.477$\pm$0.084\\
10032.71 & \ion{C}{iv}\ 1550.77 &  0.697$\pm$0.087\\
10809.16 & \ion{Al}{ii}\ 1670.79 &  0.499$\pm$0.079\\
15165.89 & \ion{Fe}{ii}\ 2344.21 &  0.342$\pm$0.118\\
15415.30 & \ion{Fe}{ii}\ 2382.76 &  0.469$\pm$0.136\\
16734.33 & \ion{Fe}{ii}\ 2586.65 &  0.774$\pm$0.145\\
16821.82 & \ion{Fe}{ii}\ 2600.17 &  0.588$\pm$0.152\\
18091.00 & \ion{Mg}{ii}\ 2796.35 &  2.264$\pm$0.135\\
18137.44 & \ion{Mg}{ii}\ 2803.53 &  1.734$\pm$0.164\\
\noalign{\smallskip}
\hline                           
\end{tabular}
\end{table}

\FloatBarrier

\begin{figure}
   \centering
   \includegraphics[clip, trim=0.cm 0cm 0cm 0cm, width=0.9\hsize]{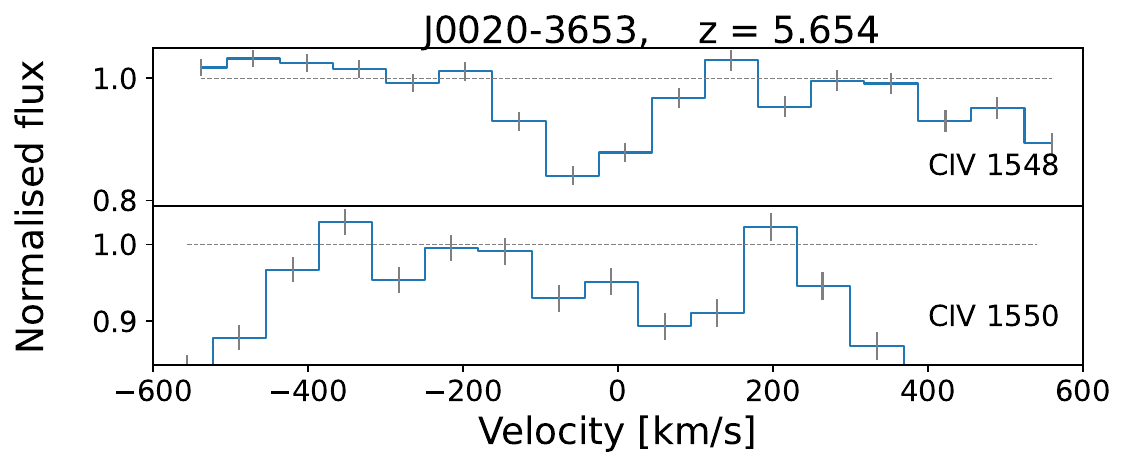}
    \caption{Absorption system at $z=5.654$ toward J0020--3653. }
    \label{fig:J0020_5.654}
\end{figure}
\begin{table}
\centering
\caption{J0020--3653 $z=5.654$} 
\label{table:J0020_5.654}  
\begin{tabular}{cccc}   
\hline\hline \noalign{\smallskip}
$\lambda_{\mathrm{obs}}$ & ID & $W_{\mathrm{obs}}$ \\   
\noalign{\smallskip} \hline \noalign{\smallskip}     
10301.69 & \ion{C}{iv}\ 1548.19 &  0.805$\pm$0.093\\
10318.82 & \ion{C}{iv}\ 1550.77 &  0.706$\pm$0.102\\
\noalign{\smallskip}
\hline                           
\end{tabular}
\end{table}

\begin{figure}
   \centering
   \includegraphics[clip, trim=0.cm 1.5cm 0cm 2cm, width=\hsize]{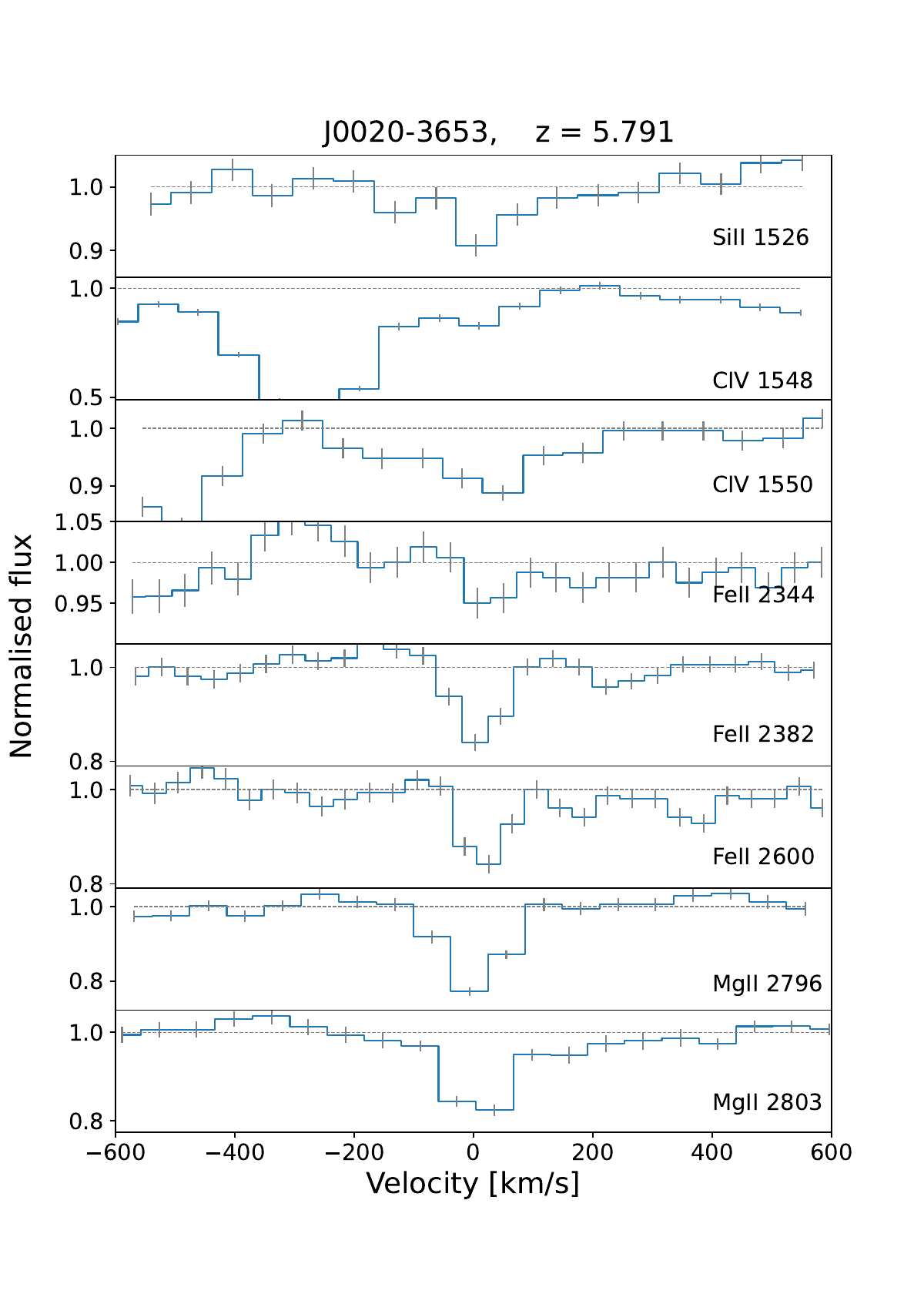}
    \caption{Absorption system at $z=5.791$ toward J0020--3653. }
    \label{fig:J0020_5.791}
\end{figure}
\begin{table}
\centering
\caption{J0020--3653  $z=5.791$} 
\label{table:J0020_5.791}  
\begin{tabular}{cccc}   
\hline\hline \noalign{\smallskip}
$\lambda_{\mathrm{obs}}$ & ID & $W_{\mathrm{obs}}$ \\   
\noalign{\smallskip} \hline \noalign{\smallskip}
10367.86 & \ion{Si}{ii}\ 1526.71 &  0.501$\pm$0.093\\
10513.79 & \ion{C}{iv}\ 1548.19 &  2.426$\pm$0.095$^\dagger$ \\
10531.28 & \ion{C}{iv}\ 1550.77 &  0.936$\pm$0.097\\
15919.56 & \ion{Fe}{ii}\ 2344.21 &  0.317$\pm$0.132\\
16181.36 & \ion{Fe}{ii}\ 2382.76 &  0.405$\pm$0.131\\
17657.77 & \ion{Fe}{ii}\ 2600.17 &  1.029$\pm$0.149\\
18990.03 & \ion{Mg}{ii}\ 2796.35 &  1.684$\pm$0.168\\
19038.78 & \ion{Mg}{ii}\ 2803.53 &  1.926$\pm$0.145\\
\noalign{\smallskip}
\hline                           
\end{tabular}
\begin{tablenotes}
\item  $^\dagger$ Line partly blended with \ion{Fe}{ii} $\lambda$2382 at $z=3.4087$
\end{tablenotes}
\end{table}

\begin{figure}
   \centering
   \includegraphics[clip, trim=0.cm 2cm 0cm 2cm, width=\hsize]{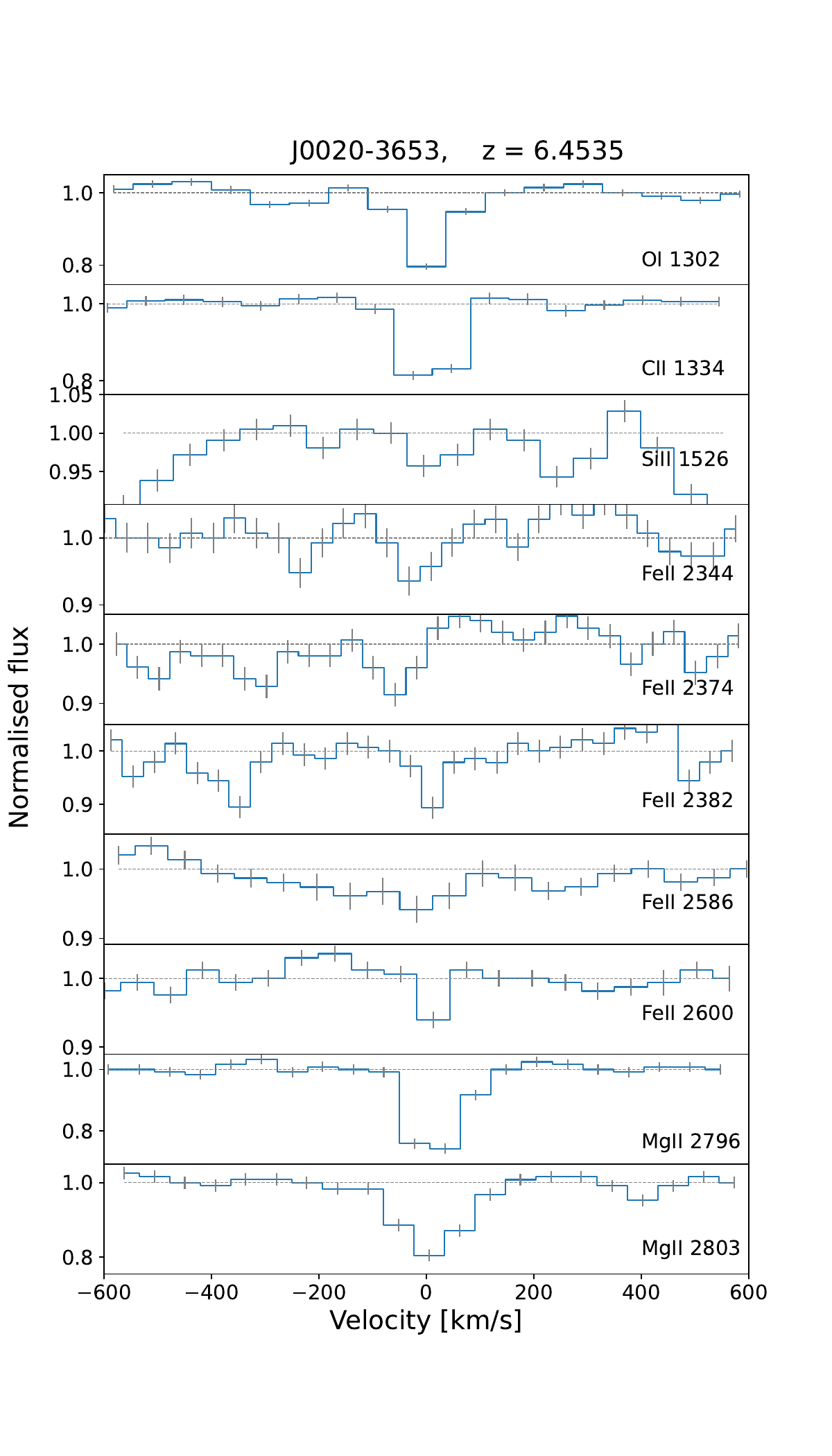}
    \caption{Absorption system at $z=6.4535$ toward J0020--3653. }
    \label{fig:J0020_6.4535}
\end{figure}
\begin{table}
\centering
\caption{J0020--3653,  $z=6.4535$} 
\label{table:J0020_6.4535}
\begin{tabular}{cccc}   
\hline\hline \noalign{\smallskip}
$\lambda_{\mathrm{obs}}$ & ID & $W_{\mathrm{obs}}$ \\   
\noalign{\smallskip} \hline \noalign{\smallskip}
9705.71 & \ion{O}{i}\ 1302.17 &  0.680$\pm$0.050\\
9946.94 & \ion{C}{ii}\ 1334.53 &  0.768$\pm$0.077\\
11379.31 & \ion{Si}{ii}\ 1526.71 &  0.212$\pm$0.089\\
17472.60 & \ion{Fe}{ii}\ 2344.21 &  0.083$\pm$0.159\\
17698.05 & \ion{Fe}{ii}\ 2374.46 &  0.094$\pm$0.148\\
17759.94 & \ion{Fe}{ii}\ 2382.76 &  0.406$\pm$0.159\\
19279.60 & \ion{Fe}{ii}\ 2586.65 &  0.741$\pm$0.187\\
19380.39 & \ion{Fe}{ii}\ 2600.17 &  0.139$\pm$0.094\\
20842.61 & \ion{Mg}{ii}\ 2796.35 &  2.301$\pm$0.175\\
20896.12 & \ion{Mg}{ii}\ 2803.53 &  1.956$\pm$0.170\\
\noalign{\smallskip}
\hline                           
\end{tabular}
\end{table}

\FloatBarrier

\begin{figure}
   \centering
   \includegraphics[width=\hsize]{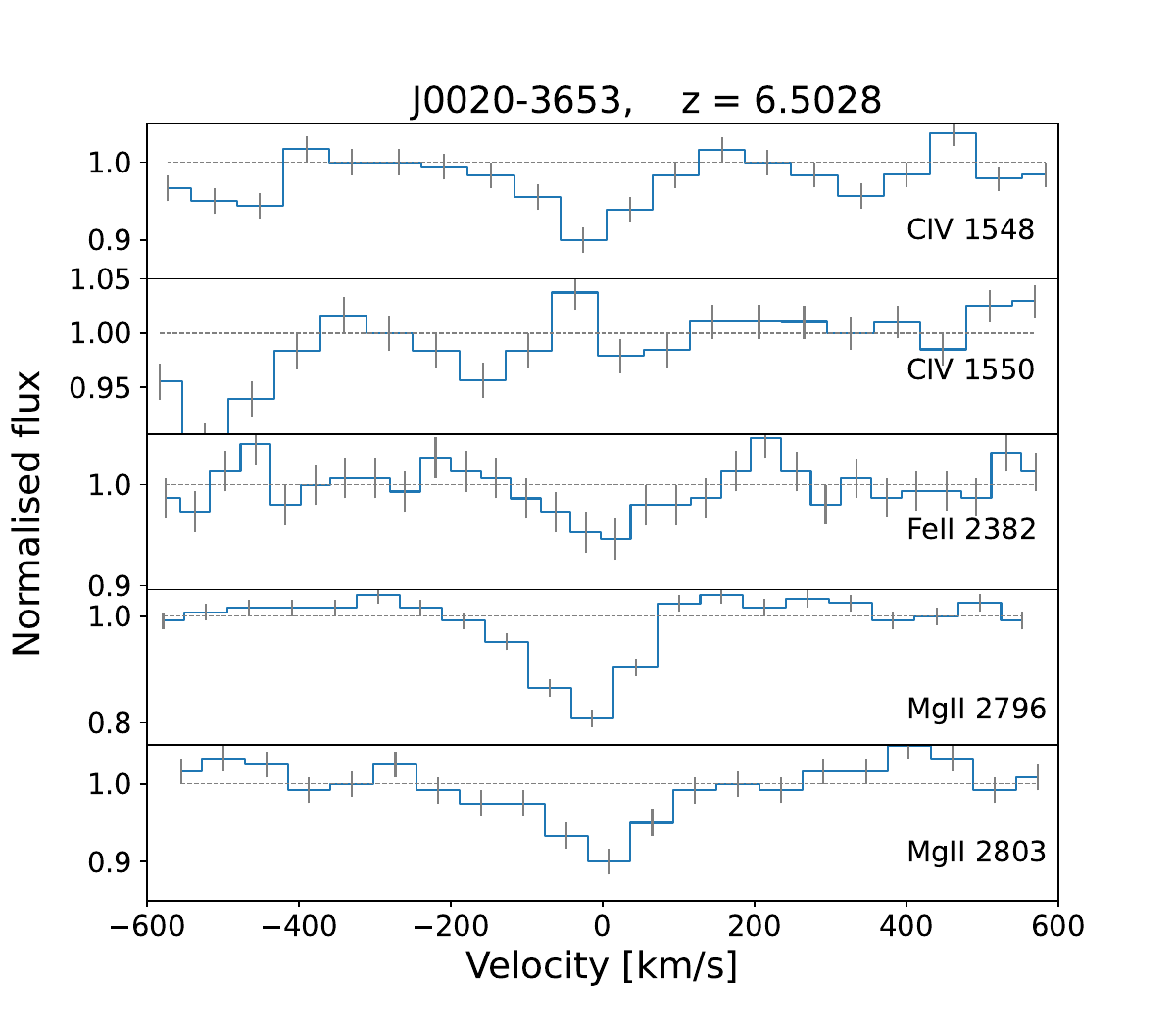}
    \caption{Absorption system at $z=6.5028$ toward J0020--3653. }
    \label{fig:J0020_6.5028}
\end{figure}
\begin{table}
\centering
\caption{J0020--3653,  $z=6.5028$} 
\label{table:J0020_6.5028 }
\begin{tabular}{cccc}   
\hline\hline \noalign{\smallskip}
$\lambda_{\mathrm{obs}}$ & ID & $W_{\mathrm{obs}}$ \\   
\noalign{\smallskip} \hline \noalign{\smallskip}
11615.80 & \ion{C}{iv}\ 1548.19 &  0.524$\pm$0.096\\
11635.12 & \ion{C}{iv}\ 1550.77 &  0.114$\pm$0.092\\
17877.41 & \ion{Fe}{ii}\ 2382.76 &  0.383$\pm$0.150\\
20980.47 & \ion{Mg}{ii}\ 2796.35 &  1.642$\pm$0.167\\
21034.33 & \ion{Mg}{ii}\ 2803.53 &  1.086$\pm$0.175\\
\noalign{\smallskip}
\hline                           
\end{tabular}
\end{table}

\begin{figure}
   \centering
   \includegraphics[width=\hsize]{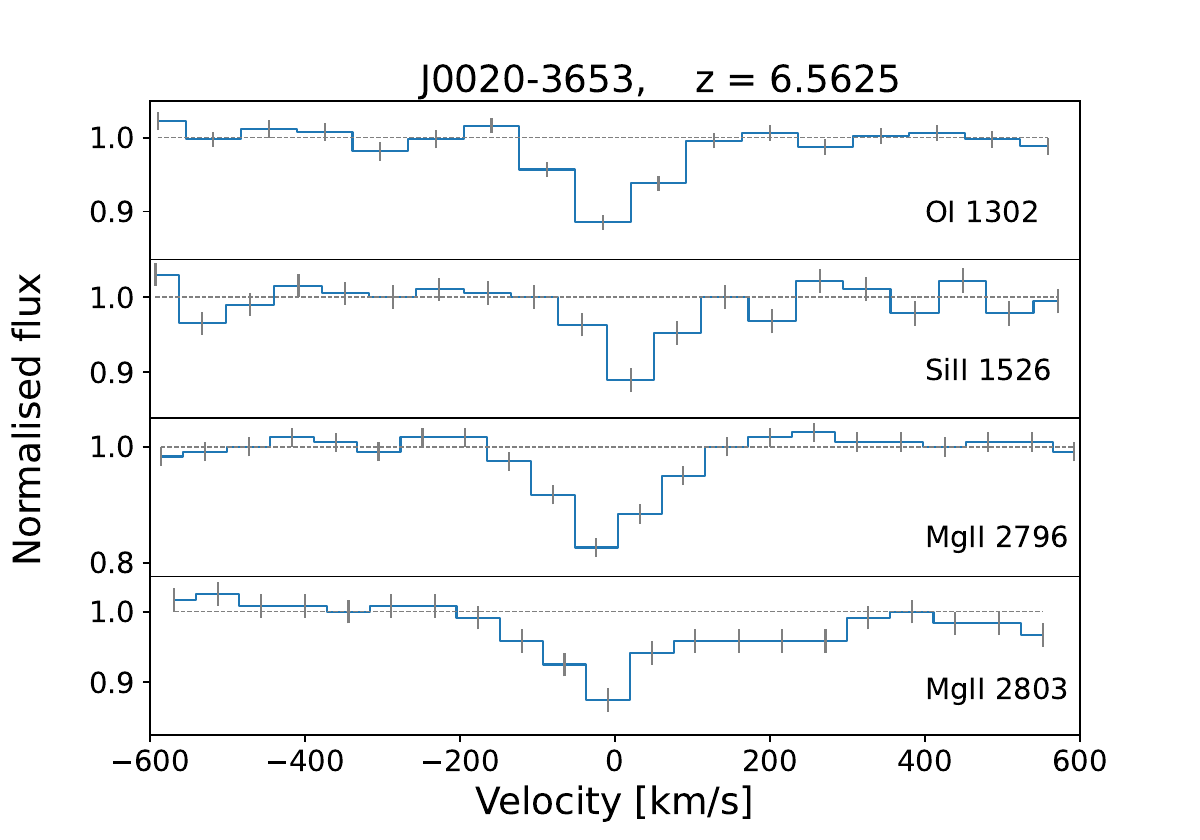}
    \caption{Absorption system at $z=6.5625$ toward J0020--3653. }
    \label{fig:J0020_6.5625}
\end{figure}
\begin{table}
\centering
\caption{J0020--3653, $z=6.5625$} 
\label{table:J0020_6.5625 }
\begin{tabular}{cccc}   
\hline\hline \noalign{\smallskip}
$\lambda_{\mathrm{obs}}$ & ID & $W_{\mathrm{obs}}$ \\   
\noalign{\smallskip} \hline \noalign{\smallskip}
9847.65  & \ion{O}{i}\ 1302.17 &  0.472$\pm$0.059\\
11545.72 & \ion{Si}{ii}\ 1526.71 &  0.445$\pm$0.091\\
10092.39 & \ion{C}{ii}\ 1334.53  & < 0.4$^\dagger$ \\
21147.41 & \ion{Mg}{ii}\ 2796.35 &  1.633$\pm$0.185\\
21201.70 & \ion{Mg}{ii}\ 2803.53 &  1.553$\pm$0.175\\
\noalign{\smallskip}
\hline                           
\end{tabular}
\begin{tablenotes}
    \item $^\dagger$\ion{C}{ii} is not clearly detected in this system. A 3$\sigma$ upper limit is listed because the limit is used in Section~\ref{sect:abundances}.
\end{tablenotes}
\end{table}

\FloatBarrier

\begin{figure}
   \centering
   \includegraphics[clip, trim=0.cm 3.5cm 0cm 4.5cm, width=\hsize]{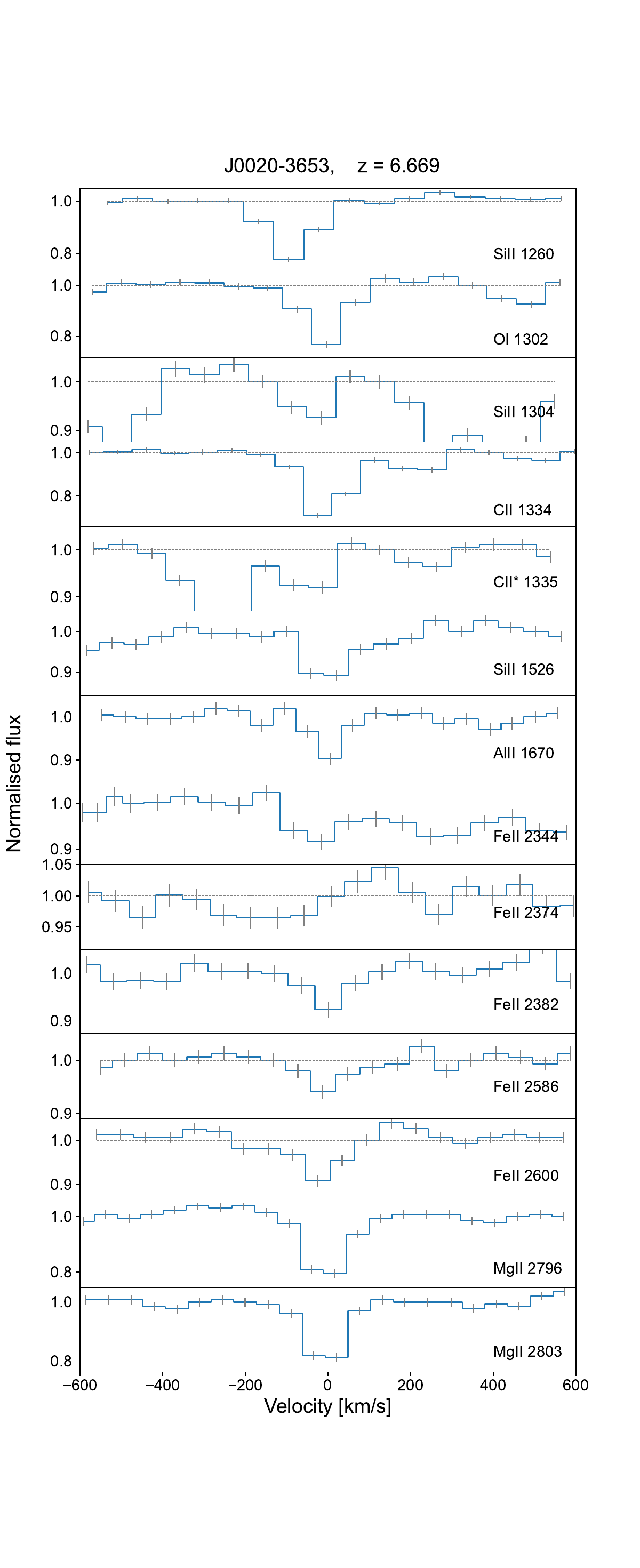}
    \caption{Absorption system at $z=6.669$ toward J0020--3653}
    \label{fig:J0020_6.669}
\end{figure}
\begin{table}
\centering
\caption{J0020--3653, $z=6.669$} 
\label{table:J0020_6.669 }
\begin{tabular}{cccc}   
\hline\hline
\noalign{\smallskip}
$\lambda_{\mathrm{obs}}$ & ID & $W_{\mathrm{obs}}$ \\   
\noalign{\smallskip} \hline \noalign{\smallskip}
9668.71 & \ion{Si}{ii}\ 1260.75 &  0.964$\pm$0.055\\
9986.33 & \ion{O}{i}\ 1302.17 &  0.884$\pm$0.074\\
10003.22 & \ion{Si}{ii}\ 1304.37 &  0.368$\pm$0.081\\
10234.53 & \ion{C}{ii}\ 1334.53 &  1.292$\pm$0.078\\
10243.54 & \ion{C}{ii*}\ 1335.71 & 0.447$\pm$0.073\\
11708.31 & \ion{Si}{ii}\ 1526.71 &  0.707$\pm$0.078\\
12813.27 & \ion{Al}{ii}\ 1670.79 &  0.318$\pm$0.090\\
17977.78$^\dagger$ & \ion{Fe}{ii}\ 2344.21 &  0.947$\pm$0.171\\ 
18209.74 & \ion{Fe}{ii}\ 2374.46 &  0.139$\pm$0.175\\
18273.42 & \ion{Fe}{ii}\ 2382.76 &  0.362$\pm$0.180\\
19837.02 & \ion{Fe}{ii}\ 2586.65 &  0.469$\pm$0.137\\
19940.73 & \ion{Fe}{ii}\ 2600.17 &  0.595$\pm$0.128\\
21445.22 & \ion{Mg}{ii}\ 2796.35 &  1.857$\pm$0.167\\
21500.28 & \ion{Mg}{ii}\ 2803.53 &  1.729$\pm$0.157\\
\noalign{\smallskip}
\hline                           
\end{tabular}
\begin{tablenotes}
\item $^\dagger$ Line blended with unidentified feature.
\end{tablenotes}
\end{table}

\begin{figure}
   \centering
   \includegraphics[width=\hsize]{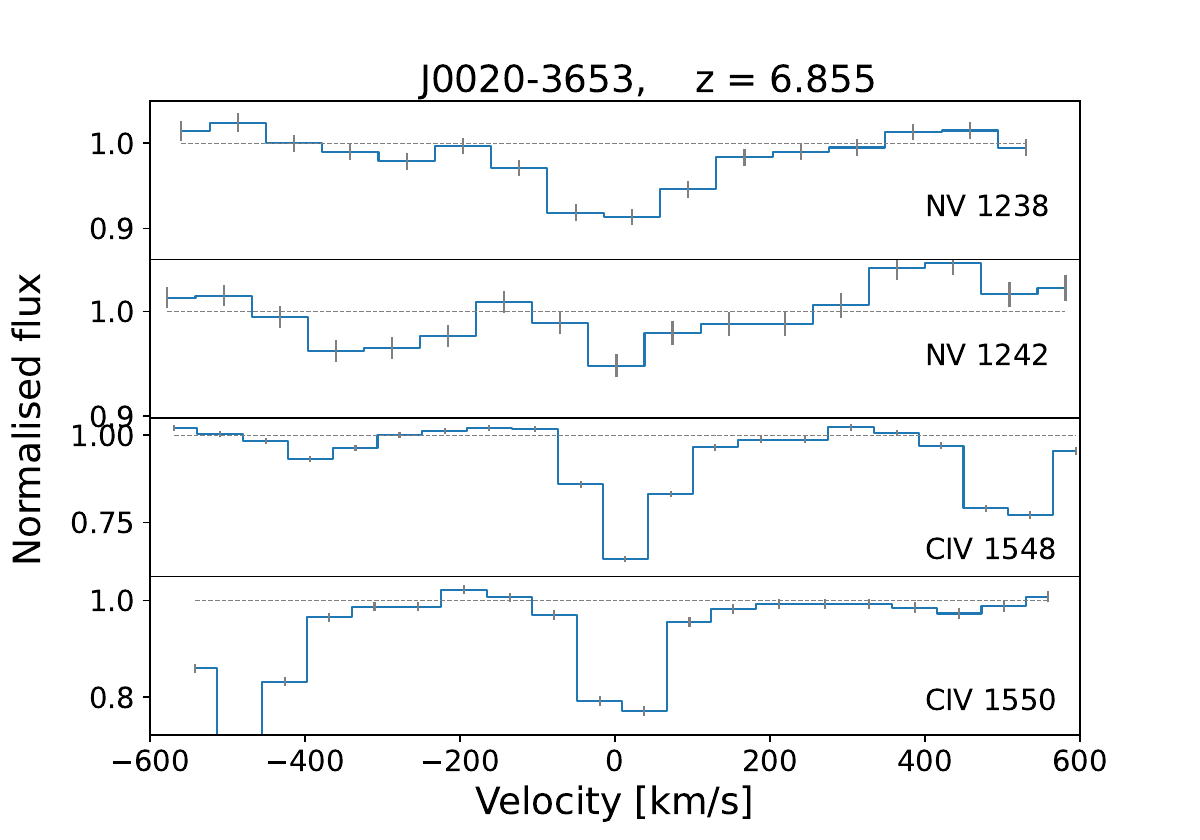}
    \caption{Absorption system at $z=6.855$ toward J0020--3653}
    \label{fig:J0020_6.855}
\end{figure}
\begin{table}
\centering
\caption{J0020--3653, $z=6.855$} 
\label{table:J0020_6.855 }
\begin{tabular}{cccc}   
\hline\hline
\noalign{\smallskip}
$\lambda_{\mathrm{obs}}$ & ID & $W_{\mathrm{obs}}$ \\   
\noalign{\smallskip} \hline \noalign{\smallskip}
9730.94  & \ion{N}{v}\,1238.82  & 0.638$\pm$0.056\\ 
9762.23  & \ion{N}{v}\,1242.80  & 0.202$\pm$0.058\\
12161.07 & \ion{C}{iv}\,1548.19 & 1.582$\pm$0.058\\
12181.30 & \ion{C}{iv}\,1550.77 & 1.176$\pm$0.063\\
\noalign{\smallskip}
\hline                           
\end{tabular}
\end{table}


\newpage
\mbox{~}
\clearpage

\begin{figure}
   \centering
   \includegraphics[width=\hsize]{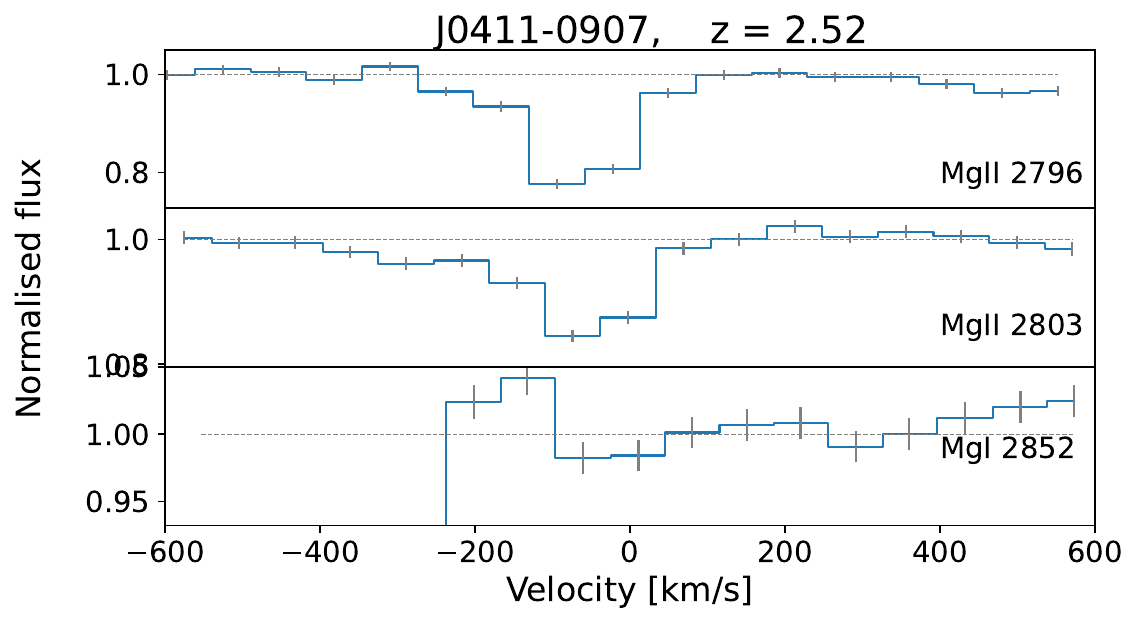}
    \caption{Absorption system at $z=2.52$ toward J0411--0907. Note that the y-axis range changes according to the strengths of the absorption lines in this and all other panels.}
    \label{fig:J0411_abs2.52}
\end{figure}
\begin{table}
\centering
\caption{J0411--0907,  $z=2.52$} 
\label{table:J0411_2.52}  
\begin{tabular}{cccc}   
\hline\hline
\noalign{\smallskip}
$\lambda_{\mathrm{obs}}$ & ID & $W_{\mathrm{obs}}$ \\   
\noalign{\smallskip} \hline \noalign{\smallskip}             
9843.16 & \ion{Mg}{ii}\ 2796.35 &  1.315$\pm$0.067\\
9868.43$^\dagger$ & \ion{Mg}{ii}\ 2803.53 &  0.967$\pm$0.072\\
10042.43 & \ion{Mg}{i}\ 2852.96 &  0.216$\pm$0.084\\
\noalign{\smallskip}\hline                           
\end{tabular}
\begin{tablenotes}
\item $^\dagger$ Line blended with \ion{Fe}{ii} $\lambda$2374 at $z=3.156$
\end{tablenotes}
\end{table}

\begin{figure}
   \centering
   \includegraphics[width=\hsize]{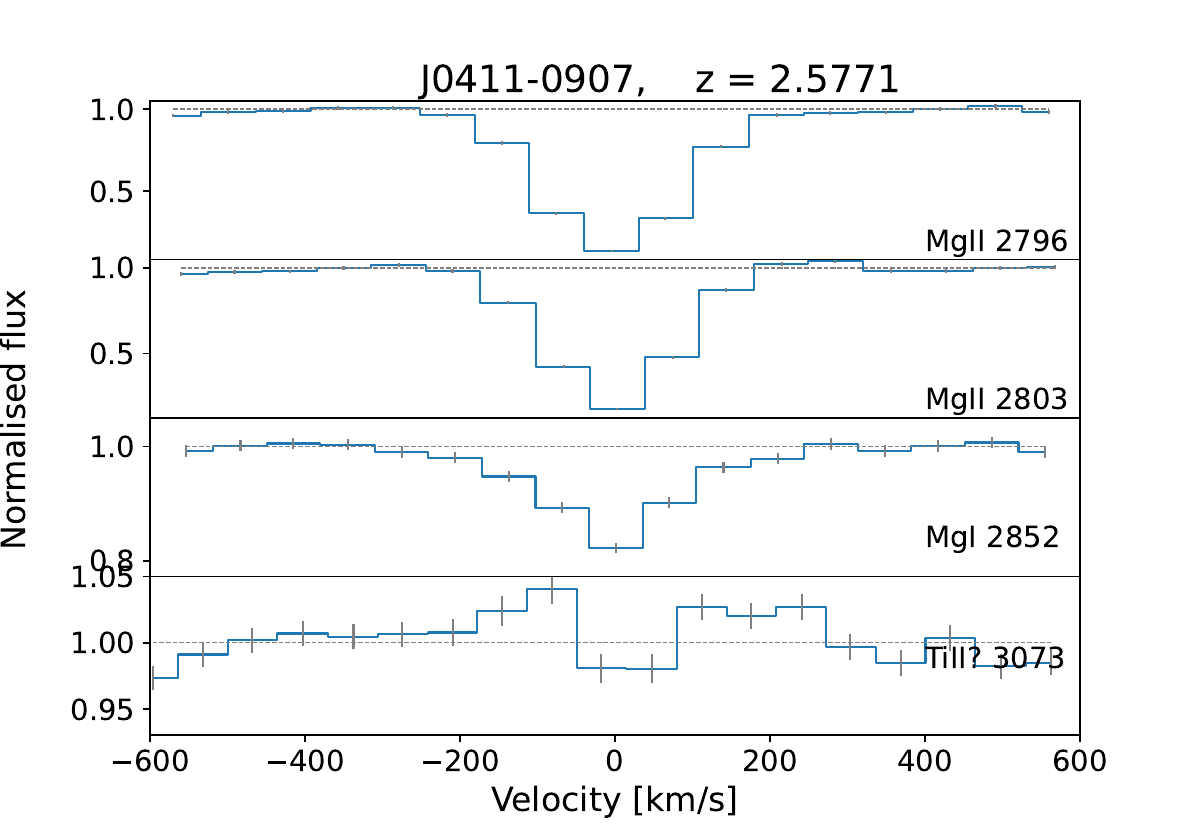}
    \caption{Absorption system at $z=2.5771$ toward J0411--0907.}
    \label{fig:J0411_abs2.5771}
\end{figure}
\begin{table}
\centering
\caption{J0411--0907,  $z=2.5771$}       
\label{table:J0411_2.5771} 
\begin{tabular}{cccc}   
\hline\hline
\noalign{\smallskip}
$\lambda_{\mathrm{obs}}$ & ID & $W_{\mathrm{obs}}$ \\   
\noalign{\smallskip} \hline \noalign{\smallskip}         
10002.83 & \ion{Mg}{ii}\ 2796.35 &  6.342$\pm$0.068\\
10028.51 & \ion{Mg}{ii}\ 2803.53 &  5.164$\pm$0.075\\
10205.34 & \ion{Mg}{i}\ 2852.96 &  1.218$\pm$0.068\\
10995.57 & \ion{Ti}{ii}\ 3073.88 &  1.340$\pm$0.068\\
\noalign{\smallskip}
\hline                           
\end{tabular}
\end{table}

\begin{figure}
   \centering
   \includegraphics[width=\hsize]{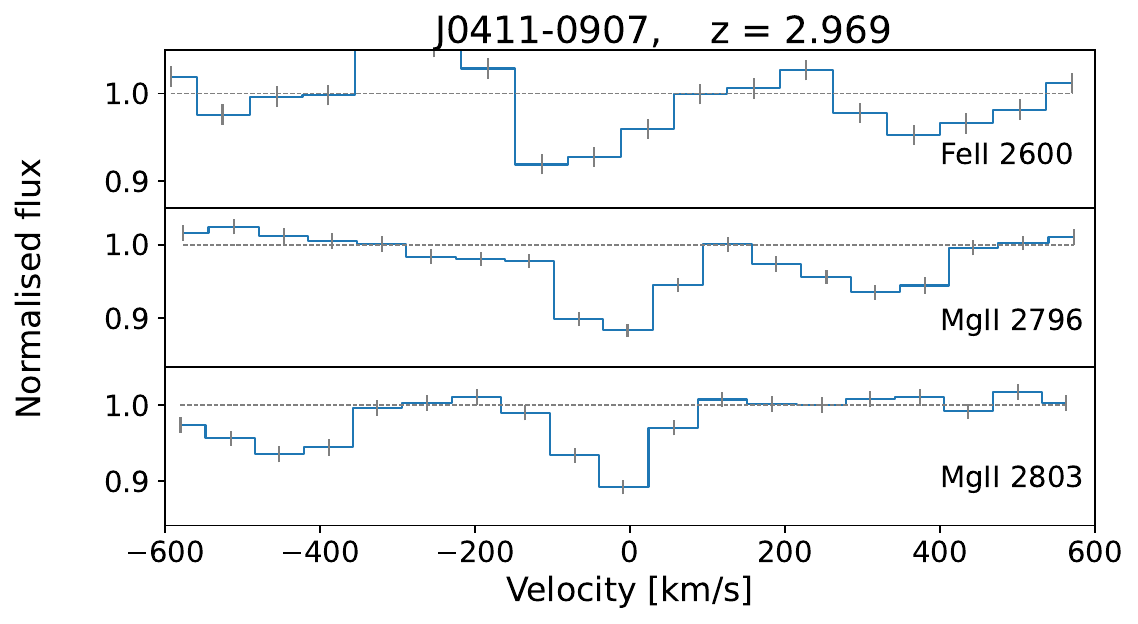}
    \caption{Absorption system at $z=2.969$ toward J0411--0907. }
    \label{fig:J0411_abs2.969}
\end{figure}
\begin{table}
\centering
\caption{J0411--0907, $z=2.969$}       
\label{table:J0411_2.969}  
\begin{tabular}{cccc}   
\noalign{\smallskip} \hline\hline \noalign{\smallskip}
$\lambda_{\mathrm{obs}}$ & ID & $W_{\mathrm{obs}}$ \\
\noalign{\smallskip} \hline \noalign{\smallskip}  
10320.09 & \ion{Fe}{ii}\,2600.17 & 0.370$\pm$0.067 \\
11098.72 & \ion{Mg}{ii}\,2796.35 & 0.804$\pm$0.061 \\
11127.21 & \ion{Mg}{ii}\,2803.53 & 0.459$\pm$0.062 \\
\noalign{\smallskip} \hline          
\end{tabular} 
\end{table}

\begin{figure}
   \centering
   \includegraphics[width=\hsize]{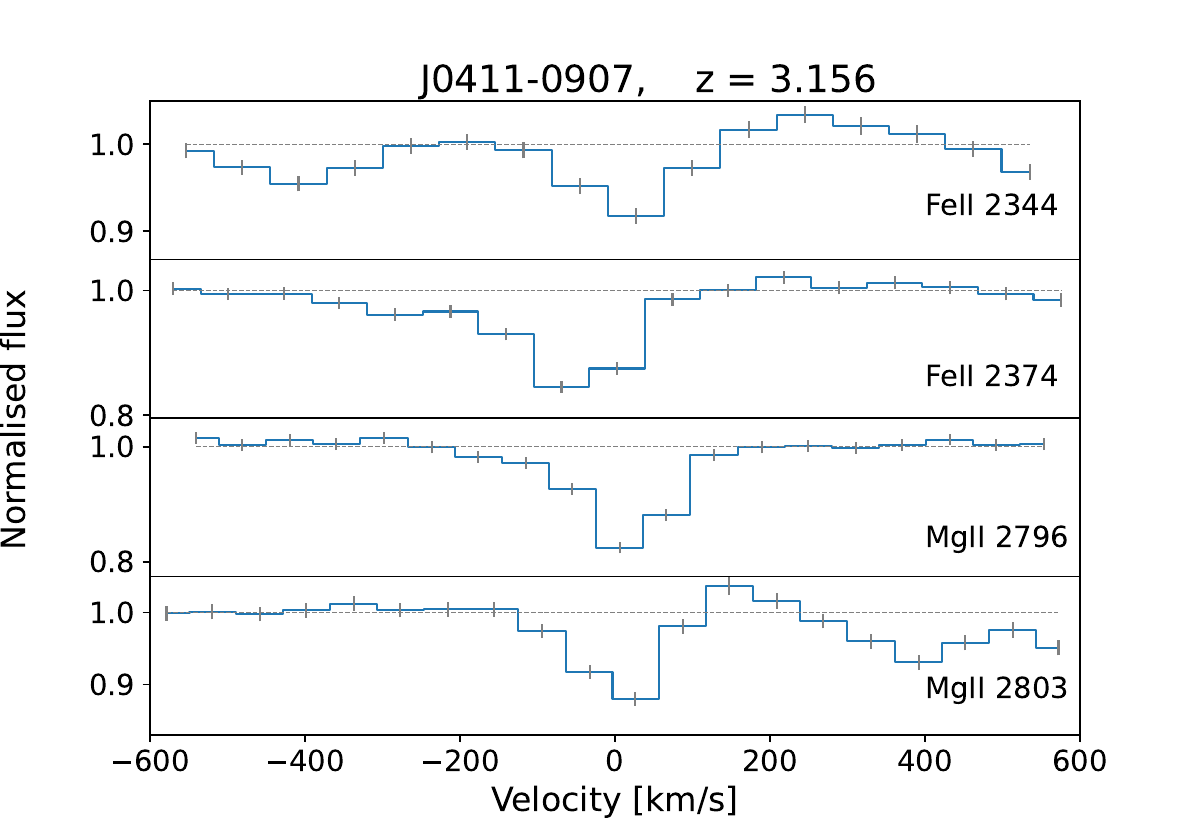}
    \caption{Absorption system at $z=3.1560$ toward J0411--0907. }
    \label{fig:J0411_abs3.1560}
\end{figure}
\begin{table}
\centering
\caption{J0411--0907,  $z=3.1560$.}  
\label{table:J0411_3.1560}   
\begin{tabular}{cccc}   
\hline\hline
\noalign{\smallskip}
$\lambda_{\mathrm{obs}}$ & ID & $W_{\mathrm{obs}}$\\
\noalign{\smallskip} \hline \noalign{\smallskip}         
9742.55 & \ion{Fe}{ii}\ 2344.21 &  0.344$\pm$0.054\\
9868.26 & \ion{Fe}{ii}\ 2374.46 &  0.857$\pm$0.053$^\dagger$ \\
11621.64 & \ion{Mg}{ii}\ 2796.35 &  1.012$\pm$0.065\\
11651.47 & \ion{Mg}{ii}\ 2803.53 &  0.485$\pm$0.061\\
\noalign{\smallskip}
\hline                           
\end{tabular}
\begin{tablenotes}
\item $^\dagger$ Line blended with \ion{Mg}{ii} $\lambda$2803 at $z=2.520$
\end{tablenotes}
\end{table}

\newpage
\mbox{~}
\clearpage

\begin{figure}
   \centering
   \includegraphics[width=\hsize]{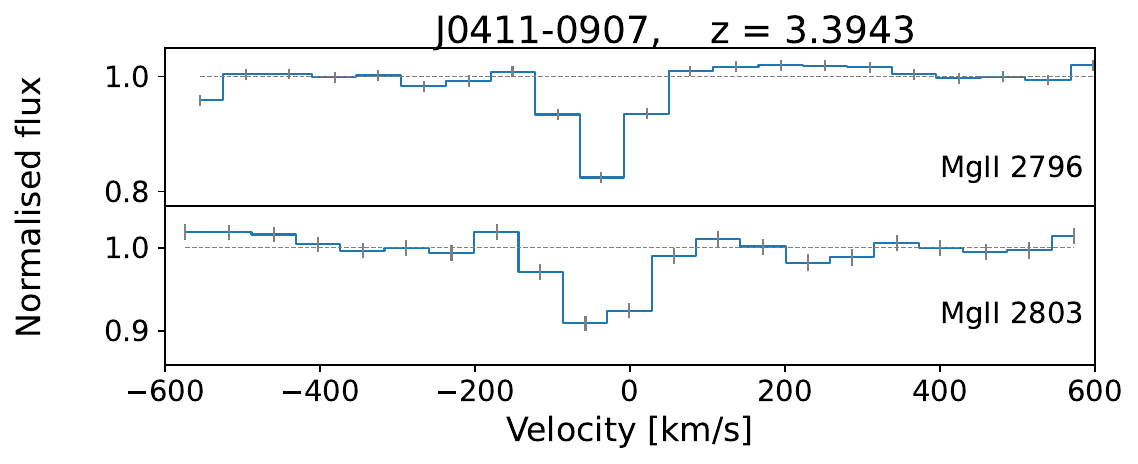}
    \caption{Absorption system at $z=3.3943$ toward J0411--0907. }
    \label{fig:J0411_abs3.3943}
\end{figure}
\begin{table}
\centering
\caption{J0411--0907, $z=3.3943$}       
\label{table:J0411_3.3943}   
 \begin{tabular}{cccc}   
\hline\hline
\noalign{\smallskip}
$\lambda_{\mathrm{obs}}$ & ID & $W_{\mathrm{obs}}$ \\   
\noalign{\smallskip} \hline \noalign{\smallskip}
12288.01 & \ion{Mg}{ii}\ 2796.35 &  0.596$\pm$0.057\\
12319.56 & \ion{Mg}{ii}\ 2803.53 &  0.412$\pm$0.059\\
\noalign{\smallskip}
\hline                           
\end{tabular}
\end{table}

\begin{figure}
   \centering
   \includegraphics[clip, trim=0.cm 1cm 0cm 0cm, width=\hsize]{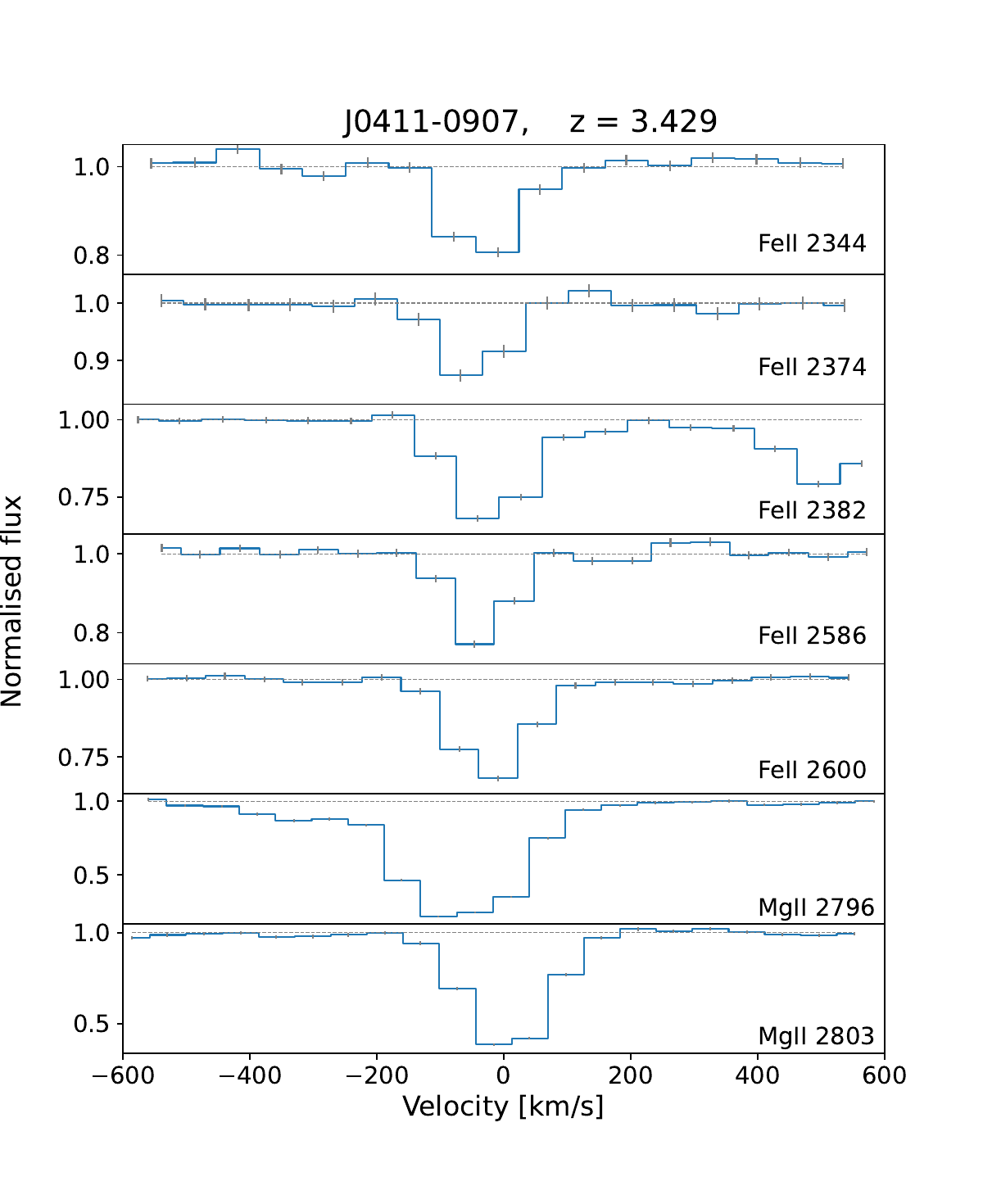}
    \caption{Absorption system at $z=3.429$ toward J0411--0907. }
    \label{fig:J0411_abs3.429}
\end{figure}
\begin{table}
\centering
\caption{J0411--0907, $z=3.429$}       
\label{table:J0411_3.429}    
\begin{tabular}{cccc}   
\hline\hline
\noalign{\smallskip}
$\lambda_{\mathrm{obs}}$ & ID & $W_{\mathrm{obs}}$\\   
\noalign{\smallskip} \hline \noalign{\smallskip} 
10382.52 & \ion{Fe}{ii}\ 2344.21 &  0.930$\pm$0.066$^{\dagger}$\\
10516.49 & \ion{Fe}{ii}\ 2374.46 &  0.518$\pm$0.060\\
10553.27 & \ion{Fe}{ii}\ 2382.76 &  1.827$\pm$0.063\\
11456.27 & \ion{Fe}{ii}\ 2586.65 &  1.013$\pm$0.057\\
11516.17 & \ion{Fe}{ii}\ 2600.17 &  1.764$\pm$0.060\\
12385.04 & \ion{Mg}{ii}\ 2796.35 &  7.222$\pm$0.048\\
12416.84 & \ion{Mg}{ii}\ 2803.53 &  4.273$\pm$0.068$^{\ddagger}$\\
\noalign{\smallskip}
\hline                           
\end{tabular}
\begin{tablenotes}
\item $^{\dagger}$ Line blended with \ion{Mg}{ii} $\lambda$2796 at $z=4.2814$
\item $^{\ddagger}$ Line blended with \ion{Fe}{ii} $\lambda$2600 at $z=3.776$
\end{tablenotes}
\end{table}

\begin{figure}
   \centering
   \includegraphics[clip, trim=0.cm 1.5cm 0cm 0cm, width=\hsize]{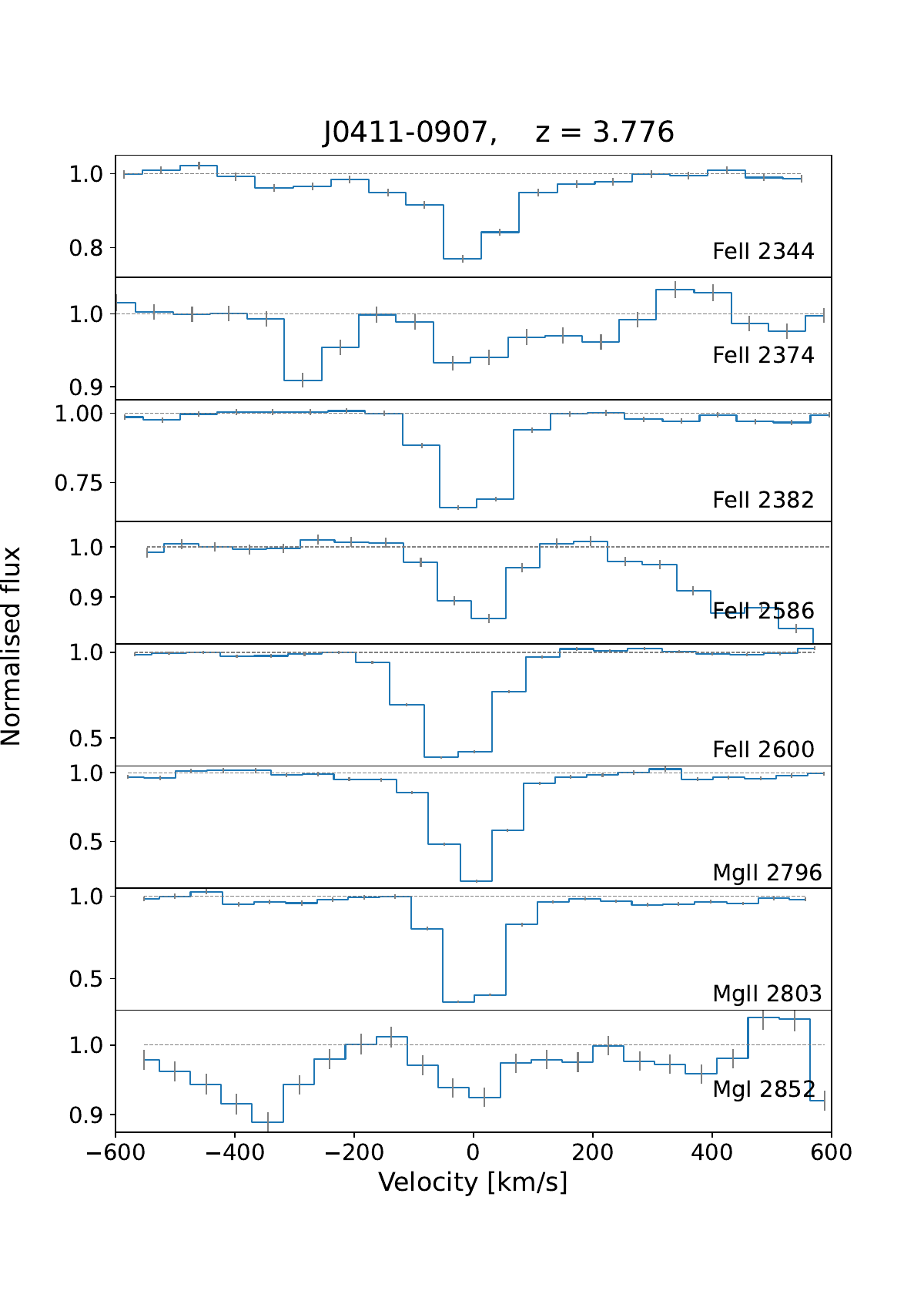}
    \caption{Absorption system at $z=3.776$ toward J0411--0907. }
    \label{fig:J0411_abs3.776}
\end{figure}
\begin{table}
\centering
\caption{J0411--0907, $z=3.776$}       
\label{table:J0411_3.776}   
\begin{tabular}{cccc}   
\hline\hline
\noalign{\smallskip}
$\lambda_{\mathrm{obs}}$ & ID & $W_{\mathrm{obs}}$ \\   
\noalign{\smallskip} \hline \noalign{\smallskip}
11195.97 & \ion{Fe}{ii}\ 2344.21 &  1.418$\pm$0.061\\
11340.43 & \ion{Fe}{ii}\ 2374.46 &  0.475$\pm$0.062\\
11380.09 & \ion{Fe}{ii}\ 2382.76 &  1.948$\pm$0.053\\
12353.84 & \ion{Fe}{ii}\ 2586.65 &  0.694$\pm$0.060\\
12418.43 & \ion{Fe}{ii}\ 2600.17 &  4.225$\pm$0.054$^\dagger$ \\
13355.38 & \ion{Mg}{ii}\ 2796.35 &  4.777$\pm$0.079\\
13389.66 & \ion{Mg}{ii}\ 2803.53 &  3.952$\pm$0.072\\
13625.76 & \ion{Mg}{i}\ 2852.96 &  0.532$\pm$0.095\\
\noalign{\smallskip}
\hline               
\end{tabular}
\begin{tablenotes}
\item $^\dagger$ Line blended with \ion{Mg}{ii} $\lambda$2803 at $z$=3.429
\end{tablenotes}
\end{table}

\newpage
\mbox{~}
\clearpage 

\begin{figure}
   \centering
   \includegraphics[width=\hsize]{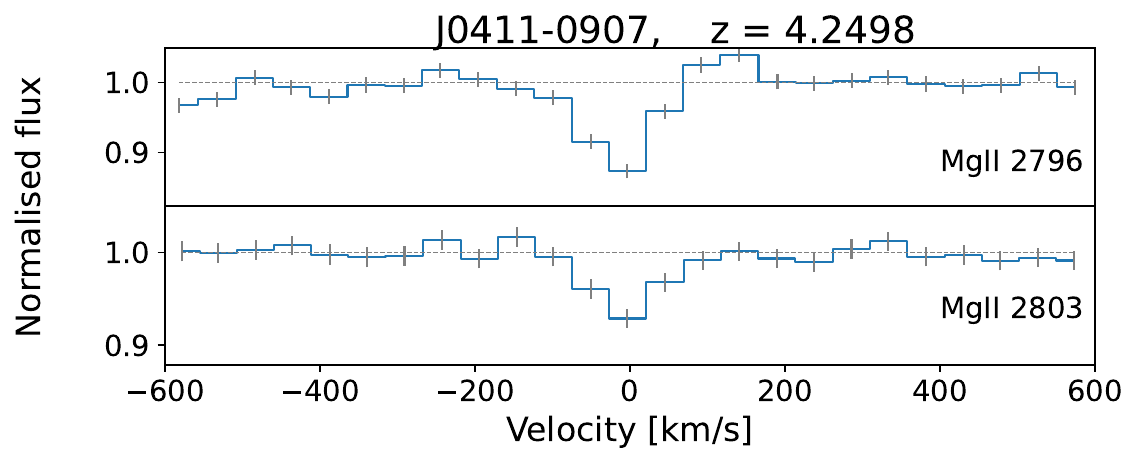}
    \caption{Absorption system at $z=4.2498$ toward J0411--0907. }
    \label{fig:J0411_abs4.2498}
\end{figure}
\begin{table}
\centering
\caption{J0411--0907, $z=4.2498$}       
\label{table:J0411_4.2498} 
\begin{tabular}{cccc}   
\hline\hline
\noalign{\smallskip}
$\lambda_{\mathrm{obs}}$ & ID & $W_{\mathrm{obs}}$ \\   
\noalign{\smallskip} \hline \noalign{\smallskip}         
14680.29 & \ion{Mg}{ii}\ 2796.35 &  0.493$\pm$0.075\\
14717.98 & \ion{Mg}{ii}\ 2803.53 &  0.359$\pm$0.074\\
\noalign{\smallskip}
\hline                           
\end{tabular}
\end{table}

\begin{figure}[t!]
   \centering
   \includegraphics[clip, trim=0.cm 7cm 0cm 7.2cm,width=0.95\hsize]{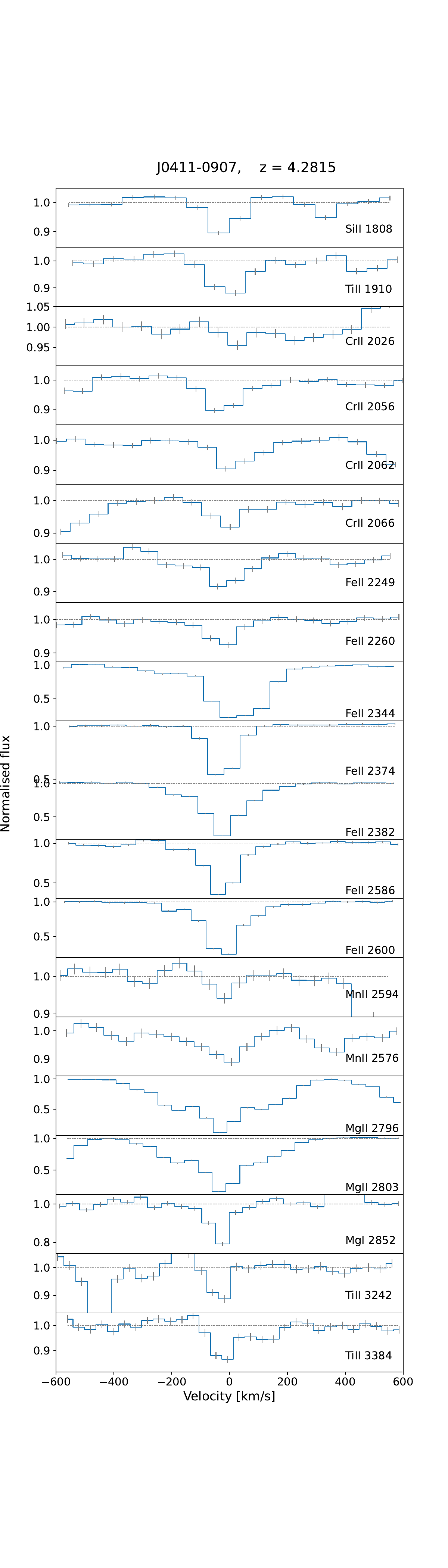}
    \caption{Absorption system at $z=4.2815$ toward J0411--0907. }
    \label{fig:J0411_abs4.815}
\end{figure}
\begin{table}
\centering
\caption{J0411--0907, $z=4.2815$}       
\label{table:J0411_4.2815}  
\begin{tabular}{cccc}   
\hline\hline
\noalign{\smallskip}
$\lambda_{\mathrm{obs}}$ & ID & $W_{\mathrm{obs}}$ \\   
\noalign{\smallskip} \hline \noalign{\smallskip}
9549.02 & \ion{Si}{ii}\ 1808.01 &  0.261$\pm$0.053\\
10091.63 & \ion{Ti}{ii}\ 1910.75 &  0.543$\pm$0.083\\
10701.74 & \ion{Cr}{ii}\ 2026.27 &  0.365$\pm$0.086\\
10860.10 & \ion{Cr}{ii}\ 2056.25 &  0.590$\pm$0.068\\
10891.69 & \ion{Cr}{ii}\ 2062.23 &  0.598$\pm$0.067\\
10912.43 & \ion{Cr}{ii}\ 2066.16 &  0.465$\pm$0.069\\
11882.72 & \ion{Fe}{ii}\ 2249.88 &  0.453$\pm$0.070\\
11940.31 & \ion{Fe}{ii}\ 2260.78 &  0.445$\pm$0.065\\
12380.97 & \ion{Fe}{ii}\ 2344.21 &  8.400$\pm$0.065$^{\dagger}$ \\
12540.72 & \ion{Fe}{ii}\ 2374.46 &  2.378$\pm$0.072$^{\ddagger}$ \\
12584.57 & \ion{Fe}{ii}\ 2382.76 &  6.032$\pm$0.063\\
13661.39 & \ion{Fe}{ii}\ 2586.65 &  3.979$\pm$0.117\\
13732.81 & \ion{Fe}{ii}\ 2600.17 &  6.267$\pm$0.105\\
13702.85 & \ion{Mn}{ii}\ 2594.50 &  0.139$\pm$0.118\\
13609.78 & \ion{Mn}{ii}\ 2576.88 &  0.983$\pm$0.113\\
14768.93 & \ion{Mg}{ii}\ 2796.35 &  13.433$\pm$0.066\\
14806.85 & \ion{Mg}{ii}\ 2803.53 &  10.878$\pm$0.073\\
15067.93 & \ion{Mg}{i}\  2852.96 &  0.875$\pm$0.094\\
17127.53 & \ion{Ti}{II}\ 3242.93 &  0.210$\pm$0.131\\
17876.50 & \ion{Ti}{ii}\ 3384.74 &  0.818$\pm$0.136\\
\noalign{\smallskip}
\hline                           
\end{tabular}
\begin{tablenotes}
\item $^{\dagger}$ Line blended with \ion{Mg}{ii} $\lambda$2796 at $z=3.429$
\item $^{\ddagger}$ Line blended with \ion{Si}{ii} $\lambda$1808 at $z=5.935$
\end{tablenotes}
\end{table}

\newpage
\mbox{~}
\clearpage

\begin{figure}
   \centering
   \includegraphics[clip, trim=0.cm 0.cm 0cm 0cm, width=\hsize]{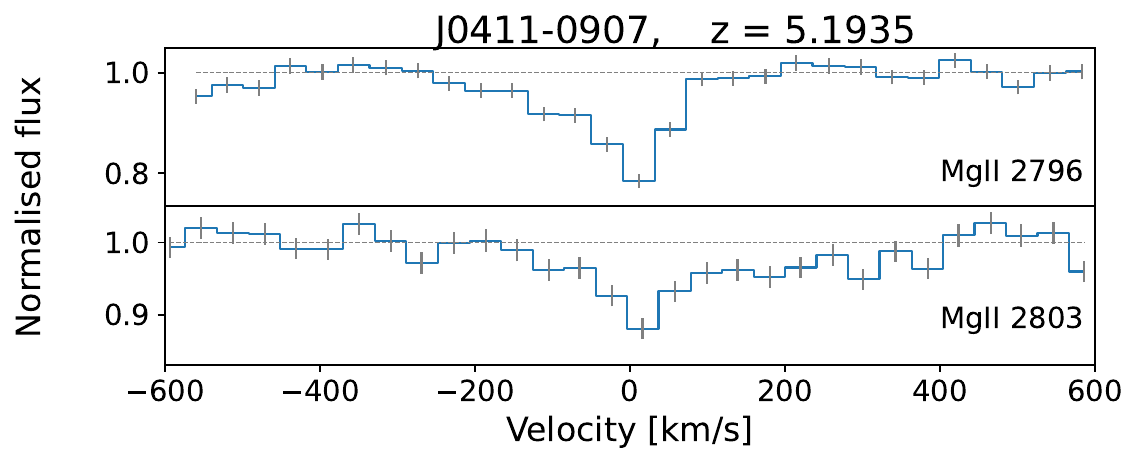}
    \caption{Absorption system at $z=5.1935$ toward J0411--0907. }
    \label{fig:J0411_abs5.1935}
\end{figure}
\begin{table}
\centering
\caption{J0411--0907, $z=5.1935$}       
\label{table:J0411_5.1935} 
\begin{tabular}{c c c c }   
\hline\hline
\noalign{\smallskip}
$\lambda_{\mathrm{obs}}$ & ID & $W_{\mathrm{obs}}$ \\   
\noalign{\smallskip} \hline \noalign{\smallskip}
17319.21 & \ion{Mg}{ii} 2796.35 & 1.740$\pm$ 0.108 \\
17363.67 & \ion{Mg}{ii} 2803.53 & 1.107$\pm$0.112 \\
\noalign{\smallskip}
\hline                           
\end{tabular}
\end{table}

\begin{figure}
   \centering
   \includegraphics[width=\hsize]{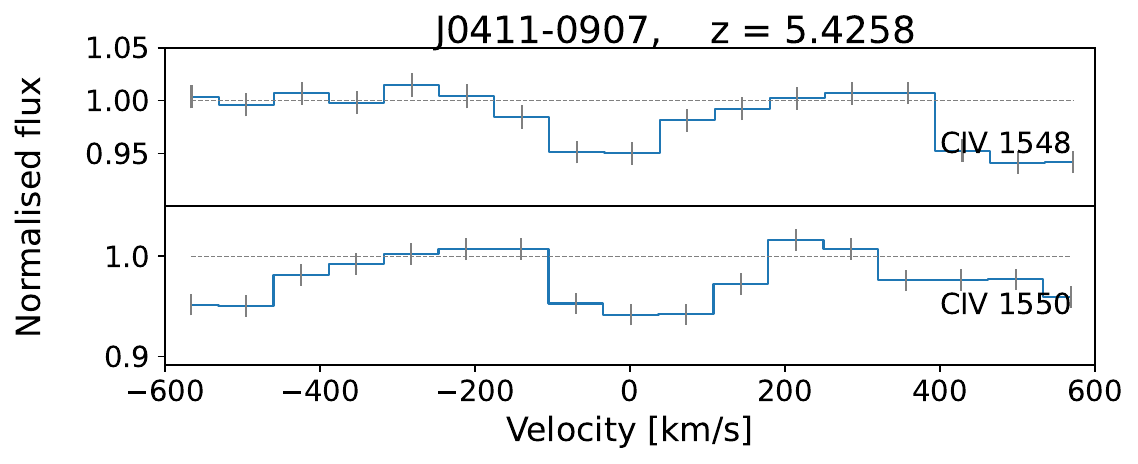}
    \caption{Absorption system at $z=5.4258$ toward J0411--0907. }
    \label{fig:J0411_abs5.4258}
\end{figure}
\begin{table}
\centering
\caption{J0411--0907, $z=5.4258$}       
\label{table:J0411_5.4258} 
\begin{tabular}{c c c c }   
\hline\hline
\noalign{\smallskip}
$\lambda_{\mathrm{obs}}$ & ID & $W_{\mathrm{obs}}$ \\   
\noalign{\smallskip} \hline \noalign{\smallskip}
9948.39 & \ion{C}{iv} 1548.19 & 0.329$\pm$0.057\\ 
9964.94 & \ion{C}{iv} 1550.77 & 0.434$\pm$0.056 \\
\noalign{\smallskip}
\hline                           
\end{tabular}
\end{table}

\begin{figure}
   \centering
   \includegraphics[clip, trim=0.cm 4cm 0cm 4.5cm, width=\hsize]{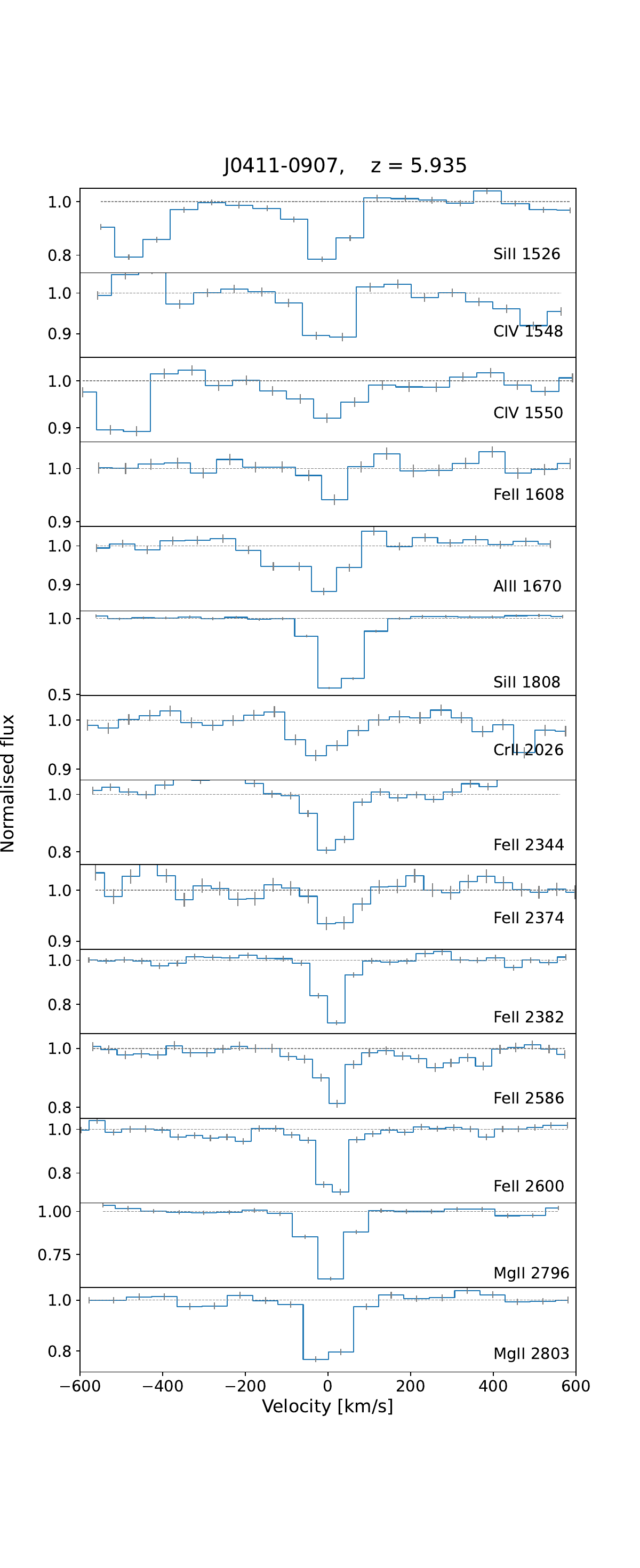}
    \caption{Absorption system at $z=5.935$ toward J0411--0907. }
    \label{fig:J0411_abs5.935}
\end{figure}
\begin{table}
\centering
\caption{J0411--0907, $z=5.935$}       
\label{table:J0411_5.935} 
\begin{tabular}{c c c c }   
\hline\hline
\noalign{\smallskip}
$\lambda_{\mathrm{obs}}$ & ID & $W_{\mathrm{obs}}$ \\
\noalign{\smallskip} \hline \noalign{\smallskip}
10587.71 & \ion{Si}{ii}\ 1526.71 &  0.990$\pm$0.068\\
10736.73 & \ion{C}{iv}\ 1548.19 &  0.464$\pm$0.061\\
10754.59 & \ion{C}{iv}\ 1550.77 &  0.491$\pm$0.065\\
11154.61 & \ion{Fe}{ii}\ 1608.45 &  0.084$\pm$0.061$^{\dagger}$ \\
11586.91 & \ion{Al}{ii}\ 1670.79 &  0.606$\pm$0.066\\
12538.57 & \ion{Si}{ii}\ 1808.01 &  2.485$\pm$0.056$^{\ddagger}$ \\
14052.18 & \ion{Cr}{ii}\ 2026.27 &  0.361$\pm$0.075\\
16257.12 & \ion{Fe}{ii}\ 2344.21 &  0.974$\pm$0.090\\
16466.89 & \ion{Fe}{ii}\ 2374.46 &  0.370$\pm$0.096\\
16524.48 & \ion{Fe}{ii}\ 2382.76 &  1.182$\pm$0.097\\
17938.42 & \ion{Fe}{ii}\ 2586.65 &  1.068$\pm$0.108\\
18032.20 & \ion{Fe}{ii}\ 2600.17 &  1.655$\pm$0.107\\
19392.70 & \ion{Mg}{ii}\ 2796.35 &  2.620$\pm$0.130\\
19442.49 & \ion{Mg}{ii}\ 2803.53 &  1.823$\pm$0.123\\
\noalign{\smallskip}
\hline  
\end{tabular}
\begin{tablenotes}
\item $^{\dagger}$ Line blended with \ion{Mg}{ii} $\lambda$2803 at $z=2.9791$
\item $^{\ddagger}$ Line blended with \ion{Fe}{ii} $\lambda$2374 at $z=4.2815$
\end{tablenotes}
\end{table}

\FloatBarrier

\begin{figure}
   \centering
   \includegraphics[clip, trim=0.cm 3cm 0cm 2.cm, width=\hsize]{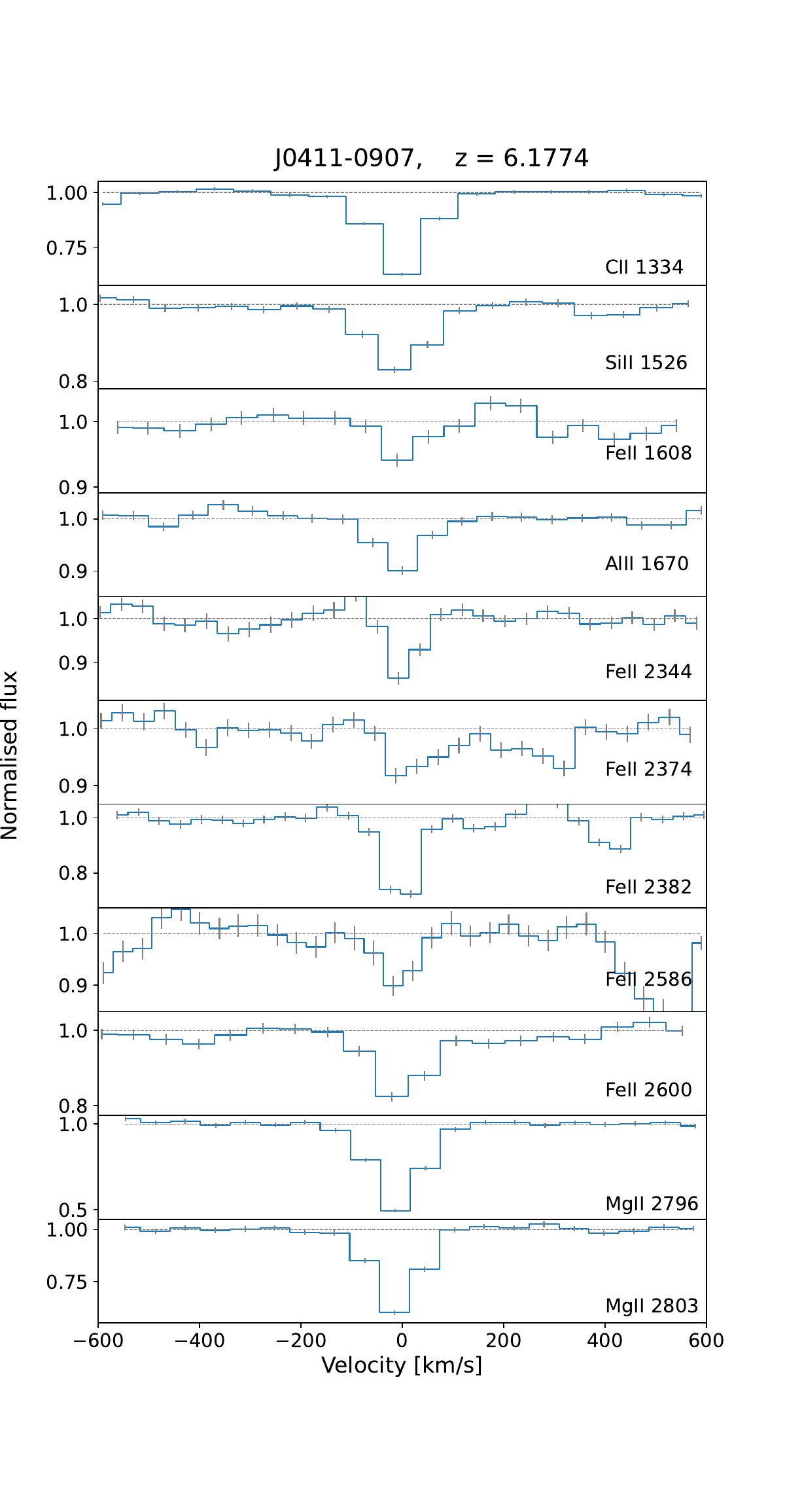}
    \caption{Absorption system at $z=6.1774$ toward J0411--0907. }
    \label{fig:J0411_abs6.1774}
\end{figure}
\begin{table}
\centering
\caption{J0411--0907, $z=6.1774$}       
\label{table:J0411_6.1774} 
\begin{tabular}{cccc}   
\hline\hline
\noalign{\smallskip}
$\lambda_{\mathrm{obs}}$ & ID & $W_{\mathrm{obs}}$  \\   
\noalign{\smallskip} \hline \noalign{\smallskip}
9578.47 & \ion{C}{ii}\ 1334.53 &  1.541$\pm$0.040\\
10957.78 & \ion{Si}{ii}\ 1526.71 &  0.912$\pm$0.057\\
11544.50 & \ion{Fe}{ii}\ 1608.45 &  0.135$\pm$0.066\\
11991.91 & \ion{Al}{ii}\ 1670.79 &  0.416$\pm$0.053\\
16825.36 & \ion{Fe}{ii}\ 2344.21 &  0.256$\pm$0.116\\
17042.46 & \ion{Fe}{ii}\ 2374.46 &  0.670$\pm$0.102\\
17102.06 & \ion{Fe}{ii}\ 2382.76 &  1.538$\pm$0.111\\
18565.42 & \ion{Fe}{ii}\ 2586.65 &  0.558$\pm$0.174\\
18662.48 & \ion{Fe}{ii}\ 2600.17 &  1.655$\pm$0.133\\
20070.54 & \ion{Mg}{ii}\ 2796.35 &  4.041$\pm$0.128\\
20122.06 & \ion{Mg}{ii}\ 2803.53 &  3.008$\pm$0.132\\
\noalign{\smallskip}
\hline                           
\end{tabular}
\end{table}

\newpage
\mbox{~}
\clearpage

\begin{figure}
   \centering
   \includegraphics[clip, trim=0.cm 0cm 0cm 0cm,width=\hsize]{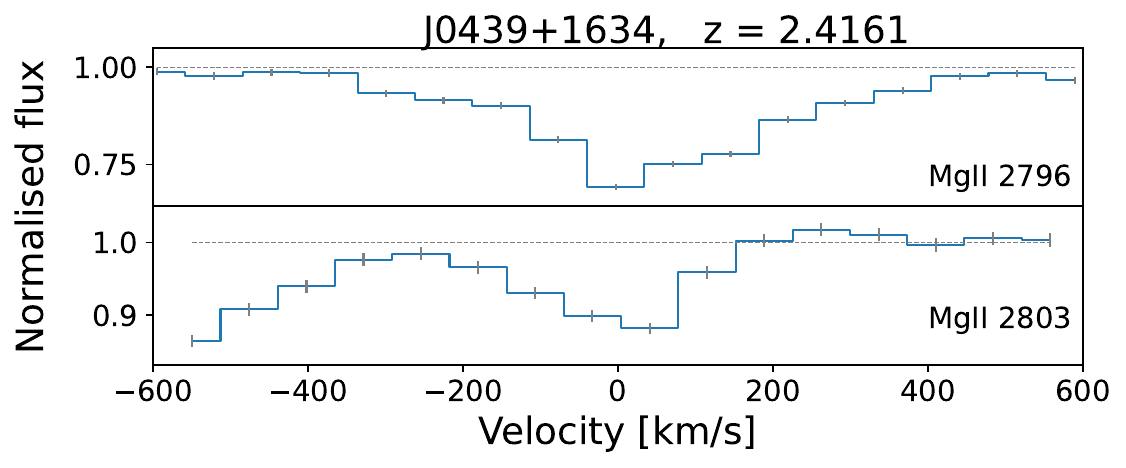}
    \caption{Absorption system at $z=2.4161$ toward J0439+1634. }
    \label{fig:J0429_abs2.4161}
\end{figure}
\begin{table}
\centering
\caption{J0439+1634, $z=2.4161$}       
\label{table:J0439_2.4161} 
\begin{tabular}{ccccl}   
\hline\hline
\noalign{\smallskip}
$\lambda_{\mathrm{obs}}$ & ID & $W_{\mathrm{obs}}$   \\   
\noalign{\smallskip} \hline \noalign{\smallskip}
9552.62 & \ion{Mg}{ii}\ 2796.35 & 3.401$\pm$0.061 \\
9577.14 & \ion{Mg}{ii}\ 2803.53 & 0.843$\pm$0.058 \\
\noalign{\smallskip}
\hline                           
\end{tabular}
\end{table}

\begin{figure}
   \centering
   \includegraphics[clip, trim=0.cm 6cm 0cm 6.2cm,width=\hsize]{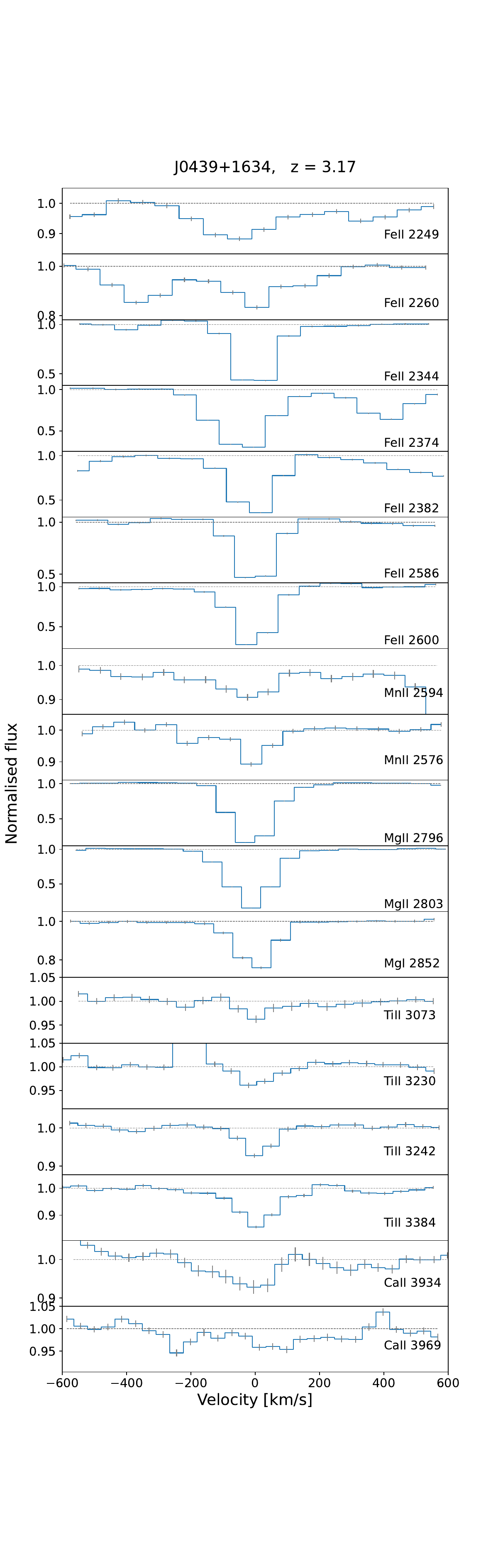}
    \caption{Absorption system at $z=3.170$ toward J0439+1634. }
    \label{fig:J0429_abs3.17}
\end{figure}
\begin{table}
\centering
\caption{J0439+1634,  $z=3.170$}       
\label{table:J0439_3.170} 
\begin{tabular}{cccc}   
\hline\hline
\noalign{\smallskip}
$\lambda_{\mathrm{obs}}$ & ID & $W_{\mathrm{obs}}$ \\
\noalign{\smallskip} \hline \noalign{\smallskip}
9381.99 & \ion{Fe}{ii}\ 2249.88 &  1.041$\pm$0.044\\
9427.45 & \ion{Fe}{ii}\ 2260.78 &  1.163$\pm$0.044\\
9775.37 & \ion{Fe}{ii}\ 2344.21 &  3.118$\pm$0.042\\
9901.50 & \ion{Fe}{ii}\ 2374.46 &  5.010$\pm$0.035$^\ddagger$ \\
9936.13 & \ion{Fe}{ii}\ 2382.76 &  3.662$\pm$0.045\\
10786.33 & \ion{Fe}{ii}\ 2586.65 &  2.915$\pm$0.042\\
10842.72 & \ion{Fe}{ii}\ 2600.17 &  4.057$\pm$0.054\\
10819.06 & \ion{Mn}{ii}\ 2594.50 &  0.764$\pm$0.060\\
10745.58 & \ion{Mn}{ii}\ 2576.88 &  0.489$\pm$0.041\\
11660.79 & \ion{Mg}{ii}\ 2796.35 &  5.468$\pm$0.029\\
11690.72 & \ion{Mg}{ii}\ 2803.53 &  5.410$\pm$0.031\\
11896.86 & \ion{Mg}{i}\ 2852.96 &  1.438$\pm$0.042\\
12818.07 & \ion{Ti}{ii}\ 3073.88 &  0.175$\pm$0.052\\
13469.65 & \ion{Ti}{ii}\ 3230.13 &  0.020$\pm$0.040\\
13523.01 & \ion{Ti}{ii}\ 3242.93 &  0.334$\pm$0.038\\
14114.37 & \ion{Ti}{ii}\ 3384.74 &  1.089$\pm$0.036\\
16407.99 & \ion{Ca}{ii}\ 3934.77 &  0.721$\pm$0.124\\
\noalign{\smallskip}
\hline                           
\end{tabular}
\begin{tablenotes}
\item $^\ddagger$ Line blended with \ion{C}{iv} $\lambda$1548  at $z=5.3945$
\end{tablenotes}
\end{table}

\newpage
\mbox{~}
\clearpage 

\begin{figure}
   \centering
   \includegraphics[width=\hsize]{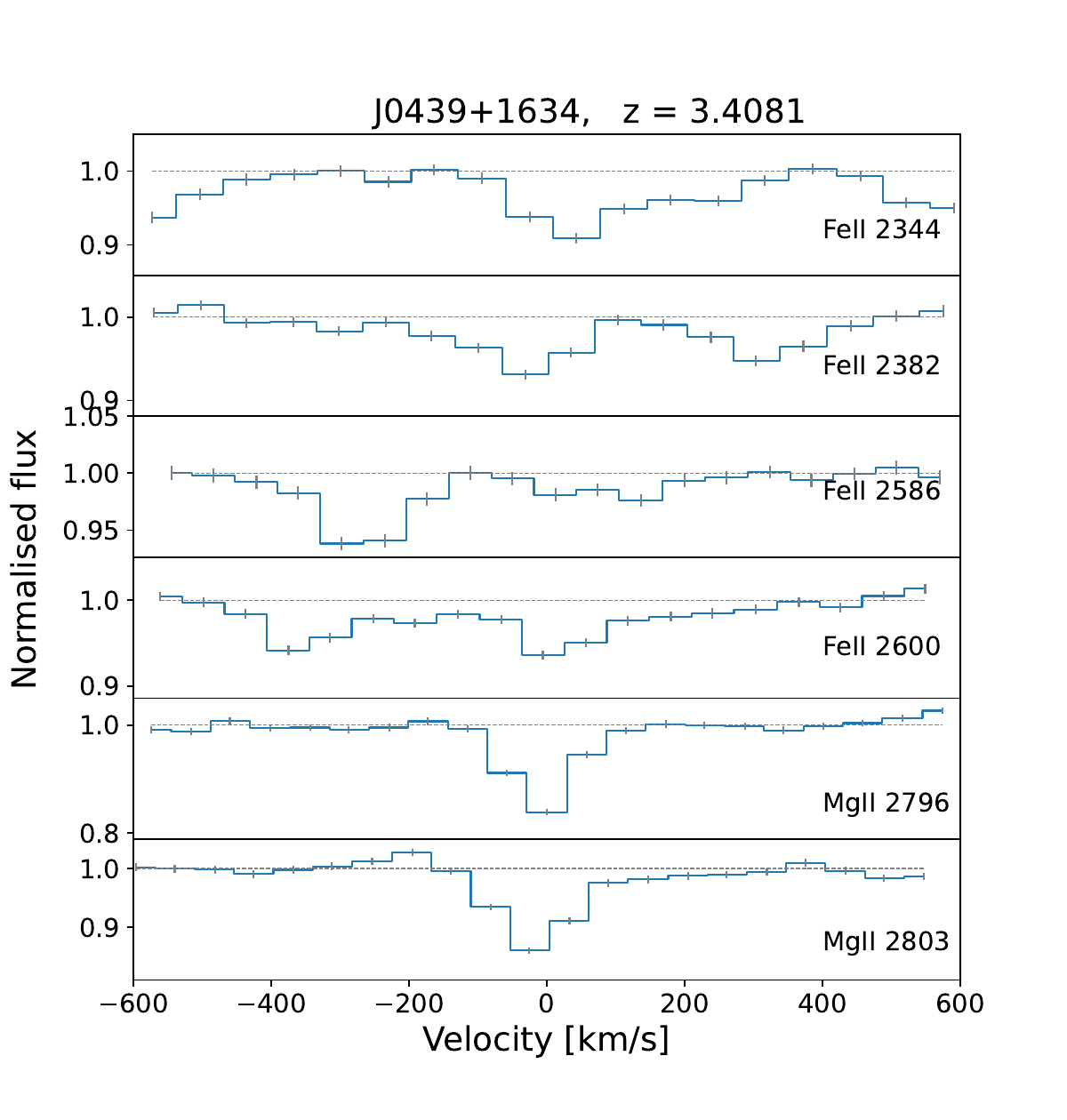}
    \caption{Absorption system at $z=3.4081$ toward J0439+1634. }
    \label{fig:J0439_abs3.4081}
\end{figure}
\begin{table}
\centering
\caption{J0439+1634, $z=3.4081$}       
\label{table:J0439_3.4081} 
\begin{tabular}{cccc}   
\hline\hline
\noalign{\smallskip}
$\lambda_{\mathrm{obs}}$ & ID & $W_{\mathrm{obs}}$  \\   
\noalign{\smallskip} \hline \noalign{\smallskip}
10333.53 & \ion{Fe}{ii}\ 2344.21 &  0.593$\pm$0.043\\
10503.47 & \ion{Fe}{ii}\ 2382.76 &  0.434$\pm$0.035\\
11402.21 & \ion{Fe}{ii}\ 2586.65 &  0.217$\pm$0.036\\
11461.82 & \ion{Fe}{ii}\ 2600.17 &  0.526$\pm$0.034\\
12326.60 & \ion{Mg}{ii}\ 2796.35 &  0.743$\pm$0.042\\
12358.25 & \ion{Mg}{ii}\ 2803.53 &  0.743$\pm$0.037\\
\noalign{\smallskip}
\hline                           
\end{tabular}
\end{table}

\begin{figure}
   \centering
   \includegraphics[clip, trim=0.cm 2cm 0cm 0.7cm, width=\hsize]{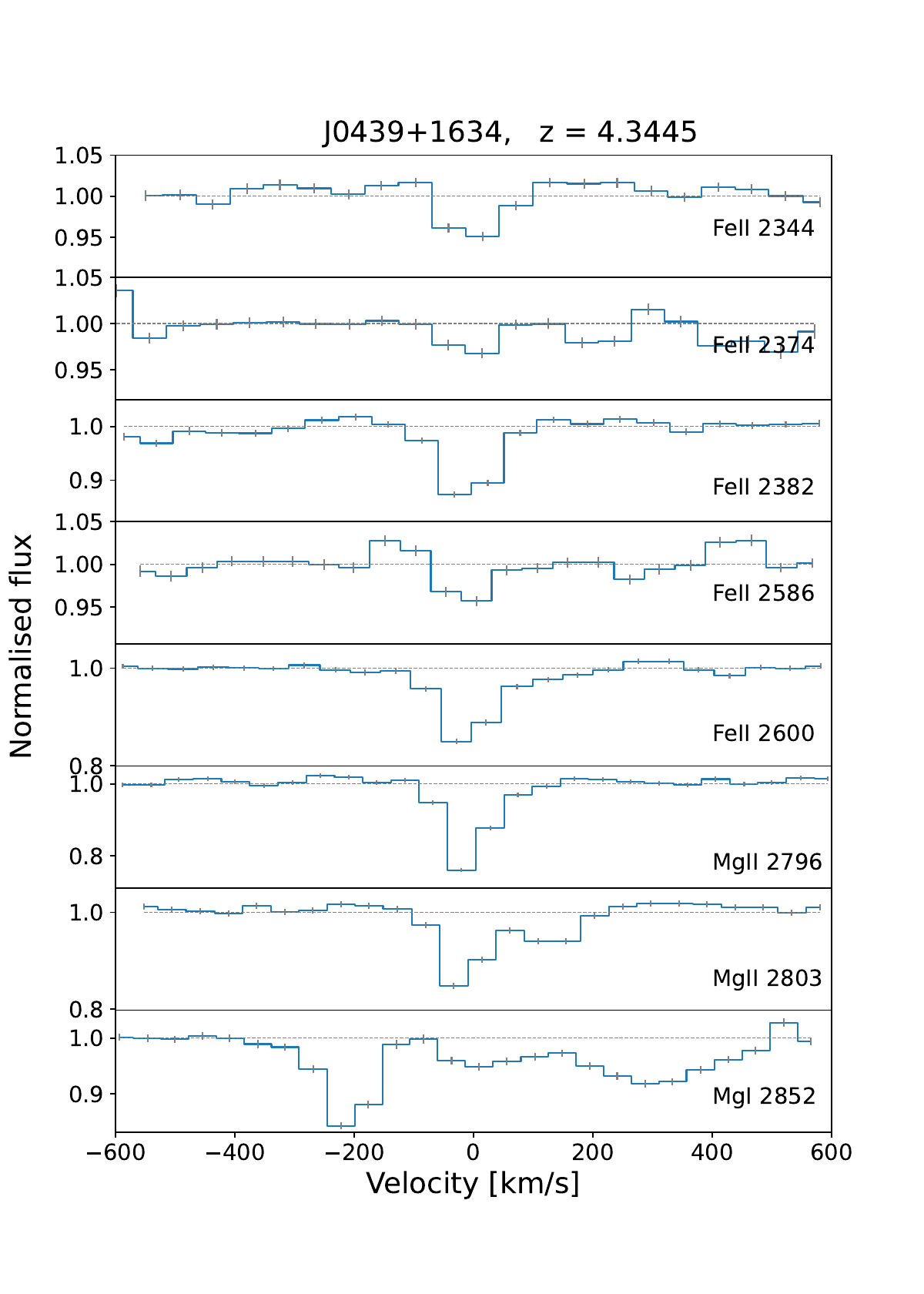}
    \caption{Absorption system at $z=4.3445$ toward J0439+1634. }
    \label{fig:J0439_abs4.3445}
\end{figure}
\begin{table}
\centering
\caption{J0439+1634, $z=4.3445$}       
\label{table:J0411_4.3445} 
\begin{tabular}{cccc}   
\hline\hline
\noalign{\smallskip}
$\lambda_{\mathrm{obs}}$ & ID & $W_{\mathrm{obs}}$  \\   
\noalign{\smallskip} \hline \noalign{\smallskip}
12528.65 & \ion{Fe}{ii}\ 2344.21 &  0.154$\pm$0.031\\
12690.30 & \ion{Fe}{ii}\ 2374.46 &  0.138$\pm$0.030\\ 
12734.69 & \ion{Fe}{ii}\ 2382.76 &  0.595$\pm$0.034\\
13824.35 & \ion{Fe}{ii}\ 2586.65 &  0.098$\pm$0.035\\
13896.62 & \ion{Fe}{ii}\ 2600.17 &  0.874$\pm$0.032\\
14945.10 & \ion{Mg}{ii}\ 2796.35 &  1.052$\pm$0.037\\
14983.47 & \ion{Mg}{ii}\ 2803.53 &  0.756$\pm$0.043\\
15247.67 & \ion{Mg}{i}\ 2852.96 &  0.558$\pm$0.038\\
\noalign{\smallskip}
\hline                           
\end{tabular}
\end{table}

\newpage
\mbox{~}
\clearpage 

\begin{figure}
   \centering
   \includegraphics[width=\hsize]{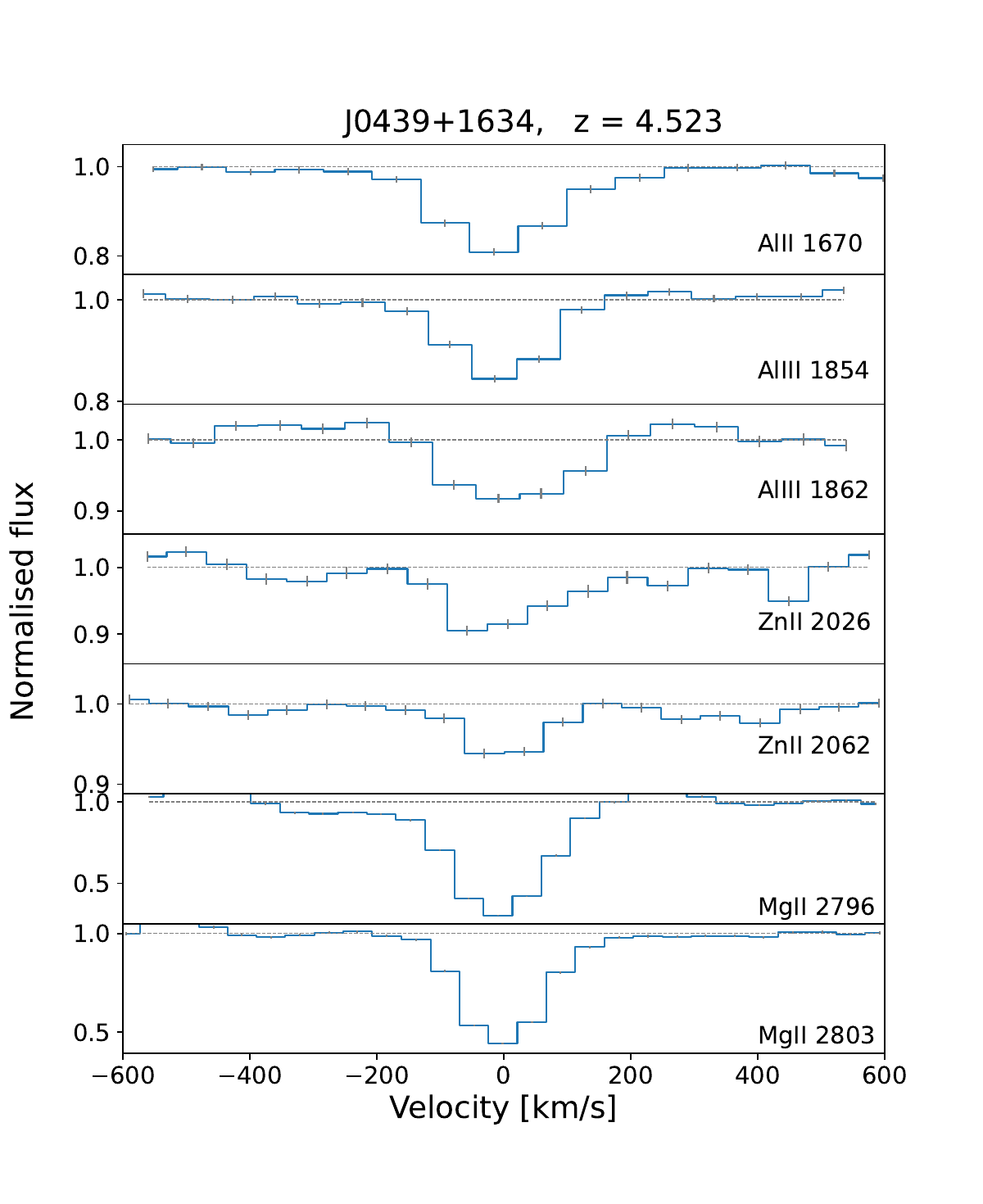}
    \caption{Absorption system at $z=4.523$ toward J0439+1634. }
    \label{fig:J0439_abs4.523}
\end{figure}
\begin{table}
\centering
\caption{J0439+1634,  $z=4.523$}       
\label{table:J0439_4.523} 
\begin{tabular}{cccc}   
\hline\hline
\noalign{\smallskip}
$\lambda_{\mathrm{obs}}$ & ID & $W_{\mathrm{obs}}$  \\   
\noalign{\smallskip} \hline \noalign{\smallskip}
9227.76 & \ion{Al}{ii}\ 1670.79 &  1.304$\pm$0.049\\
10243.60 & \ion{Al}{iii}\ 1854.72 &  0.946$\pm$0.046\\
10288.19 & \ion{Al}{iii}\ 1862.79 &  0.561$\pm$0.044$^{\dagger}$ \\
11190.35 & \ion{Zn}{ii}\ 2026.14 &  0.745$\pm$0.053\\
11392.07 &\ion{Zn}{ii}\ 2062.66 &  0.414$\pm$0.040$^{\ddagger}$ \\
15444.25 & \ion{Mg}{ii}\ 2796.35 &  6.397$\pm$0.042\\
15483.90 & \ion{Mg}{ii}\ 2803.53 &  4.731$\pm$0.047\\
\noalign{\smallskip}
\hline                           
\end{tabular}
\begin{tablenotes}
\item $^{\dagger}$ Line blended with \ion{Fe}{ii} $\lambda$1608 at $z=5.3945$
\item $^{\ddagger}$ Line blended with \ion{Al}{ii} $\lambda$1670 at $z=5.8190$
\end{tablenotes}
\end{table}

\begin{figure}
   \centering
   \includegraphics[width=\hsize]{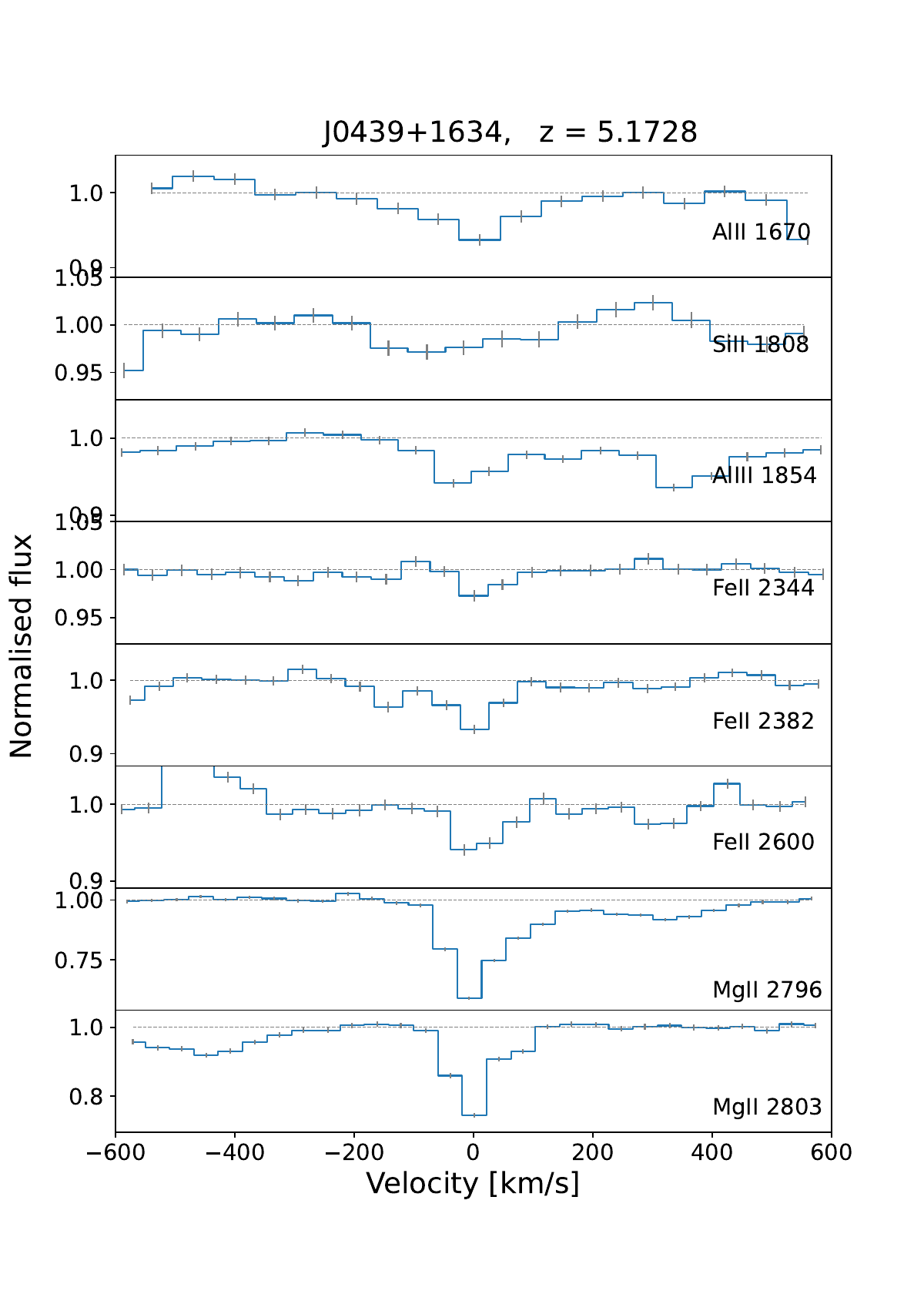}
    \caption{Absorption system at $z=5.1728$ toward J0439+1634. }
    \label{fig:J0439_abs5.1728}
\end{figure}
\begin{table}
\centering
\caption{J0439+1634,  $z=5.1728$}       
\label{table:J0439_5.1728} 
\begin{tabular}{cccc}   
\hline\hline
\noalign{\smallskip}
$\lambda_{\mathrm{obs}}$ & ID & $W_{\mathrm{obs}}$  \\   
\noalign{\smallskip} \hline \noalign{\smallskip}
10313.43 & \ion{Al}{ii} 1670.79 &  0.404$\pm$0.046\\
11160.50 & \ion{Si}{ii}\ 1808.01 &  0.245$\pm$0.047\\
11448.79 & \ion{Al}{iii}\ 1854.72 &  0.399$\pm$0.032\\
14470.36 & \ion{Fe}{ii}\ 2344.21 &  0.139$\pm$0.040\\
14708.33 & \ion{Fe}{ii}\ 2382.76 &  0.502$\pm$0.046\\
16050.35 & \ion{Fe}{ii}\ 2600.17 &  0.382$\pm$0.053\\
17261.32 & \ion{Mg}{ii}\ 2796.35 &  2.945$\pm$0.054\\
17305.64 & \ion{Mg}{ii}\ 2803.53 &  1.278$\pm$0.051\\
\noalign{\smallskip}
\hline                           
\end{tabular}
\end{table}

\newpage
\mbox{~}
\clearpage

\begin{figure}
   \centering
   \includegraphics[width=\hsize]{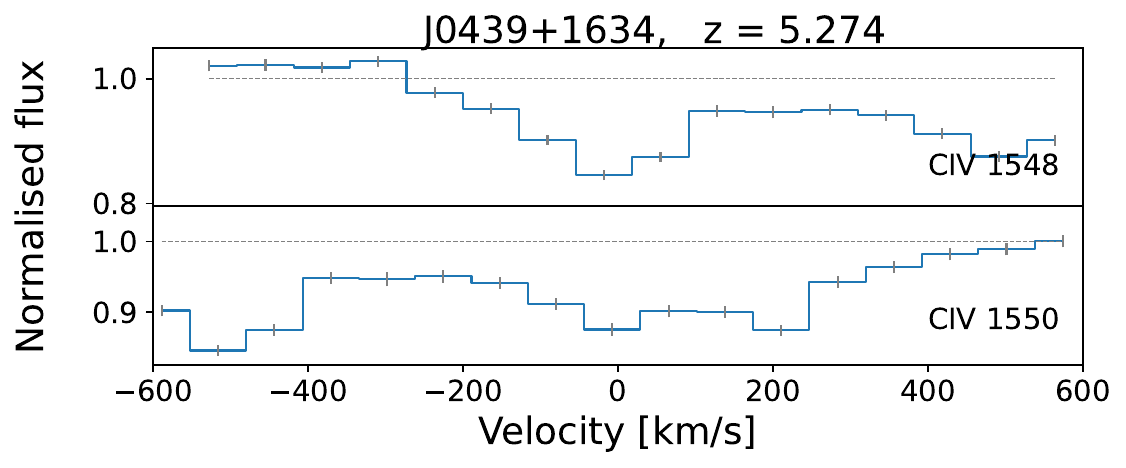}
    \caption{Absorption system at $z=5.274$ toward J0439+1634. }
    \label{fig:J0439_abs5.274}
\end{figure}
\begin{table}
\centering
\caption{J0439+1634, $z=5.274$}       
\label{table:J0439_5.274} 
\begin{tabular}{cccc}   
\hline\hline
\noalign{\smallskip}
$\lambda_{\mathrm{obs}}$ & ID & $W_{\mathrm{obs}}$  \\   
\noalign{\smallskip} \hline \noalign{\smallskip}
9713.38 &\ion{C}{iv}\,1548.19 & 1.124$\pm$0.045 \\
9729.53 &\ion{C}{iv}\,1550.77 & 1.107$\pm$0.046$^\dagger$\\
\noalign{\smallskip}
\hline                           
\end{tabular}
\begin{tablenotes}
\item  $^{\dagger}$ Line blended with \ion{C}{ii} $\lambda$1334 at $z=6.288$
\end{tablenotes}
\end{table}

\begin{figure}
   \centering
   \includegraphics[clip, trim=0.cm 3cm 0cm 4cm, width=\hsize]{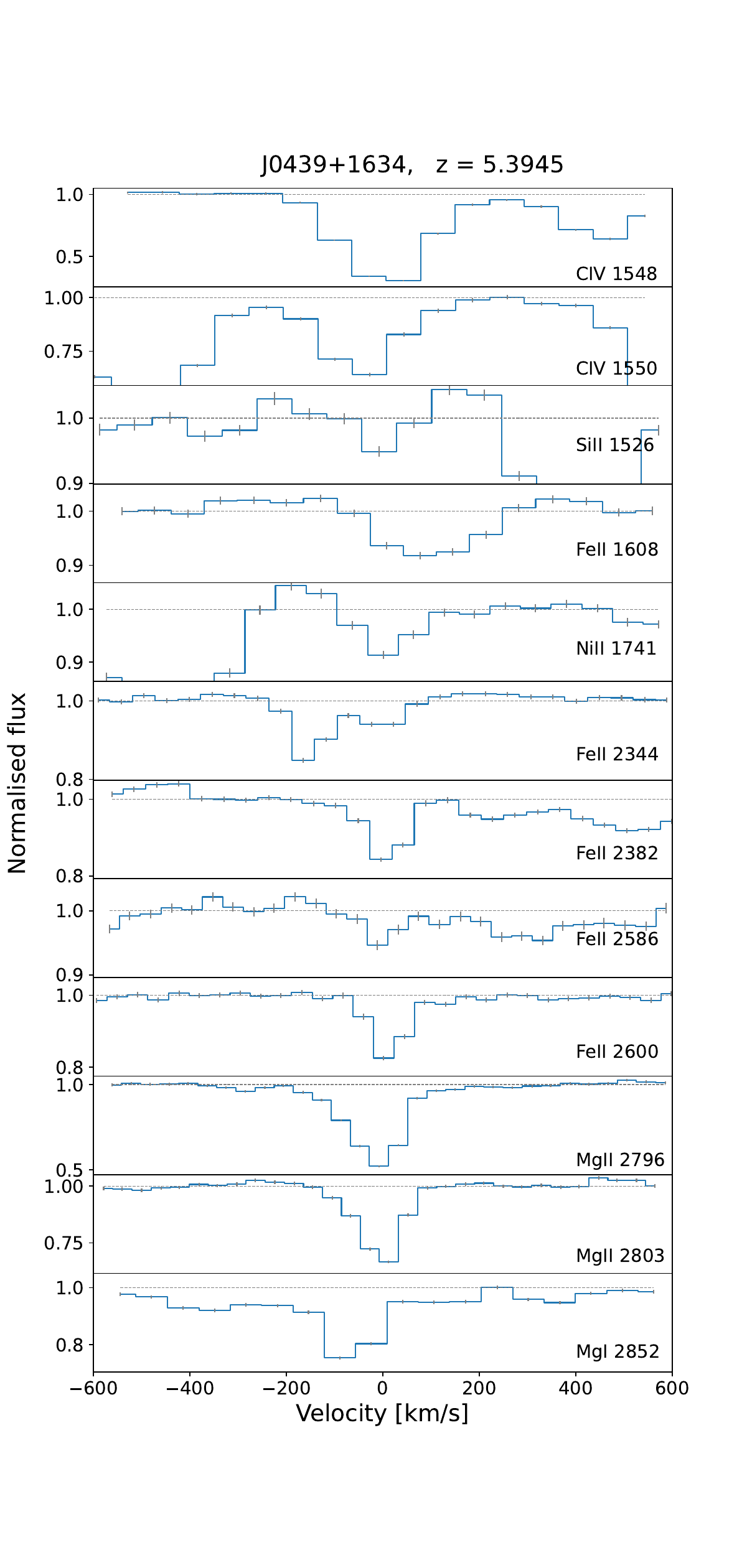}
    \caption{Absorption system at $z=5.3945$ toward J0439+1634. }
    \label{fig:J0439_abs5.3945}
\end{figure}
\begin{table}[h!]
\centering
\caption{J0439+1634,  $z=5.3945$}       
\label{table:J0439_5.3945} 
\begin{tabular}{cccc}   
\hline\hline
\noalign{\smallskip}
$\lambda_{\mathrm{obs}}$ & ID & $W_{\mathrm{obs}}$  \\   
\noalign{\smallskip} \hline \noalign{\smallskip}
9762.53 & \ion{Si}{ii}\ 1526.71 & 0.062$\pm$0.050\\ 
9899.93 & \ion{C}{iv}\ 1548.19 &  5.166$\pm$0.040$^{\dagger}$\\
9916.40 & \ion{C}{iv}\ 1550.77 &  2.331$\pm$0.053\\
10285.24 & \ion{Fe}{ii}\ 1608.45 &  0.437$\pm$0.041$^{\ddagger}$\\
11136.36 & \ion{Ni}{ii}\ 1741.55 &  0.251$\pm$0.052\\
14990.08 & \ion{Fe}{ii}\ 2344.21 &  0.378$\pm$0.032\\
15236.59 & \ion{Fe}{ii}\ 2382.76 &  0.969$\pm$0.050\\
16540.33 & \ion{Fe}{ii}\ 2586.65 &  0.252$\pm$0.053\\
16626.81 & \ion{Fe}{ii}\ 2600.17 &  0.945$\pm$0.053\\
17881.27 & \ion{Mg}{ii}\ 2796.35 &  4.005$\pm$0.054\\
17927.18 & \ion{Mg}{ii}\ 2803.53 &  2.164$\pm$0.055\\
18243.28 & \ion{Mg}{i}\  2852.96 &  2.691$\pm$0.057$^{\mathsection}$\\ 
\noalign{\smallskip}
\hline                           
\end{tabular}
\begin{tablenotes}
\item  $^{\dagger}$ Line blended with \ion{Fe}{ii} $\lambda$2374 at $z=3.170$
\item  $^{\ddagger}$ Line blended with \ion{Al}{iii} $\lambda$1862 at $z=4.523$
\item  $^{\mathsection}$ Line blended with \ion{Mg}{ii} $\lambda$2797 at $z=5.5228$
\end{tablenotes}
\end{table}

\newpage
\mbox{~}
\clearpage

\begin{figure}
   \centering
   \includegraphics[clip, trim=0.cm 2cm 0cm 3cm, width=\hsize]{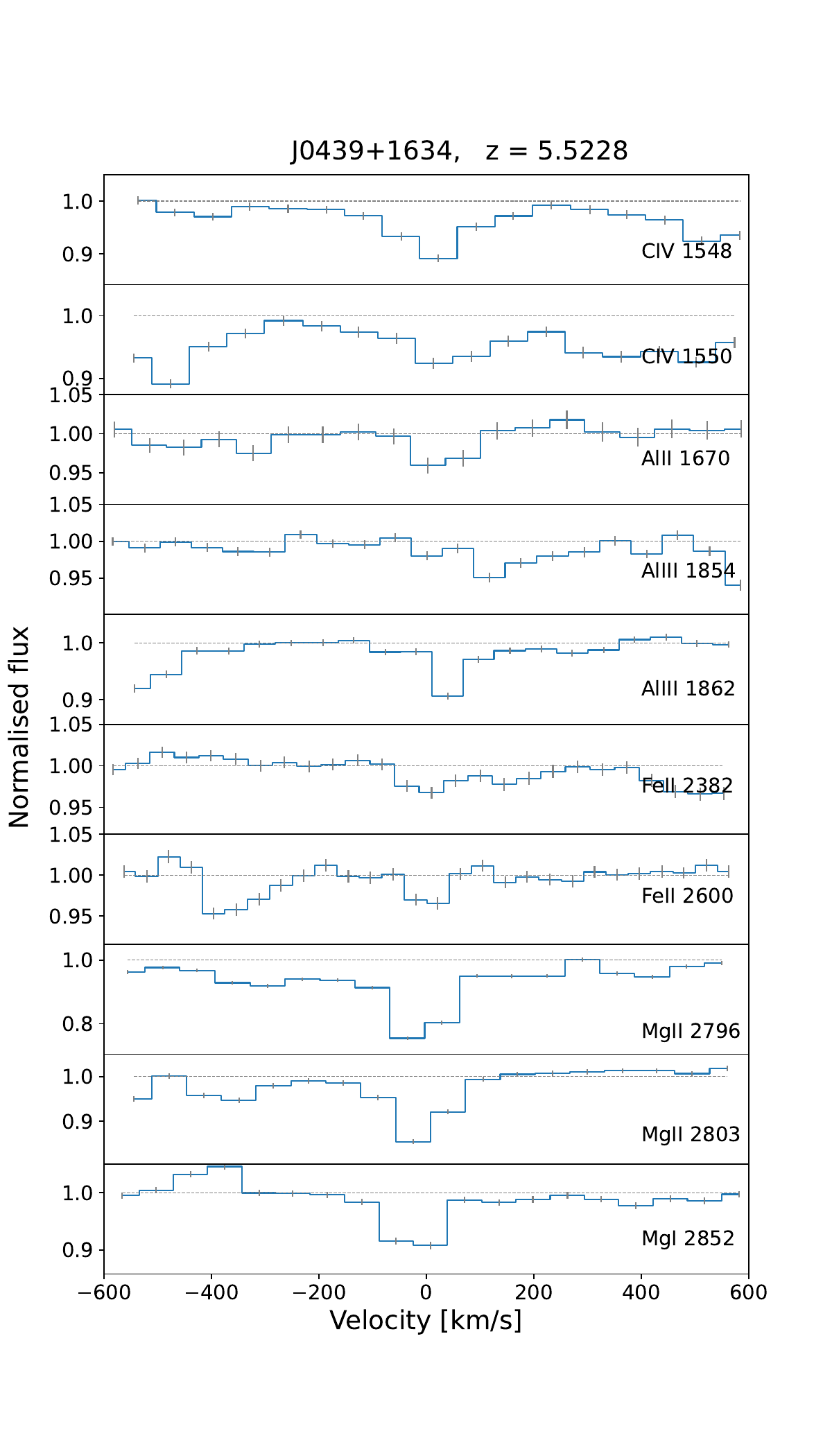}
    \caption{Absorption system at $z=5.5228$ toward J0439+1634. }
    \label{fig:J0439_abs5.5228}
\end{figure}
\begin{table}[h!]
\centering
\caption{J0439+1634, $z=5.5228$}       
\label{table:J0439_5.5228} 
\begin{tabular}{cccc}   
\hline\hline
\noalign{\smallskip}
$\lambda_{\mathrm{obs}}$ & ID & $W_{\mathrm{obs}}$  \\   
\noalign{\smallskip} \hline \noalign{\smallskip}
10098.57 & \ion{C}{iv}\,1548.19 & 0.702$\pm$0.043\\
10115.36 & \ion{C}{iv}\,1550.77 & 0.612$\pm$0.050\\
10898.21 & \ion{Al}{ii}\,1670.79 & 0.152$\pm$0.067\\
12097.94 & \ion{Al}{iii}\,1854.72 & 0.260$\pm$0.039\\
12150.60 & \ion{Al}{iii}\,1862.79 & 0.380$\pm$0.037\\
15542.30 & \ion{Fe}{ii}\,2382.76 & 0.268$\pm$0.052\\
16960.41 & \ion{Fe}{ii}\,2600.17 & 0.126$\pm$0.055 \\
17881.27 & \ion{Mg}{ii}\,2796.35 & 2.737$\pm$0.057$^\dagger$\\
17927.18 & \ion{Mg}{ii}\,2803.53 & 1.131$\pm$0.054\\
18243.28 & \ion{Mg}{i}\,2852.96  & 0.942$\pm$0.061\\ 
\noalign{\smallskip}
\hline                           
\end{tabular}
\begin{tablenotes}
\item  $^{\dagger}$ Line blended with \ion{Mg}{i} $\lambda$2852 at $z=5.3945$
\end{tablenotes}
\end{table}

\FloatBarrier

\begin{figure}
   \centering
   \includegraphics[width=\hsize]{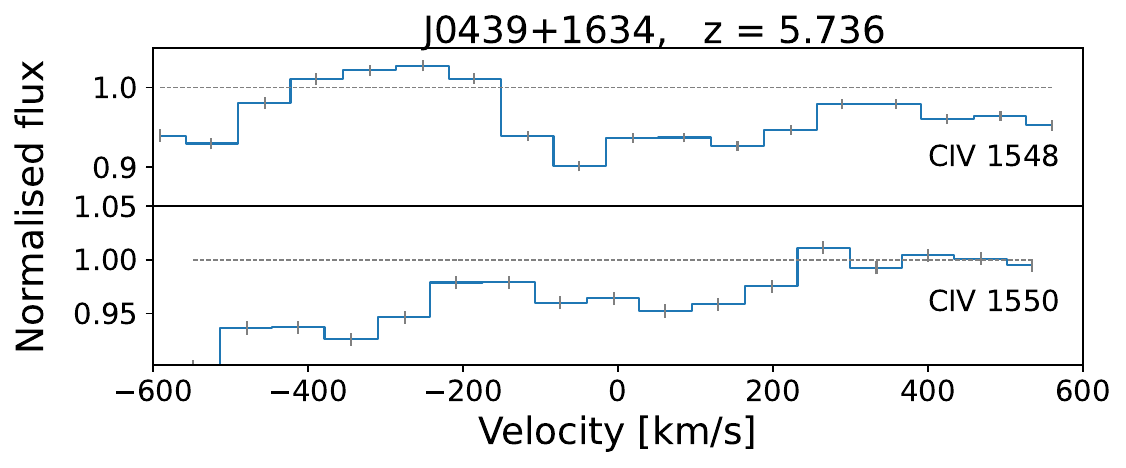}
    \caption{Absorption system at $z=5.736$ toward J0439+1634. }
    \label{fig:J0439_abs5.736}
\end{figure}
\begin{table}[h!]
\centering
\caption{J0439+1634, $z=5.736$}       
\label{table:J0439_5.736} 
\begin{tabular}{cccc}   
\hline\hline
\noalign{\smallskip}
$\lambda_{\mathrm{obs}}$ & ID & $W_{\mathrm{obs}}$  \\   
\noalign{\smallskip} \hline \noalign{\smallskip}
10428.64 & \ion{C}{iv}\ 1548.19 & 0.826$\pm$0.037\\
10445.99 & \ion{C}{iv}\ 1550.77 & 0.496$\pm$0.036\\
\noalign{\smallskip}
\hline                           
\end{tabular}
\end{table}

\FloatBarrier

\newpage
\mbox{~}
\clearpage

\begin{figure}[!h]
   \centering
   \includegraphics[clip, trim=0.cm 2.5cm 0cm 3cm, width=\hsize]{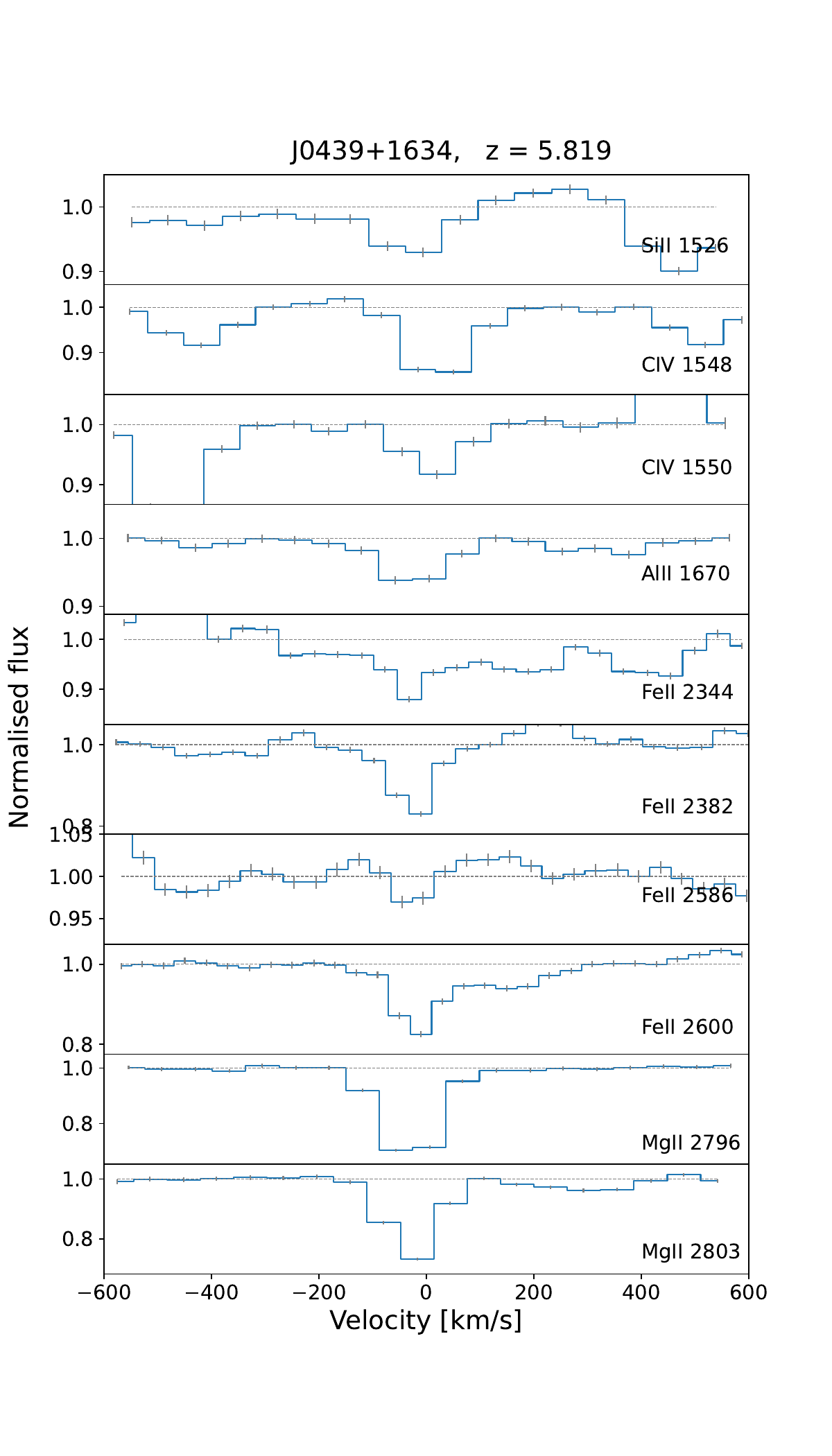}
    \caption{Absorption system at $z=5.819$ toward J0439+1634. }
    \label{fig:J0439_abs5.819}
\end{figure}
\begin{table}[h!]
\centering
\caption{J0439+1634,  $z=5.819$}       
\label{table:J0439_5.819} 
\begin{tabular}{cccc}   
\hline\hline
\noalign{\smallskip}
$\lambda_{\mathrm{obs}}$ & ID & $W_{\mathrm{obs}}$ \\
\noalign{\smallskip}
\hline     
\noalign{\smallskip}
10410.61 & \ion{Si}{ii}\ 1526.71 &  0.323$\pm$0.043\\
10557.14 & \ion{C}{iv}\ 1548.19 &  0.757$\pm$0.037\\
10574.70 & \ion{C}{iv}\ 1550.77 &  0.387$\pm$0.043$^{\dagger}$\\
11393.10 & \ion{Al}{ii}\ 1670.79 &  0.409$\pm$0.037$^{\ddagger}$ \\
15985.20 & \ion{Fe}{ii}\ 2344.21 &  1.265$\pm$0.047\\
16248.07 & \ion{Fe}{ii}\ 2382.76 &  0.891$\pm$0.049\\
17638.37 & \ion{Fe}{ii}\ 2586.65 &  0.220$\pm$0.061\\
17730.58 & \ion{Fe}{ii}\ 2600.17 &  1.582$\pm$0.061\\
19068.32 & \ion{Mg}{ii}\ 2796.35 &  2.860$\pm$0.072\\
19117.28 & \ion{Mg}{ii}\ 2803.53 &  2.062$\pm$0.061\\
\noalign{\smallskip}
\hline                           
\end{tabular}
\begin{tablenotes}
\item $^{\dagger}$ Line blended with \ion{Si}{ii} $\lambda$1526 at $z=5.928$
\item $^{\ddagger}$ Line blended with \ion{Zn}{ii} $\lambda$2062 at $z=4.523$
\end{tablenotes}
\end{table}

\begin{figure}
   \centering
   \includegraphics[width=\hsize]{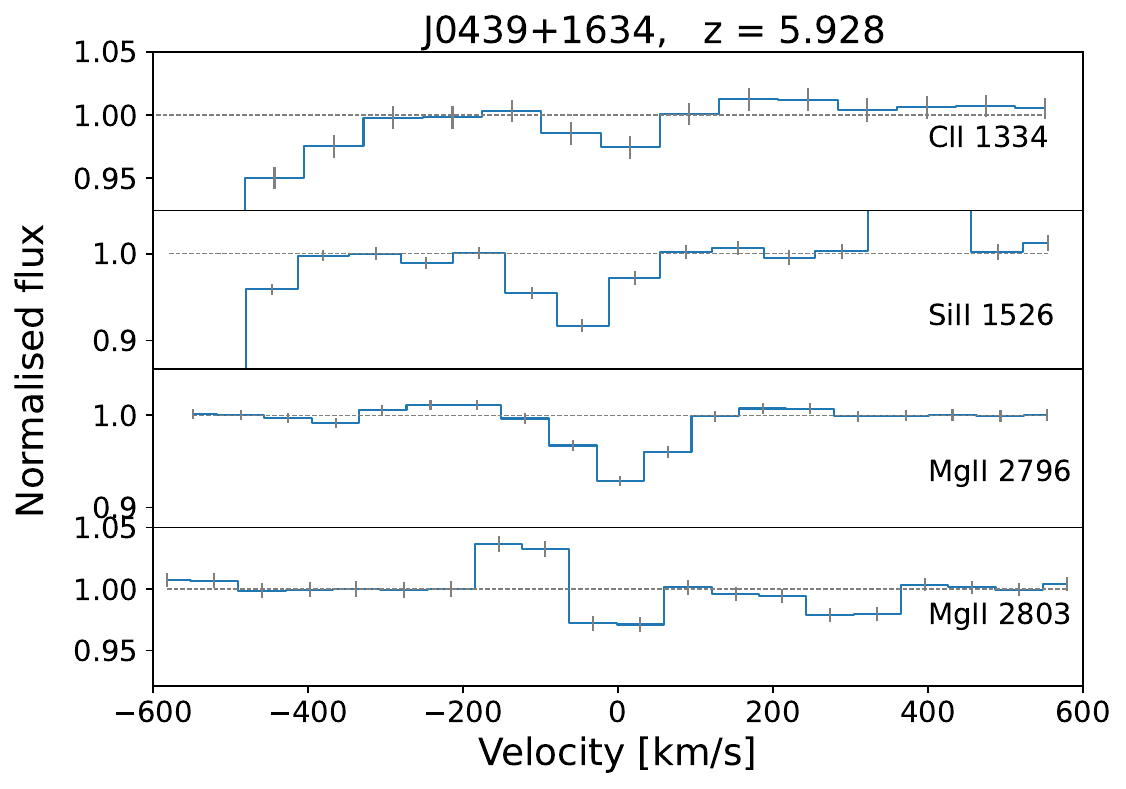}
    \caption{Absorption system at $z=5.928$ toward J0439+1634. }
    \label{fig:J0439_abs5.928}
\end{figure}
\begin{table}[h!]
\centering
\caption{J0439+1634,  $z=5.928$}       
\label{table:J0439_5.928} 
\begin{tabular}{cccc}   
\hline\hline
\noalign{\smallskip}
$\lambda_{\mathrm{obs}}$ & ID & $W_{\mathrm{obs}}$ \\
\noalign{\smallskip}
\hline     
\noalign{\smallskip}
9245.64 & \ion{C}{ii}\ 1334.53 & 0.055$\pm$0.047 \\
10577.02& \ion{Si}{ii} 1526.71 & 0.346$\pm$0.044$^\dagger$ \\
19373.13& \ion{Mg}{ii} 2796.35 & 0.518$\pm$0.061 \\
19422.86& \ion{Mg}{ii} 2803.53 &0.206$\pm$0.042\\
\noalign{\smallskip}
\hline                           
\end{tabular}
\begin{tablenotes}
\item $^\dagger$ Line blended with \ion{C}{iv} $\lambda$1550 at $z=5.819$
\end{tablenotes}
\end{table}

\begin{figure}
   \centering
   \includegraphics[width=\hsize]{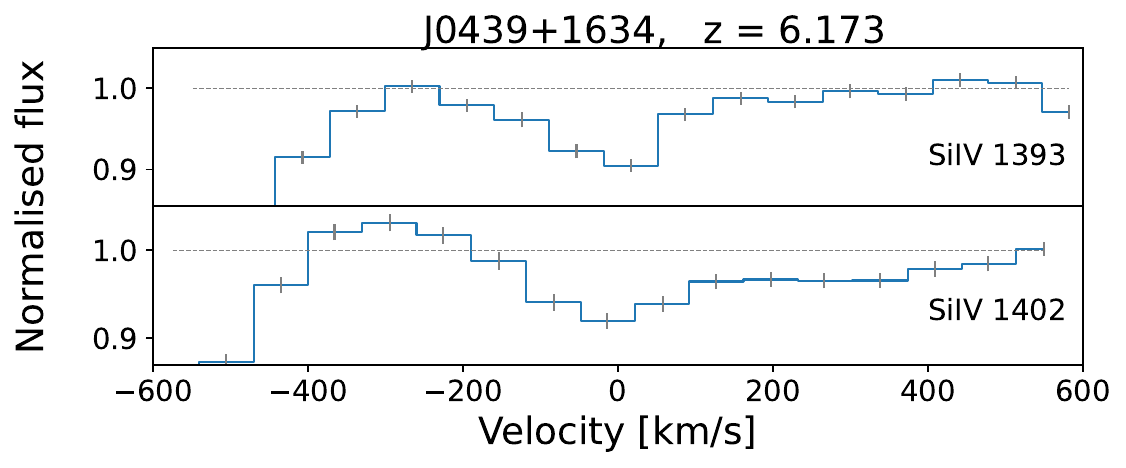}
    \caption{Absorption system at $z=6.173$ toward J0439+1634. }
    \label{fig:J0439_abs6.173}
\end{figure}
\begin{table}[h!]
\centering
\caption{J0439+1634, $z=6.173$}       
\label{table:J0439_6.173} 
\begin{tabular}{cccc}   
\hline\hline
\noalign{\smallskip}
$\lambda_{\mathrm{obs}}$ & ID & $W_{\mathrm{obs}}$  \\   
\noalign{\smallskip} \hline \noalign{\smallskip}
9997.40  & \ion{Si}{iv}\,1393.76 & 0.690$\pm$0.052\\ 
10062.07 & \ion{Si}{iv}\,1402.77 & 0.617$\pm$0.058\\
\noalign{\smallskip}
\hline                           
\end{tabular}
\end{table}

\newpage
\mbox{~}
\clearpage 
\FloatBarrier

\begin{figure}
   \centering
   \includegraphics[width=\hsize]{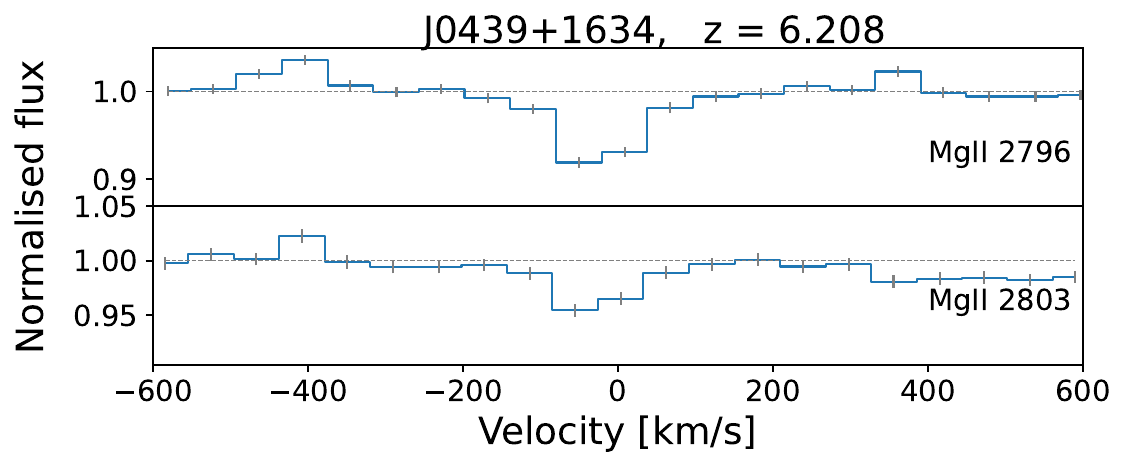}
    \caption{Absorption system at $z=6.208$ toward J0439+1634.}
    \label{fig:J0439_abs6.208}
\end{figure}
\begin{table}[hp!]
\centering
\caption{J0439+1634, $z=6.208$}       
\label{table:J0439_6.208} 
\begin{tabular}{cccc}   
\hline\hline
\noalign{\smallskip}
$\lambda_{\mathrm{obs}}$ & ID & $W_{\mathrm{obs}}$  \\   
\noalign{\smallskip} \hline \noalign{\smallskip}
20156.11 & \ion{Mg}{ii} 2796.35 & 0.812$\pm$0.062 \\
20207.85 & \ion{Mg}{ii} 2803.53 & 0.434$\pm$0.061\\
\noalign{\smallskip}
\hline                           
\end{tabular}
\end{table}

\begin{figure}
   \centering
   \includegraphics[width=\hsize]{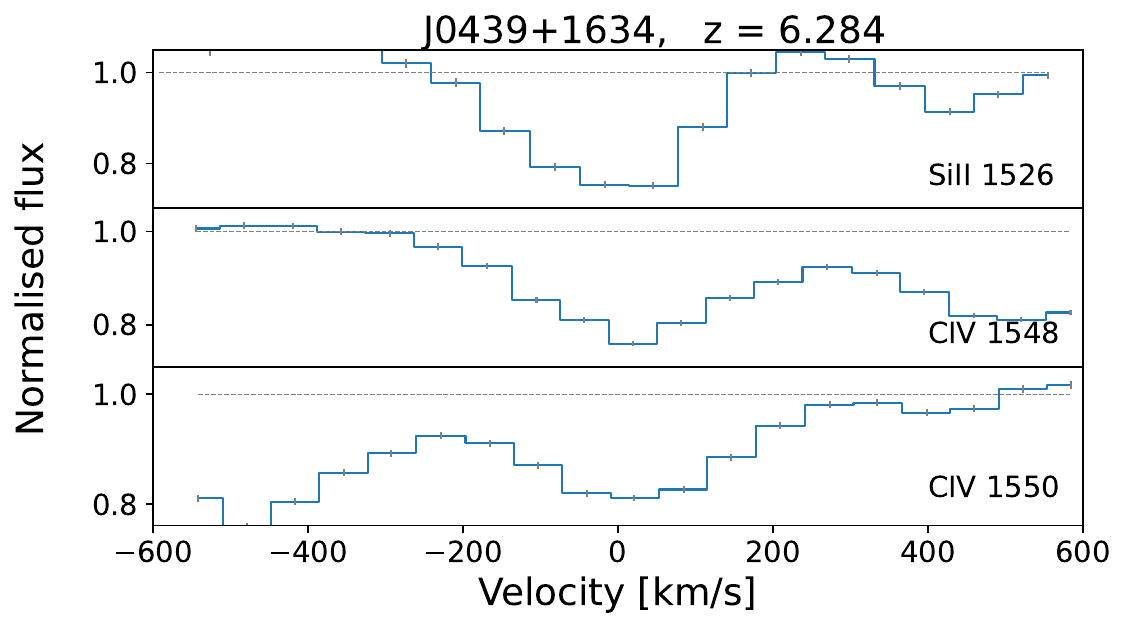}
    \caption{Absorption system at $z=6.284$ toward J0439+1634. }
    \label{fig:J0439_abs6.284}
\end{figure}
\begin{table}[h!]
\centering
\caption{J0439+1634, $z=6.284$}       
\label{table:J0439_6.284} 
\begin{tabular}{cccc}   
\hline\hline
\noalign{\smallskip}
$\lambda_{\mathrm{obs}}$ & ID & $W_{\mathrm{obs}}$  \\   
\noalign{\smallskip} \hline \noalign{\smallskip}
11120.53 & \ion{Si}{ii}\ 1526.71 &  2.083$\pm$0.066\\
11277.05 & \ion{C}{iv}\ 1548.19 &  2.843$\pm$0.049\\
11295.81 & \ion{C}{iv}\ 1550.77 &  2.677$\pm$0.047\\
\noalign{\smallskip}
\hline                           
\end{tabular}
\end{table}

\FloatBarrier

\begin{figure}
   \centering
   \includegraphics[clip, trim=0.cm 0.5cm 0cm 1cm, width=\hsize]{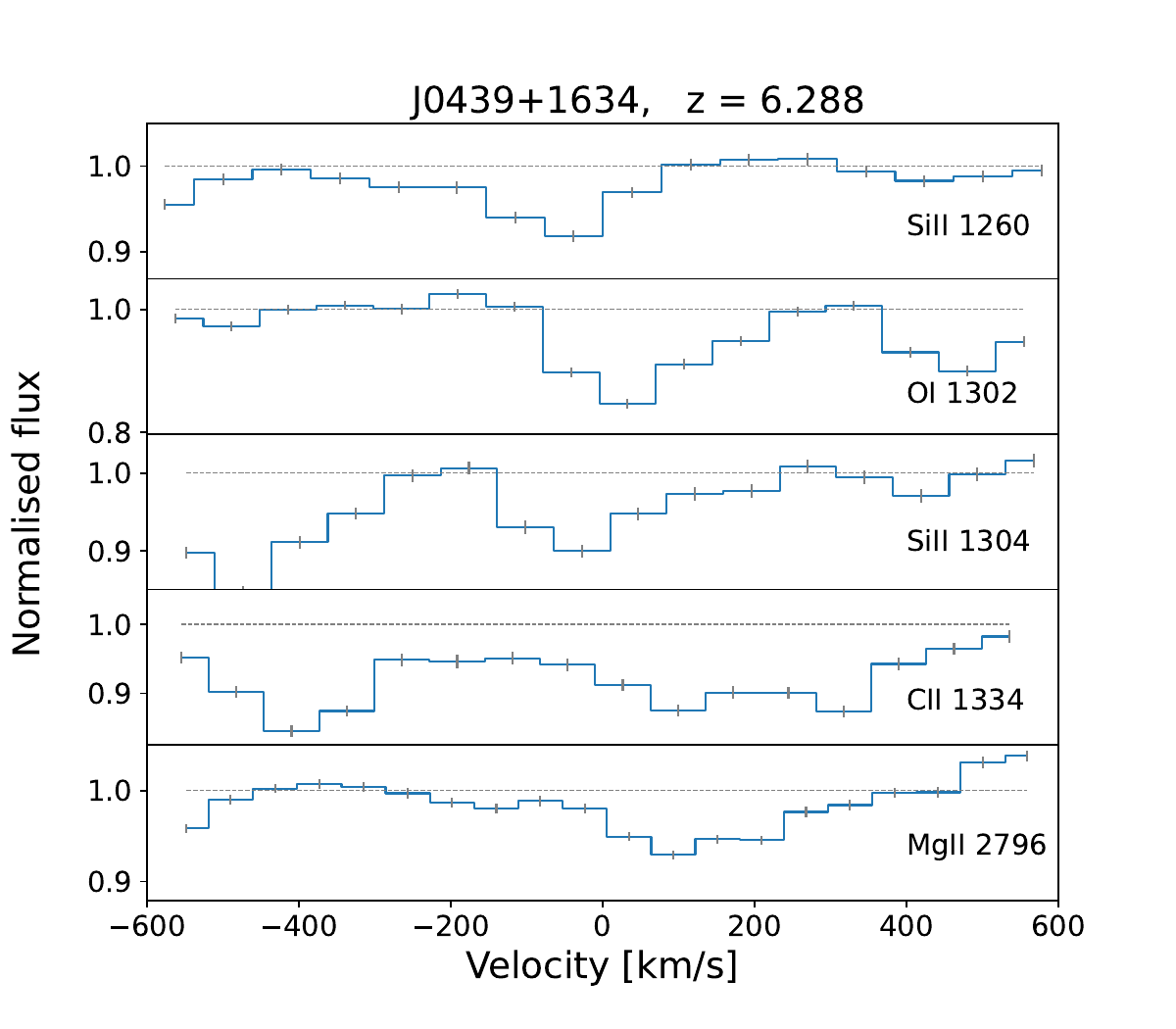}
    \caption{Absorption system at $z=6.288$ toward J0439+1634. }
    \label{fig:J0439_abs6.288}
\end{figure}
\begin{table}[h!]
\centering
\caption{J0439+1634, $z=6.288$}       
\label{table:J0439_6.288} 
\begin{tabular}{cccc}   
\hline\hline
\noalign{\smallskip}
$\lambda_{\mathrm{obs}}$ & ID & $W_{\mathrm{obs}}$ \\   
\noalign{\smallskip} \hline \noalign{\smallskip}
9188.37 & \ion{Si}{ii}\,1260.75 &  0.444$\pm$0.040\\
9490.20 & \ion{O}{i}\,1302.17 &  0.864$\pm$0.047\\
9506.25 & \ion{Si}{ii}\,1304.37 &  0.628$\pm$0.049\\
9726.07 & \ion{C}{ii}\,1334.53 &  1.114$\pm$0.052$^\dagger$ \\
20379.81 & \ion{Mg}{ii}\,2796.35  & 1.161$\pm$0.059 \\
\noalign{\smallskip}
\hline                           
\end{tabular}
\begin{tablenotes}
\item $^\dagger$ Blended with \ion{C}{iv} 1550 at $z=5.274$
\end{tablenotes}
\end{table}

\begin{figure}
   \centering
   \includegraphics[width=\hsize]{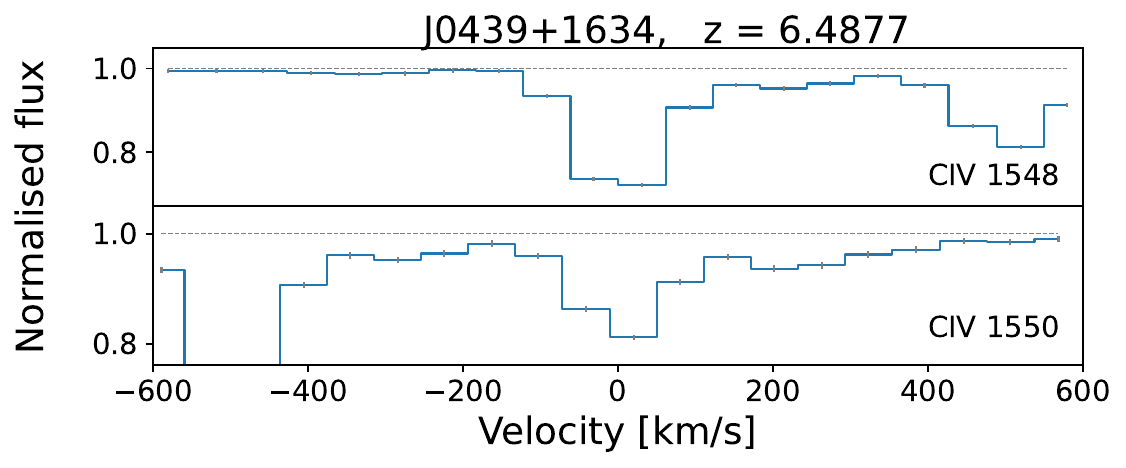}
    \caption{Absorption system at $z=6.4877$ toward J0439+1634. }
    \label{fig:J0439_abs6.4877}
\end{figure}
\begin{table}[h!]
\centering
\caption{J0439+1634, $z=6.4877$}       
\label{table:J0439_6.4877} 
\begin{tabular}{cccc}   
\hline\hline
\noalign{\smallskip}
$\lambda_{\mathrm{obs}}$ & ID & $W_{\mathrm{obs}}$  \\   
\noalign{\smallskip} \hline \noalign{\smallskip}
11592.42 & \ion{C}{iv}\ 1548.19 &  1.784$\pm$0.030\\
11611.70 & \ion{C}{iv}\ 1550.77 &  1.206$\pm$0.031\\
\noalign{\smallskip}
\hline                           
\end{tabular}
\end{table}

\FloatBarrier

\newpage
\mbox{~}
\clearpage

\begin{figure}
   \centering
   \includegraphics[width=\hsize]{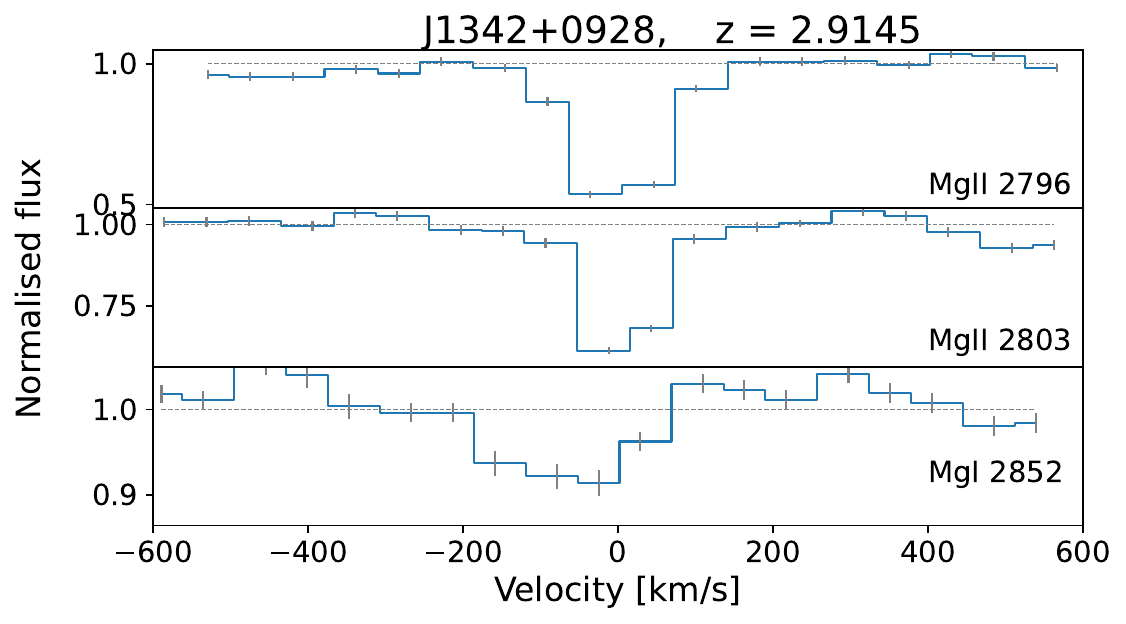}
    \caption{Absorption system at $z=2.9145$ toward J1342+0928. }
    \label{fig:J1342_abs2.9145}
\end{figure}
\begin{table}
\centering
\caption{J1342+0928,  $z=2.9145$}       
\label{table:J1342_2.9145} 
\begin{tabular}{cccc}   
\hline\hline
\noalign{\smallskip}
$\lambda_{\mathrm{obs}}$ & ID & $W_{\mathrm{obs}}$  \\   
\noalign{\smallskip} \hline \noalign{\smallskip}
10946.32 & \ion{Mg}{ii}\ 2796.35 &  2.797$\pm$0.076\\
10974.42 & \ion{Mg}{ii}\ 2803.53 &  1.776$\pm$0.084\\
11167.93 & \ion{Mg}{i}\ 2852.96 &  0.524$\pm$0.077$^\dagger$ \\
\noalign{\smallskip}
\hline                           
\end{tabular}
\begin{tablenotes}
\item  $^\dagger$ Line blended with \ion{C}{ii}\ $\lambda$1334 at $z=7.368$
\end{tablenotes}
\end{table}

\begin{figure}
   \centering
   \includegraphics[width=\hsize]{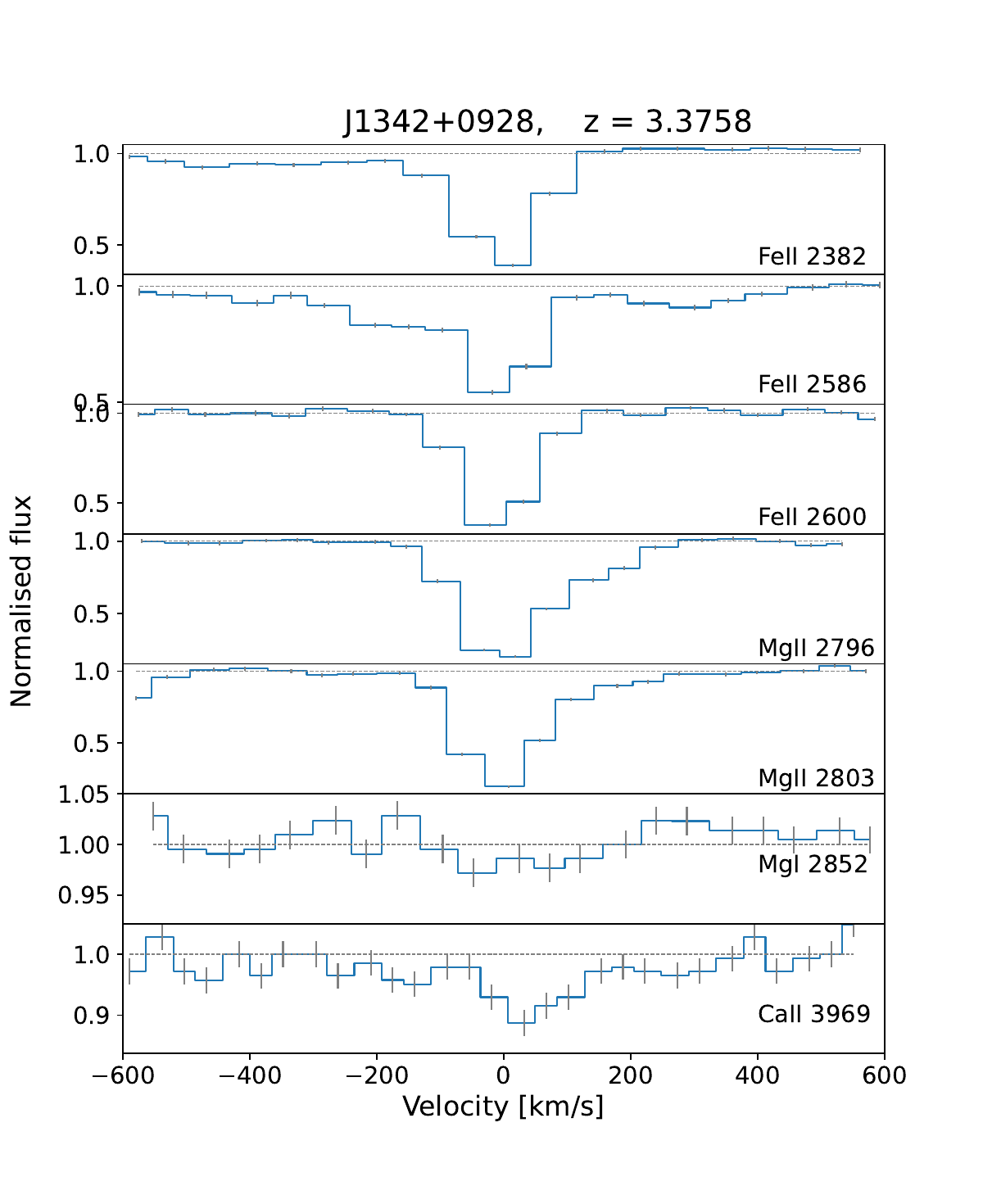}
    \caption{Absorption system at $z=3.3758$ toward J1342+0928. }
    \label{fig:J1342_abs3.3758}
\end{figure}
\begin{table}
\centering
\caption{J1342+0928,  $z=3.3758$}       
\label{table:J1342_3.3758} 
\begin{tabular}{cccc}   
\hline\hline
\noalign{\smallskip}
$\lambda_{\mathrm{obs}}$ & ID & $W_{\mathrm{obs}}$  \\   
\noalign{\smallskip} \hline \noalign{\smallskip}
10426.50 & \ion{Fe}{ii}\ 2382.76 &  3.166$\pm$0.081\\
11318.66 & \ion{Fe}{ii}\ 2586.65 &  3.864$\pm$0.081\\
11377.84 & \ion{Fe}{ii}\ 2600.17 &  3.007$\pm$0.089\\
12236.28 & \ion{Mg}{ii}\ 2796.35 &  6.490$\pm$0.097\\
12267.69 & \ion{Mg}{ii}\ 2803.53 &  5.665$\pm$0.102\\
12484.00 & \ion{Mg}{i} 2852.96 &  0.612$\pm$0.105 \\
17370.14 & \ion{Ca}{ii} 3969.59 &  1.520$\pm$0.195\\
\noalign{\smallskip}
\hline                           
\end{tabular}
\end{table}

\begin{figure}
   \centering
   \includegraphics[width=\hsize]{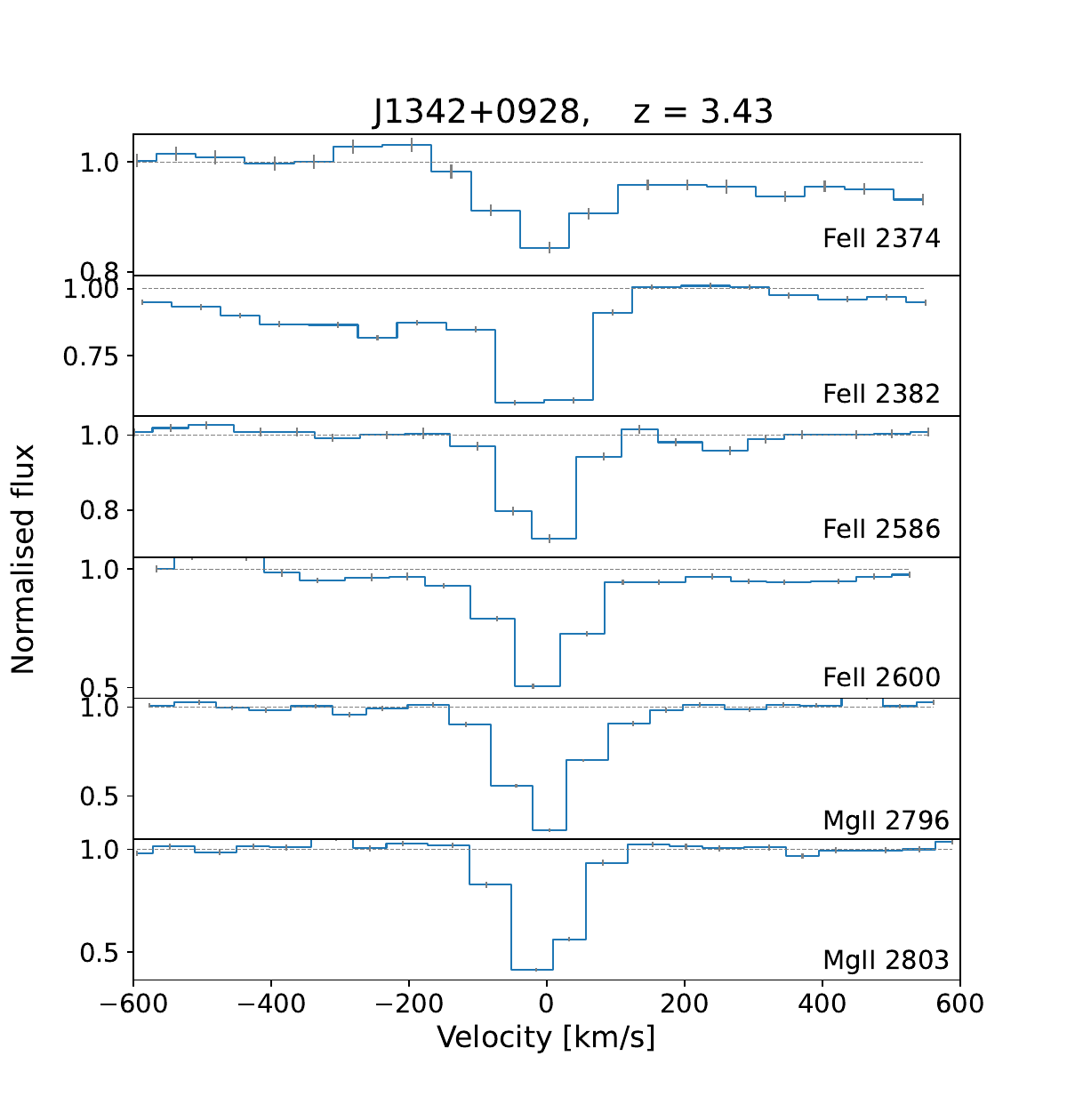}
    \caption{Absorption system at $z=3.430$ toward J1342+0928. }
    \label{fig:J1342_abs3.430}
\end{figure}
\begin{table}
\centering
\caption{J1342+0928,  $z=3.430$}       
\label{table:J1342_3.430} 
\begin{tabular}{cccc}   
\hline\hline
\noalign{\smallskip}
$\lambda_{\mathrm{obs}}$ & ID & $W_{\mathrm{obs}}$ \\   
\noalign{\smallskip} \hline \noalign{\smallskip}
10518.86 & \ion{Fe}{ii}\ 2374.46 &  1.000$\pm$0.068\\
10555.65 & \ion{Fe}{ii}\ 2382.76 &  3.305$\pm$0.074$^\dagger$ \\
11458.86 & \ion{Fe}{ii}\ 2586.65 &  1.424$\pm$0.084\\
11518.77 & \ion{Fe}{ii}\ 2600.17 &  3.062$\pm$0.084\\
12387.84 & \ion{Mg}{ii}\ 2796.35 &  3.637$\pm$0.094\\
12419.64 & \ion{Mg}{ii}\ 2803.53 &  2.563$\pm$0.091\\
\noalign{\smallskip}
\hline                           
\end{tabular}
\begin{tablenotes}
\item  $^\dagger$ Line partly blended with \ion{Si}{ii} $\lambda$1260 at $z=7.360$
\end{tablenotes}
\end{table}

\newpage
\mbox{~}
\clearpage

\begin{figure}
   \centering
   \includegraphics[width=\hsize]{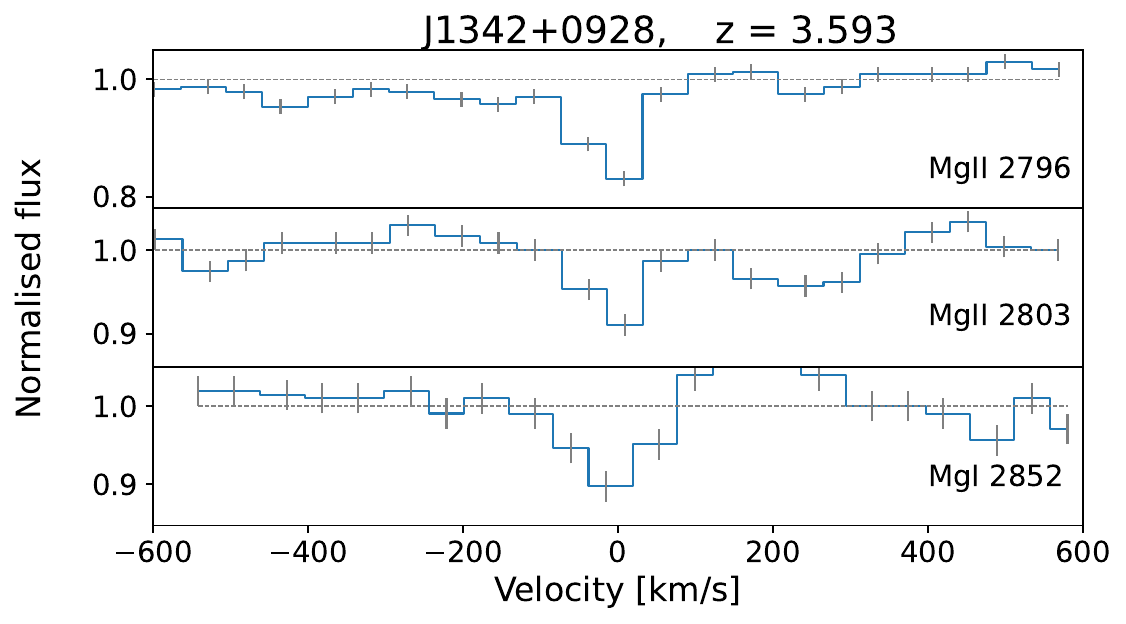}
    \caption{Absorption system at $z=3.593$ toward J1342+0928. }
    \label{fig:J1342_abs3.593}
\end{figure}
\begin{table}
\centering
\caption{J1342+0928,  $z=3.593$}       
\label{table:J1342_3.593} 
\begin{tabular}{cccc}   
\hline\hline
\noalign{\smallskip}
$\lambda_{\mathrm{obs}}$ & ID & $W_{\mathrm{obs}}$  \\   
\noalign{\smallskip} \hline \noalign{\smallskip}
12843.64 & \ion{Mg}{ii}\,2796.35 & 0.754$\pm$0.084\\ 
12876.62 & \ion{Mg}{ii}\,2803.53 & 0.397$\pm$0.084 \\
13103.66 & \ion{Mg}{ii}\,2852.96 & 0.336$\pm$0.110$^\dagger$ \\
\noalign{\smallskip}
\hline                           
\end{tabular}
\begin{tablenotes}
\item  $^\dagger$ Line blended with \ion{Al}{ii} $\lambda$1670 at $z=6.8427$
\end{tablenotes}
\end{table}

\begin{figure}
   \centering
   \includegraphics[clip, trim=0.cm 2.5cm 0cm 2.5cm, width=\hsize]{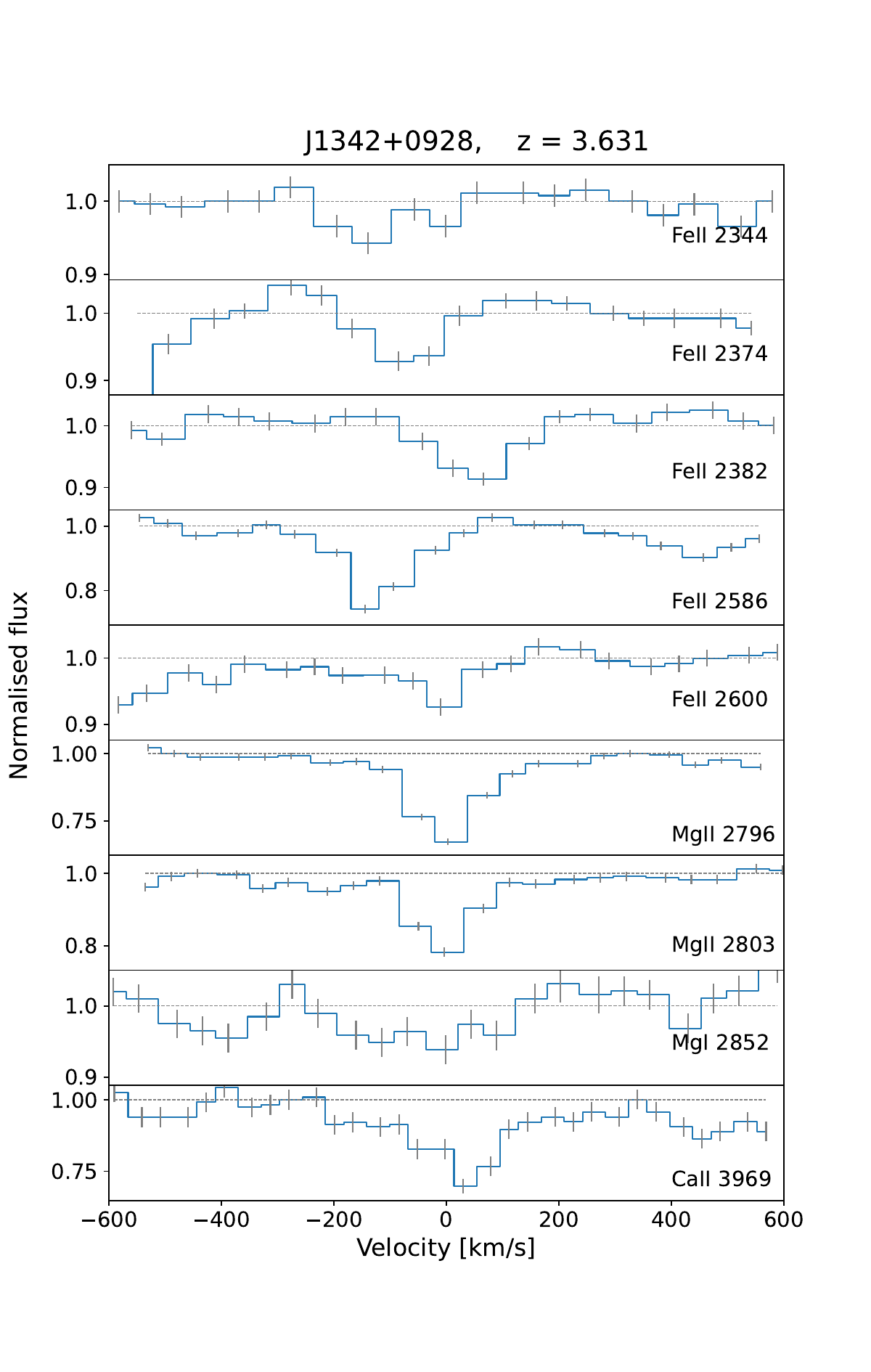}
    \caption{Absorption system at $z=3.631$ toward J1342+0928. }
    \label{fig:J1342_abs3.631}
\end{figure}
\begin{table}
\centering
\caption{J1342+0928,  $z=3.631$}       
\label{table:J1342_3.631} 
\begin{tabular}{cccc}   
\hline\hline
\noalign{\smallskip}
$\lambda_{\mathrm{obs}}$ & ID & $W_{\mathrm{obs}}$ \\   
\noalign{\smallskip} \hline \noalign{\smallskip}
10856.06 & \ion{Fe}{ii}\ 2344.21 &  0.260$\pm$0.094\\
10996.13 & \ion{Fe}{ii}\ 2374.46 &  0.229$\pm$0.091\\
11034.58 & \ion{Fe}{ii}\ 2382.76 &  0.403$\pm$0.080\\
11978.78 & \ion{Fe}{ii}\ 2586.65 &  1.333$\pm$0.090$^\dagger$ \\
12041.40 & \ion{Fe}{ii}\ 2600.17 &  0.422$\pm$0.085\\
12949.91 & \ion{Mg}{ii}\ 2796.35 &  2.333$\pm$0.087\\
12983.15 & \ion{Mg}{ii}\ 2803.53 &  1.508$\pm$0.086\\
13212.08 & \ion{Mg}{i}\ 2852.96 &  0.475$\pm$0.149\\
18383.18 & \ion{Ca}{ii}\ 3969.59 &  3.564$\pm$0.268\\
\noalign{\smallskip}
\hline        
\end{tabular}
\begin{tablenotes}
\item $^\dagger$ Line blended with \ion{Si}{ii} $\lambda$1526 at $z=6.8427$
\end{tablenotes}
\end{table}

\begin{figure}
   \centering
   \includegraphics[width=\hsize]{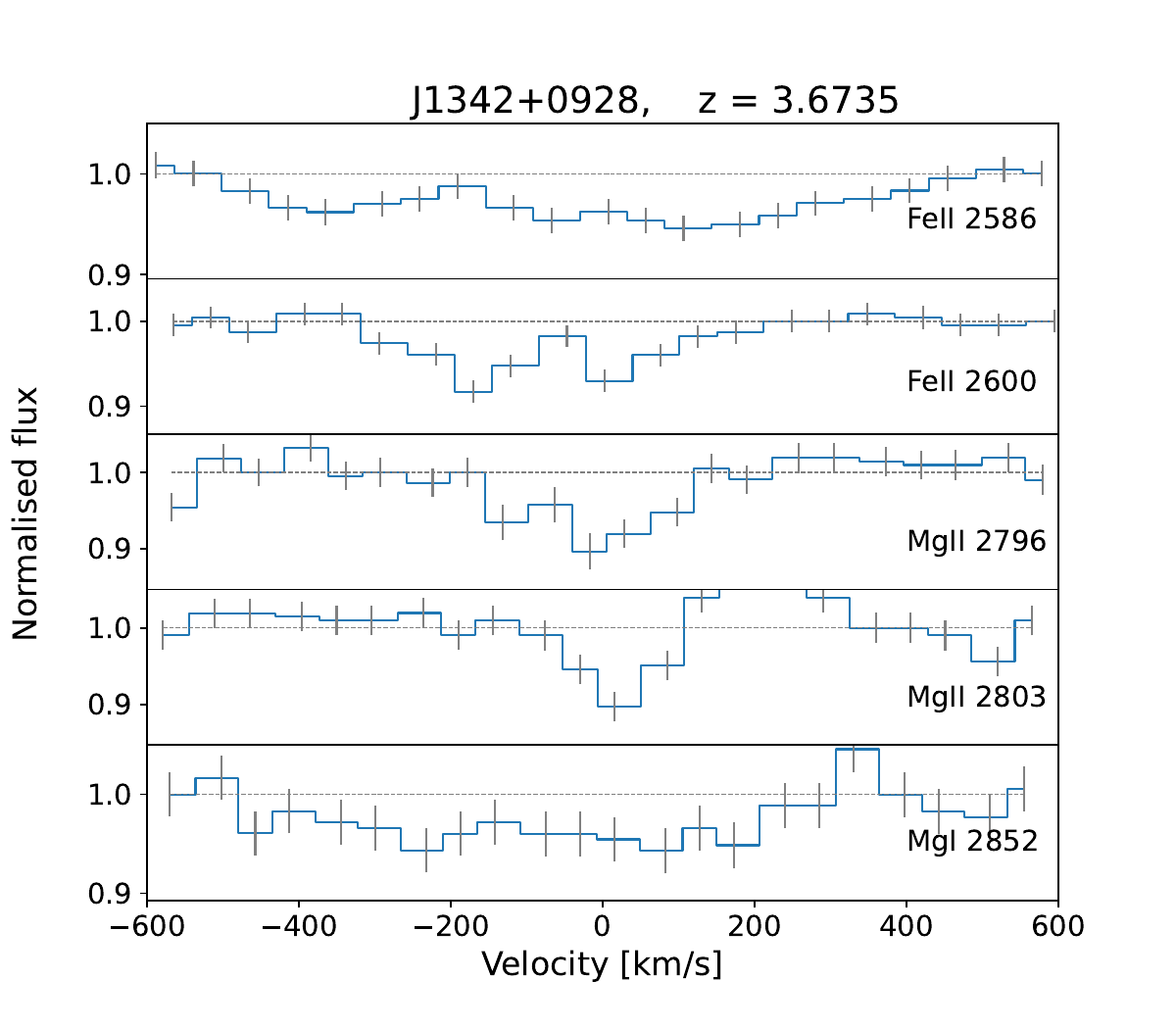}
    \caption{Absorption system at $z=3.6735$ toward J1342+0928. }
    \label{fig:J1342_abs3.6735}
\end{figure}
\begin{table}
\centering
\caption{J1342+0928,  $z=3.6735$}       
\label{table:J1342_3.6735} 
\begin{tabular}{cccc}   
\hline\hline
\noalign{\smallskip}
$\lambda_{\mathrm{obs}}$ & ID & $W_{\mathrm{obs}}$ \\   
\noalign{\smallskip} \hline \noalign{\smallskip}
12088.71 & \ion{Fe}{ii}\,2586.65 & 0.675$\pm$0.083 \\
12151.91 & \ion{Fe}{ii}\,2600.17 & 0.723$\pm$0.086 \\
13068.75 & \ion{Mg}{ii}\,2796.35 & 0.855$\pm$0.142 \\
13102.30 & \ion{Mg}{ii}\,2803.53 & 0.195$\pm$0.134$^\dagger$ \\
13333.33 & \ion{Mg}{i}\,2852.96 & 0.797$\pm$0.156 \\
\noalign{\smallskip}
\hline        
\end{tabular}
\begin{tablenotes}
\item $^\dagger$ Line blended with \ion{Al}{ii} $\lambda$1670 at $z=6.8427$ and \ion{Mg}{i} $\lambda$2852 at $z=3.5933$
\end{tablenotes}
\end{table}

\newpage
\mbox{~}
\clearpage

\begin{figure}
   \centering
   \includegraphics[clip, trim=0.cm 3cm 0cm 3cm, width=\hsize]{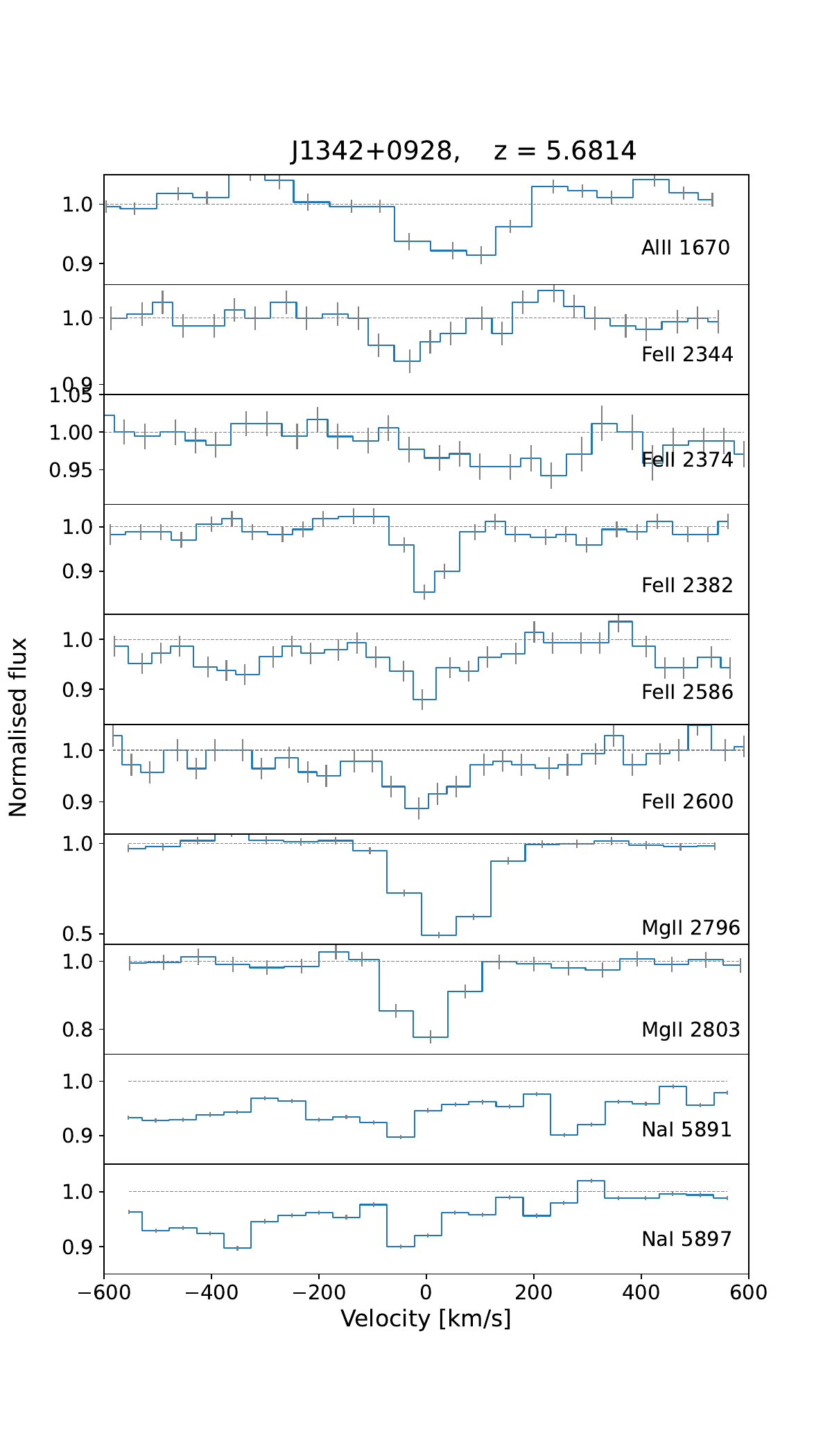}
    \caption{Absorption system at $z=5.6814$ toward J1342+0928. }
    \label{fig:J1342_abs5.6814}
\end{figure}
\begin{table}
\centering
\caption{J1342+0928,  $z=5.6814$}       
\label{table:J1342_5.6814} 
\begin{tabular}{cccc}   
\hline\hline
\noalign{\smallskip}
$\lambda_{\mathrm{obs}}$ & ID & $W_{\mathrm{obs}}$  \\   
\noalign{\smallskip} \hline \noalign{\smallskip}
11163.20 & \ion{Al}{ii}\ 1670.79 &  0.644$\pm$0.077$^\dagger$\\
15662.63 & \ion{Fe}{ii}\ 2344.21 &  0.361$\pm$0.126\\
15864.73 & \ion{Fe}{ii}\ 2374.46 &  0.523$\pm$0.123\\
15920.21 & \ion{Fe}{ii}\ 2382.76 &  0.559$\pm$0.132\\
17282.44 & \ion{Fe}{ii}\ 2586.65 &  1.058$\pm$0.152\\
17372.80 & \ion{Fe}{ii}\ 2600.17 &  1.231$\pm$0.163\\
18683.55 & \ion{Mg}{ii}\ 2796.35 &  5.253$\pm$0.188\\ 
18731.51 & \ion{Mg}{ii}\ 2803.53 &  1.739$\pm$0.196\\ 
39362.4  & \ion{Na}{i}\ 5891.58  &  3.301$\pm$0.295\\
39405.3  & \ion{Na}{i}\ 5897.56  &  2.507$\pm$0.292\\
\noalign{\smallskip}
\hline                           
\end{tabular}
\begin{tablenotes}
\item $^\dagger$ Line partly blended with \ion{C}{ii} $\lambda$1334 at $z=7.368$
\end{tablenotes}
\end{table}

\begin{figure}
   \centering
   \includegraphics[width=\hsize]{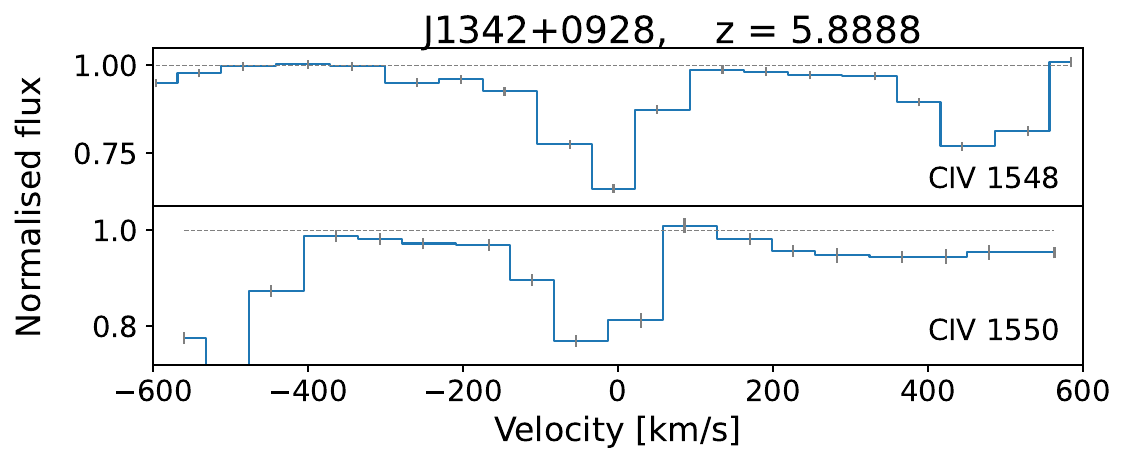}
    \caption{Absorption system at $z=5.8888$ toward J1342+0928. }
    \label{fig:J1342_abs5.8888}
\end{figure}
\begin{table}
\centering
\caption{J1342+0928,  $z=5.8888$}       
\label{table:J1342_5.8888} 
\begin{tabular}{cccc}   
\hline\hline
\noalign{\smallskip}
$\lambda_{\mathrm{obs}}$ & ID & $W_{\mathrm{obs}}$  \\   
\noalign{\smallskip} \hline \noalign{\smallskip}
10665.21 & \ion{C}{iv}\ 1548.19 &  1.813$\pm$0.072\\
10682.94 & \ion{C}{iv}\ 1550.77 &  1.344$\pm$0.079$^\dagger$\\
\noalign{\smallskip}
\hline                           
\end{tabular}
\begin{tablenotes}
\item  $^\dagger$ Line blended with \ion{Si}{ii} $\lambda$1260 at $z=7.476$
\end{tablenotes}
\end{table}

\begin{figure}
   \centering
   \includegraphics[width=\hsize]{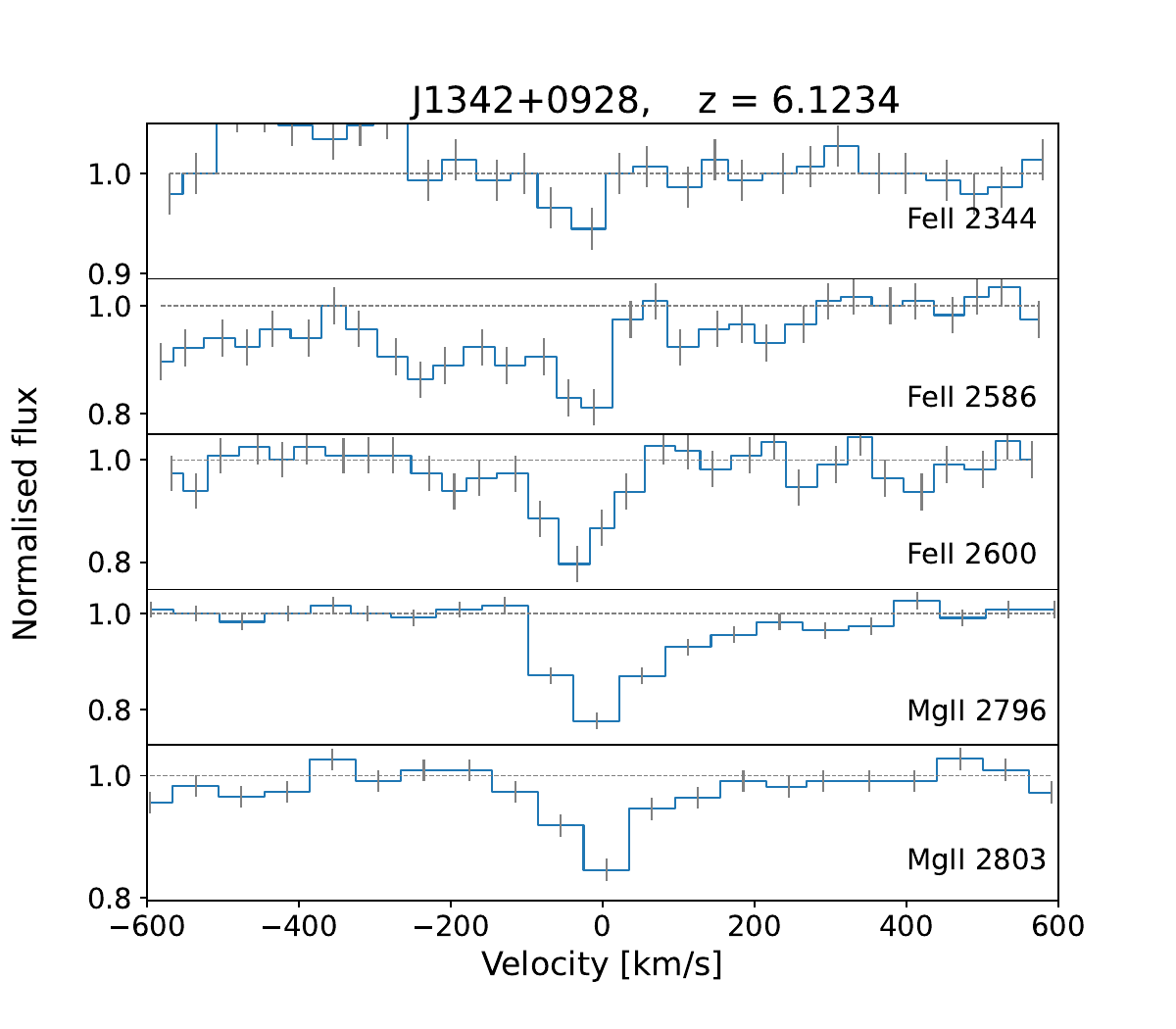}
    \caption{Absorption system at $z=6.1234$ toward J1342+0928. }
    \label{fig:J1342_abs6.1234}
\end{figure}
\begin{table}
\centering
\caption{J1342+0928,  $z=6.1234$}       
\label{table:J1342_6.1234} 
\begin{tabular}{cccc}   
\hline\hline \noalign{\smallskip}
\noalign{\smallskip}
$\lambda_{\mathrm{obs}}$ & ID & $W_{\mathrm{obs}}$  \\   
\noalign{\smallskip} \hline \noalign{\smallskip}
16698.77 & \ion{Fe}{ii}\ 2344.21 &  0.200$\pm$0.171\\
18425.74 & \ion{Fe}{ii}\ 2586.65 &  2.812$\pm$0.300\\
18522.07 & \ion{Fe}{ii}\ 2600.17 &  1.412$\pm$0.301\\
19919.53 & \ion{Mg}{ii}\ 2796.35 &  2.390$\pm$0.208\\
19970.67 & \ion{Mg}{ii}\ 2803.53 &  1.437$\pm$0.212\\
\noalign{\smallskip}
\hline                           
\end{tabular}
\end{table}

\newpage
\mbox{~}
\clearpage 

\begin{figure}
   \centering
   \includegraphics[clip, trim=0.cm 1cm 0cm 1.5cm, width=\hsize]{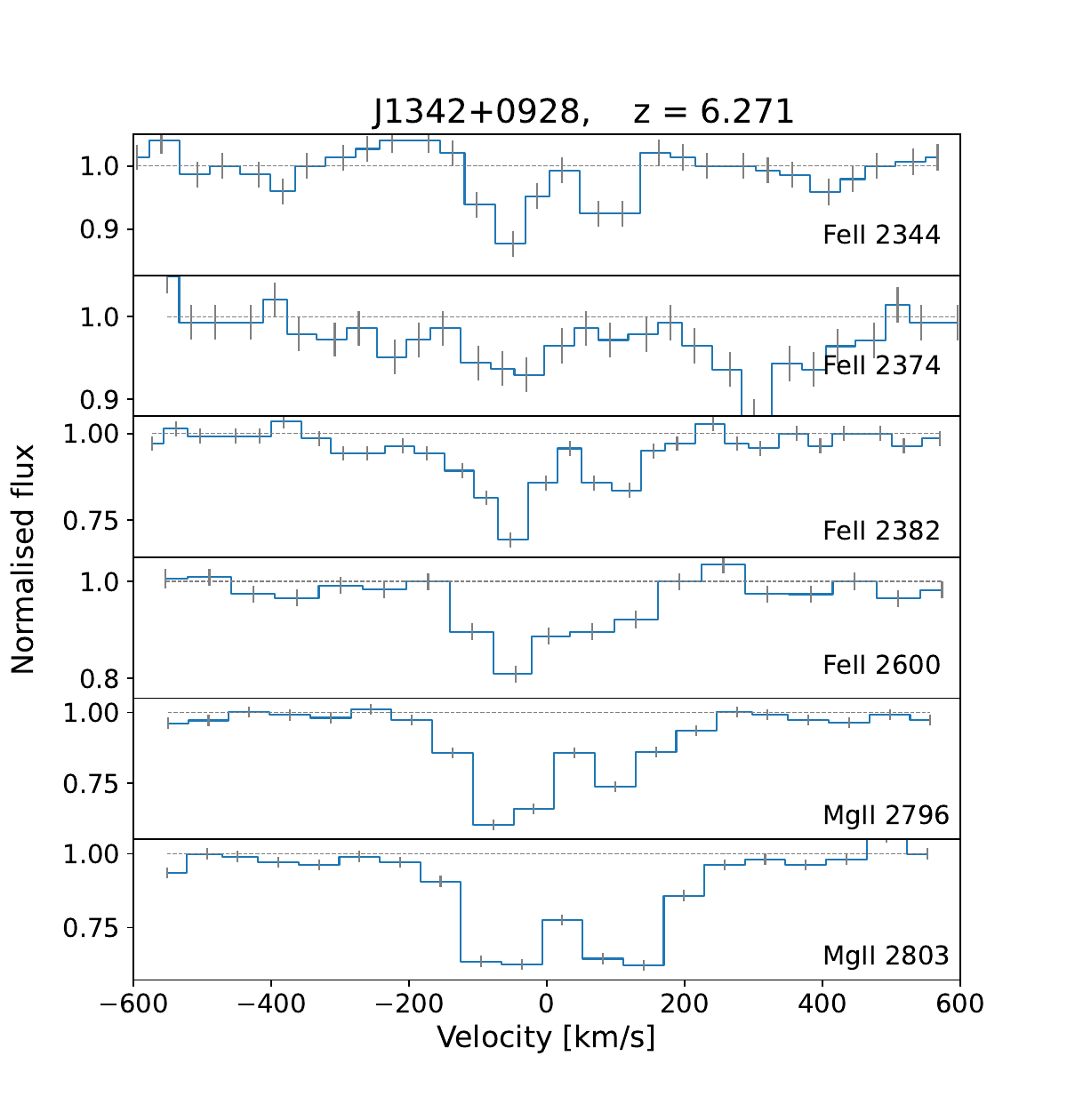}
    \caption{Absorption system at $z=6.271$ toward J1342+0928. }
    \label{fig:J1342_abs6.271}
\end{figure}
\begin{table}
\centering
\caption{J1342+0928,  $z=6.271$}       
\label{table:J1342_6.271} 
\begin{tabular}{cccc}   
\hline\hline 
\noalign{\smallskip}
$\lambda_{\mathrm{obs}}$ & ID & $W_{\mathrm{obs}}$  \\   
\noalign{\smallskip} \hline \noalign{\smallskip}
17044.78 & \ion{Fe}{ii}\ 2344.21 &  0.528$\pm$0.189\\
17264.71 & \ion{Fe}{ii}\ 2374.46 &  1.160$\pm$0.190\\
17325.08 & \ion{Fe}{ii}\ 2382.76 &  3.350$\pm$0.197\\
18905.86 & \ion{Fe}{ii}\ 2600.17 &  2.140$\pm$0.215\\
20332.28 & \ion{Mg}{ii}\ 2796.35 &  6.020$\pm$0.233\\
20384.47 & \ion{Mg}{ii}\ 2803.53 &  8.021$\pm$0.237$^\dagger$ \\
\noalign{\smallskip}
\hline        
\end{tabular}
\begin{tablenotes}
\item  $^\dagger$ Line blended with \ion{Fe}{ii} $\lambda$2600 at $z=6.8427$
\end{tablenotes}
\end{table}

\begin{figure}
   \centering
   \includegraphics[width=\hsize]{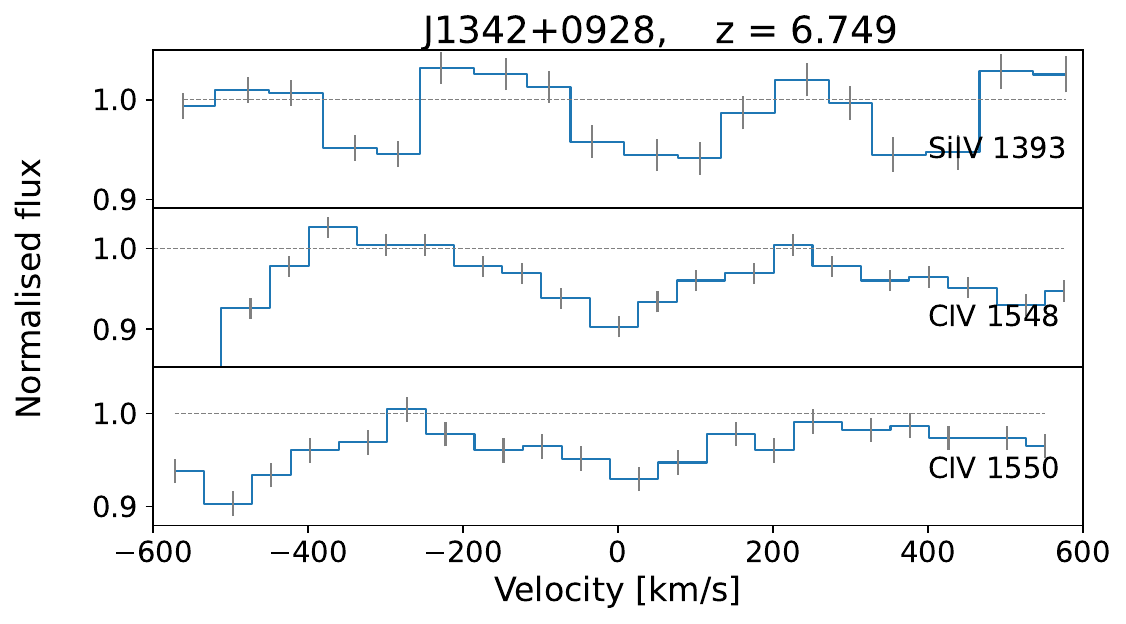}
    \caption{Absorption system at $z=6.749$ toward J1342+0928. }
    \label{fig:J1342_abs6.749}
\end{figure}
\begin{table}
\centering
\caption{J1342+0928,  $z=6.749$}       
\label{table:J1342_6.749} 
\begin{tabular}{cccc}   
\hline\hline
\noalign{\smallskip}
$\lambda_{\mathrm{obs}}$ & ID & $W_{\mathrm{obs}}$  \\   
\noalign{\smallskip} \hline \noalign{\smallskip}
10800.21 & \ion{Si}{iv}\,1393.76&  0.316$\pm$0.095\\ 
11996.96 & \ion{C}{iv}\,1548.19 &  0.797$\pm$0.081\\
12016.92 & \ion{C}{iv}\,1550.77 &  0.637$\pm$0.077\\
\noalign{\smallskip}
\hline                           
\end{tabular}
\end{table}

\begin{figure}
   \centering
   \includegraphics[clip, trim=0.cm 2cm 0cm 1.5cm, width=\hsize]{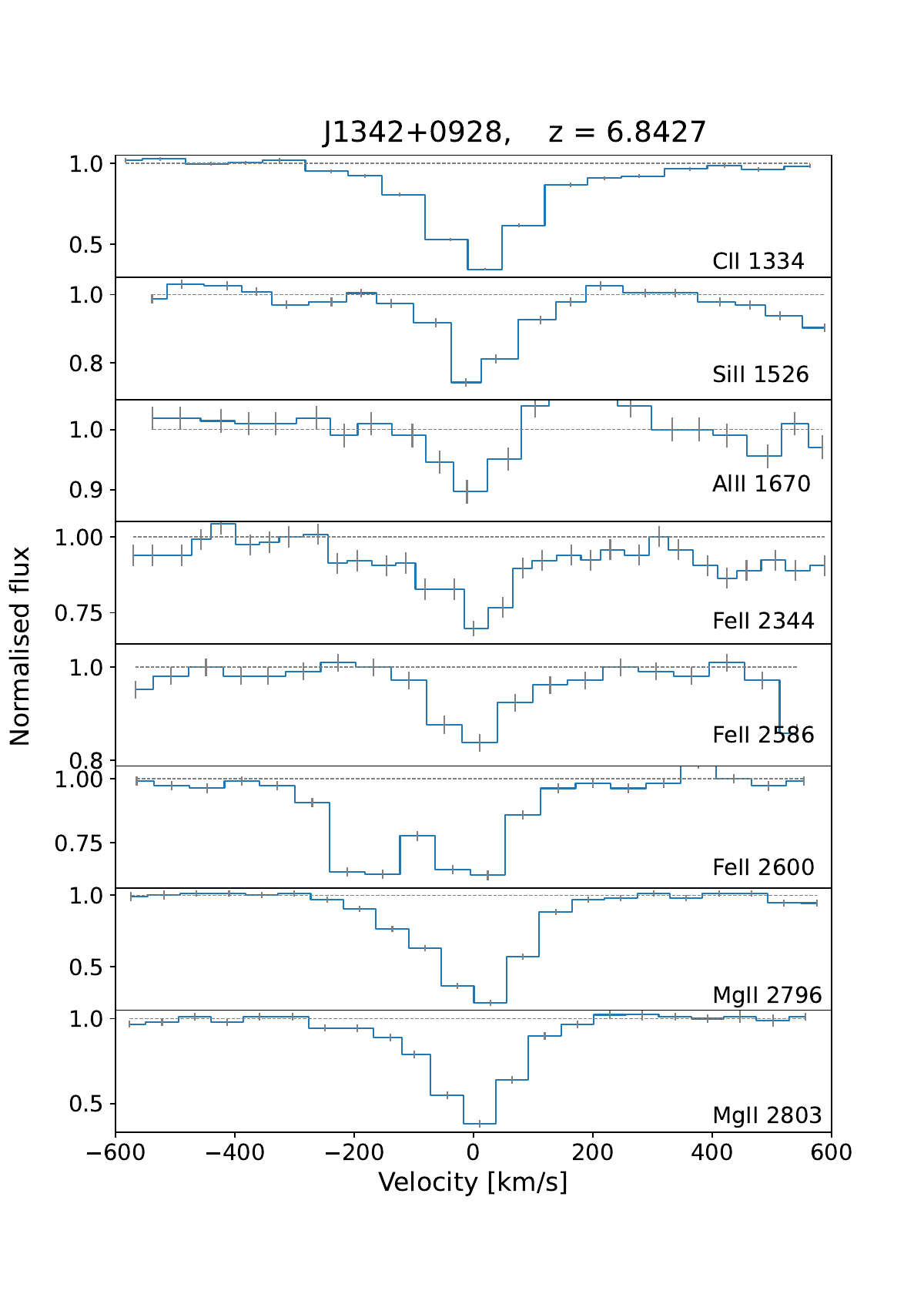}
    \caption{Absorption system at $z=6.8427$ toward J1342+0928. }
    \label{fig:J1342_abs6.8427}
\end{figure}
\begin{table}
\centering
\caption{J1342+0928,  $z=6.8427$}       
\label{table:J1342_6.8427} 
\begin{tabular}{cccc}   
\hline\hline
\noalign{\smallskip}
$\lambda_{\mathrm{obs}}$ & ID & $W_{\mathrm{obs}}$  \\   
\noalign{\smallskip} \hline \noalign{\smallskip}
10466.74 & \ion{C}{ii}\ 1334.53 &  4.913$\pm$0.085\\
11973.96 & \ion{Si}{ii}\ 1526.71 &  1.458$\pm$0.097$^{\dagger}$ \\
13103.98 & \ion{Al}{ii}\ 1670.79 &  0.317$\pm$0.117$^{\ddagger}$\\
18385.67 & \ion{Fe}{ii}\ 2344.21 &  3.948$\pm$0.318\\
20287.10 & \ion{Fe}{ii}\ 2586.65 &  1.867$\pm$0.253\\
20393.16 & \ion{Fe}{ii}\ 2600.17 &  8.098$\pm$0.238$^{\mathsection}$ \\
21931.79 & \ion{Mg}{ii}\ 2796.35 &  10.851$\pm$0.285\\
21988.09 & \ion{Mg}{ii}\ 2803.53 &  7.690$\pm$0.310\\
\noalign{\smallskip}\hline                    
\end{tabular}
\begin{tablenotes}
\item $^{\dagger}$ Line blended with \ion{Fe}{ii} $\lambda$2586 at $z=3.631$
\item $^{\ddagger}$ Line blended with \ion{Mg}{i} $\lambda$2852 at $z=3.593$
\item $^{\mathsection}$ Line blended with \ion{Mg}{ii} $\lambda$2803 at $z=6.271$
\end{tablenotes}
\end{table}

\newpage
\mbox{~}
\clearpage

\begin{figure}
   \centering
   \includegraphics[width=\hsize]{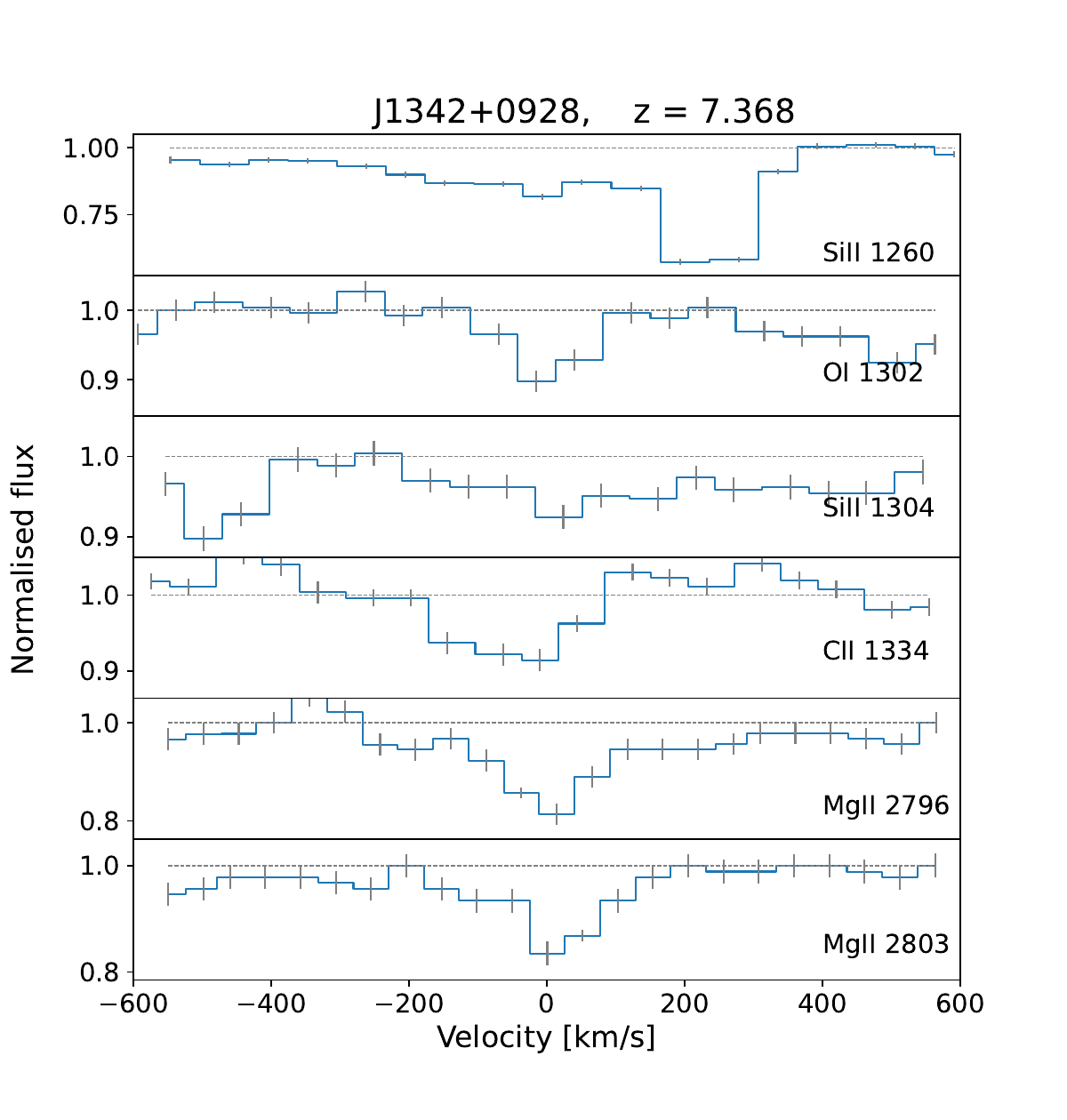}
    \caption{Absorption system at $z=7.368$ toward J1342+0928. }
    \label{fig:J1342_abs7.368}
\end{figure}
\begin{table}
\centering
\caption{J1342+0928,  $z=7.368$}       
\label{table:J1342_7.368} 
\begin{tabular}{cccc}   
\hline\hline
\noalign{\smallskip}
$\lambda_{\mathrm{obs}}$ & ID & $W_{\mathrm{obs}}$  \\   
\noalign{\smallskip} \hline \noalign{\smallskip}
10547.21 & \ion{Si}{ii}\ 1260.42 &  2.997$\pm$0.066$^{\dagger}$ \\
10896.54 & \ion{O}{i}\ 1302.17  &  0.508$\pm$0.088 \\
10914.13 & \ion{Si}{ii}\ 1304.37 &  0.653$\pm$0.088 \\
11167.37 & \ion{C}{ii}\ 1334.53 &  0.531$\pm$0.080$^{\ddagger, \mathsection}$ \\
23399.87 & \ion{Mg}{ii}\ 2796.35 &  2.851$\pm$0.236 \\
23459.95 & \ion{Mg}{ii}\ 2803.53 &  2.243$\pm$0.220 \\
\noalign{\smallskip}
\hline  
\end{tabular}
\begin{tablenotes}
\item $^\dagger$ Line blended with  \ion{Mg}{i} 2852 at $z=2.9145$
\item $^{\ddagger}$ Line partly blended with \ion{Fe}{ii} $\lambda$2382 at $z=3.430$
\item $^{\mathsection}$ Line partly blended with \ion{Al}{ii} $\lambda$1670 at $z=5.6814$
\end{tablenotes}
\end{table}

\begin{figure}
   \centering
   \includegraphics[clip, trim=0.cm 2cm 0cm 1cm, width=\hsize]{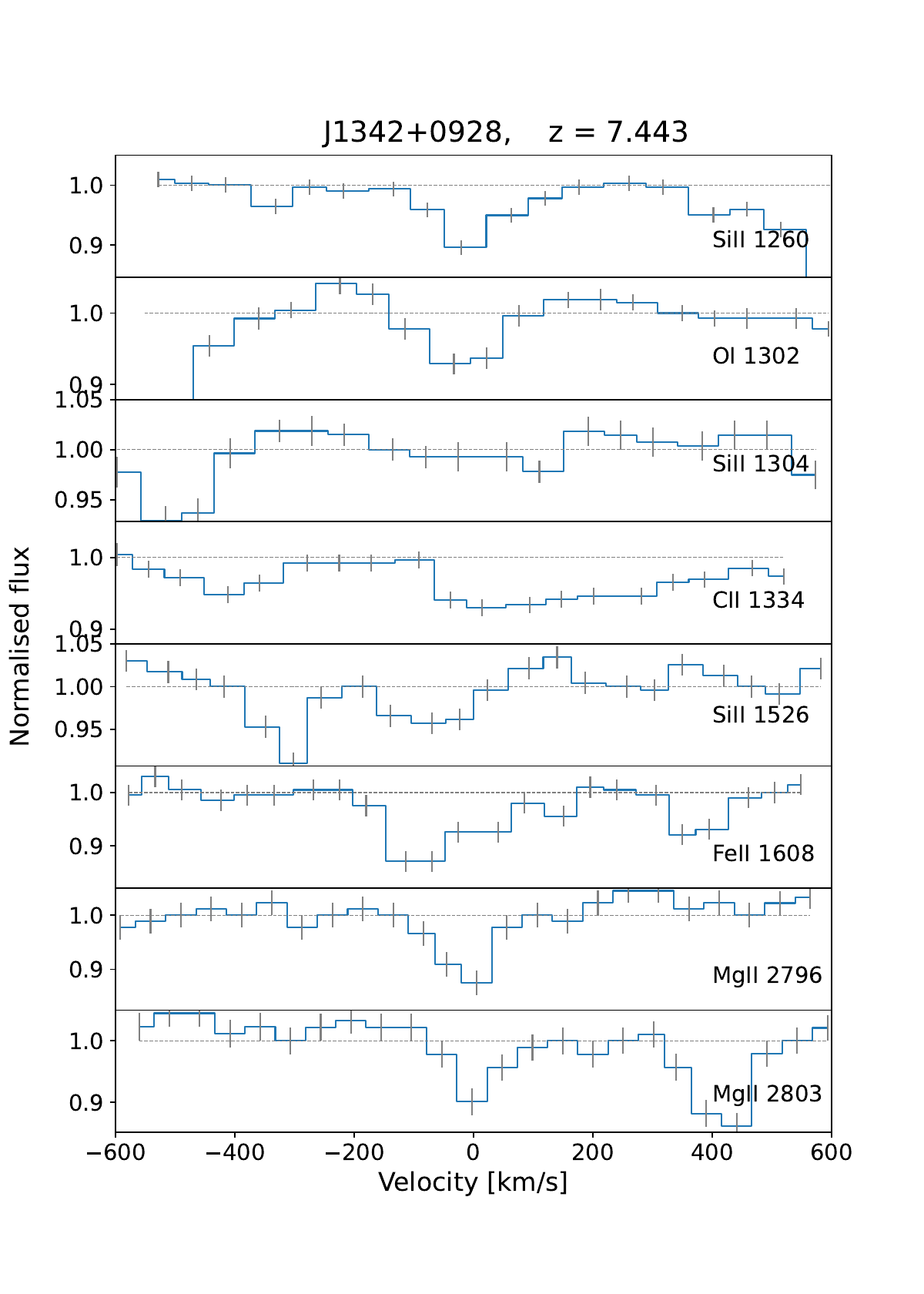}
    \caption{Absorption system at $z=7.443$ toward J1342+0928. }
    \label{fig:J1342_abs7.443}
\end{figure}
\begin{table}
\centering
\caption{J1342+0928,  $z=7.443$}       
\label{table:J1342_7.443} 
\begin{tabular}{cccc}   
\hline\hline
\noalign{\smallskip}
$\lambda_{\mathrm{obs}}$ & ID & $W_{\mathrm{obs}}$  \\   
\noalign{\smallskip} \hline \noalign{\smallskip}  
10641.74 & \ion{Si}{ii}\ 1260.42 &  0.561$\pm$0.073\\
10994.20 & \ion{O}{i}\ 1302.17 &  0.258$\pm$0.085\\
11011.95 & \ion{Si}{ii}\ 1304.37 &  0.081$\pm$0.076\\
11267.46 & \ion{C}{ii}\ 1334.53 &  0.609$\pm$0.077\\
12889.98 & \ion{Si}{ii}\ 1526.71 &  0.154$\pm$0.088\\
13580.15 & \ion{Fe}{ii}\ 1608.45 &  1.092$\pm$0.136\\
23609.60 & \ion{Mg}{ii}\ 2796.35 &  1.057$\pm$0.250\\
23670.21 & \ion{Mg}{ii}\ 2803.53 &  0.524$\pm$0.232\\
\noalign{\smallskip}
\hline                           
\end{tabular}
\end{table}

\begin{figure}
   \centering
   \includegraphics[width=\hsize]{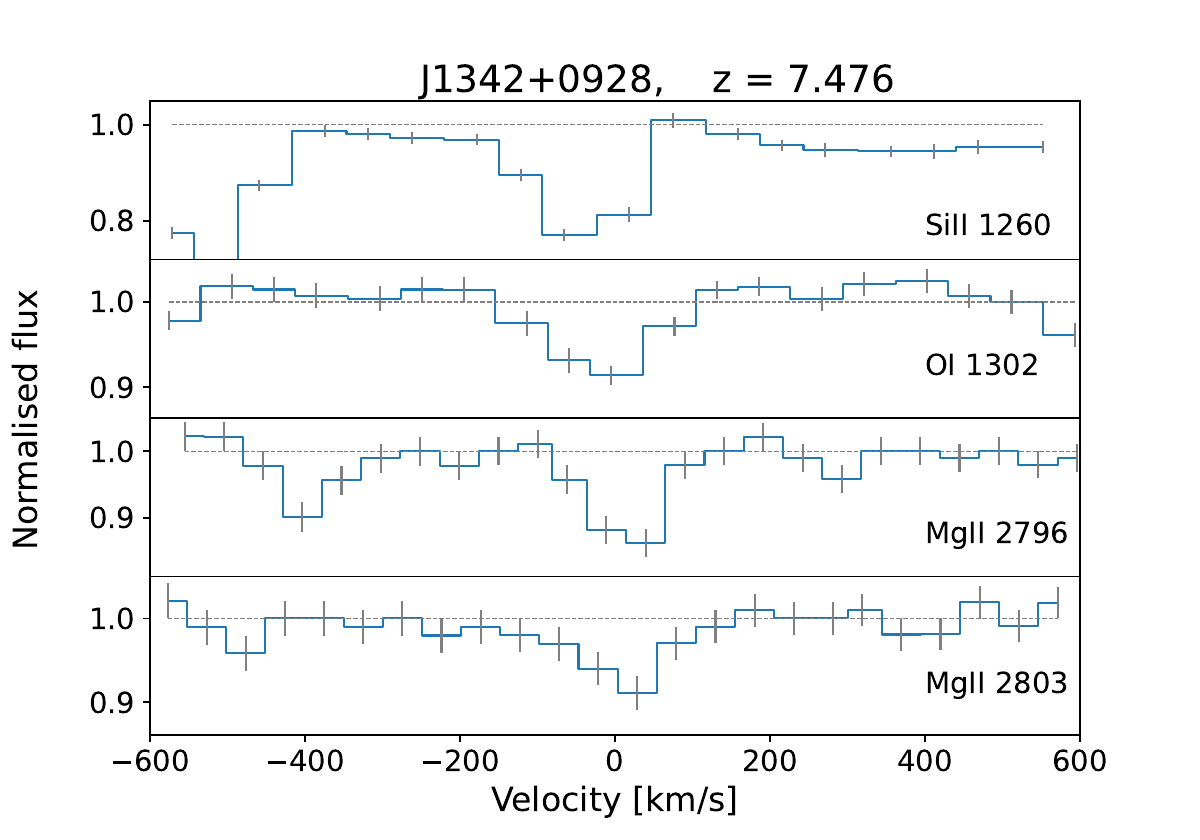}
    \caption{Absorption system at $z=7.476$ toward J1342+0928. }
    \label{fig:J1342_abs7.476}
\end{figure}
\begin{table}
\centering
\caption{J1342+0928,  $z=7.476^\dagger$}       
\label{table:J1342_7.476} 
\begin{tabular}{cccc}   
\hline\hline
\noalign{\smallskip}
$\lambda_{\mathrm{obs}}$ & ID & $W_{\mathrm{obs}}$ \\   
\noalign{\smallskip} \hline \noalign{\smallskip}  
10683.34 & \ion{Si}{ii}\ 1260.42 & 1.344$\pm$0.079$^\ddagger$ \\
11037.18 & \ion{O}{i}\ 1302.17 & 0.378$\pm$0.081 \\
23701.88 & \ion{Mg}{ii}\ 2796.35 & 1.167$\pm$0.235 \\
23762.73 & \ion{Mg}{ii}\ 2803.53 & 0.958$\pm$0.226 \\
\noalign{\smallskip}
\hline                           
\end{tabular}
\begin{tablenotes}
\item $^\dagger$This absorber lies only $\approx2000$ km~s$^{-1}$ from the quasar systemic redshift, and is considered a proximate system.
\item  $^\ddagger$ Line blended with \ion{C}{iv} $\lambda$1550 at $z=5.8888$ 
\end{tablenotes}
\end{table}


\end{appendix}
\end{document}